\documentclass{jfm}
\usepackage{graphicx}
\usepackage{epstopdf, epsfig, amsmath, bm, booktabs, multirow}

\newcommand{\be}{\begin{equation}}
\newcommand{\ee}{\end{equation}}
\newcommand{\bes}{\begin{equation*}}
\newcommand{\ees}{\end{equation*}}
\newcommand{\bea}{\begin{eqnarray}}
\newcommand{\eea}{\end{eqnarray}}
\newcommand{\bi}{\begin {itemize}}
\newcommand{\ei}{\end {itemize}}
\newcommand{\benm}{\begin{enumerate}}
\newcommand{\eenm}{\end{enumerate}}
\newcommand{\bmn}{\begin{minipage}}
\newcommand{\emn}{\end{minipage}}
\newcommand{\bfig}{\begin{figure}}
\newcommand{\efig}{\end{figure}}

\newcommand{\lnw}{\linewidth}

\newcommand{\bcls}{\begin{columns}}
\newcommand{\ecls}{\end{columns}}
\newcommand{\bcl}{\begin{column}}
\newcommand{\ecl}{\end{column}}

\newcommand{\bmat}{\begin{matrix}}
\newcommand{\emat}{\end{matrix}}
\newcommand{\bpmat}{\begin{pmatrix}}
\newcommand{\epmat}{\end{pmatrix}}
\newcommand{\bvmat}{\begin{vmatrix}}
\newcommand{\evmat}{\end{vmatrix}}
\newcommand{\bbmat}{\begin{bmatrix}}
\newcommand{\ebmat}{\end{bmatrix}}

\newcommand{\q}{\quad}
\def\bali#1\eali{\begin{align*}#1\end{align*}}
\newcommand{\ptl}{\partial}
\newcommand{\td}{\tilde}

\newcommand{\lal}{\langle}
\newcommand{\ral}{\rangle}

\newcommand{\ep}{\epsilon}
\newcommand{\al}{\alpha}
\newcommand{\da}{\delta}

\newcommand{\lm}{\lambda}
\newcommand{\om}{\omega}

\newcommand{\fF}{{\bf F}}

\newcommand{\fN}{{\bf N}}

\renewcommand{\a}{{\bf a}}
\renewcommand{\b}{{\bf b}}
\renewcommand{\c}{{\bf c}}

\renewcommand{\k}{{\bf k}}

\renewcommand{\u}{{\bf u}}
\renewcommand{\v}{{\bf v}}

\newcommand{\e}{{\bf e}}
\newcommand{\f}{{\bf f}}

\newcommand{\fp}{{\bf p}}

\newcommand{\x}{{\bf x}}

\newcommand{\y}{{\bf y}}

\newcommand{\bo}{{\bm \omega}}

\newcommand{\bxi}{{\bm \xi}}
\newcommand{\bvp}{{\bm \varphi}}
\newcommand{\blmd}{{\bm \lambda}}

\renewcommand{\SS}{{\mathcal S}}

\newcommand{\FF}{{\mathcal F}}

\newcommand{\PP}{{\mathcal P}}

\newcommand{\tA}{{\td{A}}}
\newcommand{\ts}{{\td{s}}}
\newcommand{\tog}{{\td{\omega}}}

\newcommand{\rbu}{{\rm \bf u}}
\newcommand{\rbv}{{\rm \bf v}}

\shorttitle{Reconstructing 3D turbulence by 4DVAR}
\shortauthor{Y. Li \it{et. al}}

\title{Small scale reconstruction in
three-dimensional Kolmogorov
flows using four-dimensional variational data assimilation}

\author{Yi Li\aff{1}\corresp{\email{yili@sheffield.ac.uk}},
Jianlei Zhang\aff{2}, Gang Dong\aff{2}, Naseer S. Abdullah\aff{1}
  }

\affiliation{
\aff{1}School of Mathematics and Statistics, University of
  Sheffield, Sheffield, S3 7RH, UK
  \aff{2}National Key Laboratory of Transient Physics, Nanjing University
  of Science and Technology, Nanjing 210094, P. R. China
}

\begin{document}

\maketitle

\begin{abstract} 
  Significant insights in
  computational fluid dynamics are obtained in recent years
  by adopting the data assimilation methods developed in the meteorology community.
  We apply the four dimensional variational method to reconstruct the small scales of
  three-dimensional turbulent velocity fields with a moderate Reynolds number, given a time sequence of
  measurement data on a coarse set of grid points. The problem presents new challenges because
  the evolution of the flow is dominated by the nonlinear
  vortex stretching and the energy cascade process, which are absent from
  two dimensional flows. The results show that, reconstruction is successful when the resolution of
  the measurement data, given in terms of the wavenumber, is at the order of the 
  threshold value $k_c = 0.2\eta_K^{-1}$ where $\eta_K$ is the Kolmogorov length scale of the flow. 
  When the data are available
  over a period of 
  one large eddy turn-over time scale, the filtered enstrophy and other
  small scale quantities are reconstructed with a $30\%$ or smaller normalized point-wise 
  error, and a $90\%$ point-wise correlation.
  The spectral correlation between the reconstructed and target fields is higher than $80\%$ for all
  wavenumbers.  
  Minimum volume enclosing
  ellipsoids (MVEEs) and MVEE trees are introduced to quantitatively compare
  the geometry of non-local
  structures. Results show that, for the majority samples, errors in the locations and the sizes of
  the reconstructed structures are within $15\%$, and those in the orientations is
  within $15^\circ$. 
  Overall, for this flow, satisfactory reconstruction of the scales two or more octaves
  smaller is
  possible if data at large scales are available for at least one large eddy turn-over time. 
  In comparison, a direct substitution scheme
  results in three times bigger point-wise discrepancy in enstrophy. The spectral
  difference between the reconstructed and target velocity fields is more than ten times higher than
  what is obtained with the four dimensional variational method. The results show that 
  further investigation is warranted
  to verify the efficacy of the method in flows with higher Reynolds numbers. 
\end{abstract}

%TODO: 1) update superscripts etc (use \rm \bf u and v)

%\begin{keywords}
%Authors should not enter keywords on the manuscript, as these must
%be chosen by the author during the online submission process and will
%then be added during the typesetting process (see http://journals.cambridge.
%org/data/\linebreak[3]relatedlink/jfm-\linebreak[3]keywords.pdf for the full list)
%\end{keywords}

\section{Introduction \label{sect:intro}}

Data assimilation (DA) uses
available experimental or
observational data to improve computational model predictions.
It has had a long
history in meteorology, in particular numerical weather prediction (NWP) research
(see, e.g., \citet{Courtieretal93} and \citet{Kalnay03} for reviews).
Many DA methods have been developed.
Four dimensional variational (4DVAR) methods and ensemble Kalman filter (EnKF) are among the most popular methods
\citep{Kalnay03, Kalman60, Evensen09}. 4DVAR
are based on optimal control \citep{Lions71},
where a
constrained optimization problem is
solved. Spatial as well as temporal data are assimilated
into the model by minimizing the difference between
the model prediction and the data. In recent applications, time sequences of
three dimensional spatial data have been assimilated, hence the name 4DVAR.
EnKF is a sequential method where the error of
the prediction is estimated when
the prediction (known as the \textit{a priori} estimate
\citep{BrownHwang12} or the background
\citep{Kalnay03}) is made. The
observational data are assimilated by combining
the \textit{a priori} estimate with
the measurement data, resulting in an updated prediction
called the \textit{a posteriori} estimate or the analysis. The weight for the measurement
is chosen to minimize the error of the \textit{a posteriori} estimate.

In recent years, these two methods have been applied to
fluid dynamic problems \citep{Hayase15}.
\citet{Gronskisetal13} applies the
variational method to
reconstruct the inflow and initial conditions
for two-dimensional (2D)
mixing layers and wake flows behind a cylinder.
\citet{Monsetal16} considers
2D wake flows,
where a comprehensive comparison between
different DA schemes is presented.
\citet{KatoObayashi13}, \citet{Katoetal15} and \citet{Lietal17}
use the EnKF method to improve Reynolds-Averaged Navier-Stokes (RANS)
based prediction of turbulent flows.
\citet{MeldiPoux17} applies EnKF to large eddy simulations (LES) and detached eddy
simulation (DES).
Similar problems are also investigated with variational methods
\citep{Fouresetal14}.
In \citet{Dadamoetal07}, \citet{Artanaetal12}, and
\citet{Protasetal15}, variational methods are coupled with
reduced
order models based on Galerkin
truncation.
Variational methods are also used in state
estimation
in the context of flow control \citep{BewleyProtas04, Chevalieretal06,
Colburnetal11}, to extrapolate experimental
data in a dynamically
consistent way \citep{Heitzetal10, Combesetal15} and to obtain optimal
sensor placements \citep{Monsetal17}.
\citet{Monsetal14} applies a variational method to investigate decaying isotropic turbulence, where the
eddy-damped quasi-normal Markovian model is used.
%TODO: has been updated.

%\citet{Narayanetal12} introduced a multi-model EnKF method, but it was not
%applied to fluid mechanics problems.

%\citet{Yangetal17} applied the method in \citet{Narayanetal12} to the
%simulations of sub-surface flows. A bit too far from our cases.

%\citet{Lietal17} used a RANS model coupled with EnKF. A method similar to
%\citet{KatoObayashi13}.

%\citet{Katoetal15} used ETKF to estimate flow parameters (Mach etc)

%Two-dimensional isotropic turbulence is investigated in \citet{Geshoetal16}
%where numerical simulations are compared with analytical predictions. It
%shows that the data actually required in the numerical experiments are
%significantly less than what is suggested by analytical results
%\citep{OlsonTiti03}.

The aforementioned research has demonstrated the potential of DA in turbulent
simulations as well as in understanding the physics of turbulent flows.
As having been demonstrated in NWP research, DA is unique in its ability to improve our modelling or
prediction of
\emph{instantaneous} flow fields, in addition to their \emph{statistics}.
However, few studies have explored this aspect in the simulation of 3D turbulent fields.
(see, e.g., the list of recent works on DA
tabulated in \citet{Monsetal17}).
In \citet{Yoshidaetal05}, the ability of a DA scheme to
recover the instantaneous small scales in a 3D isotropic turbulent field is investigated.
In this study, the data, given as Fourier modes, directly replace
model predictions given by the NSEs at every time step.
It is found that
the small scale instantaneous velocity field can be asymptotically recovered exactly
when
Fourier modes with wavenumber up
to a threshold value approximately equal $k_c \equiv 0.2
\eta_K^{-1}$ are provided, where $\eta_K$ is the Kolmogorov length scale.
When the amount of data
decreases towards the threshold, the time needed to recover the small scales tends to infinity
(hence an infinitely long time sequence of measurement data is needed).
The problem is investigated in \citet{Lalescuetal13} from the perspective of
chaos synchronization and a similar conclusion is reached. The DA method used in these works can be
termed `direct substitution' (DS). As far as we know, the ability of 4DVAR or EnKF to reconstruct
the small scales has not been investigated.
In this paper, we present an analysis based on the 4DVAR method.
We consider a Kolmogorov flow in a 3D periodic box.
It is assumed that a time sequence of velocity data are
given on a set of grid points. 4DVAR is employed to reconstruct the
initial velocity field such that the
velocity at later times matches given measurement data.
The time sequence of velocity fields computed from the initial field are compared with the `true'
velocity fields (the \emph{target} fields).
%In our numerical experiments, the available data are below the threshold given by
%\citet{Yoshidaetal05}.
The objective is to
ascertain how well the instantaneous
small scale velocity fields can be reconstructed with 4DVAR for a given set of parameters.
The 3D turbulence in a periodic box is the simplest turbulent flow where the nonlinear
inter-scale interaction plays the dominant role in the dynamics. The vortex stretching
process, e.g., is absent in 2D flows. Therefore, although
the mathematics is similar for 2D and 3D problems, the latter does present significant new challenges.

Using the 4DVAR method to reconstruct 3D
fully developed (i.e., statistically stationary) turbulent field is the first contribution of this investigation.
To evaluate the reconstruction, it is important
to quantify the instantaneous difference
between the reconstructed and the target velocity fields.
Much information can be learned from point-wise correlations and/or the statistics of point-wise difference.
However,
these statistics are not best suited to capture
the geometry of \emph{non-local structures} populating the
small scales of turbulent
fields (such as the vortex filaments). This deficiency becomes a significant obstacle when the distribution of
the non-local structures becomes one of our main interests.

The morphology of non-local structures have long been described in qualitative terms,
assisted by visualization. Great efforts have been made in recent years to develop methods for quantitative description.
\citet{MorenoPullin08} uses the probability density function
(PDF) of the
curvatures on the surface of a structure as its signature.
\citet{YangPullin11} describes the structures in a channel flow in terms of curvelets and angular
spectra. Indices named shapefinders, defined in terms of the Minkowski functionals, are used in
\citet{Leungetal12} to classify the structures. These methods provide very detailed
quantitative descriptions of the structures. However, they focus on the intrinsic geometry;
the information about locations, orientations and sizes of the structures sometimes is missing,
which happens to be important when we compare the geometry in two different
fields.
Therefore, as the second contribution of this investigation, 
we propose to use minimum volume enclosing ellipsoids (MVEEs)
to describe the
geometry of a nonlocal structure, and use MVEE trees where the structures are highly
non-convex.
MVEE is used widely in areas such as statistical estimate, cluster
analysis and image processing
\citep{Todd16}.
Its application in turbulence research, however, has not been reported.
The results demonstrate that MVEEs and MVEE trees are useful tool for the
analysis of the non-local geometry in turbulence.

The paper is organized as follows. In Section \ref{sect:eq}, the 4DVAR formulation of the problem is
introduced; the description of the small scales of homogeneous turbulence is reviewed, and then the
calculation of MVEEs and MVEE tree is explained. In Section \ref{sect:results}, the
simulations and the results are presented and analyzed. The conclusions are summarized
in Section \ref{sect:conclusion}.

%structure:
%1) analysis of the reconstructed fields: nonlinear and spectral
%properties
%2) Dynamics of reconstruction: what is the question? How the data affect the
%reconstructions of the small scales (large scales are straightforward)? How do
%we answer this. The adjoint field gives the gradient of the cost with
%respect to variations of velocity at all scales. We can answer the question by
%analyzing the adjoint field.
%3) Effects of nudging (???Unable to predict just yet)
%4) Reconstruction by imperfect models. (How? Using a dynamic model? Many
%questions: need to verify the model; does it enhance the prediction of
%parameters out side of the input data set? Does it matter if the model is a
%poor one? )

\section{The governing equations and the description of small scales of turbulent velocity fields \label{sect:eq}}

\subsection{The problem setup and the optimality system \label{sect:optsys}}

%In 4DVAR, the DA problem is formulated as a constrained optimization problem. In this section, we
%summarize the formulation of the problem.
Let $B$ denote
the three dimensional periodic box $[0,2\pi]^3$, and $\v(\x, t)$ be a time sequence of velocity
fields in $B$ for $t \in [0,T]$
with $T>0$. $T$ is called the optimization horizon hereafter.
It is assumed that only part of $\v(\x,t)$ is known from measurement, and this part is denoted by $\FF \v$,
where $\FF$ is a filter to be defined
later. $\FF$ plays the
same role as the measurement operator used in the meteorology community.
The question is, from given $\FF\v(\x,t)$, how to reconstruct
the full field $\v(\x,t)$ for $t \in [0,T]$.

To do so, an underlying `model' for the data $\v(\x,t)$ has to be assumed. We assume $\v(\x,t)$ is
the solution of the
Navier-Stokes equations (NSEs), from some unknown initial condition (IC).
%The reconstruction process involves combining the data $\FF\v$ with the solution of the NSEs,
%leading to a data assimilation problem.
Let $\u(\x,t)$ be a solution of the NSEs in $B$. If one
finds a
velocity field $\bvp(\x)$ such that, if $\u(\x,0) = \bvp(\x)$
then $\u(\x,t)$ agrees with $\v(\x,t)$ for $(\x,t) \in B\times [0,T]$, then
$\u(\x,t)$ is a reconstruction of $\v(\x,t)$.
It is unlikely to find $\bvp(\x)$
such that $\u(\x,t) = \v(\x,t)$ exactly. In 4DVAR,
one attempts to approximate
$\bvp(\x)$
by the solution of a constrained
optimization problem. The detail is described in what follows.

We define the inner product between two scalar or vector fields $\a(\x,t)$
and $\b(\x,t)$ as
\be
\lal \a, \b\ral \equiv \frac{1}{(2\pi)^3}\int_{\x \in B} ~ \a \cdot \b ~  d\x. 
\ee
The cost function $J$ for the optimization problem is defined as
\be \label{eq:cost}
  J = \frac{1}{2} \int_0^T \lal \FF \u -\FF \v, \FF \u - \FF\v\ral dt.
\ee
Qualitatively speaking, $J$ is the total difference between the measurements based on $\u$ and
$\v$ over the space-time domain $B \times [0,T]$.
The objective of the optimization problem is to find $\bvp(\x)$ such that
$J$ is minimized while $\u(\x,t)$ satisfies
    \be \label{eq:NSE}
    \fN(\u, p) \equiv -\ptl_t \u - \u \cdot \nabla \u - \nabla p + \nu \nabla^2 \u + \f = 0,
    \quad \nabla \cdot \u = 0,  \quad
    \u(\x, 0) = \bvp(\x),
    \ee
where $p$ is the pressure, $\nu$ is the kinematic viscosity and $\f(\x,t)$ is the forcing term. 
The flow has been assumed to be
incompressible.

The optimal solution for the problem will be found with the adjoint method.
We introduce the Lagrangian functional:
\be \label{eq:lag}
L (\u, \bxi, \sigma, \blmd, \bvp) = J (\u) - \int_0^T \lal\bxi,
\fN\ral dt - \int_0^T \lal \sigma,
\nabla \cdot \u \ral dt - \lal
\blmd, \u(\x,0) -
\bvp(\x) \ral,
\ee
where $\bxi(\x,t)$, $\sigma(\x,t)$, and $\blmd(\x)$ are the adjoint variables corresponding to the
NSEs, the continuity equation, and the IC.
$\bxi$ and $\sigma$ are also
called the adjoint velocity and the adjoint pressure.
The minima of $J$ are found at the stationary points for the
Lagrangian where
the functional derivatives of $L$ with respect to $\u$, $\bxi$,
$\sigma$, $\blmd$, and $\bvp$ are zero. These conditions give rise to
equations for the adjoint variables (i.e. the adjoint equations) in addition to
Eq. (\ref{eq:NSE}). The equations for $\bxi$ read:
    \be \label{eq:xi}
    -\ptl_t \bxi - \u \cdot \nabla \bxi + \nabla \u \cdot \bxi  +
    \nabla \sigma - \nu \nabla^2 \bxi -
    \fF = 0, \quad \nabla \cdot \bxi = 0,
    \ee
with the end-point condition:
    \be \label{eq:xi_ic}
    \quad  \bxi(\x, T) = 0.
    \ee
The forcing term $\fF$ in Eq. (\ref{eq:xi}) is given by
    \be \label{eq:adjf}
    \fF(\x,t) \equiv - \frac{\ptl J}{\ptl \u} =
      -\FF^+\FF[\u(\x,t) - \v(\x, t)]
    \ee
where $\FF^+$ is the adjoint operator of $\FF$. The forcing term $\f(\x,t)$ in Eq. (\ref{eq:NSE}) has been
assumed to be independent of $\u$ and, as a consequence, does not have a counter part in the adjoint equations. 
When $\u$, $p$, $\bxi$ and $\sigma$ satisfy Eqs.
(\ref{eq:NSE}) and (\ref{eq:xi}-\ref{eq:adjf}),
the gradient of $J$ with respect to $\bvp$, denoted by
$DJ/D\bvp$, is given by
    \be \label{eq:dj}
    \frac{D J}{D \bvp} = \frac{\partial L}{\partial \bvp} = -\bxi(\x, 0).
    \ee
Eqs. (\ref{eq:NSE}), (\ref{eq:xi}-\ref{eq:adjf}) and
the condition $D J/D\bvp =0$
constitute the optimality system of the optimization problem. 
The solution of the optimization problem is a solution
of the optimality system.

The solution method of the optimality system will be explained later.
The optimal solution for $\bvp(\x)$ will be denoted as
$\bvp^\rbu(\x)$.
The solution $\u(\x,t)$ of Eq. (\ref{eq:NSE}) with
$\bvp(\x) = \bvp^\rbu(\x)$ is the reconstruction of $\v(\x,t)$ from $\FF\v(\x,t)$.
In this investigation,
a sequence of known fully developed
turbulent velocity fields in the 3D
periodic
box $B$ is used as $\v(\x,t)$. $\FF$ is chosen as a
cutoff filter in the Fourier space with a cutoff wavenumber $k_m$,
such that only the low wavenumber Fourier modes
are used in the optimization problem. $k_m$ is an indicator of the spatial resolution of the
measurement data.
As a result, the cost
function $J$ can be written as
\be \label{eq:J1}
J = \frac{1}{2}\int_0^T \int_{k \le k_m} (\hat{\u}(\k,t) -
\hat{\v}(\k,t))\cdot (\hat{\u}^*(\k,t) - \hat{\v}^*(\k,t)) d\k,
\ee
where $k=|\k|$, and $\hat{\v}(\k,t)$ and $\hat{\u}(\k,t)$ denote the Fourier
modes of $\v(\x,t)$ and $\u(\x,t)$, respectively.
Letting $\hat{\fF}(\k,t)$ be the Fourier transform of $\fF(\x,t)$,
we obtain
\be \label{eq:adjf1}
\hat{\fF}(\k,t) =
\begin{cases}
  -[\hat{\u}(\k,t) - \hat{\v}(\k, t)]  & \text{for} \ k \le k_m , \\
  0  & \text{otherwise}.
\end{cases}
\ee

\subsection{The questions to be answered \label{sect:questions}}

We call $\u(\x,t)$ and $\v(\x,t)$ the
`reconstructed' and the
`target' fields, respectively.
The target field at $t=0$,
$\v(\x,0)$, will also be denoted by notation $\bvp^{\rm \bf v}(\x)$, corresponding to
its reconstruction $\u(\x,0) \equiv \bvp^{\rm \bf u}(\x)$.
Since $\u(\x,t)$ is the solution of the NSEs with
$\bvp^{\rm \bf u}(\x)$ as the IC, it is expected $\hat{\u}(\k,t) \approx
\hat{\v}(\k,t)$ when $k \le k_m$. The interesting question is how
closely $\hat{\u}(\k,t)$ matches $\hat{\v}(\k,t)$ when $k > k_m$; in other words,
to what extent small scales in $\v(\x,t)$ can be reconstructed.

As $\u(\x,t)$ and $\v(\x,t)$ are both solutions
of the forced NSEs, one expects their \emph{statistics} to be the same for $t\to T$ when $\u$ has fully
developed. Therefore, for $t\to T$, we are interested in the \emph{point-wise,
instantaneous} difference or correlation between $\u$ and $\v$.
However, good reconstruction at $t\to T$ is possible only when the IC
$\bvp^\rbu(\x)$ captures to some extent the features of real turbulence.
Therefore, we will also examine selected statistics of 
$\bvp^{\rm \bf u}(\x)$.
The statistics provide information on how much can be reconstructed at $t=0$ given the measurement data
$\FF \v(\x,t)$, hence is also relevant to the observability of the system.
Nevertheless, the full discussion of this aspect is left for the future.

The efficacy of 4DVAR can be assessed by comparison with other methods.
As we will show
later, the
straightforward Lagrangian interpolation fails badly.
In \citet{Yoshidaetal05}, an assimilation scheme we call `direct substitution' (DS)
is used. The authors
integrated the NSEs starting from a Gaussian random field.
The low wavenumber modes with $k\le k_m$ in the solution $\u(\x,t)$
are replaced by those in the target field $\v(\x,t)$
at every time step. \citet{Yoshidaetal05} shows that, if $k_m \ge 0.2 \eta_K^{-1}$,
the small scales in $\u(\x,t)$ approach those in
$\v(\x,t)$ asymptotically as $t\to \infty$ (see also \citet{Lalescuetal13}).
In most cases we are investigating, the resolution $k_m$ is slightly below this threshold (c.f., the parameters in Sect.
\ref{sect:DNS}), and the measurement data are available
only for $t\in[0,T]$ where $T$ is at the order of one large eddy turn-over time scale. 
As a consequence, the DS scheme will only achieve partial reconstruction.
The DS scheme will be compared with the 4DVAR method, to demonstrate the improvement provided by the
latter.

\subsection{Description of the small scales of turbulent velocity fields \label{sect:descript}}

To compare the small scales of the reconstructed field $\u$ and the target field $\v$,
the filtering approach is used to separate different scales and
the analysis is conducted mainly in the physical space.
The filtered velocity field is defined by
\be
\td{\u}(\x,t) = \int G_\Delta(\x-\y) \u(\y,t) d \y,
\ee
where $\td{\u}$ denotes the filtered velocity and $G_\Delta (\x)$ is a filter with length scale $\Delta$.
$G_\Delta$
separates $\u$ into $\td{\u}$ and the
subgrid-scale (SGS) velocity. The parameters used to
describe the structures of $\td{\u}$ and its interactions with the
SGS scales are now summarized. In the analysis of the velocity
field $\u$, we always use the Gaussian filter \citep{Pope00}.

The local structures of $\td{\u}$ are described by the filtered
velocity gradient $\td{A}_{ij} \equiv \ptl_j \td{u}_i$, the filtered strain rate tensor
$\td{s}_{ij} \equiv (\td{A}_{ij} + \td{A}_{ji})/2$, and the filtered vorticity $\td{\om}_i \equiv
\varepsilon_{ijk} \td{A}_{kj}$.
For $\td{A}_{ij}$, much insight has be learned from its tensor invariants
\be
Q \equiv -\frac{1}{2} \td{A}_{ij} \td{A}_{ji}, \q R \equiv -\frac{1}{3} \tA_{ij}\tA_{jk}\tA_{ki}.
\ee
One of the key features of turbulence is that the joint PDF of $R$ and $Q$ displays a skewed
tear-drop shape, which captures
the prevalence of straining motion in turbulent flows; see, e.g., \citet{Cantwell92}.

%The tensor invariants of $\ts_{ij}$ are related to the eigenvalues by
%\be
%Q_s \equiv -\frac{1}{2} \ts_{ij}\ts_{ij} = -\frac{(\lm_\al^{s})^2 + (\lm_\beta^s)^2 +
%(\lm_\gamma^s)^2}{2}, \q
%R_s \equiv - \frac{1}{3} \ts_{ij}\ts_{jk}\ts_{ki} = -\lm_\al^s\lm_\beta^s\lm_\gamma^s.
%\ee
%$Q_s$ measures the magnitude of $\ts_{ij}$.
%The dimensionless shape parameter $S^*_s$ defined by
%\be \label{eq:shapes}
%S^*_s \equiv \frac{- 3 \sqrt{6} \lm_\al^s \lm_\beta^s \lm_\ga^s}{[(\lm_\al^s)^2 + (\lm_\beta^s)^2 +
%(\lm_\ga^s)^2 ]^{3/2}}
%\ee
%is used to describe the geometry of $\ts_{ij}$. It can be shown mathematically that $-1\le S^*_s \le 1$
%where $S^*_s = \pm 1$ when $\lm_\al^s : \lm_\beta^s : \lm_\ga^s = \pm (1 : 1: -2)$. Thus, when $S^*_s = 1$
%(and $-1$), the strain
%rate tensor represents contraction (expansion) in one direction and axisymmetric expansion
%(contraction) in the other two
%directions. It is well-known that $S^*_s$ has a strong peak at $1$ in turbulent flows.

\begin{figure}
  \centerline{\includegraphics[width=0.4\lnw]{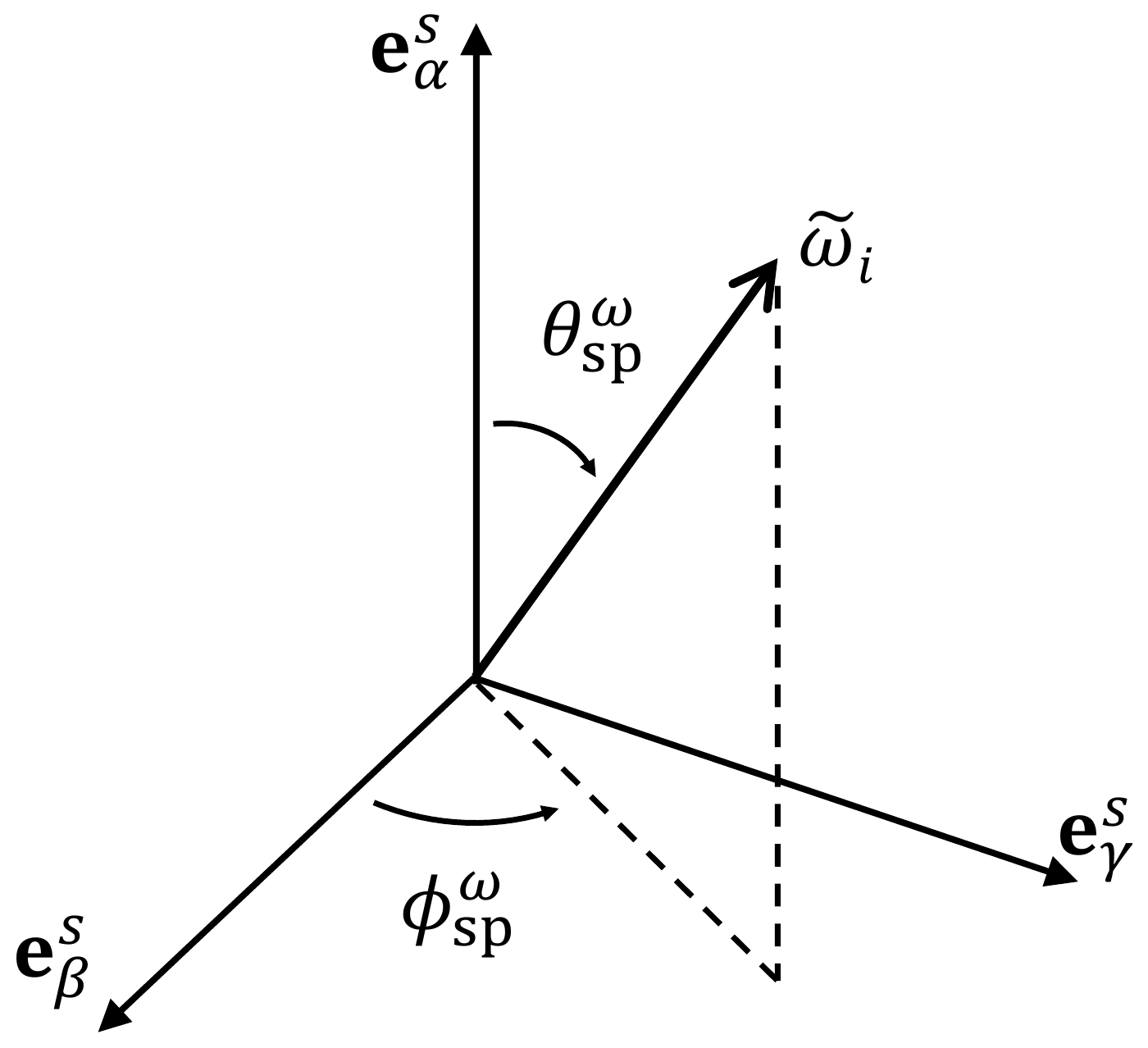}}% Images in 100% size
  \caption{Definitions of the polar and azimuthal angles used to describe the orientation
  of  $\tog_i$ in the eigen-frame of $\td{s}_{ij}$. Angles defined in a similar way for other
  vectors and eigen-frames are also used in the paper.}
\label{fig:eigenframe}
\end{figure}

The eigenvectors of $\td{s}_{ij}$ are denoted by $\e^s_\al$, $\e^s_\beta$
and $\e^s_\gamma$, with corresponding
eigenvalues $\lm^s_\al \ge \lm_\beta^s
\ge \lm_\gamma^s$. $\lm_\al^s + \lm_\beta^s + \lm_\gamma^s = 0$ due to the incompressibility of the filtered
velocity field.
The enstrophy of the resolved velocity field, defined by $\tog^2 \equiv \tog_i \tog_i$,
is a measure of the magnitude of the resolved vorticity. The main source of growth for $\tog^2$
is the vortex stretching term
$P_\omega \equiv \ts_{ij} \tog_i \tog_j$.
Evidently,
\be \label{eq:thetaos}
P_\omega = \tog^2 (\lm_\al^s \cos^2\theta^{s}_\al + \lm_\beta^s \cos^2 \theta^{s}_\beta + \lm_\gamma^s \cos^2
\theta^{s}_\gamma),
\ee
where $ \theta^{s}_\al$ is the angle between $\td{\bo}$ and $\e_\al^s$ with $\theta^{s}_\beta$ and
$\theta^{s}_\gamma$ defined in a similar way. Eq. (\ref{eq:thetaos}) shows that $P_\omega$ are
determined by the eigenvalues of $\ts_{ij}$
as well as the alignment between $\tog_i$ and the eigenvectors of $\ts_{ij}$.
The eigenvectors of $\ts_{ij}$ form a orthogonal
coordinate frame. The orientation of $\tog_i$ can be described by the polar
and azimuthal angles relative to $\e^s_\al$ in this frame,
and they are denoted by $\theta_{\rm sp}^\om$ and $\phi_{\rm
sp}^\om$. Fig. \ref{fig:eigenframe} gives an illustration of the angles. In turbulence, by examining the joint PDF of
$\cos \theta_{\rm sp}^\om$ and $\phi_{\rm sp}^\om$ or the PDFs of $\theta^s_i$ ($i=\al, \beta,
\gamma$), it has been
shown that $\tog_i$ prefers to align with $\e^s_\beta$.

The dynamics of $\td{u}_i$ is governed by the
filtered NSEs, which is unclosed due to the SGS stress tensor
$\tau_{ij} = \widetilde{u_i u}_j - \td{u}_i \td{u}_j$ (see, e.g., \citet{MeneveauKatz00}).
$\tau_{ij}$ represents the nonlinear interactions between the resolved and the SGS scales.
An important parameter related to $\tau_{ij}$ is the
SGS energy dissipation rate
\be \label{eq:pi}
\Pi_\Delta = - \tau^d_{ij} \td{s}_{ij}
\ee
where $\tau^d_{ij} \equiv \tau_{ij} - \delta_{ij} \tau_{kk}/3$ is
the deviatoric part of $\tau_{ij}$.
$\Pi_\Delta$ is the turbulent kinetic energy flux cascading from resolved scales to
the SGS scales.
Correctly reproducing the statistics of $\Pi_\Delta$ is one of the main objectives
in SGS modelling. As a consequence,
the statistics of $\tau^d_{ij}$ and $\Pi_\Delta$ and their correlations with
$\ts_{ij}$, $\tog_i$, $P_\om$ and $\tA_{ij}$
have been
extensively documented;
see, e.g. \citet{MeneveauKatz00} and \citet{Sagaut02} for
reviews on SGS modelling and large eddy simulations.

$\Pi_\Delta$ is determined by the
relative alignment between the eigenvectors of
$\tau^d_{ij}$ and $\td{s}_{ij}$ as well as their eigenvalues.
Specific preferential alignment has also been observed in turbulent
flows \citep{Taoetal02}.
The eigenvectors of $-\tau_{ij}^d$ are denoted by
$\e_\alpha^{\tau}$, $\e_\beta^{\tau}$, and $\e_\gamma^{\tau}$, with
corresponding eigenvalues
$\lm_\alpha^{\tau}\ge\lm_\beta^{\tau}\ge\lm_\gamma^{\tau}$ where
$\lm_\al^\tau + \lm_\beta^\tau + \lm_\gamma^\tau = 0$ by definition.
The vorticity $\td{\om}_i$
also displays preferential alignment with
$-\tau^d_{ij}$ \citep{Horiuti03}, specifically
$\tog_i$ tends to be perpendicular to $\e_\gamma^\tau$ while aligned with $\e_\al^\tau$ or
$\e_\beta^\tau$.

\subsection{Description of the non-local structures \label{sect:cs_intro}}

\begin{figure}
  \bmn{0.5\lnw}
  \centerline{\includegraphics[width=0.8\lnw]{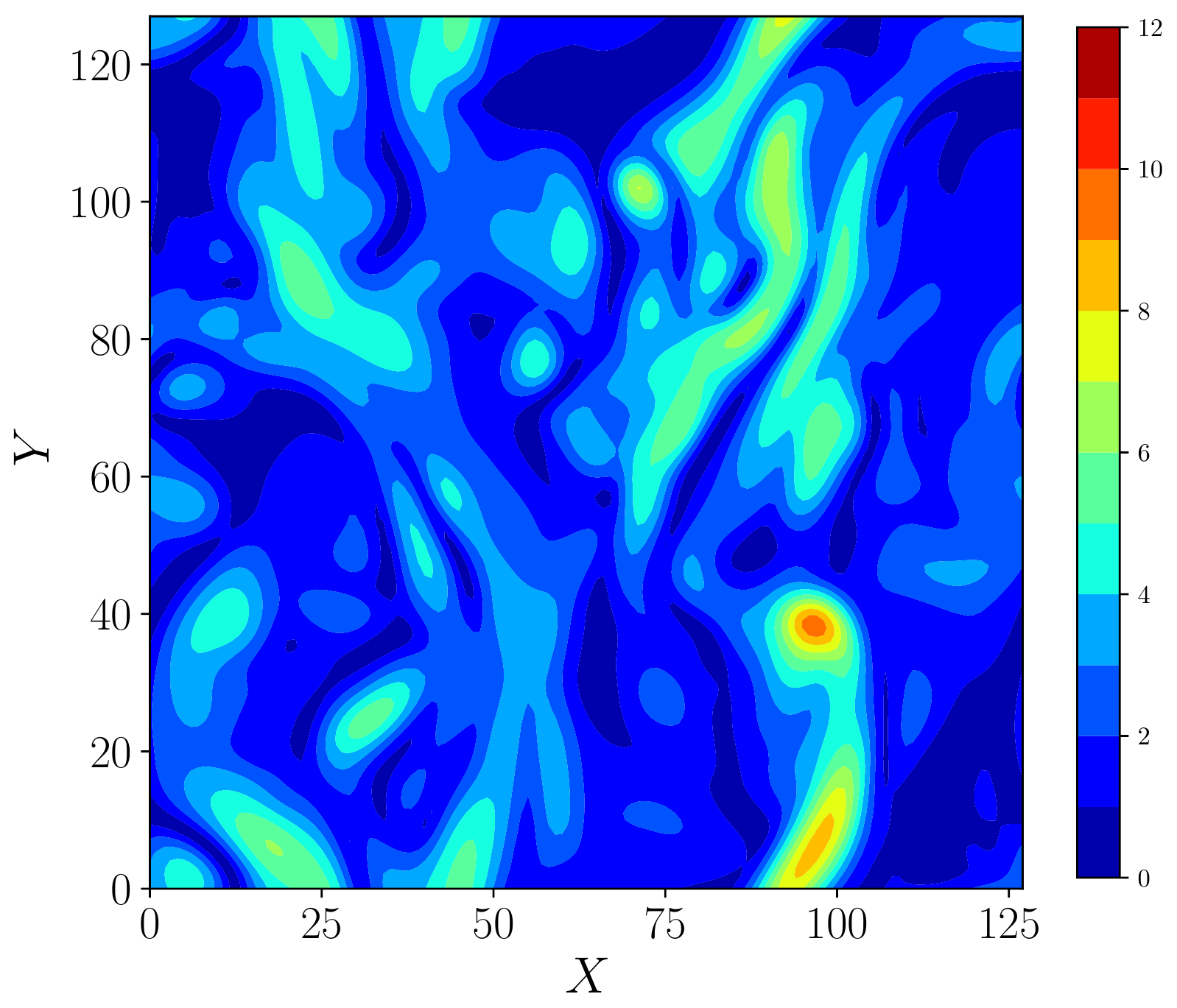}}
  \emn%
  \bmn{0.5\lnw}
  \centerline{\includegraphics[width=0.8\lnw]{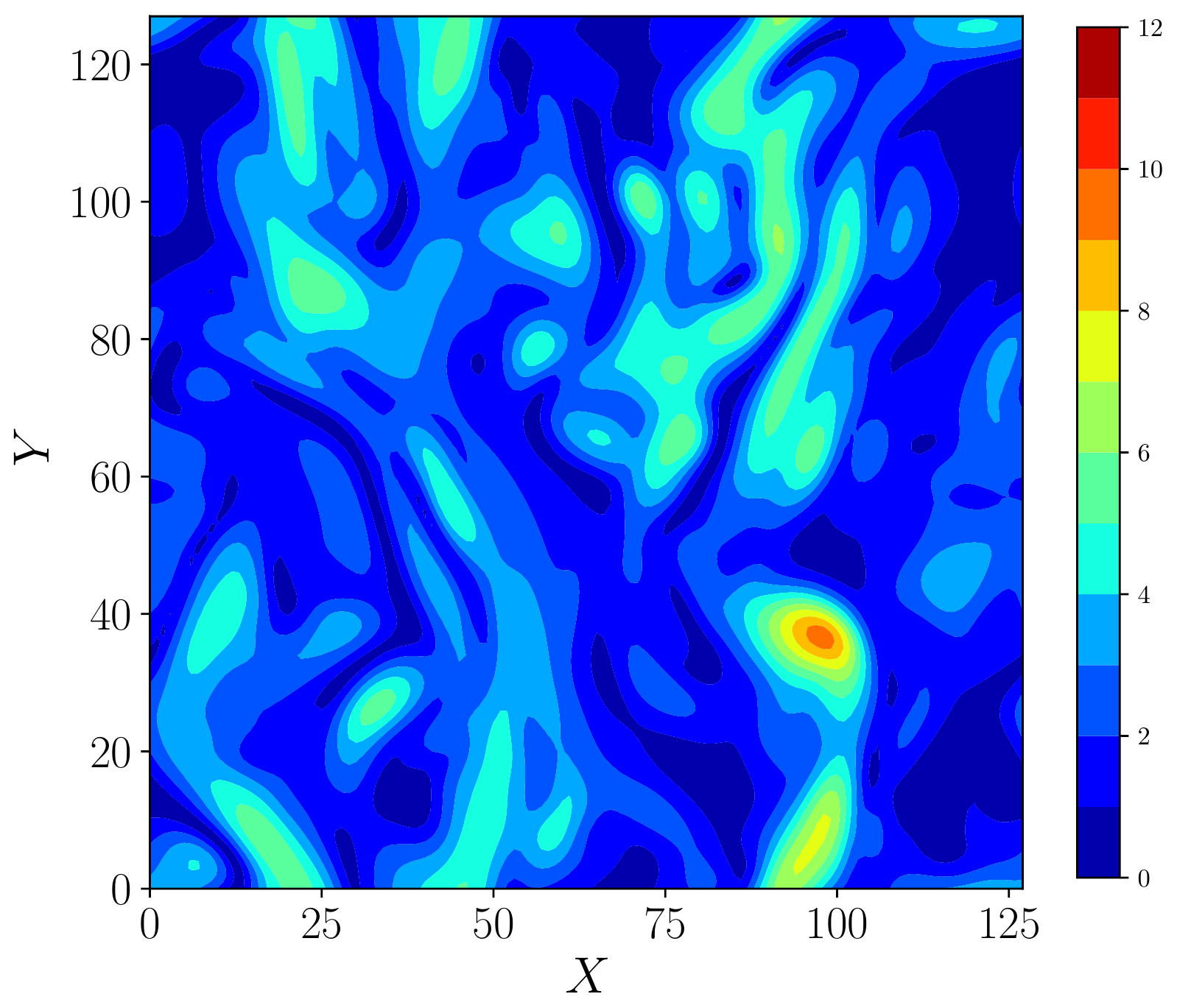}}
  \emn
  \caption{Instantaneous distributions of $\tog\equiv (\tog_i\tog_i)^{1/2}$ on a plane perpendicular
  to the $z$ axis. Left: from a target field; right: from the corresponding reconstructed field.
  $X$ and $Y$ are the labels of the grid points in the $x$ and $y$ directions, respectively. 
  }
\label{fig:oo-contour}
\end{figure}

The filtered velocity gradient $\tA_{ij}$ and related quantities describe the local structure of the
filtered velocity field in an infinitesimal neighborhood of a spatial location.
Non-local structures of finite sizes are also of great interests.
The most well-known example of such structures
is that of the vortex filaments that are characterized by strong enstrophy. The ability of the reconstructed
velocity field to capture the \emph{locations}, the \emph{dimensions}
and the \emph{orientations} of these structures is not always clear
from the point-wise statistics.
To illustrate this point, Fig. \ref{fig:oo-contour} plots selected contours
of $\tog\equiv (\tog_i\tog_i)^{1/2}$
in one target field and the corresponding reconstructed field.
Detailed comparison is given
later. However, it is already clear that
the locations, orientation and sizes of the
contours display remarkable similarity, especially in high $\tog$ regions. Usual statistics,
including joint correlations between multiple points, do not always capture this similarity.
Additional diagnostics are needed.

%In this analysis, the terminology `coherent structure' is used loosely to describe a continuous spatial domain
%with a physical parameter taking a specific range of values (typically values larger than a
%threshold value). For example, we will look into coherent vortical structures, which are defined as
%continuous spatial domains with $\tog \ge \tog_{\rm th}$ for some given threshold value $\tog_{\rm th}$.

\begin{figure}
  \bmn{0.34\lnw}
  \centerline{\includegraphics[width=1.05\lnw]{run15ww300-cntrf-crop}}
  \emn%
  \bmn{0.33\lnw}
  \centerline{\includegraphics[width=0.92\lnw]{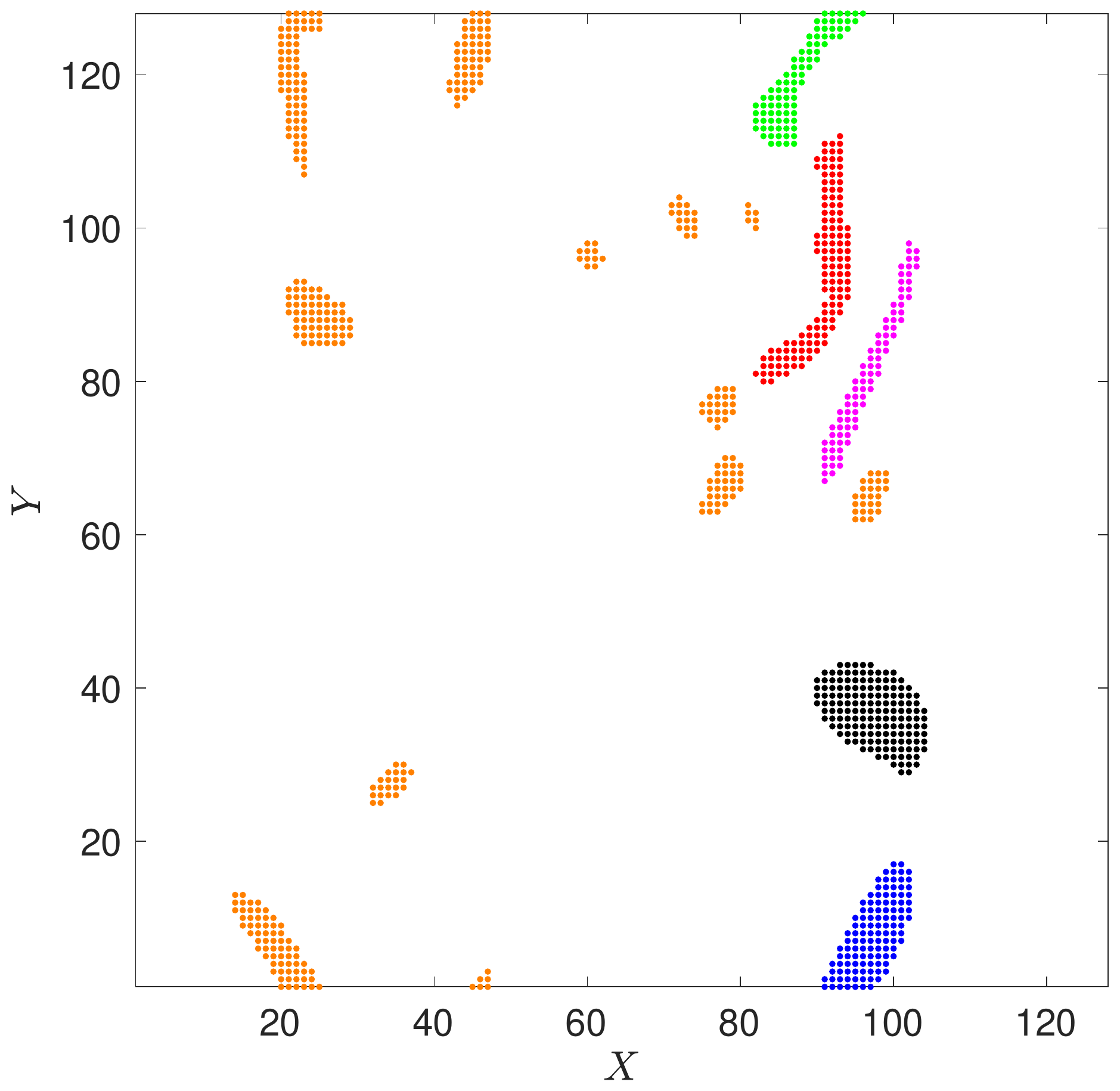}}
  \emn %
  \bmn{0.33\lnw}
  \centerline{\includegraphics[width=0.92\lnw]{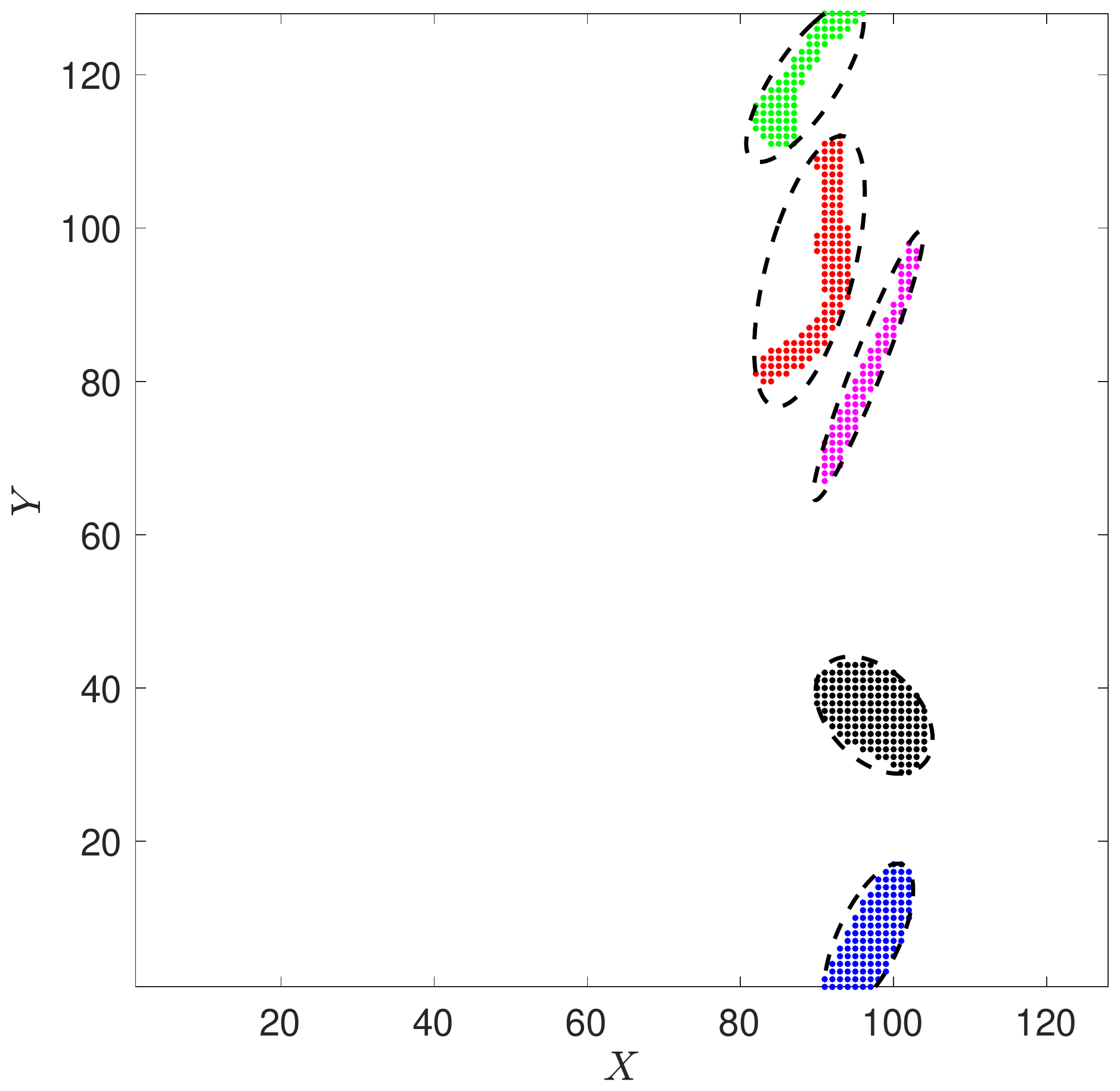}}
  \emn
  \caption{The calculation of the MVEEs. Left: the contour
  plot of $\tog$ on a 2D slice (same as the right panel of Fig.
  \ref{fig:oo-contour}).
  Middle: the 17 structures with $\tog \ge 2 \lal
  \tog_i\tog_i\ral^{1/2}$ shown with the grid points,
  which are identified by
  the DBSCAN algorithm (see text for
details). The 5 largest structures are highlighted
  with black, red, blue, green and magenta, respectively.
  Right: the MVEEs for the 5 largest structures. 
  $X$ and $Y$ are the labels of the grid points in the $x$ and $y$ directions, respectively. 
  }
\label{fig:2Ddemo}
\end{figure}

Note that, although an intuitive description of the structures
visualized by the contours
is straightforward, the precise description of their shapes, sizes and even
locations is difficult,
and actually there are no unique definitions for these concepts (also see discussions in, e.g., \citet{MorenoPullin08,
YangPullin11, Leungetal12}).
As explained in Section \ref{sect:intro}, we use the minimum volume enclosing ellipsoids
(MVEEs) of a structure to define and describe
its geometry. By definition, of all the ellipsoids that contain a
structure,
the MVEE is the one with the smallest
volume. The location, dimensions, and orientation of an MVEE are well-defined, and are given by its
center, lengths of the three axes, and the orientations of the axes. We thus \emph{define} the
location, dimensions, and orientation of a structure using those of its MVEE.
The structures in the reconstructed and the target fields are compared on this basis.
The comparison includes four steps of calculations, which are explained below and illustrated in Fig.
\ref{fig:2Ddemo} using a 2D slice.
%\begin{enumerate}
%  \item

In the first step,
the set $\PP$ of the grid points where certain threshold condition is
satisfied are extracted. $\PP$ usually consists of a number of disjoint regions. The `structures' we
    will be discussing refer to this type of
    regions; each structure corresponds to a disjoint region (c.f., the middle panel of Fig. \ref{fig:2Ddemo}).

    %\item
In the second step, individual structures in $\PP$ are extracted and distinguished from others.
      The programmatic realization of this task is non-trivial despite its innocuous appearance.
      It is accomplished by recasting it as a clustering problem, and using a density based clustering algorithm called
    DBSCAN \citep{Esteretal96, Schubertetal17}.
    In this algorithm,
    the `neighbors' of a point $p$ in $\PP$ are first identified, where a neighbor is defined as a
    point whose distance to $p$ is less than a given value
    $\ep_c$. A point with fewer than $n_c$ neighbors is considered `noise'. A cluster
    is defined according to the following rule: the neighbors of a non-noise point $p$ are in
    the same cluster to which point $p$ belong.
    Algorithmically, the points in the cluster are identified by recursively applying
    this rule until all the points identified are noise.
    The same process is repeated for other non-noise points if they
    do not belong to any clusters that have been found.

    For more details of the DBSCAN algorithm, the readers are referred to \citet{Esteretal96}.
    The outcome of applying the algorithm is that all the points in a structure are found and stored separately from
those in other
    structures, as illustrated in the middle panel of Fig. \ref{fig:2Ddemo}.
    We normalize the distance in such a way that the distance between two consecutive grid points is
    $1$. As a consequence, $\ep_c =  D^{1/2}$ and $n_c = D + 1$ are used, where $D$ is the dimension
    of the embedding
    space (i.e. $D=2$ for structures on a 2D slice whereas $D=3$ for those in the 3D space).

    %\item
    In the third step,
      the MVEE for each structure is calculated.
      An ellipsoid is defined by a symmetric positive definite matrix $E$, which specifies the
      orientations and lengths of the axes, and a vector $\c$, which specifies its center.
    Its volume is proportional to the determinant of $E^{1/2}$,
    $\text{det}\, E^{1/2}$.
      Let $P$ be the set of
      points ${\fp}_i$ $(i=1,...,N$) in the structure. The MVEE is given
      by the optimal $E$ and $\c$ that minimize
    $\text{det}\, E^{1/2}$, subject to the constraints
\be \label{eq:ell}
    (\fp_i^T -\c^T) E^{-1} (\fp_i -\c) \le 1, \q \forall \fp_i \in P.
\ee
The constraints ensure $\fp_i$ is inside the ellipsoid.
This minimization problem can be solved by the Khachiyan algorithm \citep{Khachiyan96, Todd16}.
We use the MATLAB implementation by \citet{Moshtagh09}.
%\end{enumerate}

In the fourth (and last) step, the correspondence between a structure in the target field and its
reconstruction in the reconstructed field is established.
The above three steps are applied to both the reconstructed and the target fields.
If the reconstructed field mimics the target field perfectly, one would then obtain two groups of MVEEs,
with a
one-to-one correspondence between the members. The correspondence can
be trivially established.
In reality, however,
this naive method breaks down because
  the difference between the two fields sometimes is big enough to destroy the trivial one-to-one correspondence.
The following procedure thus has been used,
where, in essence, the structures are identified by their locations and sizes.
The structures in the target field are first arranged into a list in the descending
order according to their sizes, where the \emph{size} of a structure is defined as the number of
grid points in the structure.
Let $\SS_T$ be a structure in the list. The structure in the reconstructed field
whose distance to $\SS_T$ is the minimum is then identified,
where the distance between two structures is defined as the shortest distance from one to the other.
Let $\SS_O$ be this structure.
If $\SS_O$ is not unique,
the one with the largest size is chosen. This $\SS_O$ is taken as the reconstruction of
$\SS_T$. $\SS_T$ and $\SS_O$ are called the
\emph{matching} structures. The above procedure is repeated for each structure
in the target field, starting from
the one largest in size, until all matching structures are identified.

The matching structures and corresponding MVEEs obtained after these four steps are then compared
and statistics are calculated, including relative displacement, alignment, and relative sizes,
among others.
The analysis is applied to several different physical
quantities, including high strain rate structures for which the magnitude of
$\td{s}_{ij}$ is large, vortical structures with strong $\tog$, and structures with
high SGS energy dissipation rate $\Pi_\Delta$ or negative $\Pi_\Delta$. In all cases, the analysis
is performed directly on 3D structures.

\begin{figure}
  \centerline{\includegraphics[width=0.8\lnw]{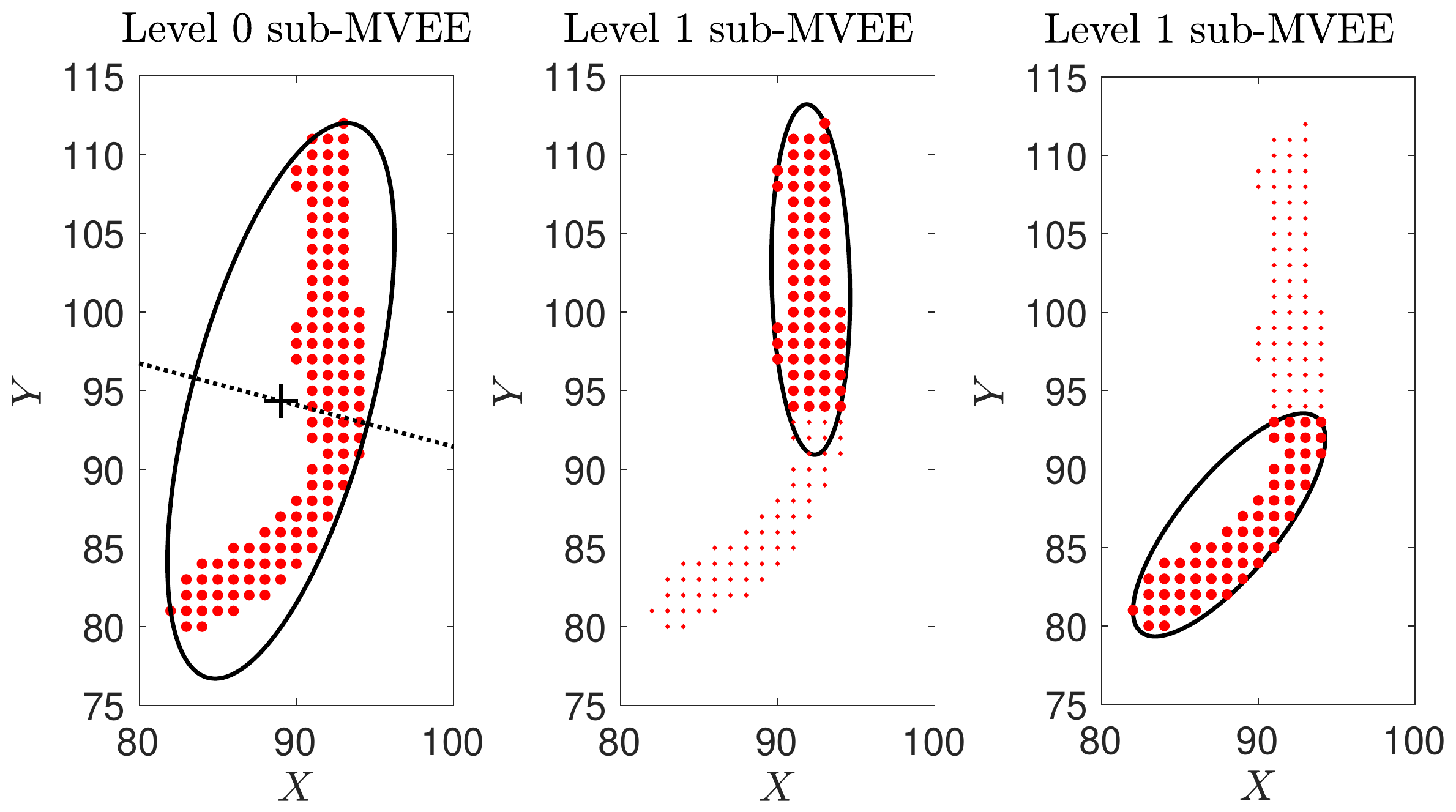}}
  \caption{Calculation of sub-MVEEs and the MVEE tree. Left: one of the
  vortical structures shown in Fig. \ref{fig:2Ddemo}, which is enclosed by the level 0 sub-MVEE and is split
  into two groups along the dotted line. The red dots are the grid points in the structure.
  Middle: the level 1 sub-MVEE for the upper half. Right: the
  level 1 sub-MVEE for the lower half.
  $X$ and $Y$ are the labels of the grid points in the $x$ and $y$ directions, respectively. 
  }
\label{fig:2DdemoTree}
\end{figure}

MVEEs provide a good description of the enclosed structure when the latter is convex.
%Luckily,
%the structures as we have defined are usually convex (see, e.g., the discussion in \citet{Leungetal12}).
A more complete description is desirable, however, if the structure is
non-convex, such as the second structure from the top in the right panel of Fig. \ref{fig:2Ddemo}.
In this case, we propose to use what we call the MVEE trees.
The construction of MVEE trees is illustrated with Fig. \ref{fig:2DdemoTree}.
It is supposed that the MVEE for the structure has been found.
In the context of MVEE trees,
this MVEE is called the level 0 (L0)  MVEE.
The L0 MVEE allows one to split the
grid points in the structure into two groups, using the symmetric plane of the MVEE perpendicular to its major
axis (see the left panel of Fig. \ref{fig:2DdemoTree}).
The MVEEs for the two parts can then be found. These are called the L1 sub-MVEEs. The
splitting can be repeated for each structure in this level, and those in each new level. The
procedure thus
forms a tree-like hierarchy of MVEEs, hence the name MVEE trees.

The MVEE trees are used to analyze strongly non-convex structures.
For this purpose, the degree of non-convexity of a structure is defined as the ratio of the volume of the structure to
the volume of its convex hull \citep{BoydVandenberghe04}. The ratio is denoted by $R_V$. Smaller
$R_V$ indicates stronger non-convexity.
Only structures with small
$R_V$ are selected when MVEE trees are calculated.

Moment of inertia ellipsoids (MoIE) have also been used in the past to characterize non-local
structures. However, MoIEs do not capture the extremal points in a structure, hence tends to miss the
details of
highly non-convex structures. In comparison, MVEE and MVEE trees are better in this regard.

%The superscript $t$ is used to denote
%quantities from the target field.
%The directions of the three axes of an MVEE in a target field are denoted by $\e^t_\al$,
%$\e^t_\beta$ and $\e^t_\gamma$, respectively, while the lengths of the axes are denoted by $\ell^t_\al$,
%$\ell^t_\beta$ and $\ell^t_\gamma$. The center of the MVEE is denoted by $\x^t_c$.
%Same notations with superscript $o$ are used for the quantities
%in the reconstructed fields.

\section{Simulations and results} \label{sect:results}

\subsection{Parameters and the computation of the target flow fields}\label{sect:DNS}

The target fields $\v(\x,t)$ are obtained by numerically integrating the NSEs with a fully
dealiased
pseudo-spectral code in a $[0,2\pi]^3$ box with periodic boundary conditions. 
A second order explicit improved Euler method based on the trapezoidal rule is used in
time-stepping.
$\v(\x, t)$ is made statistically stationary by a constant forcing term
$\f$, where
\be
\f = (0, f_a \cos(k_f x_1),0)
\ee
with $f_a = 0.15$ and $k_f = 1$ (these and the following parameters are all given in code units). 
Therefore, the flow is a 3D Kolmogorov flow with a
sinusoidal mean velocity profile \citep{BorueOrszag96, KangMeneveau05}. The Kolmogorov flow has been
chosen because the forcing term for this flow is particularly simple. 

  In all cases $128^3$ grid
  points have been used. The viscosity is $\nu =
  0.006$. Time step is $\delta t = 0.00575$.
  The time and length scales of $\v$ are estimated from $f_a$, $k_f$, 
  $\nu$, and the root-mean-square (RMS) velocity $v_{\textrm{rms}}$ which is found numerically to be
  about $0.65$.
  The large eddy
  turnover time scale is thus $\tau_L \equiv v_{\textrm{rms}}/f_a
  \approx 4.35$. The energy dissipation rate $\ep$ is
  calculated numerically from the energy spectrum and $\nu$, which gives $\ep\approx 0.08$. The
  Kolmogorov time scale is $\tau_K \equiv (\nu /
  \ep)^{1/2} \approx 0.28$. The Kolmogorov length scale is $\eta_K \equiv
  (\nu^3/\ep)^{1/4} \approx 0.04$. Therefore, $\delta x/\eta_K \approx
  1.25$ where $\delta x$ is the grid size of the simulations. The
  Reynolds number based on the Taylor micro-scale is $Re_\lambda \approx 75$, whereas the Reynolds
  number based on the forcing length scale is $Re_f \approx 340$. 

\subsection{The solution of the optimality system \label{sect:solopt}}

  The optimality system (Eq.
  (\ref{eq:NSE}) and Eqs. (\ref{eq:xi}-\ref{eq:dj})) is solved
iteratively. 
Given an initial guess for $\bvp$, the NSEs (Eq.
(\ref{eq:NSE})) are numerically integrated from $t=0$ to $t=T$ to find
$\u(\x,t)$. The cost $J$ is calculated from $\u$. Unless $J$ is
already smaller than a given tolerance, $\bvp$ has
to be updated to reduce $J$. To do so,
Eq. (\ref{eq:xi}) is integrated
backward in time from $t=T$ to $t=0$, starting from the end-point condition given by Eq.
(\ref{eq:xi_ic}). The time sequence of the forward solution $\u(\x,t)$ is used in the backward
integration.
The solution for $\bxi$ provides
$DJ/D\bvp$ according to Eq. (\ref{eq:dj}), which is then used to
update $\bvp$.
The updating step uses the nonlinear conjugate gradient method using the Polak-Ribi\`{e}re
formula \citep{NocedalWright99}.
The Brent's method in the Numerical Recipe \citep{Pressetal92} is used in the line search step.

The above several steps constitute the main loop of the solution algorithm.
Due to the chaotic nature of the solution of the NSEs, our
experience is that the iterations
may become difficult to converge when $T$
is large.
We thus solve the optimality system first with a small $T$, using the
iterations outlined above. The optimal $\bvp$ for this problem is used as the initial
guess for 
the problem with a larger $T$. The procedure is repeated until
the intended optimization horizon is reached. As the computational
time is usually very long on the computer we used (see below for more information),
we save the optimal solutions for the intermediate $T$ values on the hard
disk, which safeguards against unexpected interruptions of the
computation. Therefore, this implementation is also useful from a practical
point of view. The choices for the initial small $T$ and the increment
in $T$ are mostly based on trial and error. An initial $T =
0.25\tau_L \approx 4\tau_K $ with an increment equal $0.025\tau_L\approx 0.4\tau_K$ has been used in most cases, 
but we expect other similar choices will also work.

We now explain briefly the solution of the individual equations in the
main loop.
$\u(\x,t)$ is solved from
  the NSEs using the same numerical methods and parameters as
  those for $\v(\x,t)$.
The initial guess for $\u(\x,0) \equiv \bvp$ is usually
a divergence free Gaussian random velocity field, with
the Fourier modes with $k\le k_m$ replaced by those in the target
data $\bvp^\rbv(\x)\equiv
\v(\x,0)$ (c.f., Section \ref{sect:optsys}).
Our tests show that the initial guess has no effects on the
statistics of optimal solutions. 
It is possible to use an initial guess with the
energy spectrum matching the initial target field. However,
our tests showed that it did not improve the
results nor speed up convergence. A possible reason is that
matching the energy spectrum imposes only a weak constraint
on the control $\bvp$ since the latter has a very large number
($3\times 128^3$) of components.

As will be explained below, we run several related cases
with different parameters. Sometimes the optimal solution $\bvp^\u$ for a solved case
is then used as the initial guess for $\bvp$ in a related case.
For example, the optimal solution obtained with a larger tolerance
has been used as the initial guess for a case with a smaller tolerance; the one with a smaller $T$
used as the initial guess for a case with a larger $T$.

The adjoint equations (Eq. (\ref{eq:xi})) are solved with the same numerical schemes used for $\v$
and $\u$. The divergence free condition for $\bxi$ is used to eliminate the adjoint pressure
$\sigma$, in the same way as $p$ is eliminated from the NSEs when they are solved numerically. However, 
Eq. (\ref{eq:xi}) has to be integrated backwards in time, as it is supplemented
with the end-point condition in Eq. (\ref{eq:xi_ic}).
The time sequence of
$\u(\x, t)$ ($t\in [0,T]$) is needed for the solution of $\bxi$.
Therefore, $\u(\x,t)$ has to be
saved in the forward integration of the NSEs.
In our computation, typically the available RAM
is not large enough to hold the whole time sequence.
We thus use checkpointing,
where $\u(\x,t)$ is saved on the hard disk at selected $t$'s (the checkpoints) in the
forward integration. The sequence of $\u(\x,t)$ between two checkpoints, 
which will be short enough to store in the RAM, are computed
(using the data on a checkpoint as the initial condition) when the backward
integration of $\bxi$ reaches the time interval between the two checkpoints. 
Checkpointing thus incurs several passes of forward
integrations in one single backward integration. But it avoids having to read data at
every time step from the hard disk, which typically is a time
consuming operation.
The technique can be optimized by choosing the checkpoints
judiciously although we do not attempt to do so in this investigation. More details
about checkpointing can be found in, e.g., \citet{Wangetal09}.

  %TODO: update
\subsection{The choices of computational parameters \label{sect:param}}

%\begin{table}
%  \centering
%  \setlength{\tabcolsep}{12pt}
%  \begin{tabular}{ccccc}
%    \multirow{2}{*}{Realizations}&  \multicolumn{4}{c}{$(T,k_m)$}  \\
%    \cmidrule{2-5}
%     & $(0.5\tau_L, 2)$ & $(0.5\tau_L,4)$ & $(0.5\tau_L,8)$ & $(\tau_L,4)$ \\
%    \hline
%     R1 & -         & 2         & -         & {\bf 2}\\
%     R2 & -         & 2         & -         & {\bf 2}\\
%     R3 & -         & 2         & -         & {\bf 2}\\
%     R4 & 2, 1, 0.5 & 2, 1, 0.5 & 2, 1, 0.5 & {\bf 2}, 1.5 \\
%     R5 & 2, 1, 0.5 & 2, 1, 0.5 & 2, 1, 0.5 & {\bf 2}, 1.5 \\
%     \hline
%\end{tabular}
%\caption{\label{tab:cases0} The parameters for the cases being
%  investigated. The data entries are $10^4e_{\rm
%  tot}$ where $e_{\rm tot}$ is the tolerance used for each set of parameters. A
%  short dash indicates that the case is not computed. The numbers in
%  bold-face correspond to the baseline cases.}
%\end{table}

\begin{table}
  \centering
  \setlength{\tabcolsep}{6pt}
  \begin{tabular}{lcccc@{\hskip 1cm}lcccc}
    Case & $T/\tau_L$ &$k_m$ &  $10^4e_{\rm tot}$ & $N_R$ &
    Case & $T/\tau_L$ &$k_m$ &  $10^4e_{\rm tot}$ & $N_R$ \\
    \hline
    {\bf C1} & {\bf 1} & {\bf 4} & {\bf 2}   & {\bf 5} & C7  & 0.5& 4& 1 & 2 \\
    C2 & 1 & 4 & 1.5 & 2 & C8  & 0.5& 4& 0.5 & 2 \\
    C3 & 0.5& 2& 2   & 2 & C9  & 0.5& 8& 2 & 2 \\
    C4 & 0.5& 2& 1   & 2 & C10 & 0.5& 8& 1 & 2 \\
    C5 & 0.5& 2& 0.5 & 2 & C11 & 0.5& 8& 0.5 & 2 \\
    C6 & 0.5& 4& 2   & 5 &    &    &  &     & \\
     \hline
\end{tabular}
\caption{\label{tab:cases} The parameters for the cases being
  investigated. $N_R$ is the number of realizations. Case C1 is the
  baseline case with the parameters highlighted in bold.}
\end{table}

We assume that the data $\FF \v$ is available continuously, so that the forcing term in Eq.
(\ref{eq:adjf}) is applied at each time step. 
  We assume the optimal solution is found
  when $J/J(\u = 0)$ is less than a
  tolerance $e_{\rm tot}$, where, according to Eq. (\ref{eq:cost}),
  $J(\u = 0)$ is the total kinetic energy of the
  measurement data $\FF \v (\x, t)$ ($0\le t \le T$).
  Due to our limited computational resources, this investigation has
  used only $128^3$ grid points. However, several values of $T$, $k_m$
  and $e_{\rm tot}$ are examined, and up to five realizations are run for each
  set of parameters. The target data $\v(\x,t)$ in different
  realizations are different segments of the same time series of
  DNS data.

  The parameters are summarized in
  Table \ref{tab:cases}.
  We consider $T$ up to one large eddy
  turn-over time scale $\tau_L$ and resolutions $k_m = 2$, $4$ and $8$. However,
  for $T=\tau_L $, only cases with $k_m = 4$
  are reported for the following reasons. Firstly, for $(T, k_m) = (\tau_L,2)$, we failed to
  find convergent results (see the discussion of the results below for further comments).
  Secondly, for $(T, k_m) = (\tau_L,8)$, it turns out that the optimal solutions
  are the same as those with 
  $(T,k_m) = (0.5\tau_L,8)$ with $e_{\rm tot} = 0.5\times 10^{-4}$ (i.e., the cases with a shorter
  $T$ but more stringent tolerances).
  Therefore,
  the results for these two $k_m$'s have been omitted. 

For $(T,k_m) = (\tau_L,4)$,
optimal solutions are found for $e_{\rm tot}  = 2
\times 10^{-4}$ in five realizations (case C1). In each of these
realizations, it took around 15
days to find the
optimal solution, using up to 48G RAM
and 4 cores on an Intel Ivybridge or Westmere based CPU, with the optimal solutions from
Case 6 as the initial guesses. 
In two of the realizations, the optimal solutions are further
improved by additional iterations to reduce $J/J(\u=0)$ below
$1.5\times 10^{-4}$ (case C2). The improvement required about 15 more
days of computer time. Given the significant computational cost,
we did not attempt to reduce the cost function further.
On the other hand, for $T = 0.5\tau_L$, optimal solutions were
obtained
for $e_{\rm tot} = 10^{-4}$ and $0.5\times 10^{-4}$ in two realizations for all three $k_m$ values.
The cost decreases rapidly initially, but further improvement becomes very slow. Fig.
\ref{fig:jcst} plot the change of $J/J(\u=0)$ in the iterations for a run in case C6, which
depicts this behavior. The inset zooms into the curve to show the change at later iterations. 
Abrupt jumps in $J$ are observed when $T$ is extended. In this run, the optimal solution for $T=0.5\tau_L$ is
found after 124 iterations. In each iteration, around 5 forward integrations are performed in order
to find the conjugate gradient direction. Taking backward integration into account, $700\sim 800$
passes of integration are used, although we note that part of these integrations are over shorter optimization
horizons. 
\bfig
  \centerline{\includegraphics[width=0.5\lnw]{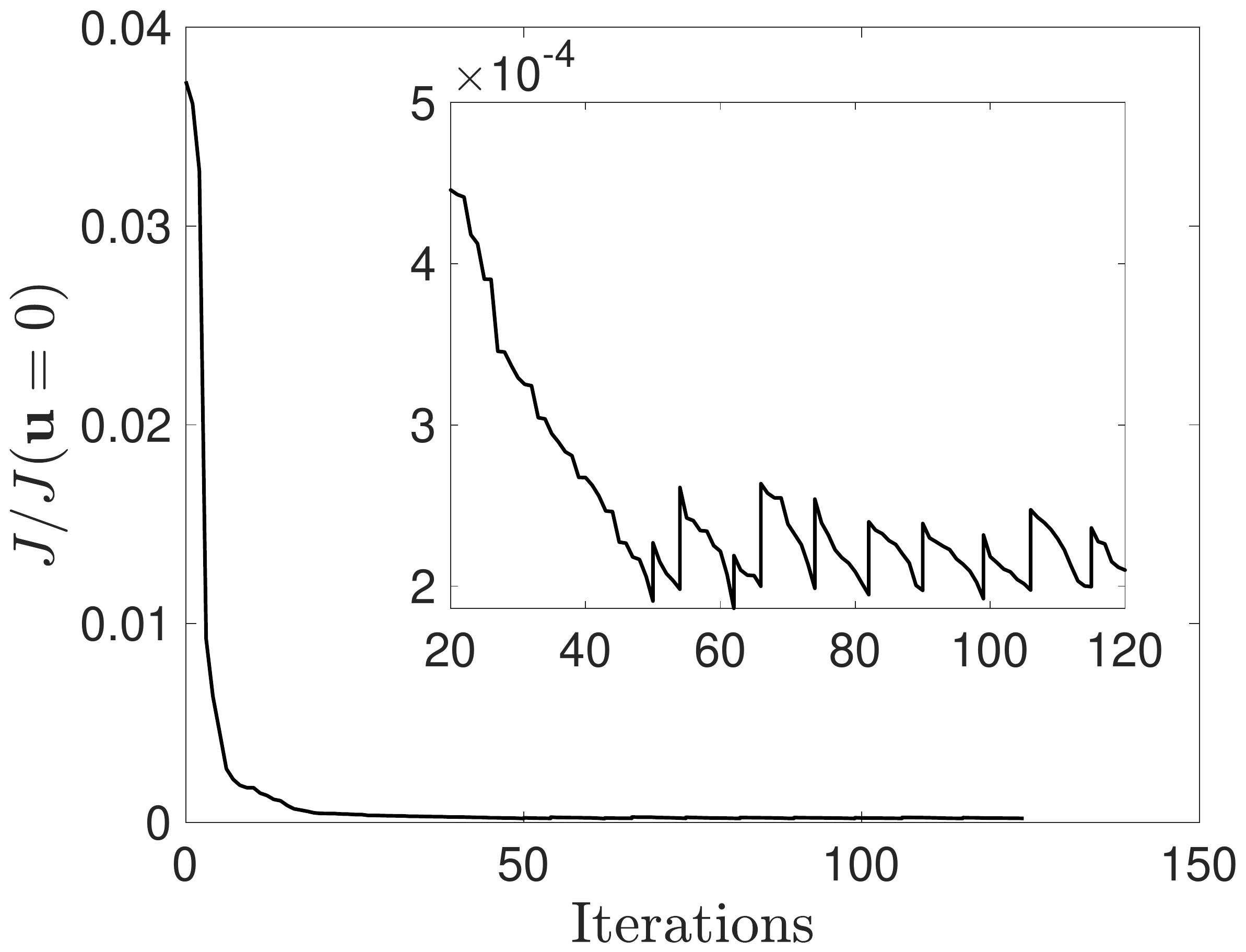}}
  \caption{$J/J(\u=0)$ as a function of iterations in one of the runs. 
  Inset: a zoom-in between the 20th and the 120th
  iterations. The jumps occur at the points where the iterations have converged and $T$ is extended.}
\label{fig:jcst}
\efig

We consider Case C1 (shown in bold in Table \ref{tab:cases}) the
baseline case.
The comparison between this and other cases will illustrate the effects of the resolution of the
measurement data $k_m$, the optimization horizon $T$
and the tolerance $e_{\rm tot}$.
However, the
majority of results is presented only for the baseline case.
The statistics are averaged over all the
realizations in each case.

%TODO: Number of forward-backward iterations? Changes in the error
%over iterations?

\subsection{The energy spectra and spectral correlation for different
computational parameters \label{sect:results_ek}}

\bfig
  \bmn{0.5\lnw}
  \centerline{\includegraphics[width=\lnw]{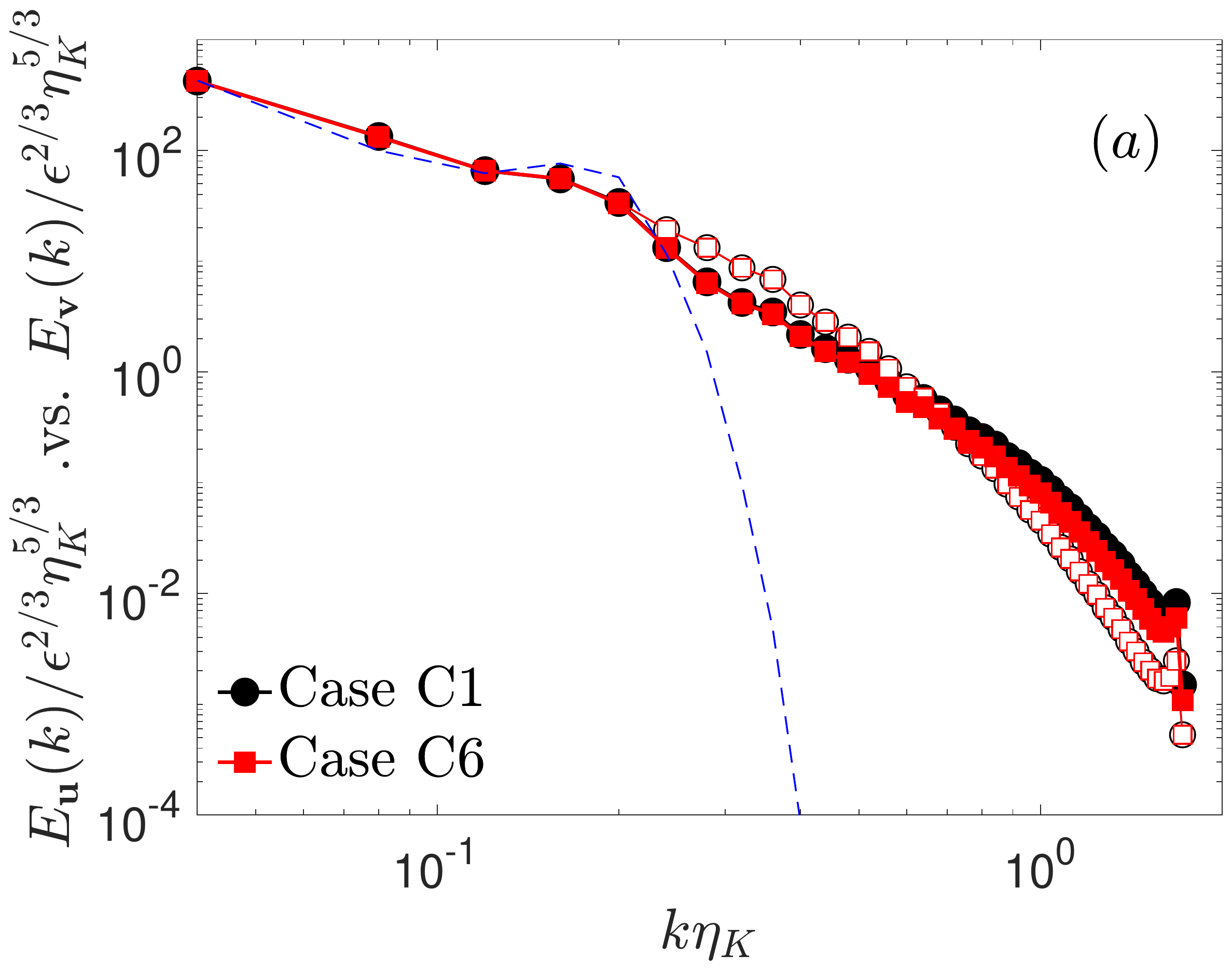}}
  \emn %
  \bmn{0.5\lnw}
  \centerline{\includegraphics[width=\lnw]{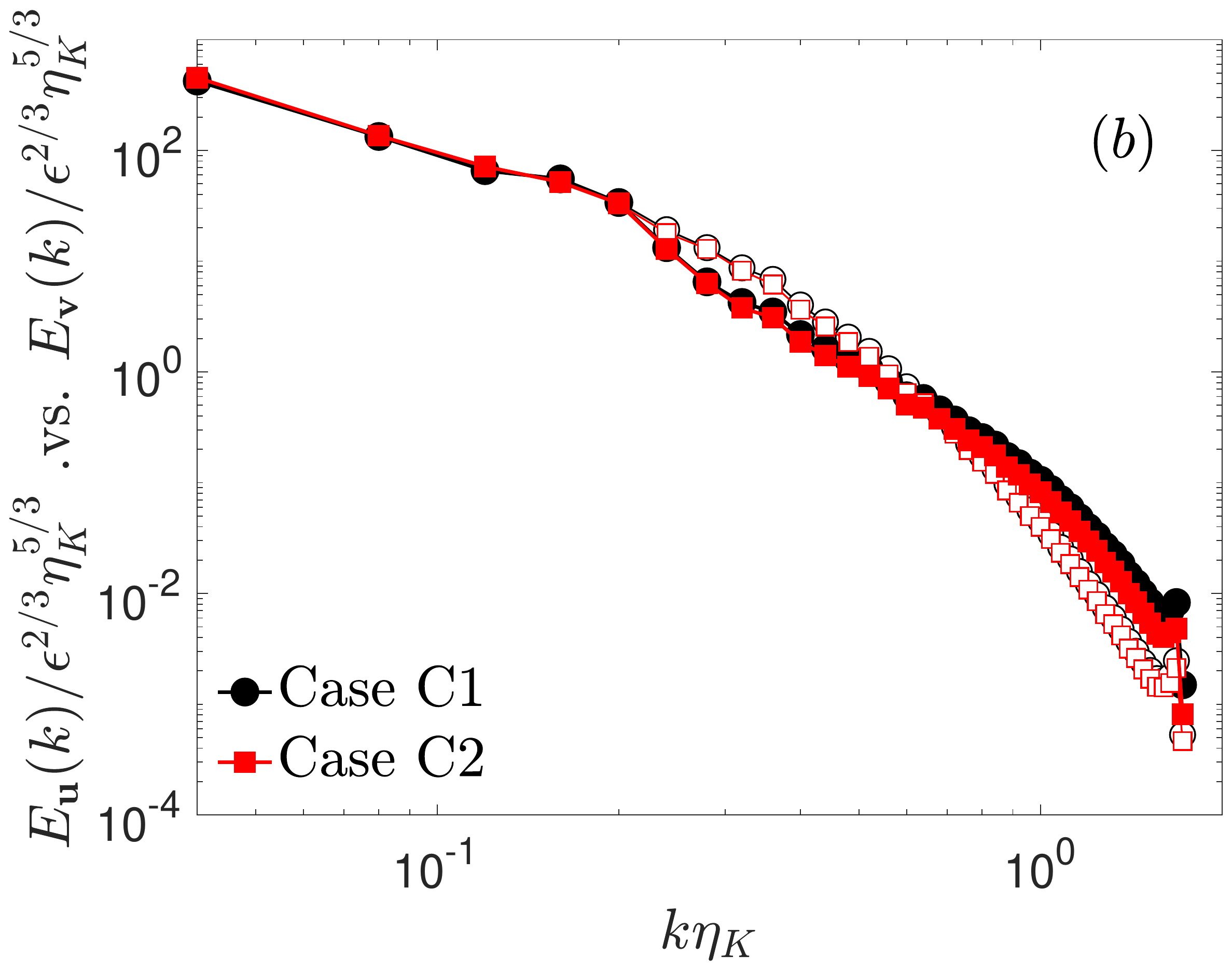}}
  \emn

  \bmn{0.5\lnw}
  \centerline{\includegraphics[width=\lnw]{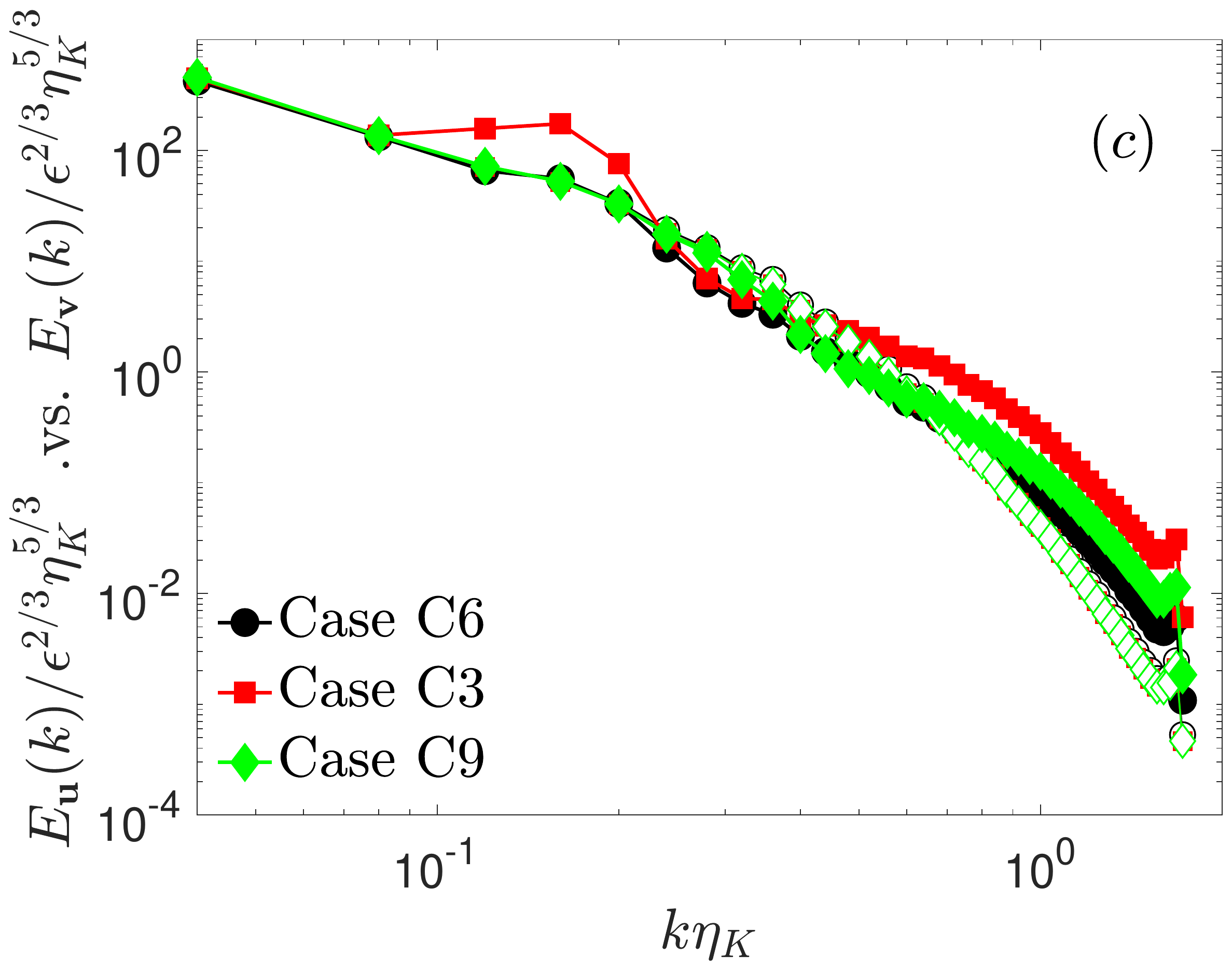}}
  \emn %
  \bmn{0.5\lnw}
  \centerline{\includegraphics[width=\lnw]{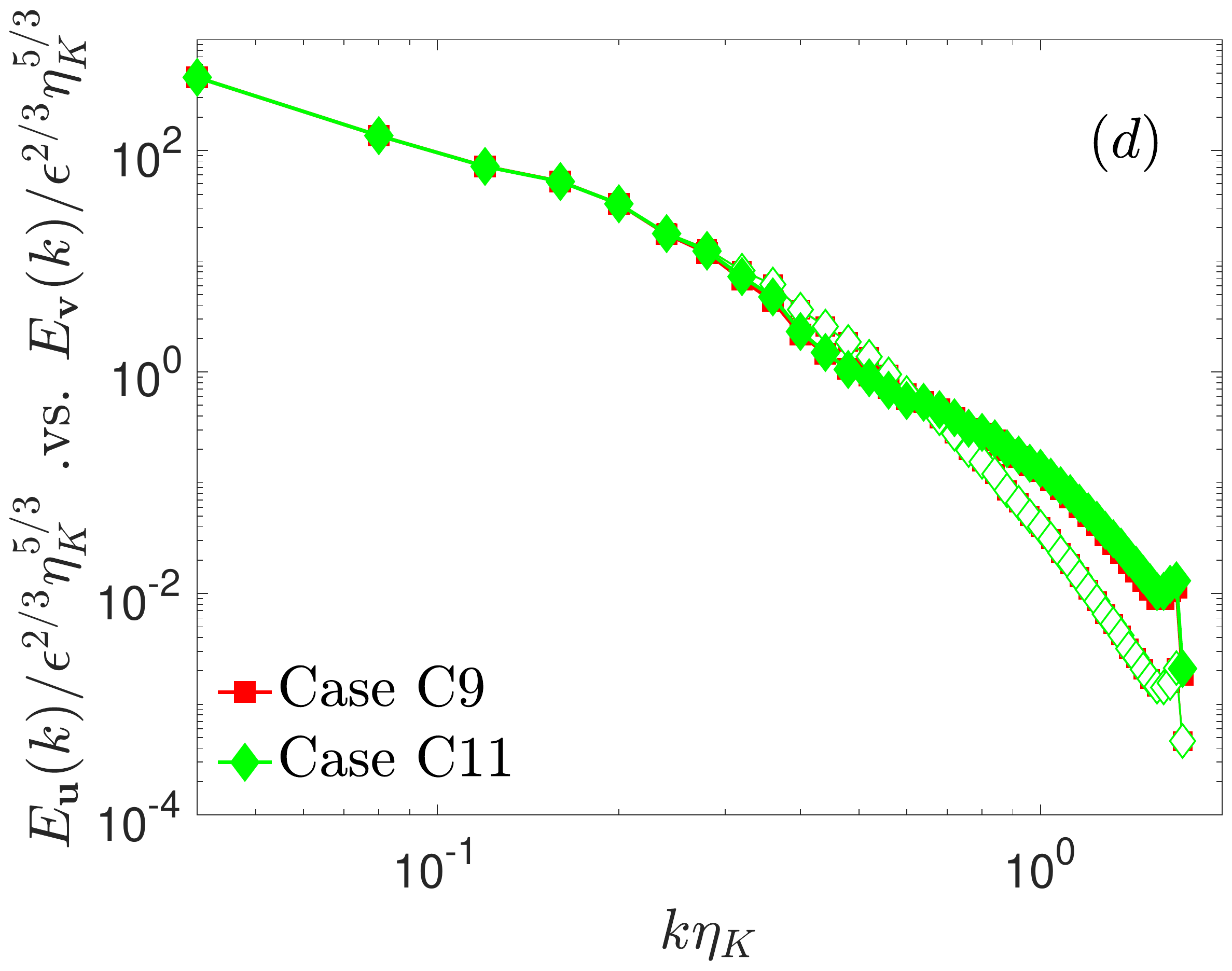}}
  \emn
  \caption{The energy spectra for $\bvp^\rbu(\x)$ (solid symbols)
  and $\bvp^\rbv(\x)$ (empty symbols of same
  shapes). (a) with different optimization horizon $T$ (the dashed line shows the energy spectrum of
  the initial guess for $\bvp^\u(\x)$ in one of the realizations); (b) with
  different tolerance $e_{\rm tot}$; (c) with different measurement resolution $k_m$;
  (d) with different tolerance for $k_m = 8$. }
\label{fig:spec-ini}
\efig

The spectral distributions of $\u$ and $\v$ obtained with different
computational parameters are compared in this
subsection. The purpose is to give an overview of the quality of the
reconstruction. More detailed investigation, focusing on the baseline
case, will be presented later. Symbols $\u$ and
$\v$ in the subscripts or superscripts are used to denote results from
$\u$ (the reconstruction) and $\v$ (the target), respectively.

Fig. \ref{fig:spec-ini} shows the energy spectra for $\bvp^\u$
and $\bvp^\v$, which are plotted with solid and empty symbols,
respectively.  The figure shows that
perfect match between
$\bvp^\u$ and $\bvp^\v$ is obtained for $k\le k_m$,
as expected.
Fig. \ref{fig:spec-ini}(a) and (b) show that, for $k_m = 4$,
$\bvp^\u$ under-predicts the spectrum for intermediate
$k$'s and over-predicts it for large $k$'s. Nevertheless, the spectra
of
$\bvp^\u$ broadly
follow those of $\bvp^\v$. The agreement does not change much with $T$
and $e_{\rm tot}$ for the ranges of the parameters we are
considering.
Note that it is non-trivial to observe the fairly good agreement.
The dashed line in Fig. \ref{fig:spec-ini}(a) plots the energy spectrum for the initial guess for
$\bvp^\u(\x)$, 
which shows that 
the small scales in $\bvp^\u$
are not present in the initial guess and are
generated by the data assimilation process.

The resolution of the measurement data $k_m$ has significant effects on $E_\u(k)$, as is
demonstrated in Fig. \ref{fig:spec-ini}(c). $E_\u(k)$ for $k_m = 2$
essentially fails to capture the features of $E_\v(k)$. On the other
hand, the spectrum for $k_m = 8$ shows similar level of
discrepancy as for
$k_m = 4$ at the high wavenumber end, which is somewhat surprising. Reducing the tolerance to
$e_{\rm tot} = 0.5\times 10^{-4}$ does not improve the agreement, as
shown in Fig. \ref{fig:spec-ini}(d).

\bfig
  \bmn{0.5\lnw}
  \centerline{\includegraphics[width=\lnw]{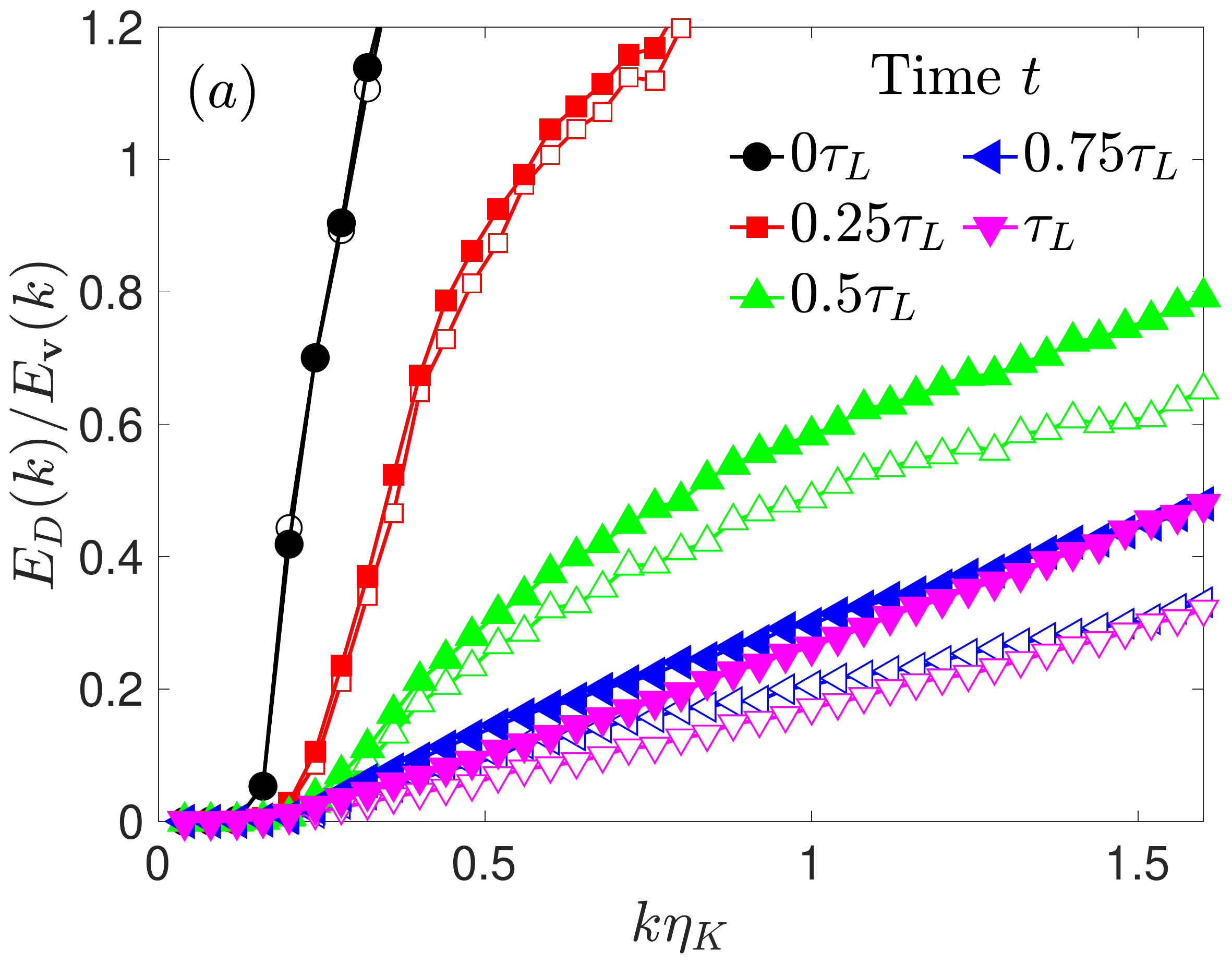}}
  \emn %
  \bmn{0.5\lnw}
  \centerline{\includegraphics[width=\lnw]{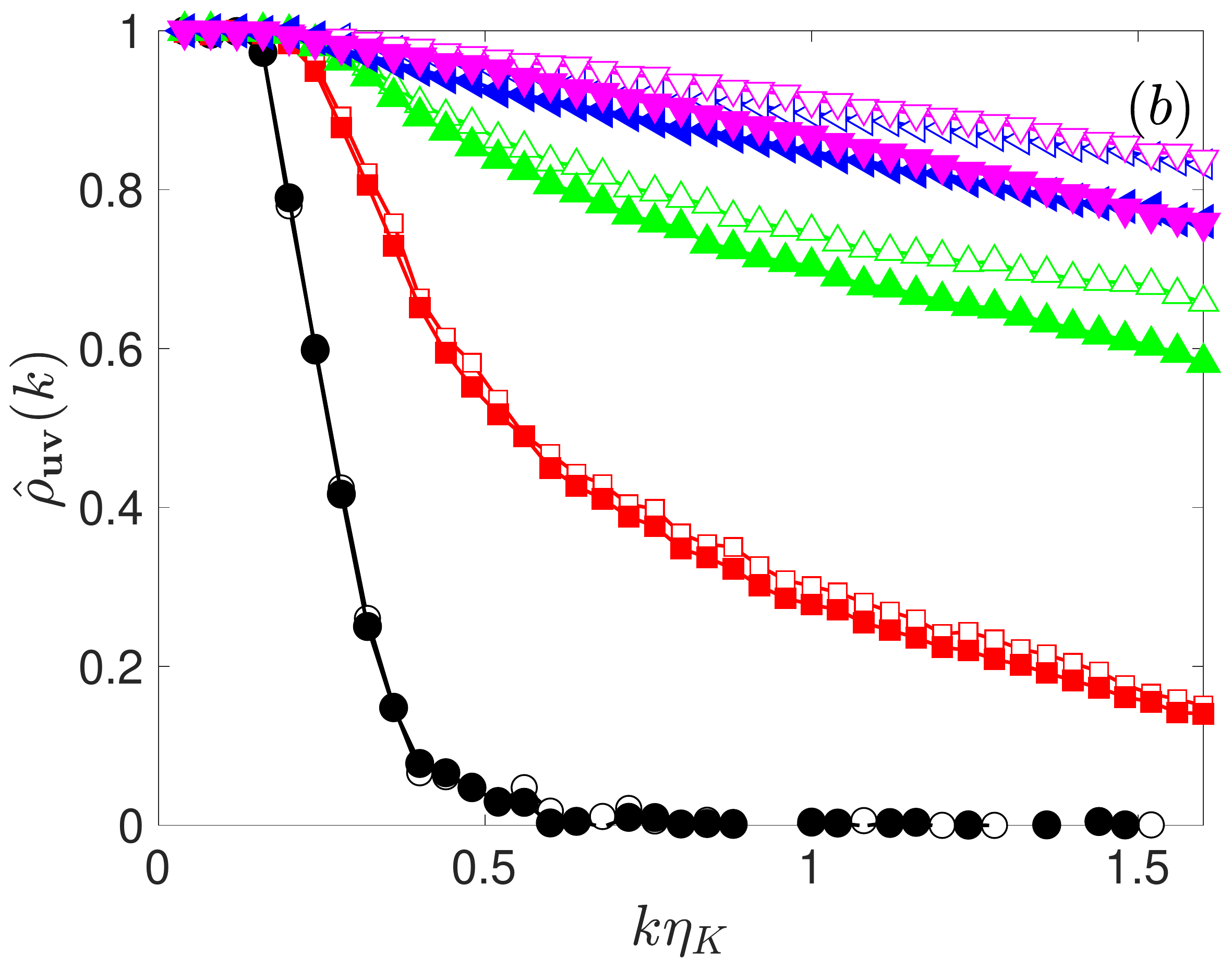}}
  \emn

  \bmn{0.5\lnw}
  \centerline{\includegraphics[width=\lnw]{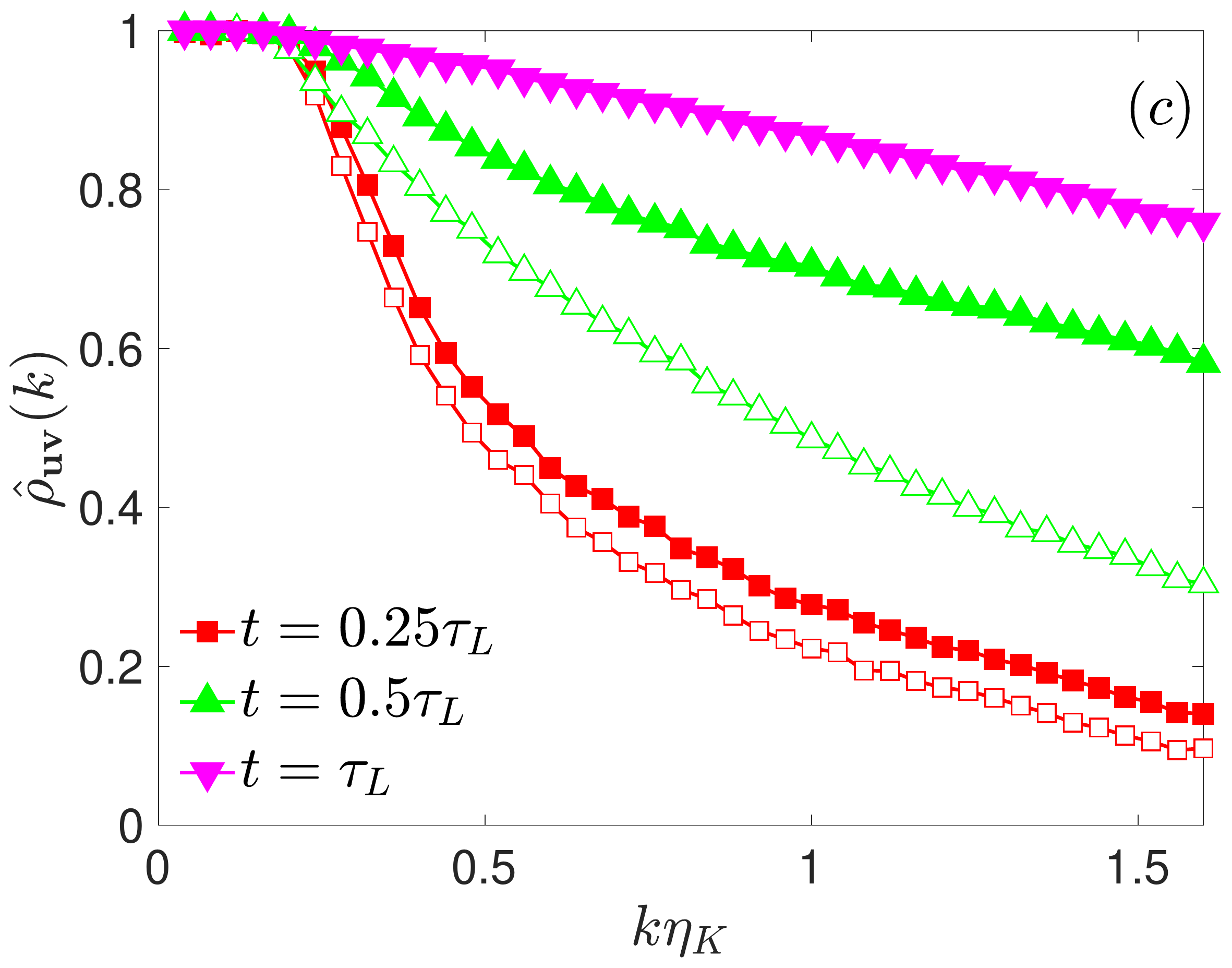}}
  \emn %
  \bmn{0.5\lnw}
  \centerline{\includegraphics[width=\lnw]{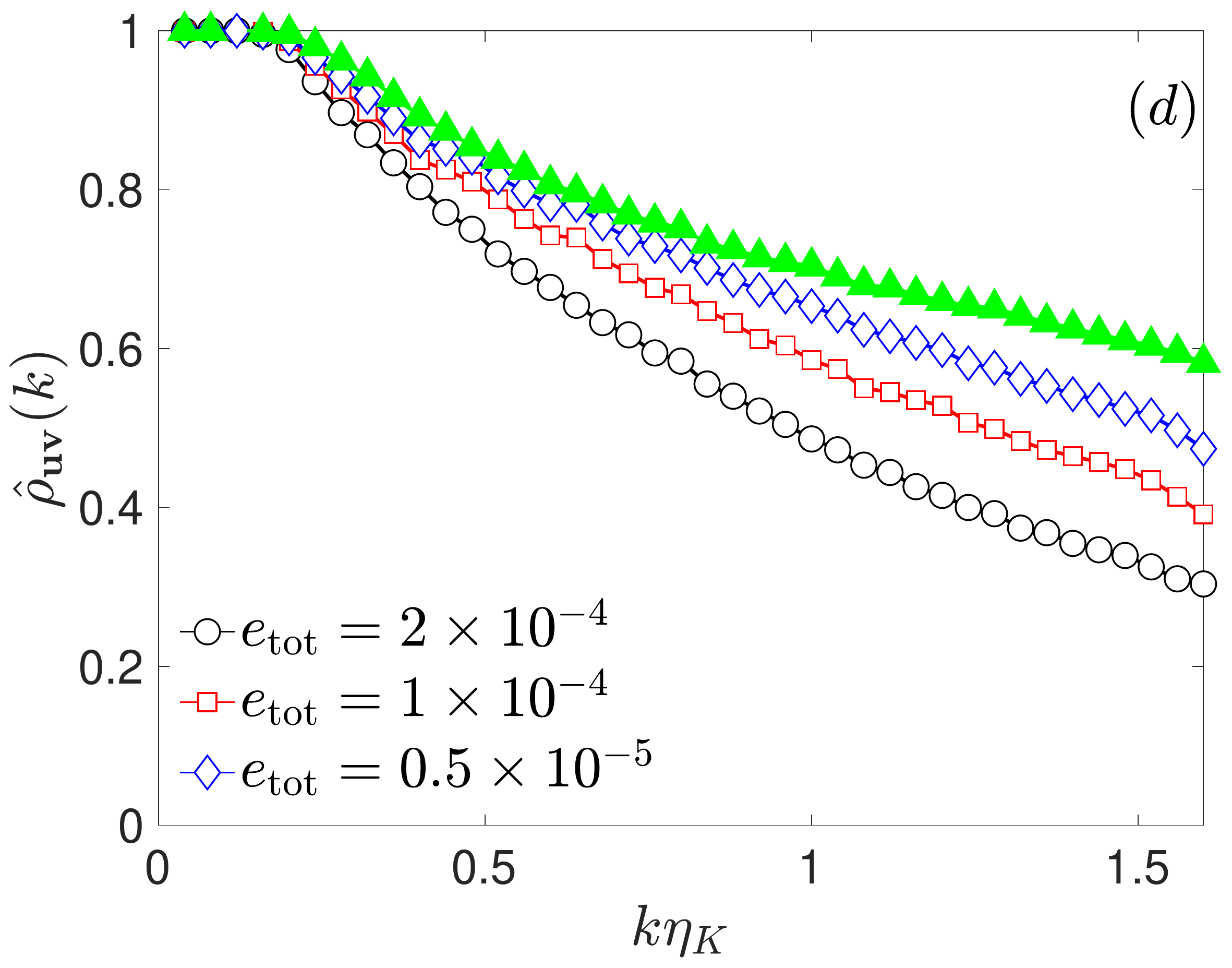}}
  \emn

  \bmn{0.5\lnw}
  \centerline{\includegraphics[width=\lnw]{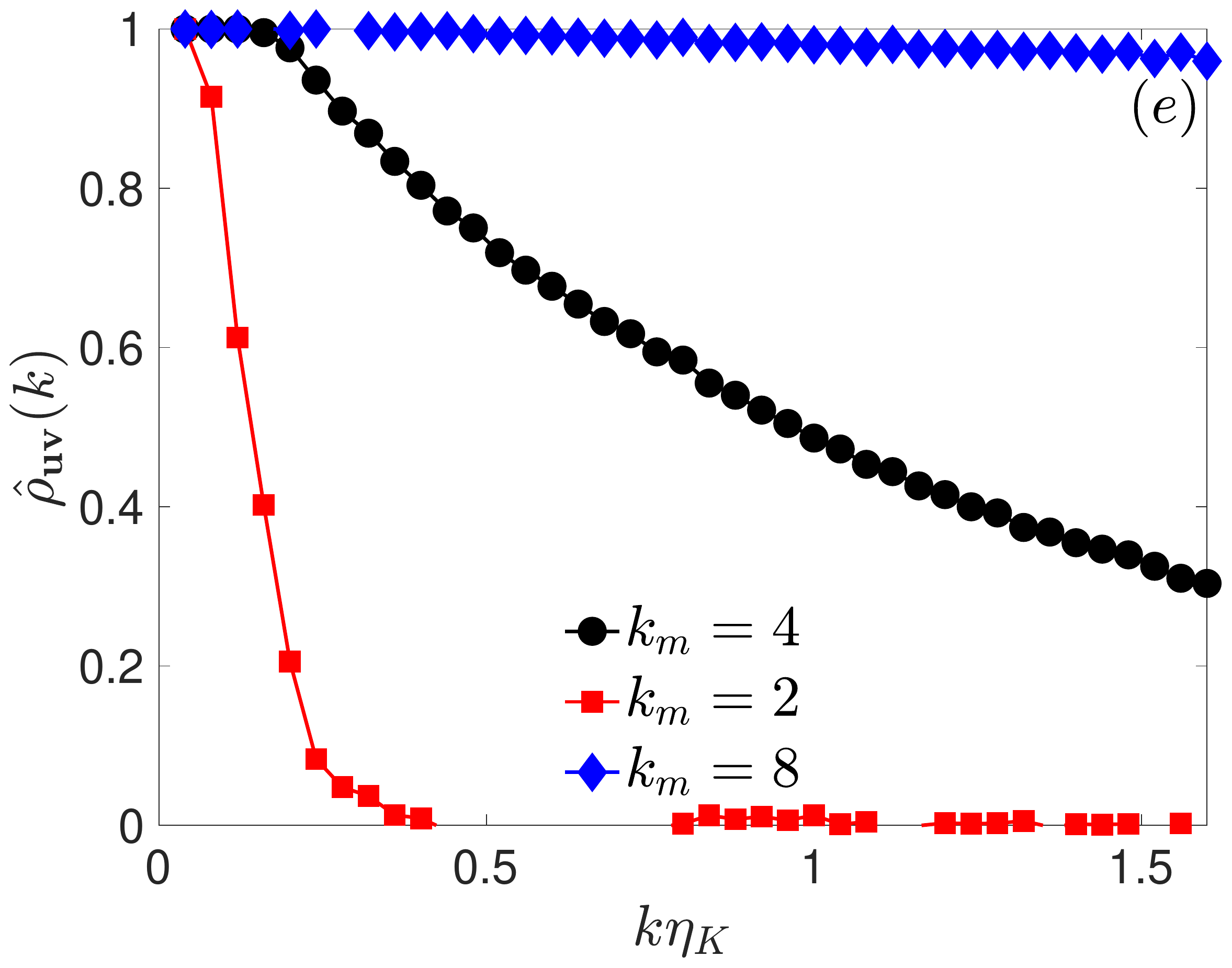}}
  \emn %
  \bmn{0.5\lnw}
  \centerline{\includegraphics[width=\lnw]{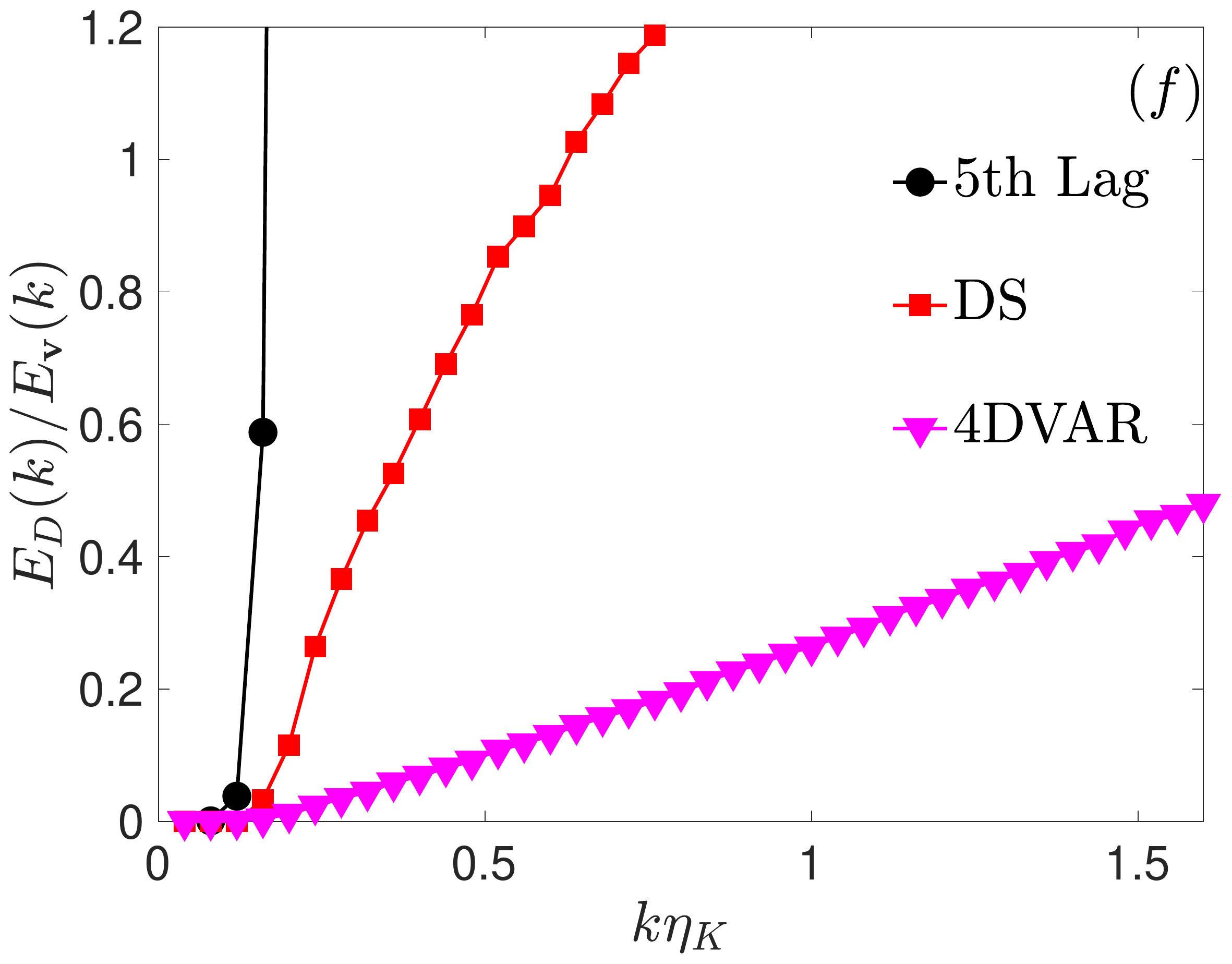}}
  \emn
  \caption{(a) and (b): $E_D(k)/E_\v(k)$ and $\hat{\rho}_{\u\v}(k)$ at
  different time $t$
  for case C1 (solid symbols) and case C2 (empty symbols), with $(T, k_m)=(\tau_L, 4)$ and $e_{\rm tot} =
  2\times 10^{-4}$ or $1.5\times 10^{-4}$. (c): $\hat{\rho}_{\u\v}(k)$
  for case C1 where $T=\tau_L$ (solid symbols) and case C6 where $T=0.5\tau_L$ (empty symbols). 
  (d): $\hat{\rho}_{\u\v}(k)$ for different $e_{\rm tot}$ (cases C6, C7, C8; empty symbols)
  at $t=0.5\tau_L$. Solid symbols: the result for case C1 as a comparison.
  (e): $\hat{\rho}_{\u \v}(k)$ at $t=0.5\tau_L$ for different $k_m$ (cases C3, C6, C9). 
  (f): $E_D(k)/E_\v(k)$ at $t=\tau_L$ obtained from 4DVAR (case C1; triangles), from DS
  (squares), and from 5th-order Lagrange interpolation (circles).}
\label{fig:spec-time}
\efig

Since both $\u(\x,t)$ and $\v(\x,t)$ are solutions of the forced NSEs, their
\emph{statistics} are expected to be the same for $t\to T$.
It is interesting, however, to investigate the instantaneous point-wise difference between
$\u(\x,t)$ and $\v(\x,t)$ at a later time $t$ for $t\in
[0,T]$.
Two quantities are used to quantify the difference.
The first one is the energy spectrum for the velocity difference $\da
\u(\x,t) = \u(\x,t) - \v(\x,t)$,
denoted by $E_D(k)$, and called spectral difference. 
Obviously, a smaller $E_D(k)$ indicates better match between
$\u(\x,t)$ and $\v(\x,t)$, hence a better reconstruction.
The second one is the spectral correlation (i.e., normalized co-spectrum) between $\u$ and $\v$,
defined by
\be
\hat{\rho}_{\u \v}(k) \equiv \frac{\oint\lal \hat{\u}^*(\k,t) \cdot
\hat{\v}(\k,t) \ral dS_k}{2\sqrt{E_\u(k) E_\v(k)}}.
\ee
The pointed brackets in the numerator denotes ensemble average whereas
the integral is a surface integral over the sphere $|\k| = k$. By definition, $0\le \hat{\rho}_{\u \v} \le 1$, and a larger
$\hat{\rho}_{\u \v}$ indicates better point-wise agreement between
$\u$ and $\v$.
Note that $E_D(k)$
and $\hat{\rho}_{\u \v}(k)$ both depend on time $t$, although the
dependence is not given explicitly in the expressions.

Fig. \ref{fig:spec-time} compares $E_D(k)$ and $\hat{\rho}_{\u \v}(k)$
from different cases in various ways.
Fig.
\ref{fig:spec-time}(a) and (b) show that, in Case C1 and C2, the point-wise agreement
between $\u$ and $\v$ improves over time within the optimization
horizon. $\hat{\rho}_{\u\v}$ reaches $0.8$ or higher at $t=\tau_L$ for all 
wavenumbers, 
whereas $E_D(k)/E_\v(k)$ can be reduced down to $30\%$ even for $k=40$. Note that the
measurement data contain approximately $8^3/2 = 256$
Fourier modes, constituting less than $0.1\%$ of the Fourier modes in
the whole flow field.

Fig.
\ref{fig:spec-time}(a) and (b) also show that
imposing a smaller tolerance $e_{\rm tot}$ can lead to significantly better
agreement at later times, even if it is only marginally smaller (i.e.,
$1.5\times 10^{-4}$ versus $2\times 10^{-4}$).
On the other hand, the effect
is quite limited at earlier times (e.g., at $t=0$), where significant
discrepancy between $\u$ and $\v$ is observed for both $e_{\rm tot}$. This observation is
interesting, suggesting that the improvements at later times are due to
effects that are not completely revealed by $E_D(k)$ and $\hat{\rho}_{\u
\v}$ at earlier times. %TODO: double check

The effects of optimization horizon $T$ are investigated in Fig.
\ref{fig:spec-time}(c), where $\hat{\rho}_{\u \v}$ for $T=0.5\tau_L$
and $T=\tau_L$
is shown with empty and filled symbols, respectively. The results for
$T=0.5\tau_L$ are shown for $t=0.25\tau_L$ (squares) and $0.5\tau_L$
(triangles). Those for $T=\tau_L$ are shown for the same times and  $t=\tau_L$ too. 
It is observed that larger $T$ improves the
point-wise agreement, and the improvement increases over time.
%For $T=0.5\tau_L$, one could argue that
%the correlation between $\u$ and $\v$ is not satisfactory for higher
%wavenumbers.
The agreement, again, can be improved by imposing a smaller tolerance.
Fig. \ref{fig:spec-time}(d) compares $\hat{\rho}_{\u\v}$ at $t=T$ 
for three $e_{\rm tot}$'s. For $e_{\rm tot} = 0.5\times
10^{-4}$, the correlation can be improved to about the same level
obtained with a longer horizon $T=\tau_L$. Nevertheless, in order to
obtain satisfactory agreement over the whole range of wavenumbers,
even smaller $e_{\rm tot}$ or $T > 0.5\tau_L$ is needed.

The parameter $k_m$ affects the results strongly. Fig.
\ref{fig:spec-time}(e) plots $\hat{\rho}_{\u\v}$ at $t= T$ for $T =
0.5\tau_L$ with $k_m = 2, 4, 8$. When $k_m = 2$, $\u$ shows little
correlation with $\v$. For $k_m = 8$, almost perfect correlation is obtained.
To understand these results, it is instructive to compare the 4DVAR method
with the DS method
\citep{Yoshidaetal05, Lalescuetal13} (c.f. Section \ref{sect:intro}), which is shown in Fig.
\ref{fig:spec-time}(f). 
In \citet{Yoshidaetal05}, the authors
integrate the NSEs with a Gaussian random field as the IC.
The low wavenumber modes with $k\le k_m$ in the solution $\u(\x,t)$
are replaced by those in the target field $\v(\x,t)$
at each time step.
They show that, if $k_m \ge k_c \equiv 0.2 \eta_K^{-1}$,
the small scales in $\u(\x,t)$ approach those in
$\v(\x,t)$ asymptotically as $t\to \infty$, assuming an infinite
sequence of data are available.
In our investigation, the threshold wavenumber is
$k_c = 5$ so $k_m = 4$ is slightly below $k_c$, and the measurement data are available
only for $t\in[0,T]$. As a consequence, the DS scheme will only achieve partial reconstruction.
The reconstruction at $t=\tau_L$ using the DS scheme with $k_m =4$
is compared with the 4DVAR method in Fig.
\ref{fig:spec-time}(f). It shows that the 4DVAR
method improves the reconstruction significantly. A more simplistic
reconstruction,
obtained with a 5th-order Lagrange interpolation, is also examined in
Fig. \ref{fig:spec-time}(f). The reconstruction has essentially no
correlation with the target field for $k> k_m$.

Fig. \ref{fig:spec-time}(e) and (f) together thus show that, for $k_m$
around or greater than $k_c$, 4DVAR can exploit a short time sequence of data to
obtain better (for $k_m = 4$) or almost perfect (for $k_m=8$) reconstruction. 
For $k_m$
much smaller than $k_c$ (e.g., for $k_m = 2$), 4DVAR will
fail. These results suggest that the threshold $k_c$ found with the DS method is also valid for 4DVAR.  
Nevertheless, a more
comprehensive investigation is needed to confirm this preliminary observation.
Simulations with other
$k_m$'s around $k_c$, as well as other Reynolds numbers, are needed.
We leave this investigation for the future.

\subsection{Statistics of the reconstructed fields at $t=0$
\label{sect:results_ini}}

%It is observed in Section \ref{sect:results_ini} that,
%initially $E_D(k)$ and $\hat{\rho}_{\u\v}(k)$ can be quite close in
%different cases, but
%become very different at later times.
%To understand this behavior, it is worth 
In this subsection, we examine
other dynamically
important statistics of the
reconstructed initial fields (i.e. $\bvp^\u(\x)$).
We consider the cases with $k_m = 4$ only and focus mostly on the baseline 
case C1 where $T=\tau_L$, $k_m = 4$ and $e_{\rm tot} = 2\times 10^{-4}$.
We conduct the analysis in the physical space, using the filtering
approach where necessary (c.f. Section \ref{sect:descript}). The Gaussian filter is used,
and the filter
scale $\Delta$ is usually taken as $\Delta_m/2$ or $\Delta_m/4$
where $\Delta_m \equiv \pi/k_m$ is the equivalent filter scale associated with $k_m$.

\begin{figure}
  \bmn{0.5\lnw}
  \centerline{\includegraphics[width=0.95\lnw]{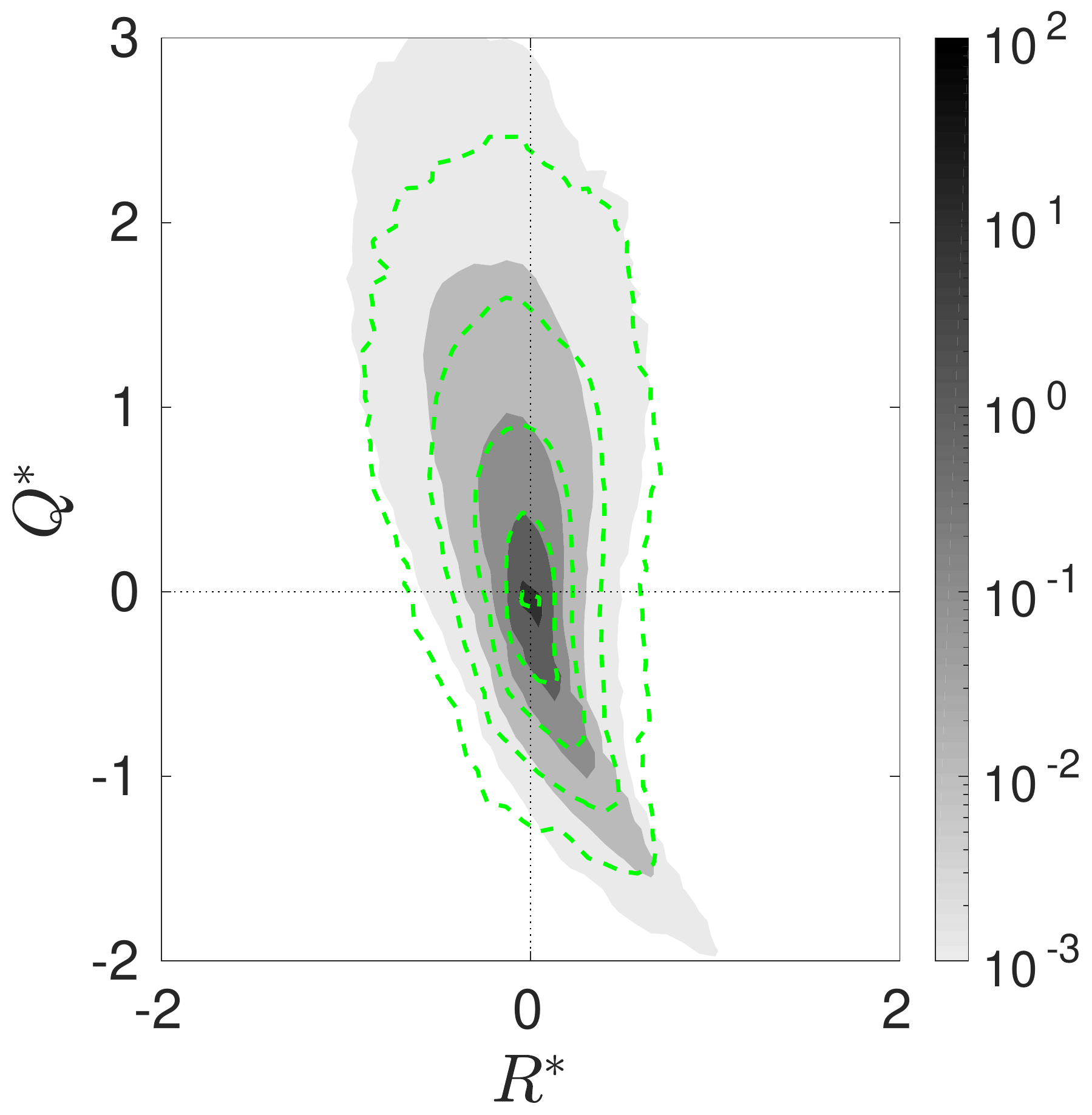}}% Images in 100% size
  \emn %
  \bmn{0.5\lnw}
  \centerline{\includegraphics[width=0.95\lnw]{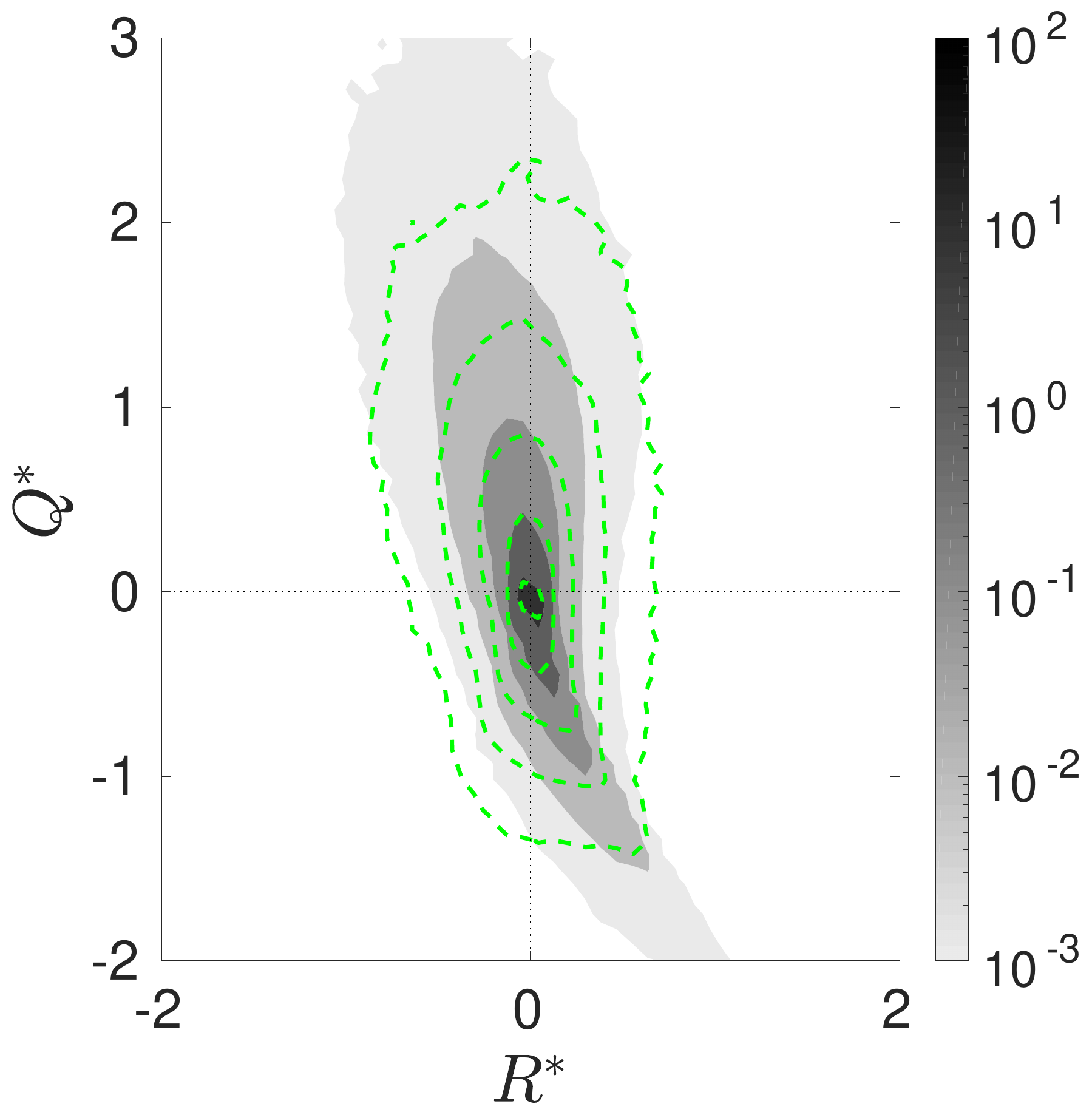}}% Images in 100% size
  \emn

  \bmn{0.5\lnw}
  \centerline{\includegraphics[width=0.95\lnw]{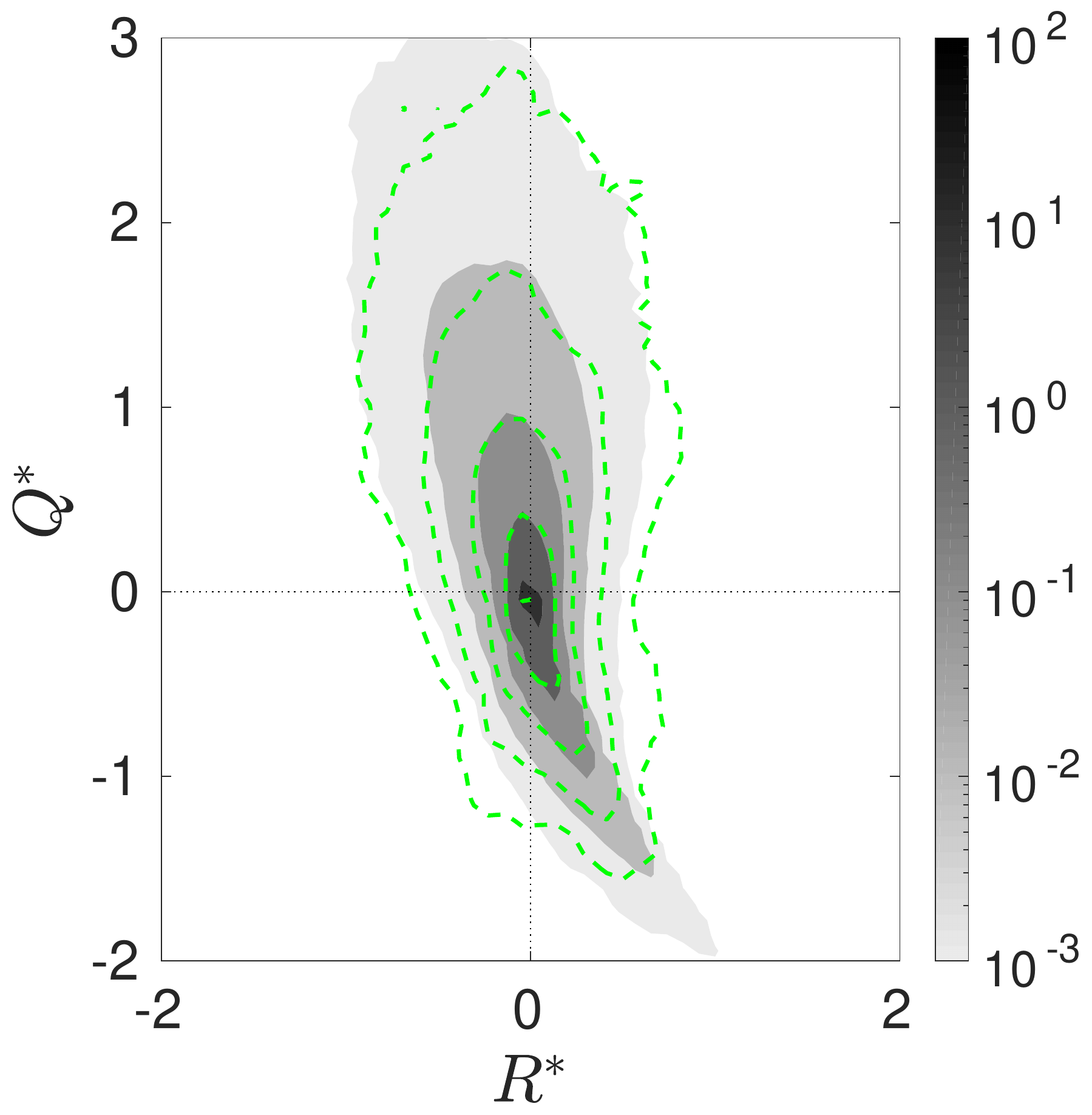}}% Images in 100% size
  \emn%
  \bmn{0.5\lnw}
  \centerline{\includegraphics[width=0.95\lnw]{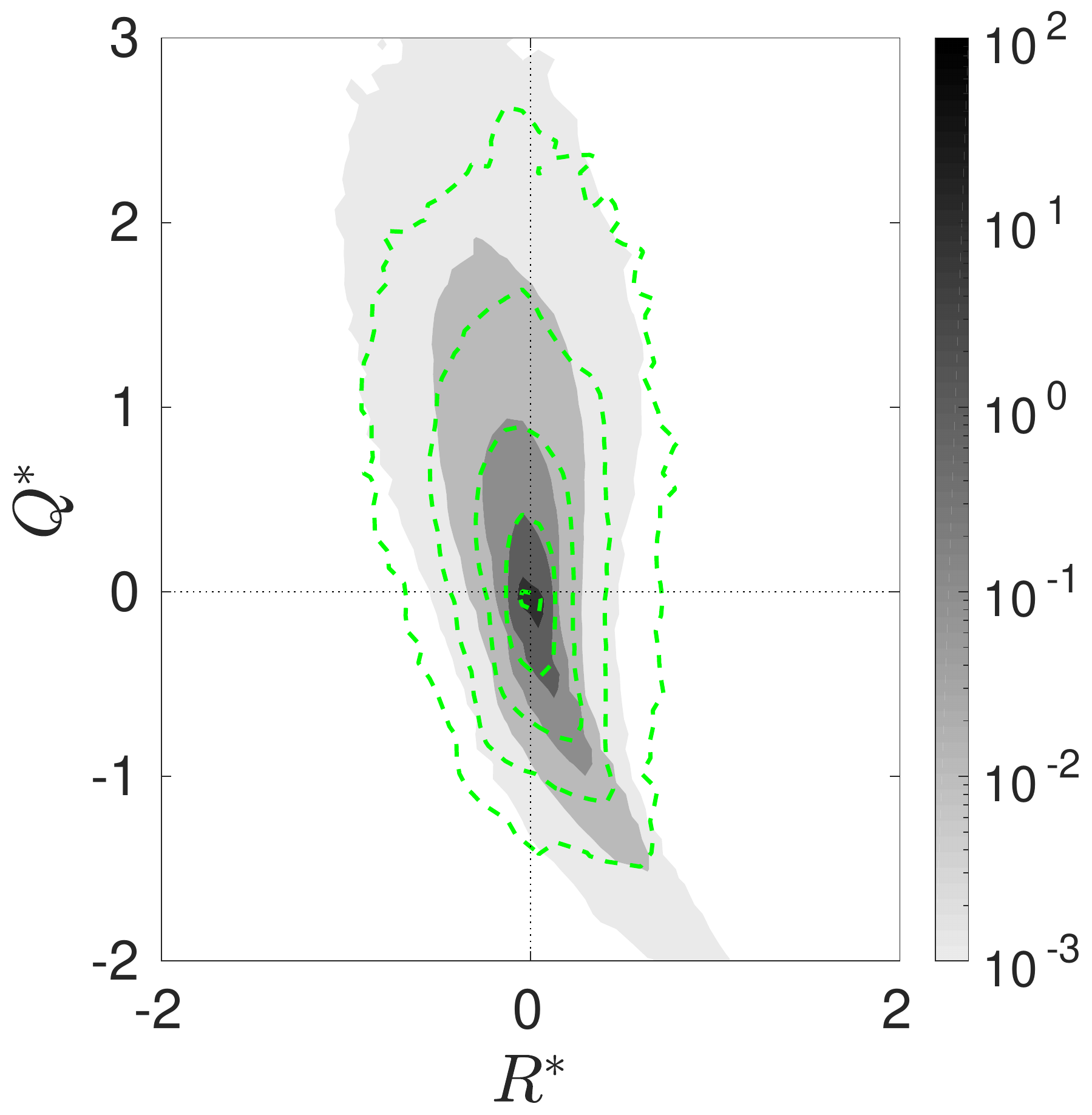}}% Images in 100% size
  \emn

  \bmn{0.5\lnw}
  \centerline{\includegraphics[width=0.95\lnw]{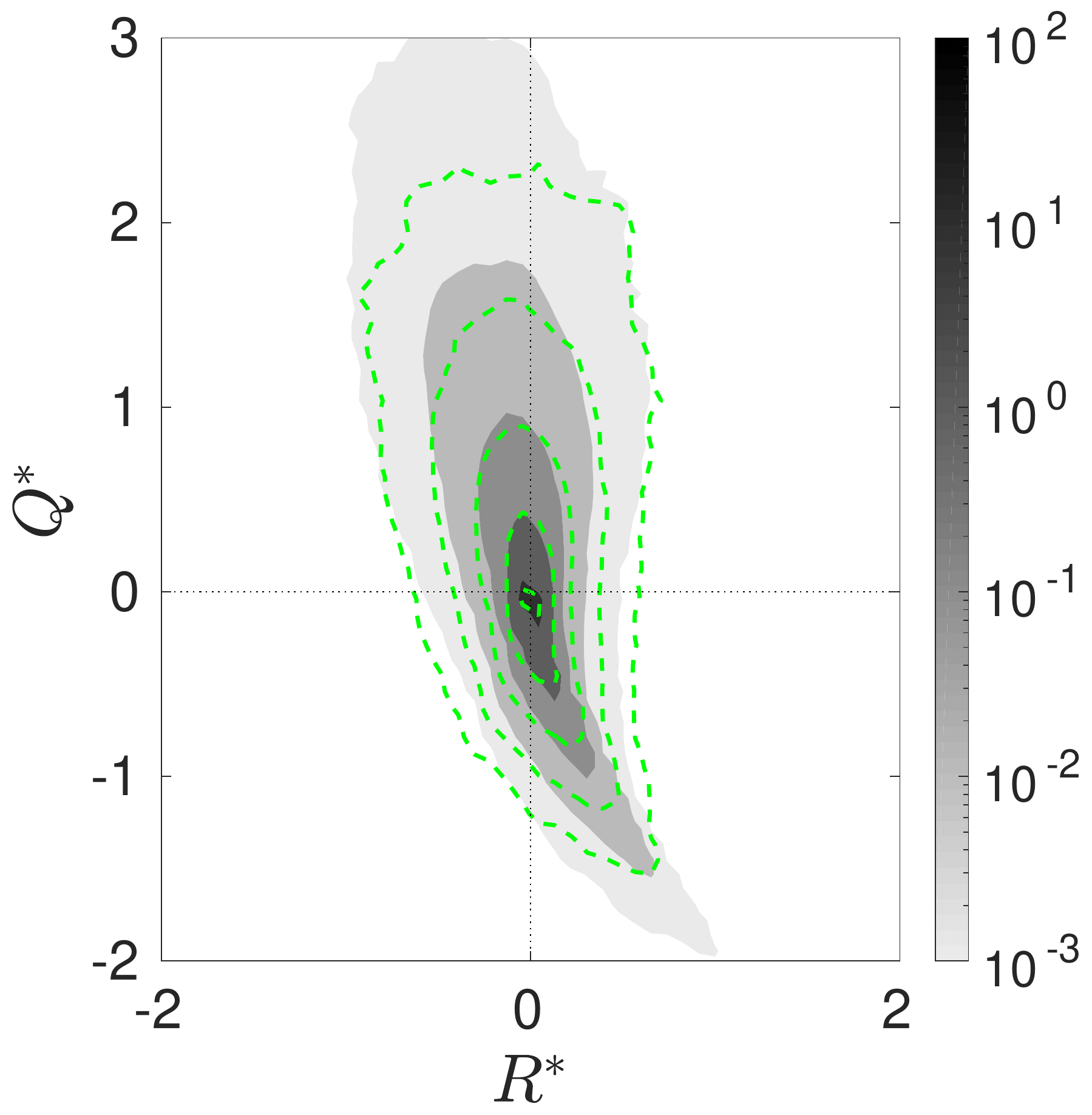}}% Images in 100% size
  \emn%
  \bmn{0.5\lnw}
  \centerline{\includegraphics[width=0.95\lnw]{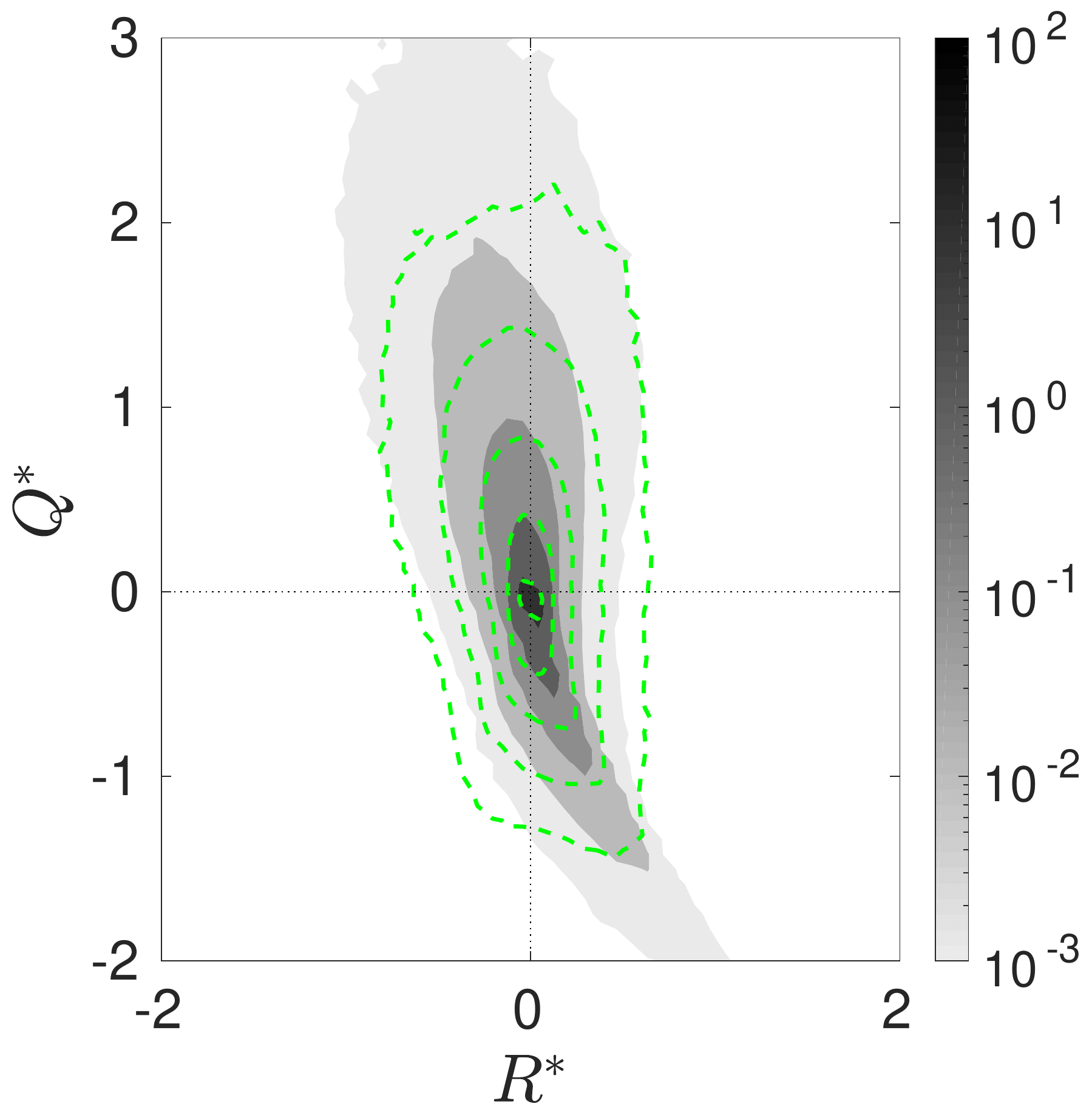}}% Images in 100% size
  \emn

  \caption{The joint PDF of $Q^*\equiv Q/\lal \tA_{ij} \tA_{ij}\ral$ and $R^* \equiv R/\lal \tA_{ij}
  \tA_{ij}\ral^{3/2}$. Left column: $\Delta
  = \Delta_m/2$; right column: $\Delta = \Delta_m/4$. Top row:
  case C1; middle row: case C2; bottom row: case C6. Grey scales: from $\bvp^\v(\x)$; dashed lines:
  from $\bvp^\u(\x)$.}
\label{fig:jpdf-aij-qr}
\end{figure}

The joint PDFs of normalized $Q$ and $R$ are plotted in Fig. \ref{fig:jpdf-aij-qr}.
The contours from
the target fields depict the well-known skewed tear-drop shape with the  Vieillefosse tail extended in the
fourth quadrant \citep{Vieillefosse84, Cantwell92}. Events in the second and the fourth quadrants
occur with higher probabilities.
These features are qualitatively reproduced by the contours of the reconstructed
fields (dashed lines). However, the Vieillefosse
tail is shorter, and the probabilities for large excursions are lower by a rather significant amount.
The lowest contours for the reconstructions with larger $T$ and smaller $e_{\rm tot}$ (the top and
middle rows) are
closer to those for the target, although the improvement is very small.
The right column shows that the discrepancy is bigger when the filter scale is smaller.

\begin{figure}
  \centerline{\includegraphics[width=0.6\lnw]{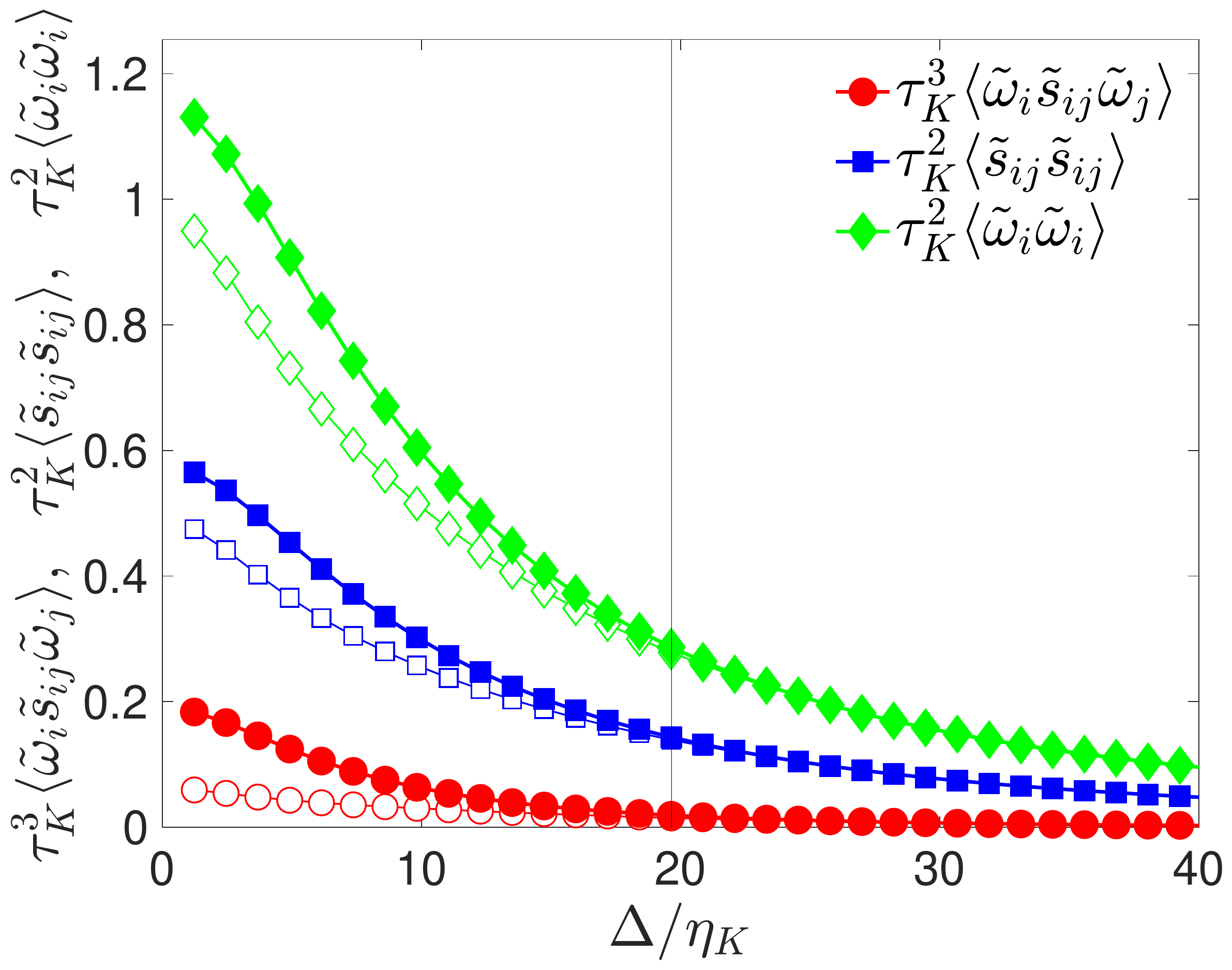}}% Images in 100% size
  \caption{$\lal
  \td{s}_{ij}\td{\om}_i \tog_j\ral$ (circles), $\lal \ts_{ji} \ts_{ij}\ral$ (squares), and $\lal \tog_i
  \tog_i \ral$ (diamonds) as functions of filter scale $\Delta$ from case C1.
 Filled
  and empty symbols: from $\bvp^\v(\x)$ and $\bvp^\u(\x)$,
  respectively. The vertical line marks the scale where $\Delta = \Delta_m$. }
\label{fig:mosomssmoo-ic}
\end{figure}

\begin{figure}
  \centerline{\includegraphics[width=0.6\lnw]{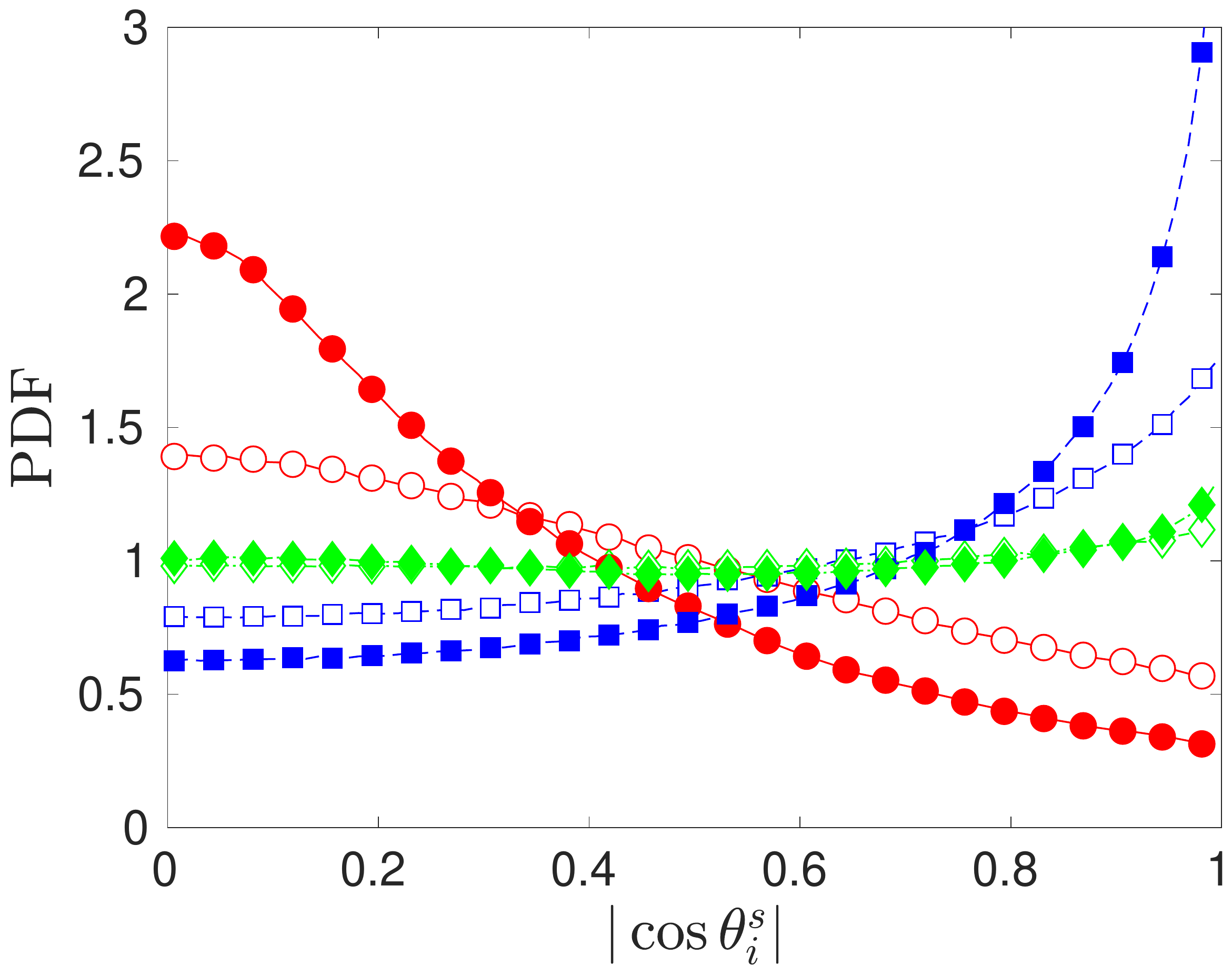}}% Images in 100% size
  \caption{PDFs of $|\cos \theta^s_i|$ with $\theta^s_i$ being the angle between $\td{\bo}$ and
  $\e^s_\alpha$ (diamonds), $\e^s_\beta$ (squares), and $\e^s_\gamma$ (circles).
  Filled symbols: $\bvp^\v(\x)$; empty symbols: $\bvp^\u(\x)$. From case C1 and $\Delta = \Delta_m/2$.}
\label{fig:alignos-1d-ic-8dx}
\end{figure}

The mean strain rate, enstrophy, and vortex stretching term, normalized by the Kolmogorov time scale
$\tau_K$, are given in Fig.
\ref{fig:mosomssmoo-ic} as functions of the filter scale. Not surprisingly, the differences between
the results in the two fields
are larger for smaller filter scales. For
$\lal \ts_{ij}\ts_{ij}\ral$ and $\lal \tog_i\tog_i\ral$ the discrepancy is below $20\%$ for all
filter scales. The discrepancy for the stretching term however is much bigger.
The latter can be partially explained by the mis-alignment between $\tog_i$ and
$\ts_{ij}$ in $\bvp^\u$, which is shown in Fig. \ref{fig:alignos-1d-ic-8dx}
with the PDFs for the cosines of the angles between $\tog_i$ and the eigenvectors of $\ts_{ij}$.
The results for $\bvp^\u$ display
qualitatively correct features, i.e., a preferred alignment between
$\tog_i$ and
$\e^s_\beta$, and the tendency for $\tog_i$ to be perpendicular to $\e^s_\gamma$.
However, the peaks of the PDFs
are lower to a significant degree. Therefore, the preferable alignment in $\bvp^\u$ 
is weaker, and this mis-alignment contributes to the observed
discrepancy in the vortex stretching term.

\begin{figure}
  \bmn{0.5\lnw}
  \centerline{\includegraphics[width=\lnw]{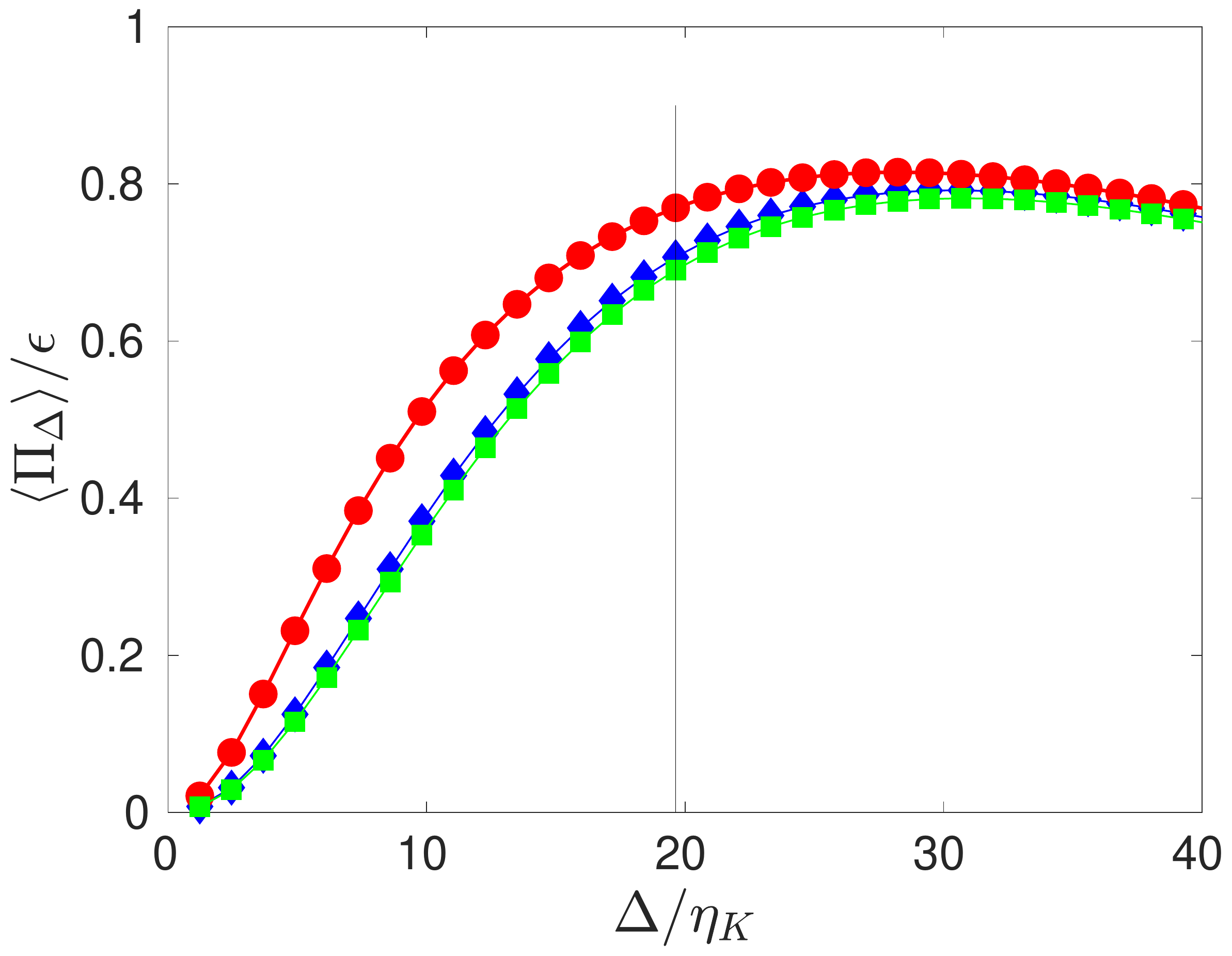}}% Images in 100% size
  \emn %
  \bmn{0.5\lnw}
 \centerline{\includegraphics[width=\lnw]{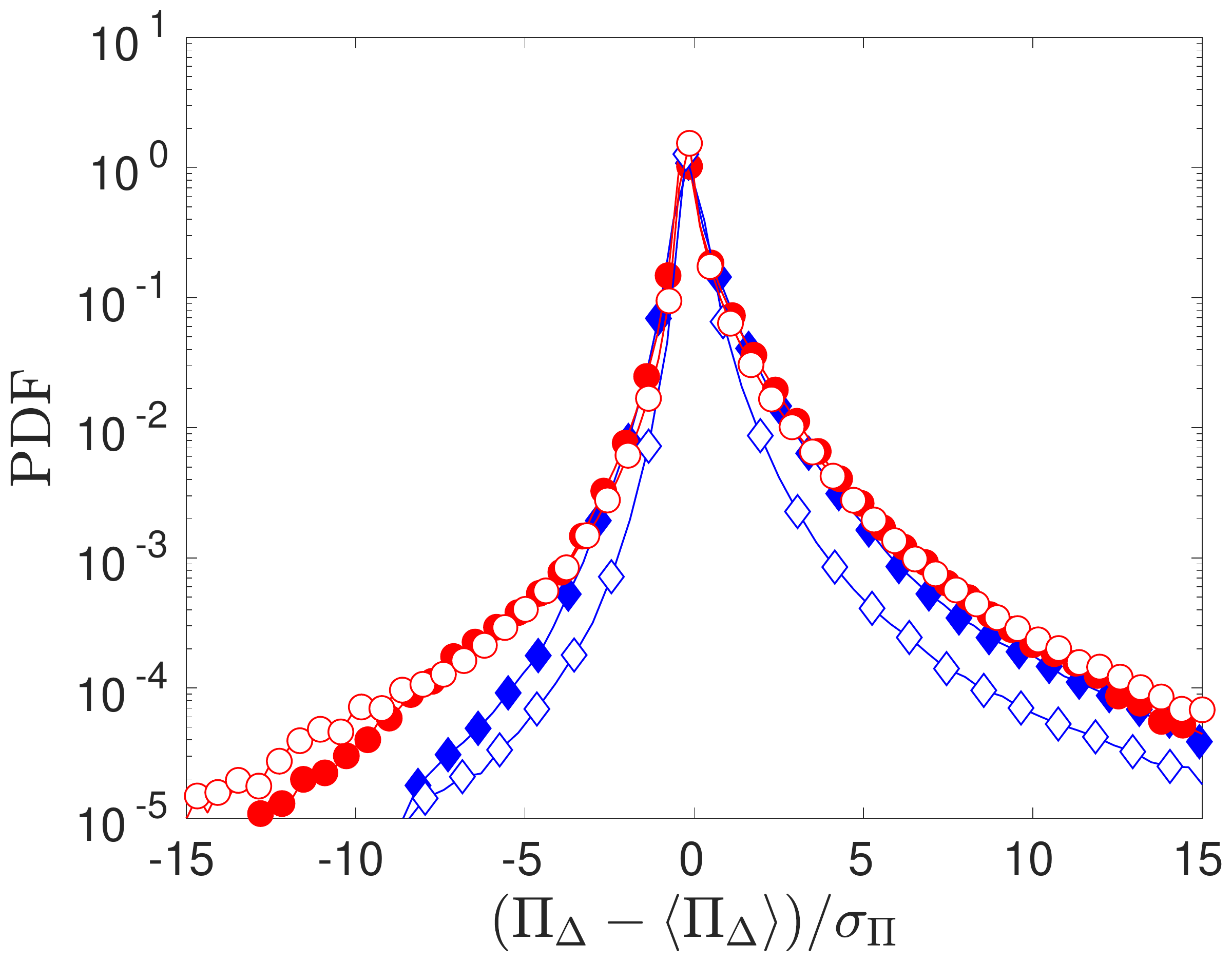}}
  \emn
  \caption{Left: the mean SGS energy dissipation $\lal \Pi_\Delta\ral/\ep$ as a function of
  $\Delta/\eta_K$
  for $\bvp^\v(\x)$ (circles) and
  $\bvp^\u(\x)$ from cases C1 (diamonds) and C6 (squares).
  The vertical line corresponds to the
  $\Delta = \Delta_m$. Right:
  PDF of normalized $\Pi_\Delta$ for $\Delta = \Delta_m/2$ (filled
  symbols) and $\Delta = \Delta_m/4$ (empty symbols), from $\bvp^\v$
  (circles) and $\bvp^\u$ (diamonds) for case C1.
  }
\label{fig:pi_horizon}
\end{figure}
%\begin{figure}
%  \caption{PDF of normalized $\Pi_\Delta$ with $\Delta = \Delta_m$
%  (left) and $\Delta = \Delta_m/2$ (right). Legends are the same as in Fig.
%  \ref{fig:spec-ic-horizon}. }
%\label{fig:pi_pdf}
%\end{figure}

%Note that the measurement data are available for $\Delta/\eta_K \ge 20$.
%Thus about $1/8$ of the volume is covered by measurement data.
The mean SGS energy dissipation $\lal \Pi_\Delta \ral$ is
shown in the left panel of Fig. \ref{fig:pi_horizon}.
The results from $\bvp^\u$ display the
same trends as those from $\bvp^\v$. This result is a direct
evidence that the nonlinear interaction between different scales are
partially reconstructed.
At $\Delta = \Delta_m/2$ ($\Delta/\eta_K \approx 10$), the value
calculated from $\bvp^\u$ is smaller by about
$25\%$.
The results for the two different $T$'s
show only very slight difference.
The PDFs for the normalized $\Pi_\Delta$
are shown in the right panel of Fig. \ref{fig:pi_horizon}.
The well-known positive skewness
is captured by $\bvp^\u(\x)$.
The agreement on the positive tail is very good, although the
negative tail is under-predicted, meaning the reconstructed
fields under-predict the probability for instantaneous energy backscattering.
%The contribution to $\Pi_\Delta$ comes mainly from the
%local interactions between eddies with sizes at the order of $\Delta$
%(TODO: Eyink reference). Therefore, for $\Delta = \Delta_m$,
%the density of measurement data is higher relative to
%the sizes of the eddies.
Only the PDFs for $T = \tau_L$ are shown, as the results with
different $T$'s have no discernible differences.
%Thus, in most cases we will use
%data from $T=0.4\tau_L$in what follows.

\begin{figure}
  \bmn{0.33\lnw}
  \centerline{\includegraphics[width=\lnw]{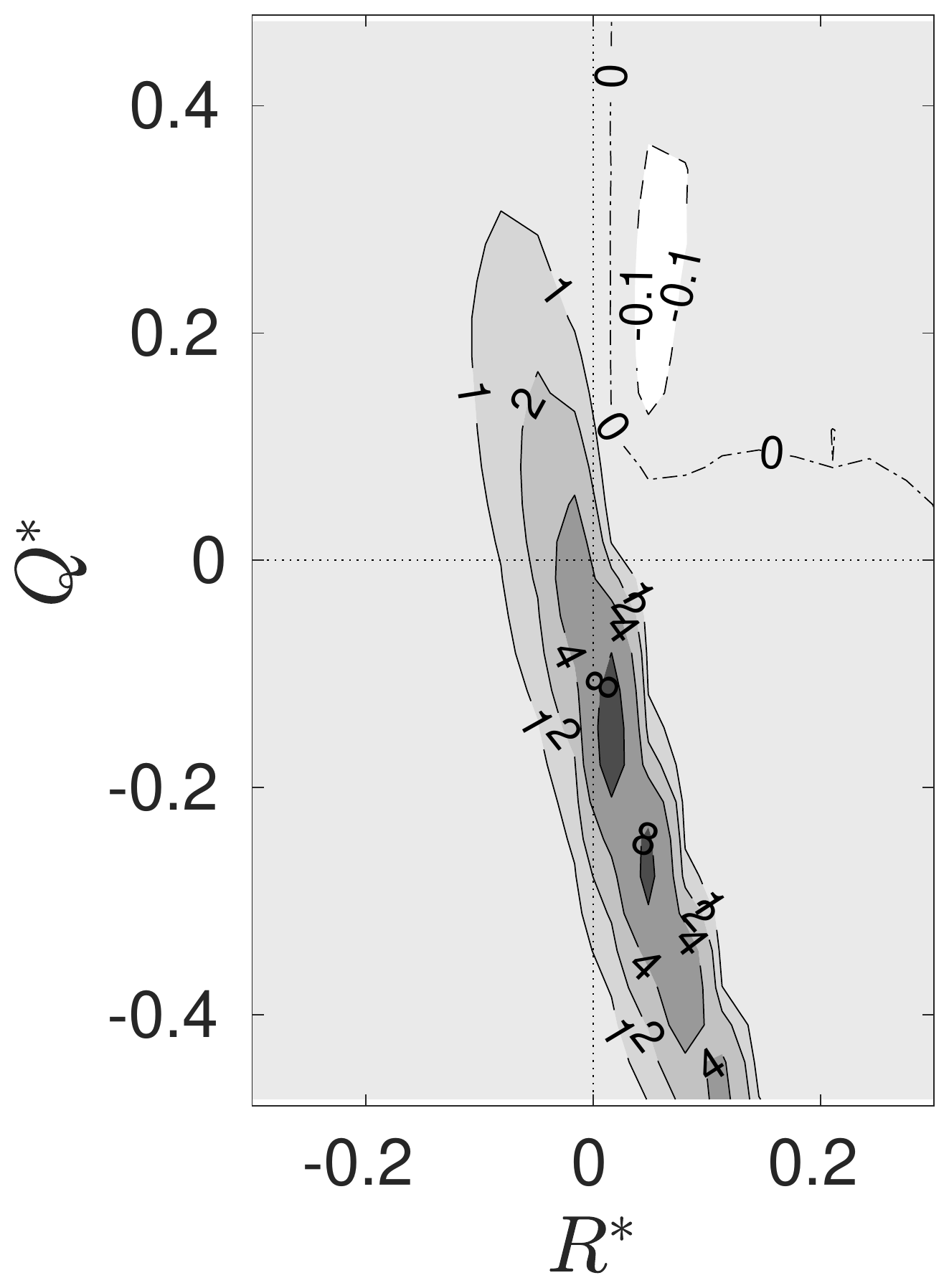}}% Images in 100% size
  \emn %
  \bmn{0.33\lnw}
  \centerline{\includegraphics[width=\lnw]{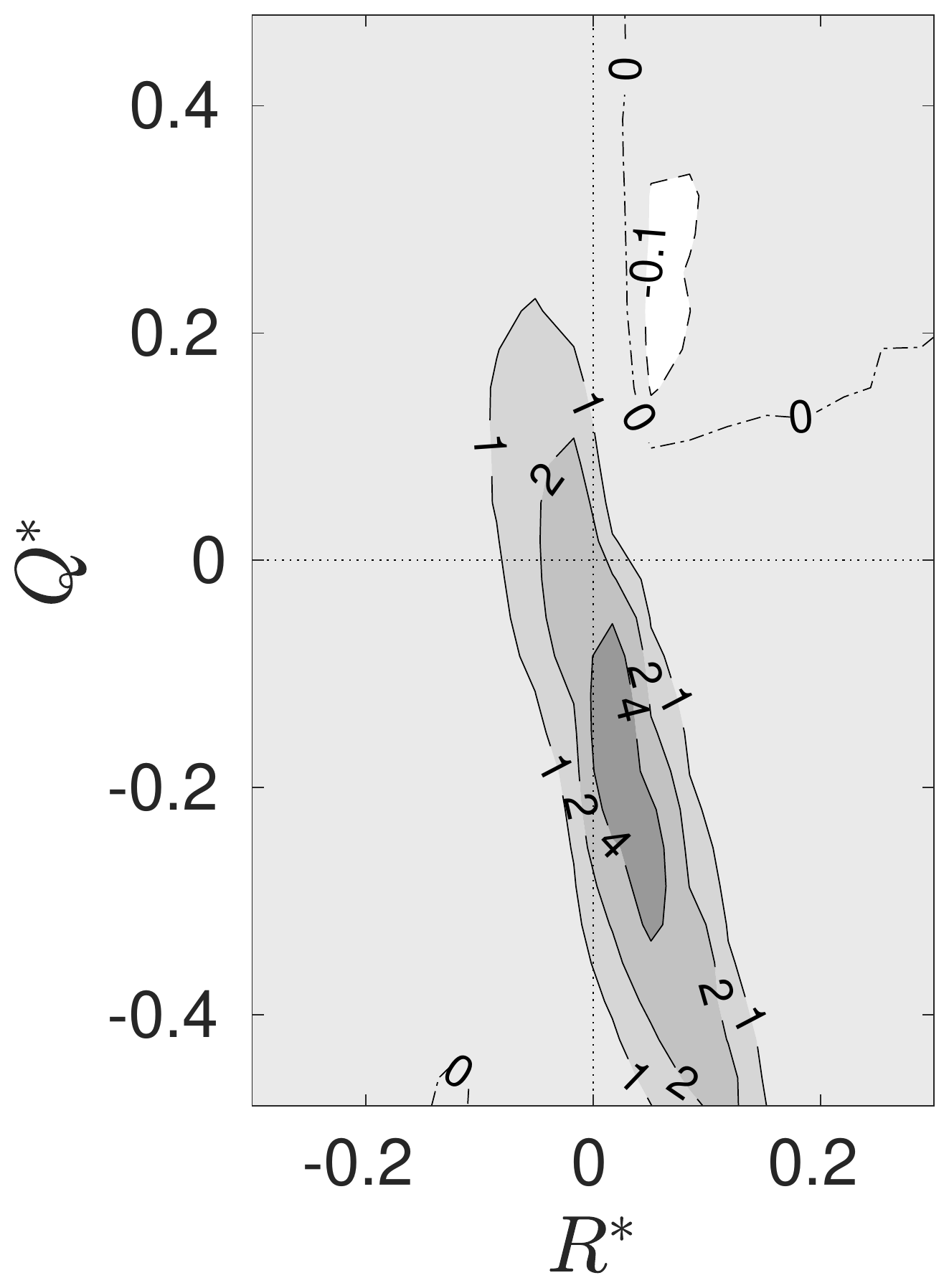}}% Images in 100% size
  \emn %
  \bmn{0.33\lnw}
  \centerline{\includegraphics[width=\lnw]{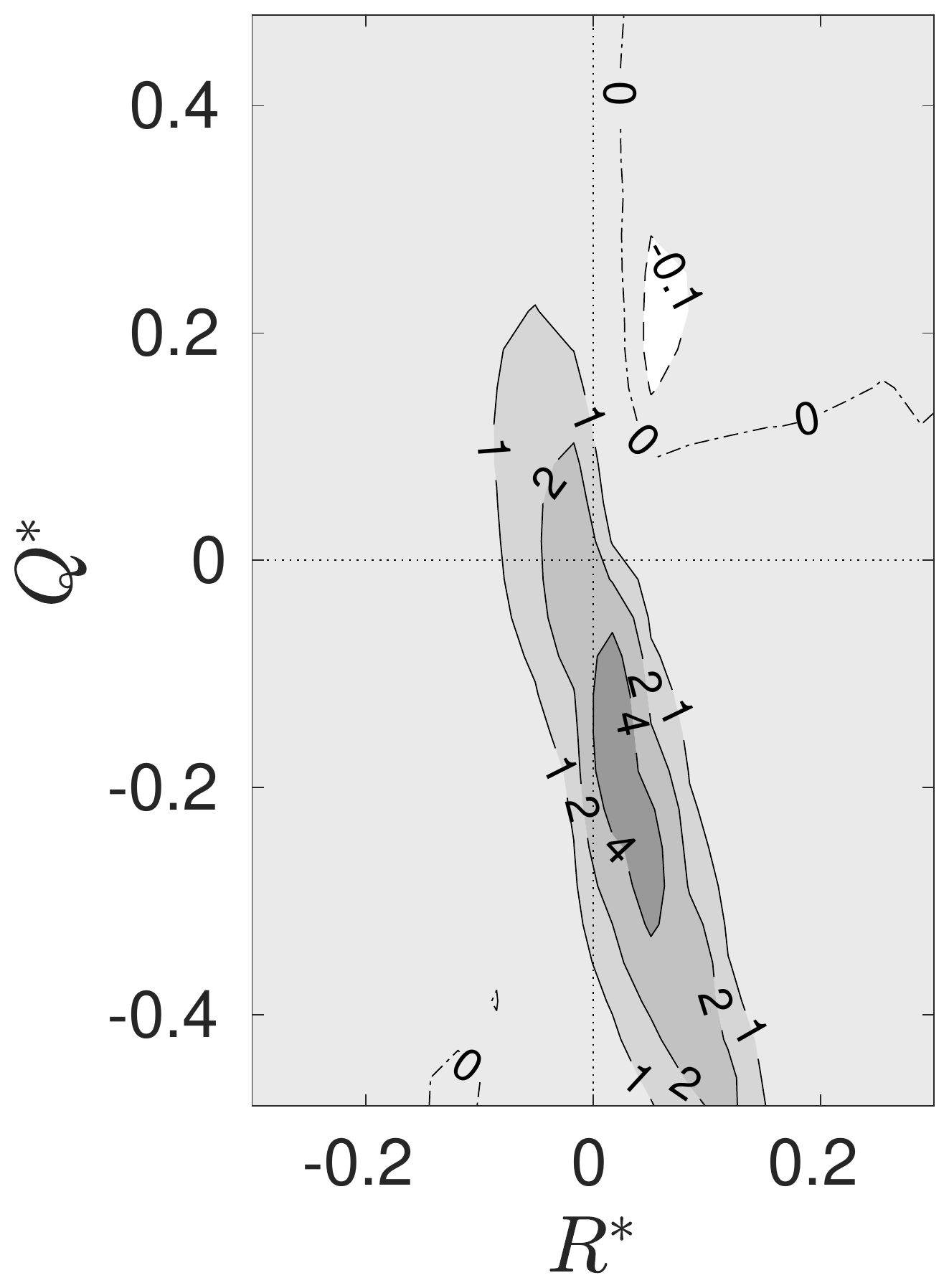}}% Images in 100% size
  \emn

  \bmn{0.33\lnw}
  \centerline{\includegraphics[width=\lnw]{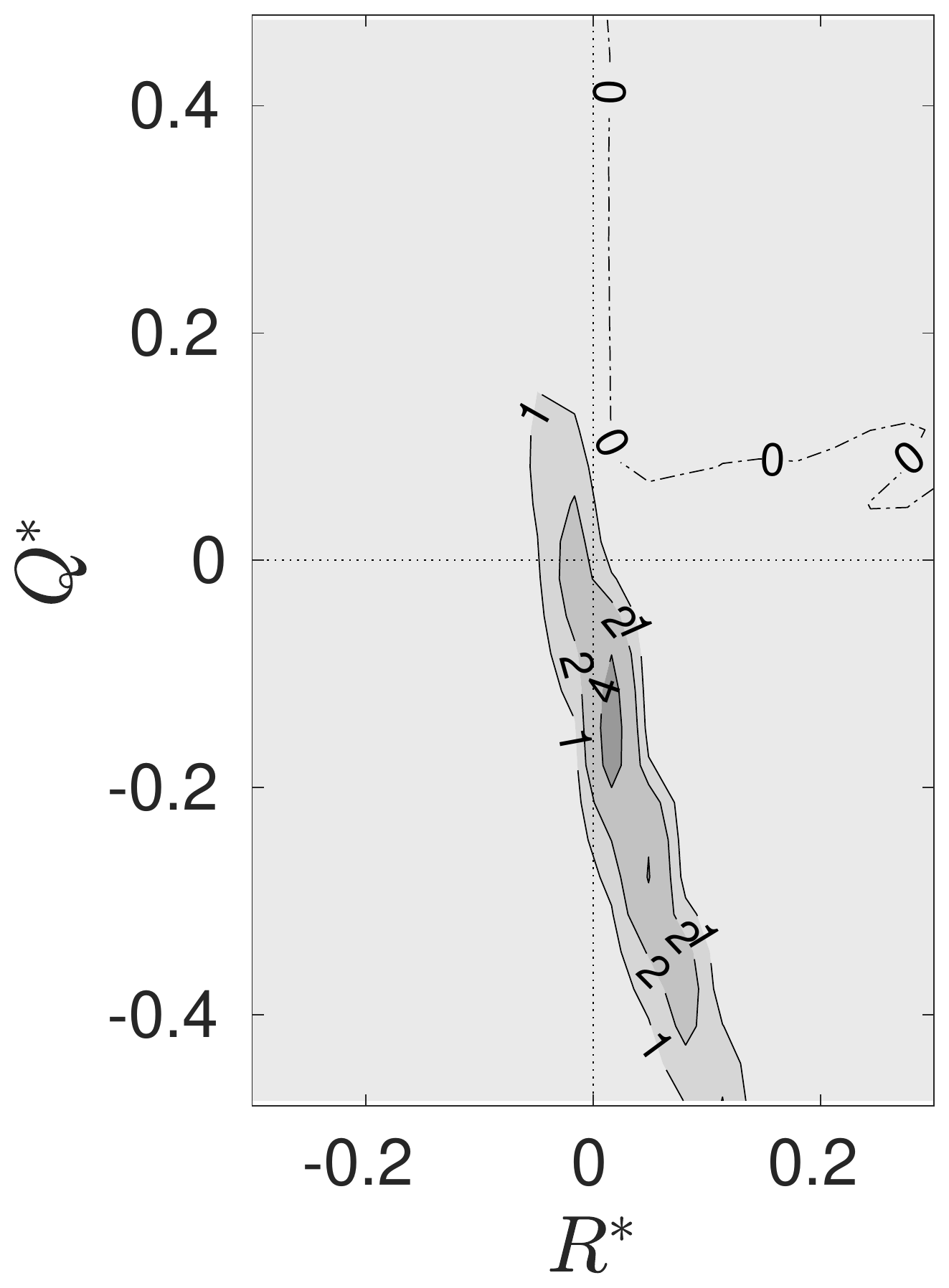}}% Images in 100% size
  \emn %
  \bmn{0.33\lnw}
  \centerline{\includegraphics[width=\lnw]{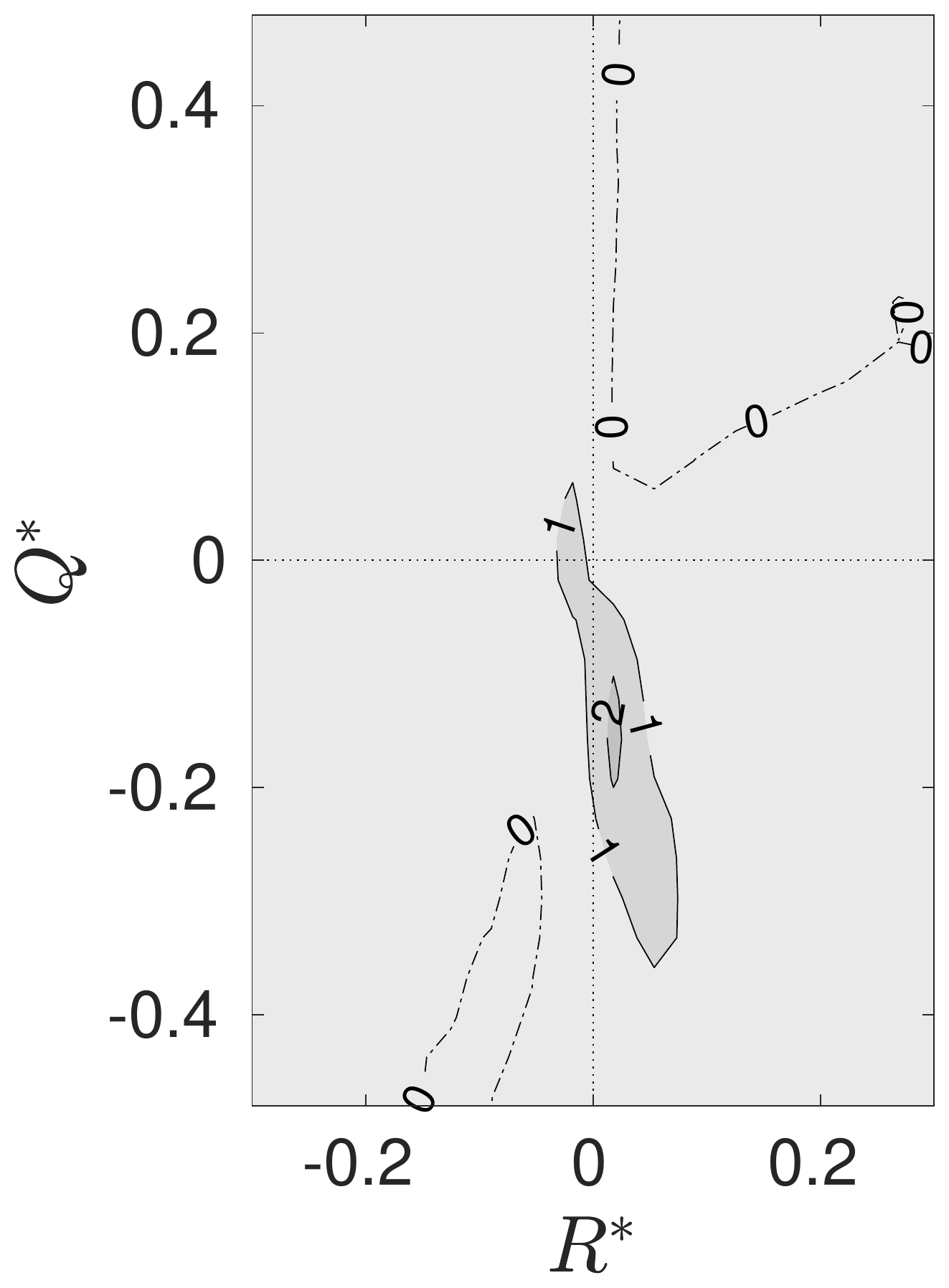}}% Images in 100% size
  \emn %
  \bmn{0.33\lnw}
  \centerline{\includegraphics[width=\lnw]{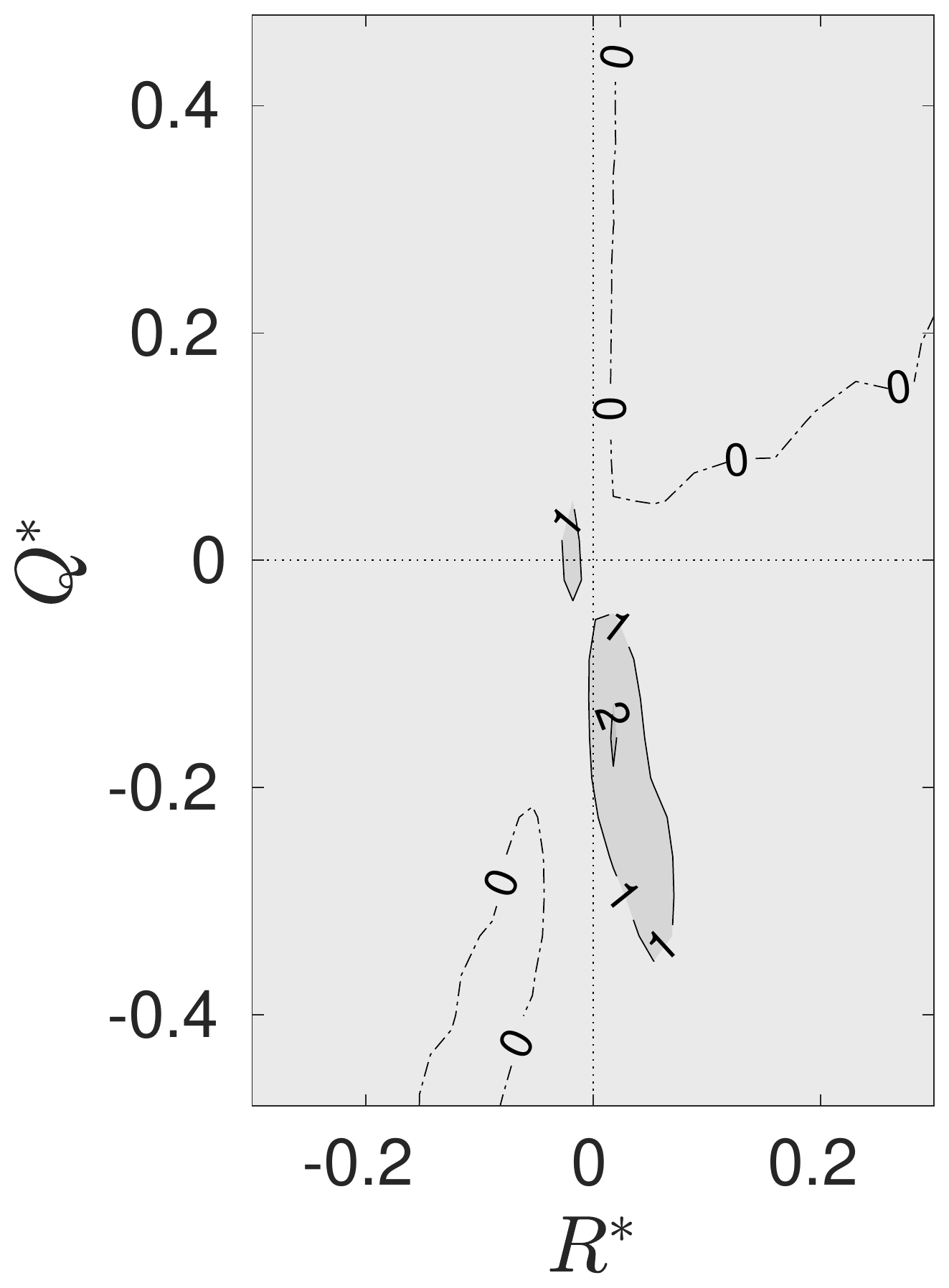}}% Images in 100% size
  \emn
  \caption{The normalized weighted conditional average of the
  SGS energy dissipation $\ep^{-1}P(R^*, Q^*)\lal \Pi_\Delta|R^*,
  Q^*\ral$. Left: from $\bvp^\v$;
  middle: from $\bvp^\u$ in case C1; right:
  from $\bvp^\u$ in case C6. Top row: $\Delta = \Delta_m/2$; bottom row: $\Delta =
  \Delta_m/4$.}
\label{fig:msgsed-cnd-qr-8dx}
\end{figure}
The conditionally averaged SGS energy dissipation $\lal \Pi_\Delta |
Q^*, R^*\ral$ has also been calculated,
where $Q^* \equiv
Q/\lal \tA_{ij} \tA_{ij}\ral$ and $R^*  \equiv R/ \lal \tA_{ij}\tA_{ij}\ral^{3/2}$.
Shown in Fig. \ref{fig:msgsed-cnd-qr-8dx}
is $\lal \Pi_\Delta | Q^*, R^*\ral$ weighted by the joint PDF $P(R^*,Q^*)$ and normalized by
the mean energy dissipation rate $\ep$.
The results from $\bvp^\v$
(the left column) show
that higher SGS dissipation
is observed around the Vieillefosse tail where high straining
motion happens.
The behavior is reproduced qualitatively by $\bvp^\u$, shown in the middle and on the right for $T=\tau_L$ and
$0.5\tau_L$, respectively.
Negative SGS dissipation,
representing energy back-scattering, is found mainly in the first quadrant, a feature also
captured by $\bvp^\u$. However, overall,
$\bvp^\u$ tends to underestimate the magnitude
of the weighted SGS energy dissipation. The results at $\Delta =
\Delta_m/2$ are closer than those at $\Delta = \Delta_m/4$. There are
some small improvements with $T=\tau_L$ compared with $T=0.5\tau_L$.

To summarize, the results in this subsection demonstrate that $\bvp^\u(\x)$ captures
the main statistical features of real turbulence with reasonable agreement. This is the basis for the
reconstruction of the instantaneous velocity fields at later times.
The discrepancy between $\bvp^\v(\x)$ and $\bvp^\u(\x)$, manifested by
the several statistics we considered, can be reduced slightly by
extending the optimization horizon from $T=0.5\tau_L$ to $T=\tau_L$ or
by using a smaller $e_{\rm tot}$. Given the chaotic nature of the flow, these small differences
might be the explanation for the significant improvement found at later time $t\to T$ shown in
Section
\ref{sect:results_ek}. This conjecture may be quantified by calculating the sensitivities of the
statistics although this is beyond the scope of this paper.

\subsection{Reconstruction of the instantaneous local structures for $t\to T$}
For the reconstructed field $\u(\x,t)$ at later times,
it is expected that its \emph{statistics} would agree well with those in the target field $\v(\x,t)$, for the
simple reason that both are the stationary solutions of the NSEs.
However, there is no \emph{a priori} reason to assert that
the $\u(\x,t)$ would have \emph{point-wise} agreement with $\v(\x,t)$.
The point-wise differences have been examined briefly using $E_D(k)$
and $\hat{\rho}_{\u \v}(k)$ in Section \ref{sect:results_ek} to
demonstrate the effects of the choices of computational parameters.
In this subsection, the base line case C1 (c.f. Table \ref{tab:cases})
is examined in further detail. Note that time $t$ is now another
parameter. We usually choose $t = T$. However, when comparing results with
different $T$'s, other values of $t$ are also used.

\subsubsection{Geometry of the local structures}

\bfig
\bmn{0.33\lnw}
  \centerline{\includegraphics[width=1.17\lnw]{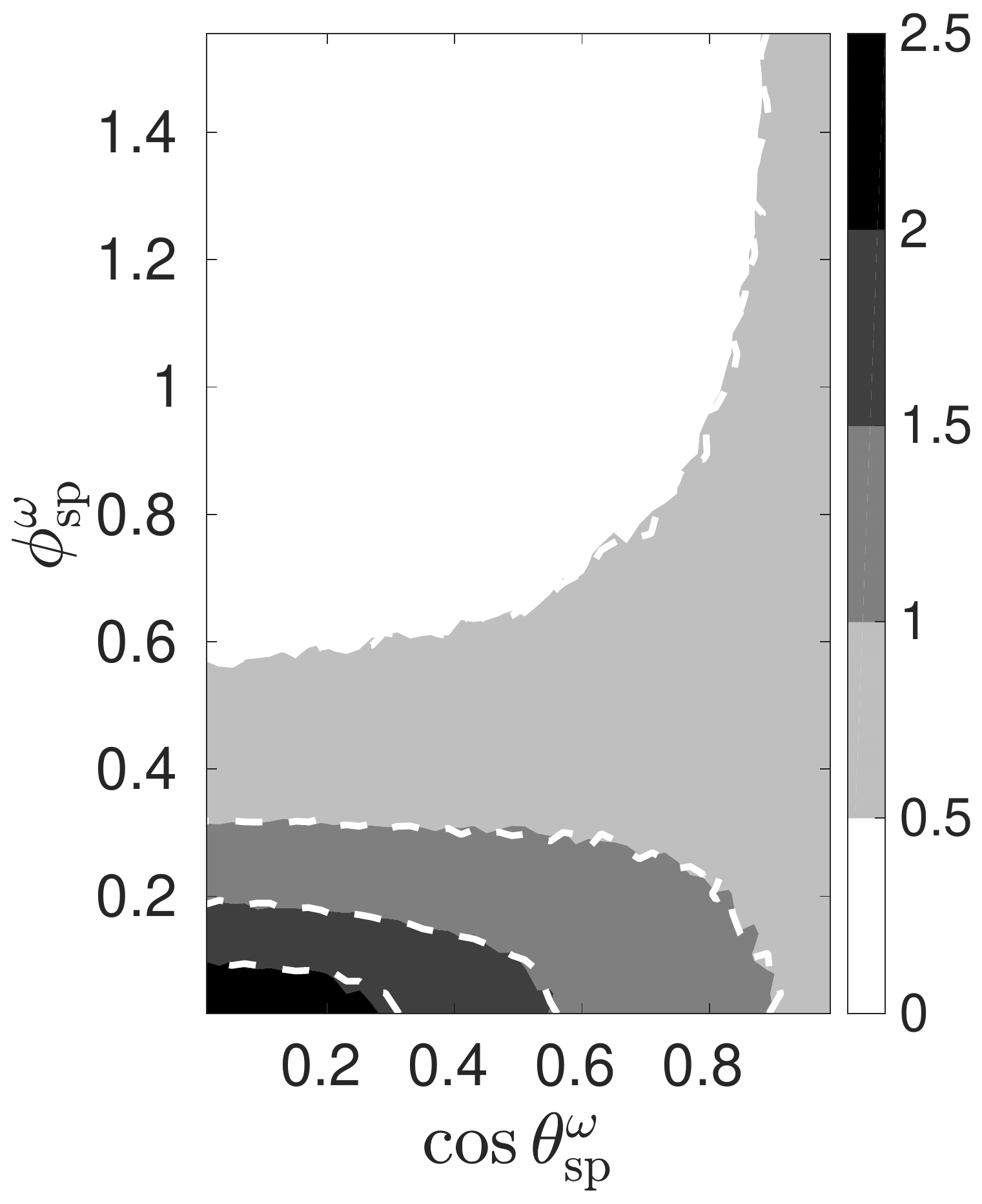}}
\emn %
\bmn{0.33\lnw}
  \centerline{\includegraphics[width=1.17\lnw]{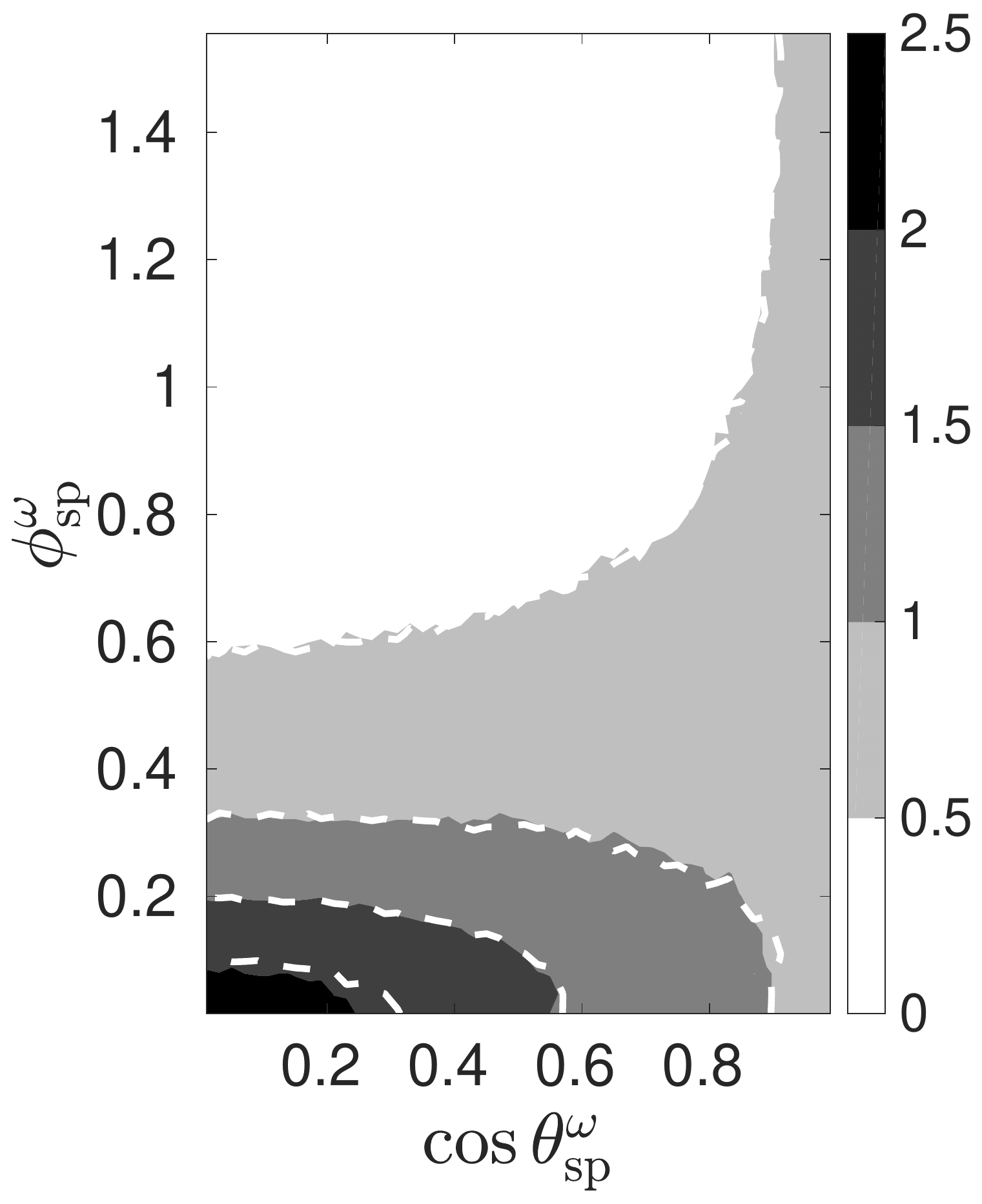}}
\emn %
\bmn{0.33\lnw}
  \centerline{\includegraphics[width=1.17\lnw]{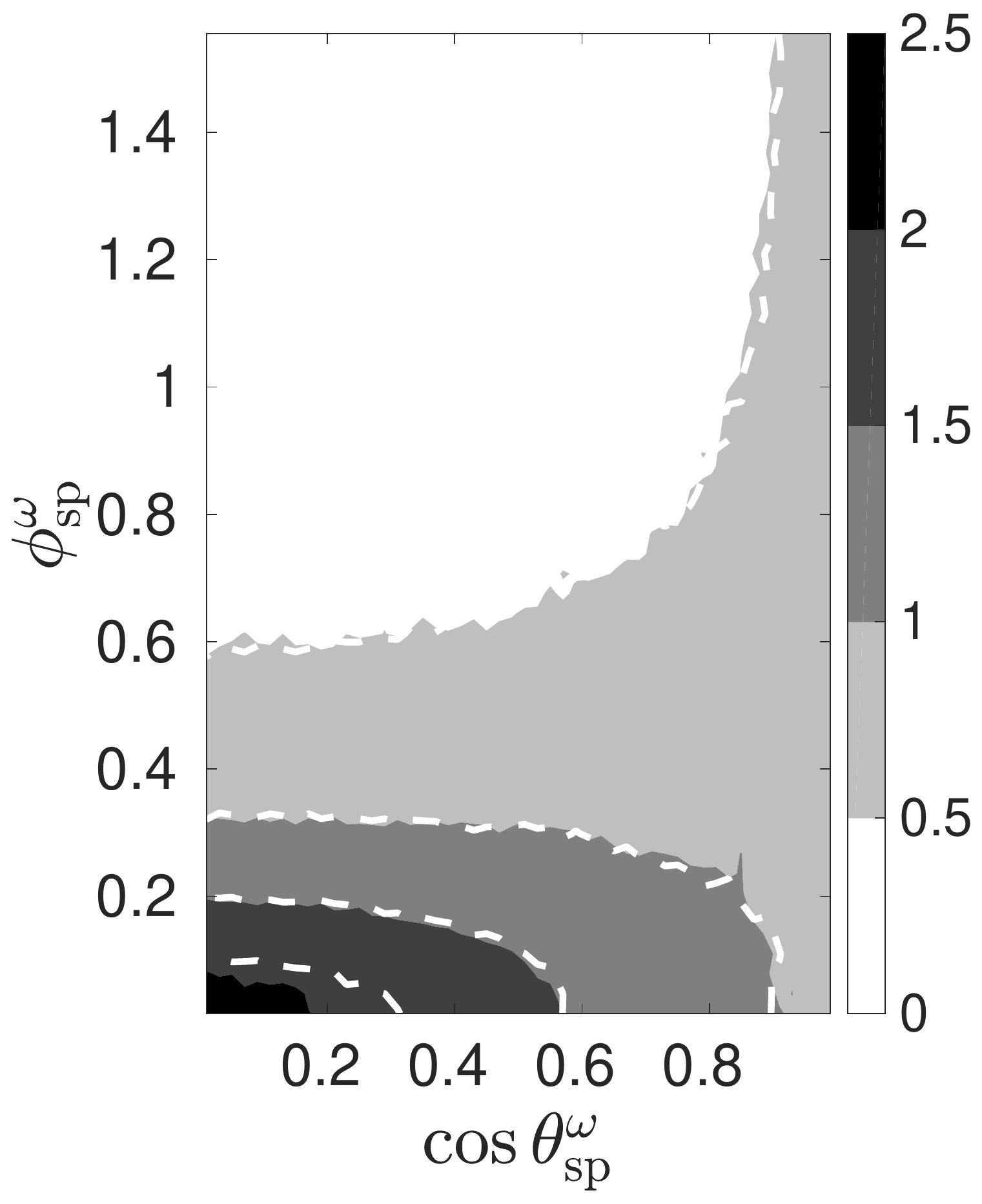}}
\emn

\bmn{0.33\lnw}
  \centerline{\includegraphics[width=1.17\lnw]{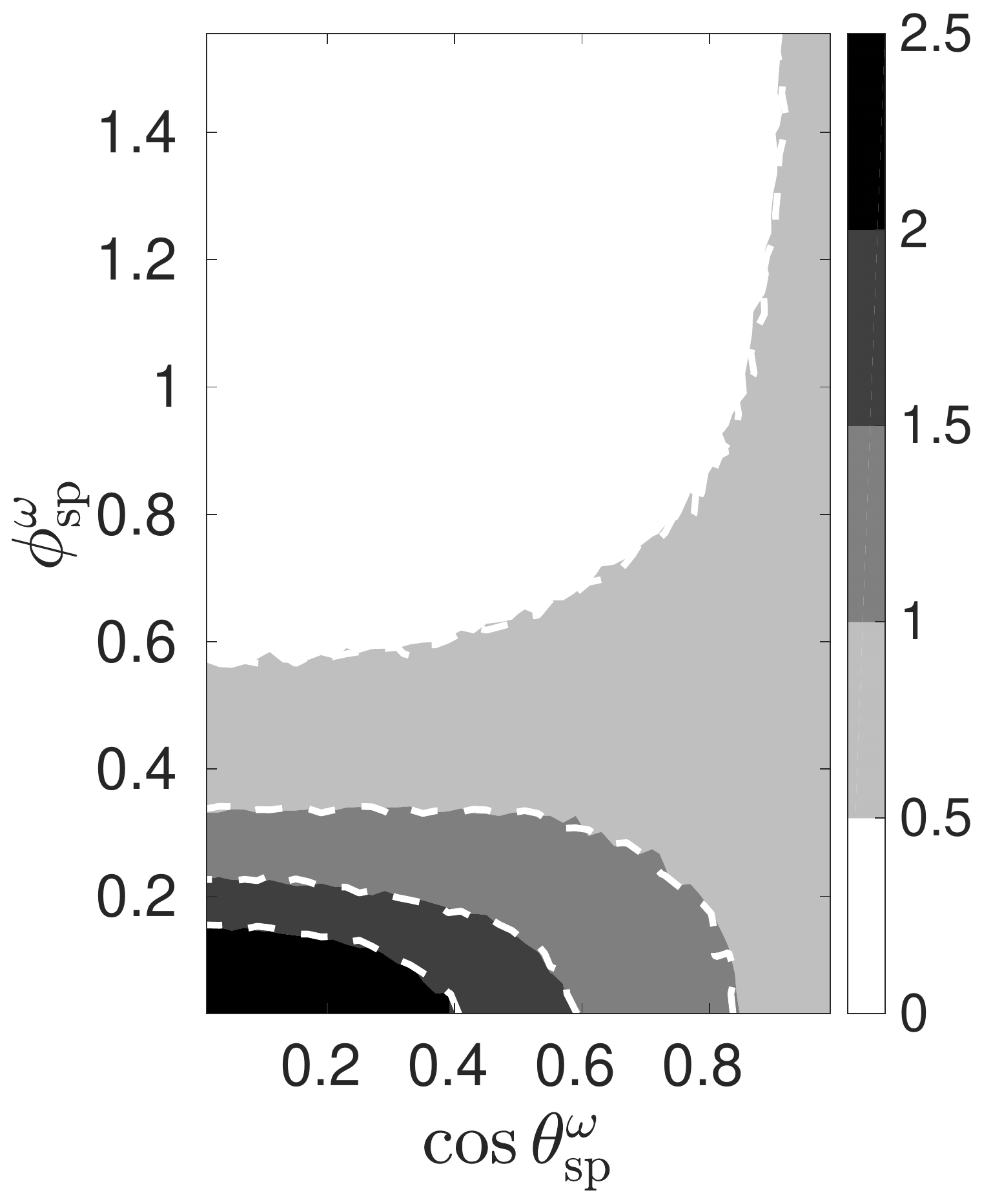}}
\emn %
\bmn{0.33\lnw}
  \centerline{\includegraphics[width=1.17\lnw]{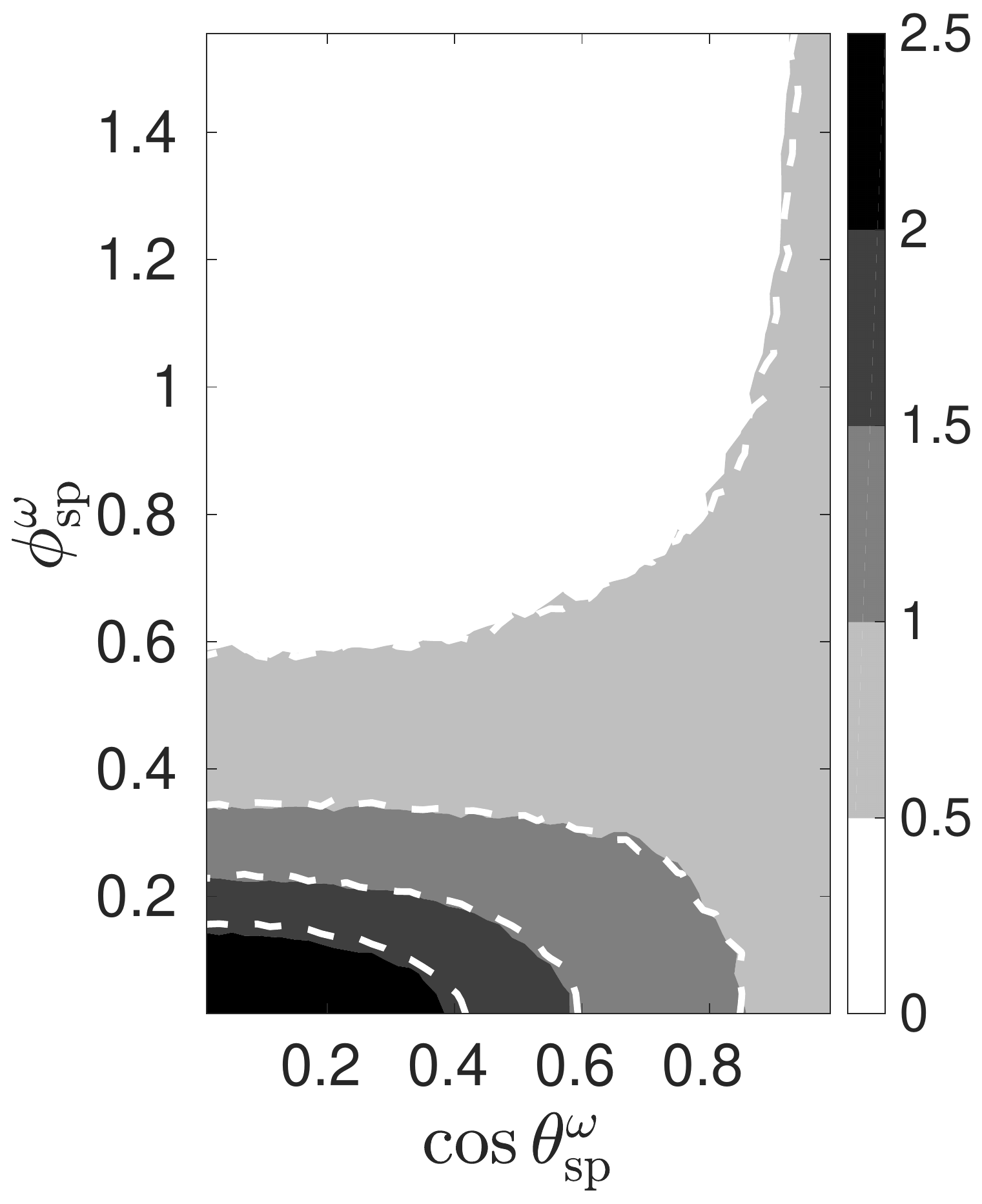}}
\emn %
\bmn{0.33\lnw}
  \centerline{\includegraphics[width=1.17\lnw]{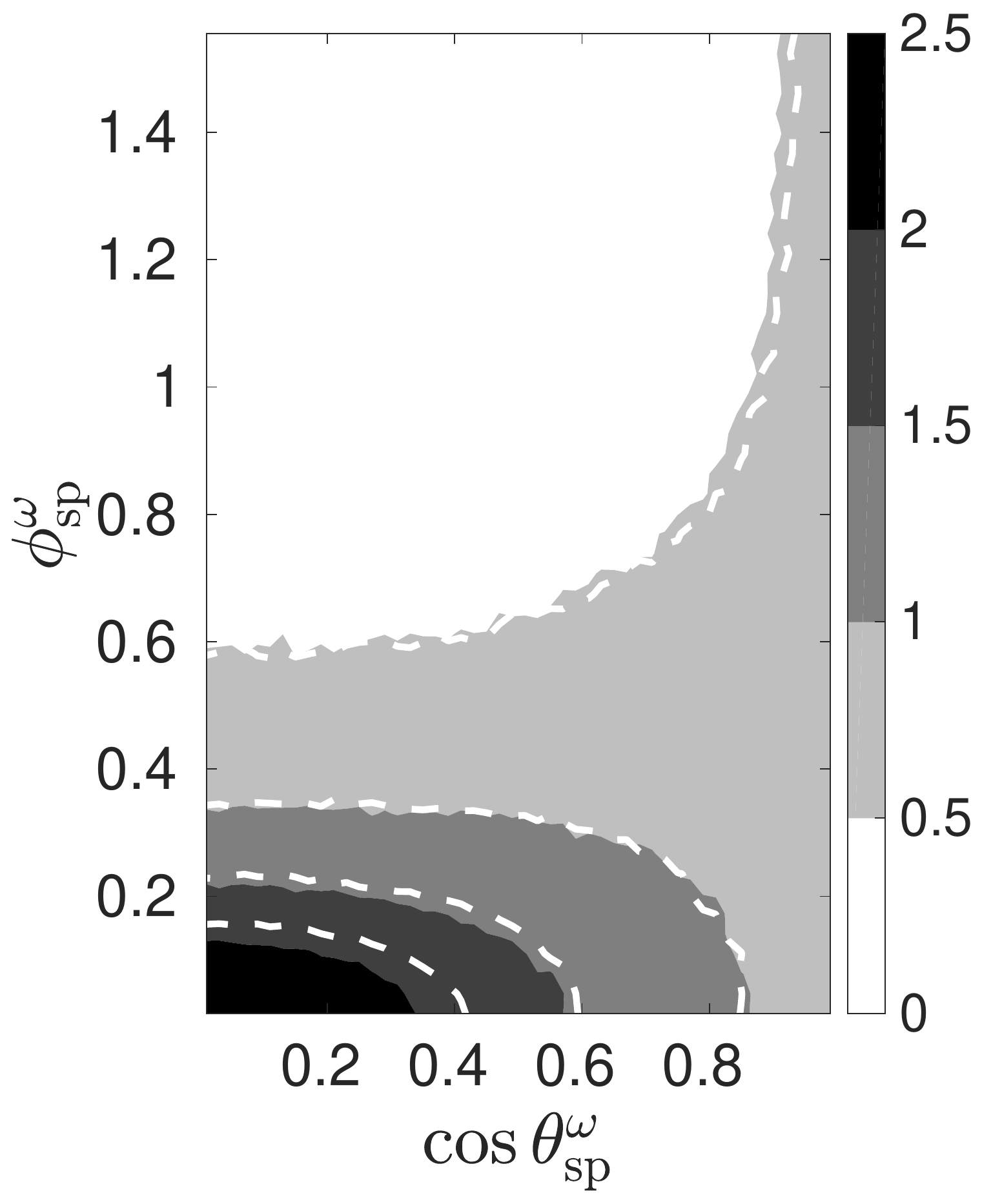}}
\emn
  \caption{The joint PDFs of $(\cos \theta^\omega_{\rm sp}, \phi^\omega_{\rm sp})$
  for angles between $\tog_i^\v$
  and $\ts_{ij}^\u$ (gray scales) and those between $\tog_i^\v$ and
  $\ts_{ij}^\v$ (dashed lines). Top row: $\Delta = \Delta_m/2$; bottom
  row: $\Delta = \Delta_m/4$.
  Left: $t=T$ for case C1; middle: $t=0.5T$ for case C1;
  right: $t=T$ for case C6.}
\label{fig:alignos-cross}
\efig

The instantaneous difference in the geometrical structures of the reconstructed fields $\u$ and the
target fields $\v$
is first investigated. The first quantity we examine is
the `cross-alignment' between
the vorticity in $\v$ ($\tog^\v_i$) and
the eigenvectors of the strain rate tensor in $\u$
($\ts_{ij}^\u$).
The alignment is measured by the joint PDF
of $\cos \theta^\omega_{\rm sp}$ and
$\phi^\omega_{\rm sp}$, where the angles define the orientation
of $\tog^\v_i$ in the eigen-frame of
$\ts_{ij}^\u$ (c.f., Fig. \ref{fig:eigenframe}). The joint PDF is shown with gray scale contours in
Fig. \ref{fig:alignos-cross}. The joint PDF for
the usual alignment between $\tog^\v_i$ and $\ts_{ij}^\v$, both from $\v$, is plotted in dashed
lines as a
comparison.
The left column of Fig. \ref{fig:alignos-cross} shows that
the joint PDF for $\tog^\v_i$ and $\ts_{ij}^\u$
is very close to the one for $\tog^\v_i$ and $\ts_{ij}^\v$ for both
$\Delta = \Delta_m/2$ and $\Delta_m/4$. Both joint
PDFs
display a high peak around $(0,0)$, which shows the well-known
preferred alignment
between the vorticity vector with $\e^s_\beta$ in turbulence.
The peak probability density for
the cross alignment is only slightly lower.
However, the fact
that replacing $\ts^\v_{ij}$ by $\ts^\u_{ij}$ still yields a
close joint PDF proves that the instantaneous orientations of the
eigenvectors of $\ts^\u_{i}$
and $\ts^\v_{ij}$ are very close to each other.  
Same behaviours are observed for
the results in the middle column, where the results at an earlier time
$t = T/2$ for the same $T=\tau_L$ are shown.
The joint PDFs
display only slightly larger discrepancy.
The result obtained with a shorter horizon
$T=0.5\tau_L$, given in
the right column, shows much
larger discrepancy, as the highest level contour is much smaller even
though the location of the peak is
correctly found at $(0,0)$.

\bfig
\bmn{0.5\lnw}
  \centerline{\includegraphics[width=0.9\lnw]{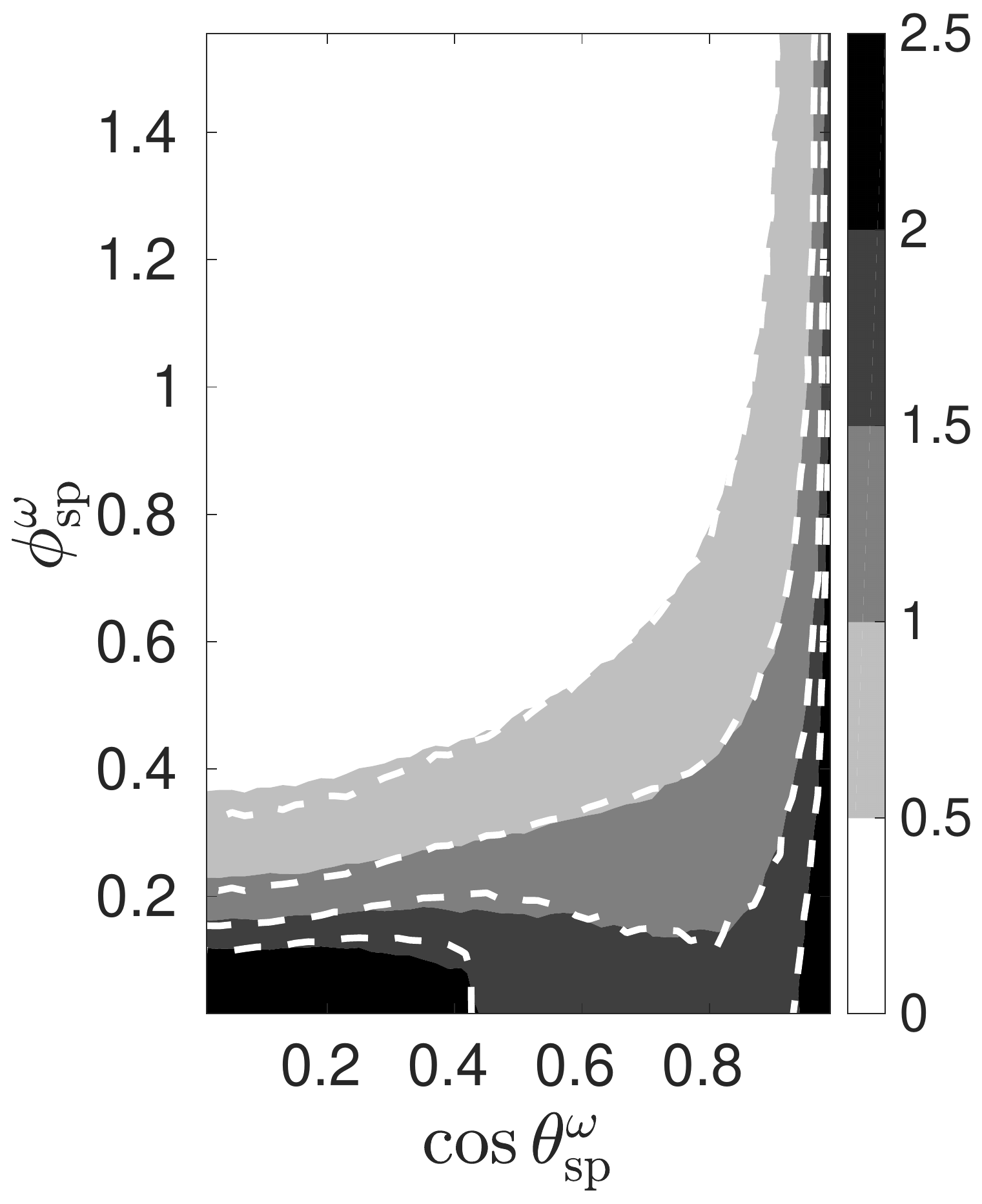}}
\emn %
\bmn{0.5\lnw}
  \centerline{\includegraphics[width=0.9\lnw]{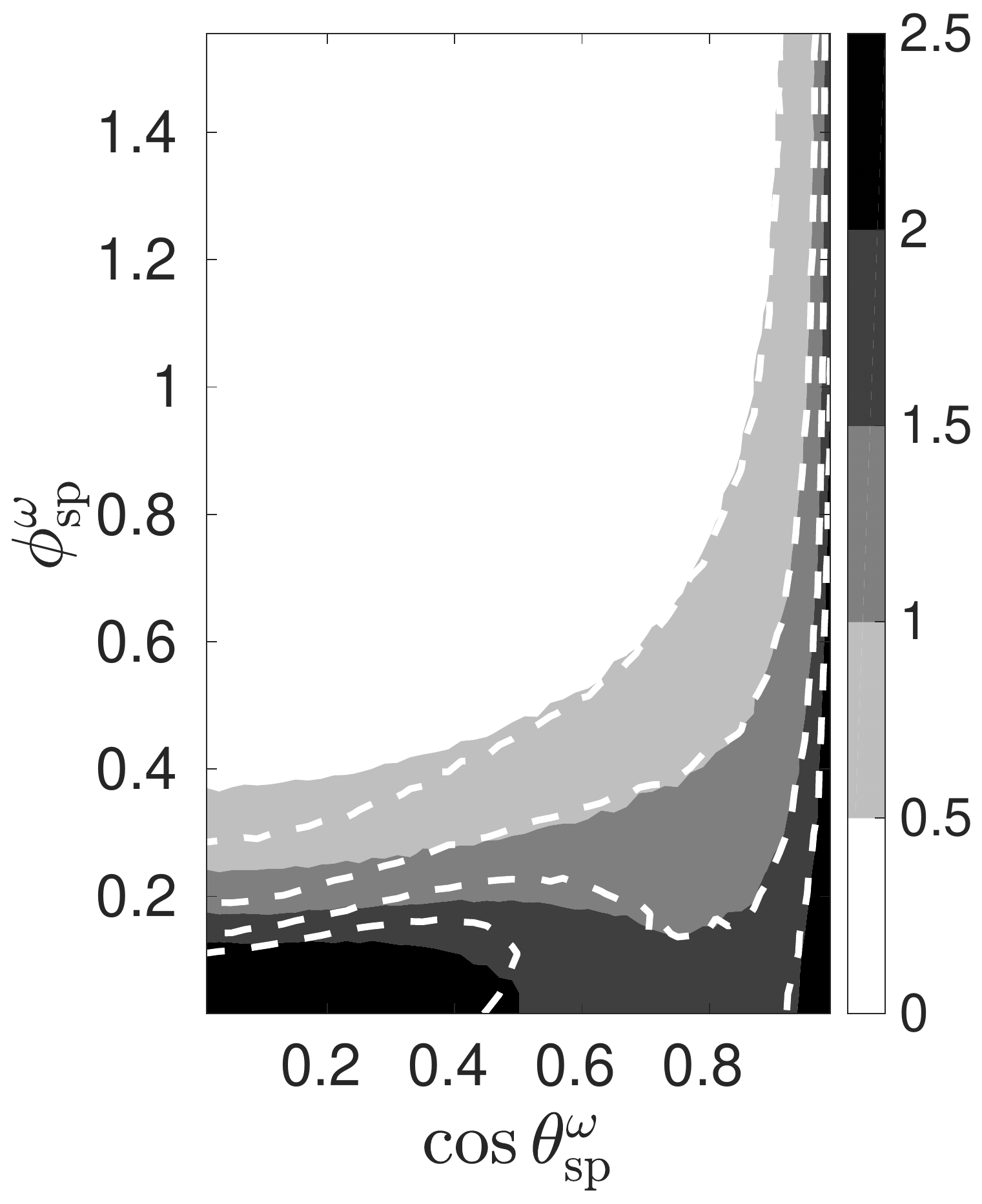}}
\emn

%\bmn{0.5\lnw}
%  \centerline{\includegraphics[width=0.9\lnw]{alignot_pdf_2d_tgt_8dx_lr1_Th-crop}}
%\emn %
%\bmn{0.5\lnw}
%  \centerline{\includegraphics[width=0.9\lnw]{alignot_pdf_2d_tgt_4dx_lr1_Th-crop}}
%\emn

%\bmn{0.5\lnw}
%  \centerline{\includegraphics[width=0.9\lnw]{alignot_pdf_2d_tgt_8dx_sr1-crop}}
%\emn %
%\bmn{0.5\lnw}
%  \centerline{\includegraphics[width=0.9\lnw]{alignot_pdf_2d_tgt_4dx_sr1-crop}}
%\emn
  \caption{The joint PDFs of $(\cos \theta^\omega_{\rm sp}, \phi^\omega_{\rm sp})$
  for angles between $\tog_i^\v$
  and $-\tau_{ij}^{d,\u}$ (gray scales) and those between $\tog_i^\v$ and
  $-\tau_{ij}^{d,\v}$ (dashed lines),
  for $\Delta = \Delta_m/2$ (left) and $\Delta_m/4$ (right) with $t=T$ in case C1.}
\label{fig:alignot-cross}
\efig

Fig. \ref{fig:alignot-cross} considers
the cross alignment between $\tog_i^\v$ and the eigenvectors of
$-\tau^{d,\u}_{ij}$.
  The dashed lines show the joint PDF for the usual alignment in $\v$. The two peaks of the joint PDF
  depict the known feature in
  turbulence, where
  $\tog_i$ tends to align with $\e_\alpha^\tau$ and $\e_\beta^\tau$.
The cross alignment between $\tog_i^\v$ and $-\tau_{ij}^{d,\u}$
qualitatively reproduces these features. The result at $\Delta =
\Delta_m/2$ show excellent quantitative agreement,
although the discrepancy becomes larger for $\Delta = \Delta_m/4$.
The comparison at $t=0.5\tau_L$ for $T = 0.5\tau_L$ and $\tau_L$ is not shown,
because it simply confirms our expectation that the results with a smaller optimization horizon $T$
display larger discrepancy.

\bfig
\bmn{0.33\lnw}
  \centerline{\includegraphics[width=\lnw]{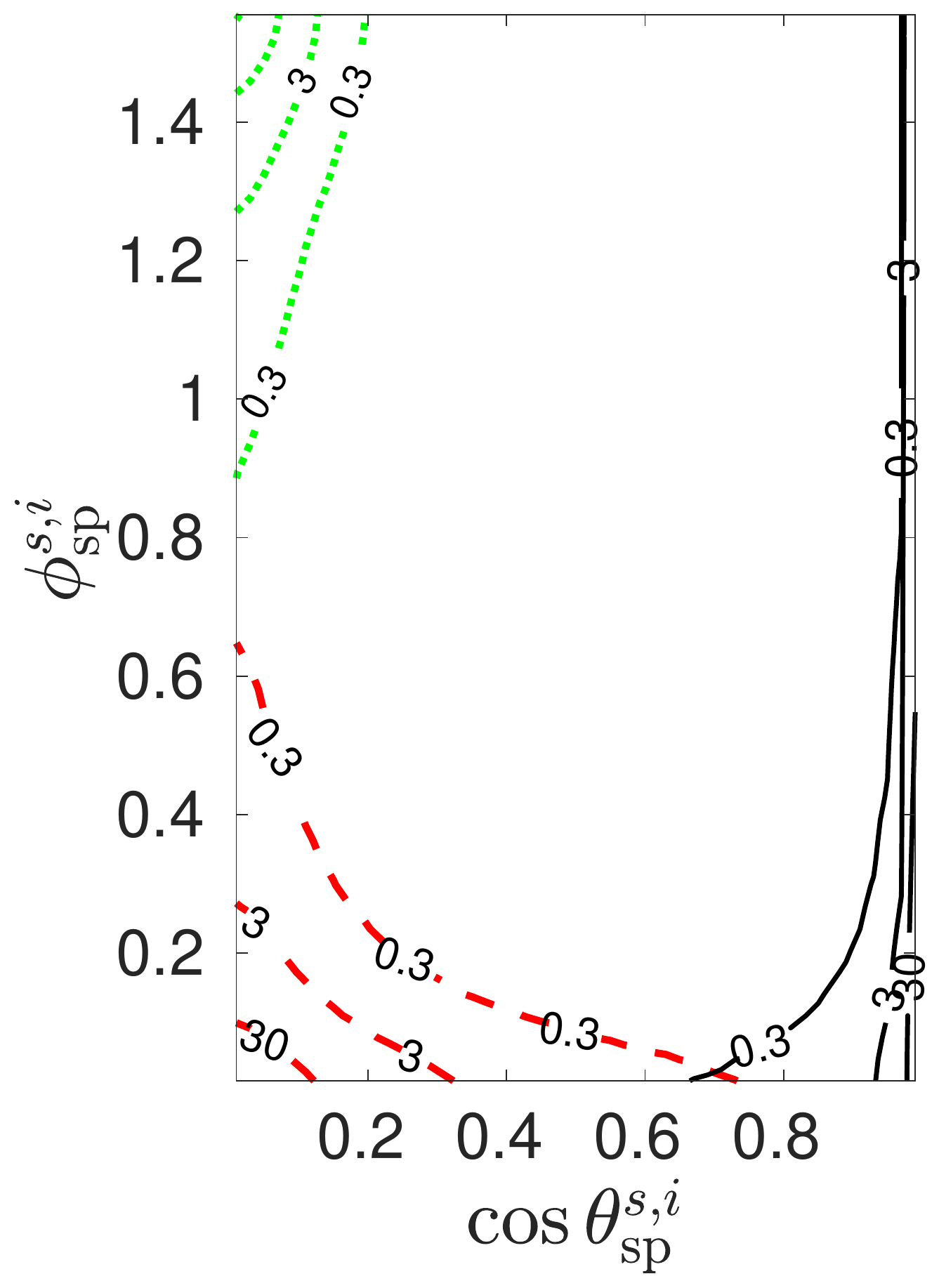}}
\emn %
\bmn{0.33\lnw}
  \centerline{\includegraphics[width=\lnw]{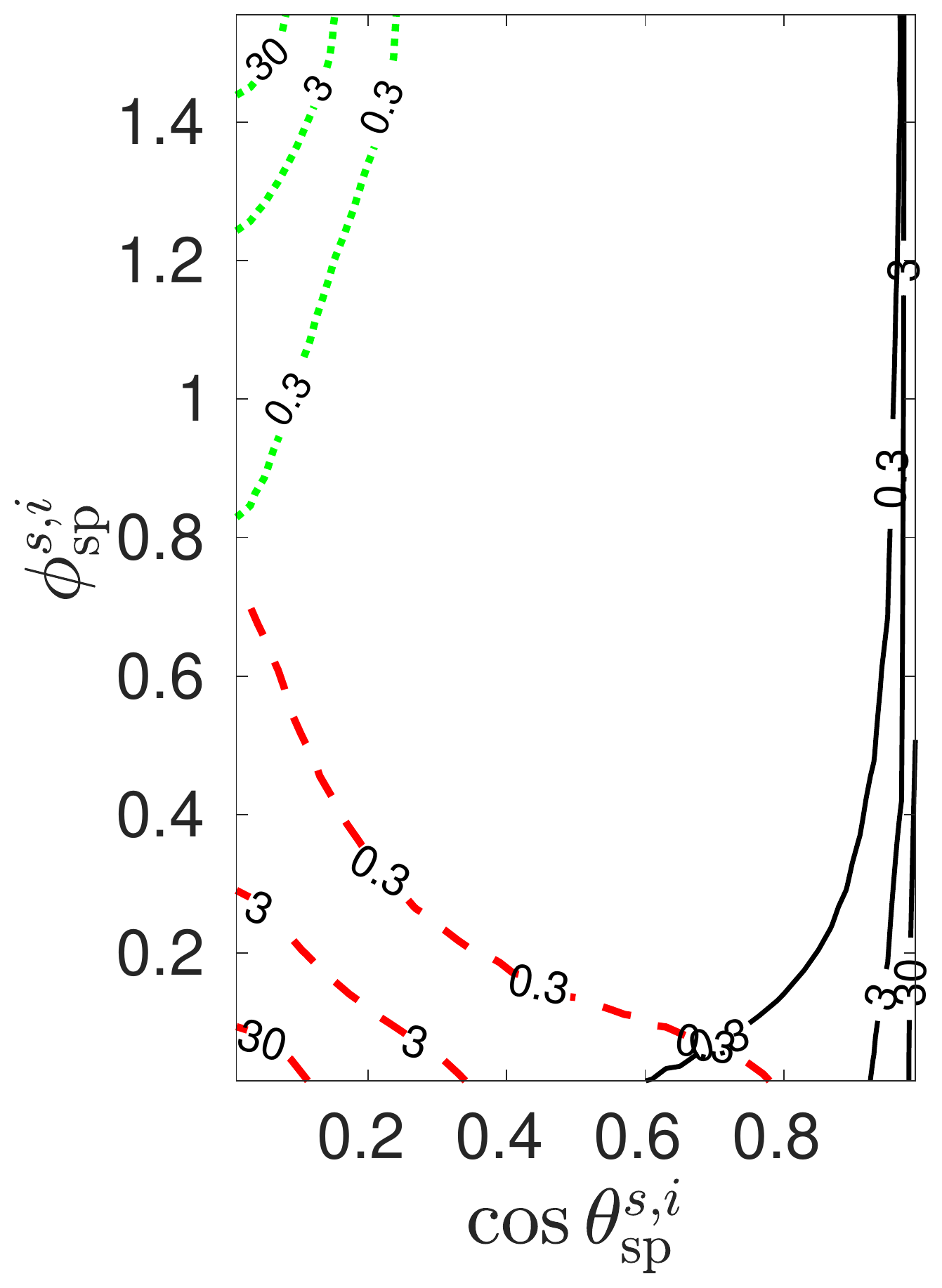}}
\emn %
\bmn{0.33\lnw}
  \centerline{\includegraphics[width=\lnw]{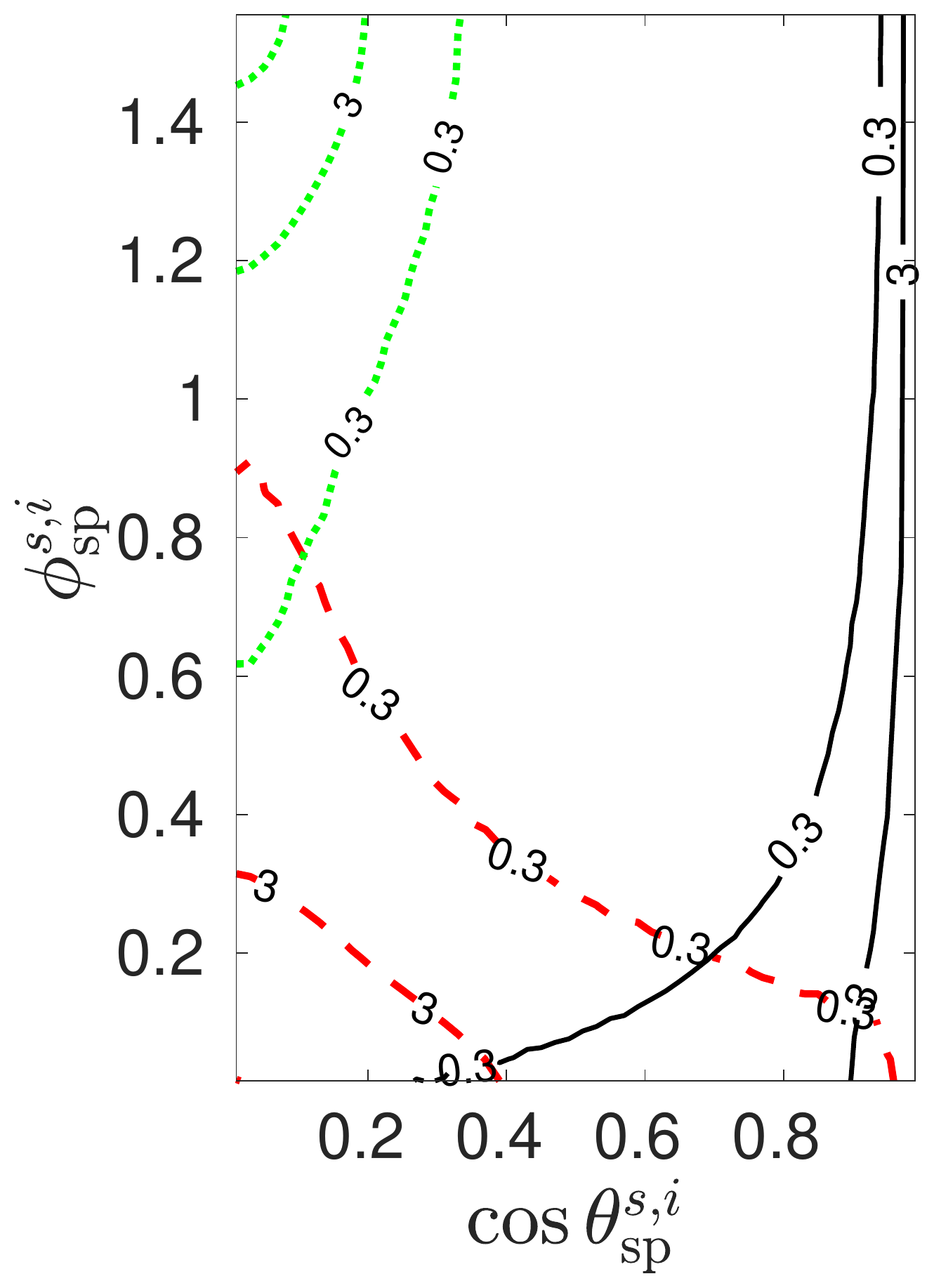}}
\emn
  \caption{The joint PDF of $(\cos\theta^{s,i}_{\rm sp},
  \phi^{s,i}_{\rm sp})$ for angles made by
  $\e^{s,\u}_i$ in the eigen-frame of $\ts^\v_{ij}$. Left: $\Delta =
  \Delta_m/2$ and $t = T$ from case C1; middle: $\Delta = \Delta_m/4$ and
  $t = T$ from case C1; right: $\Delta = \Delta_m/2$ and $t = T$ for case C6. Solid lines: $i=\alpha$;
  dashed lines: $i=\beta$; dotted lines: $i=\gamma$. The contour
  levels are $0.3$, $3$, and $30$.}
\label{fig:alignss}
\efig

\bfig
\bmn{0.33\lnw}
  \centerline{\includegraphics[width=\lnw]{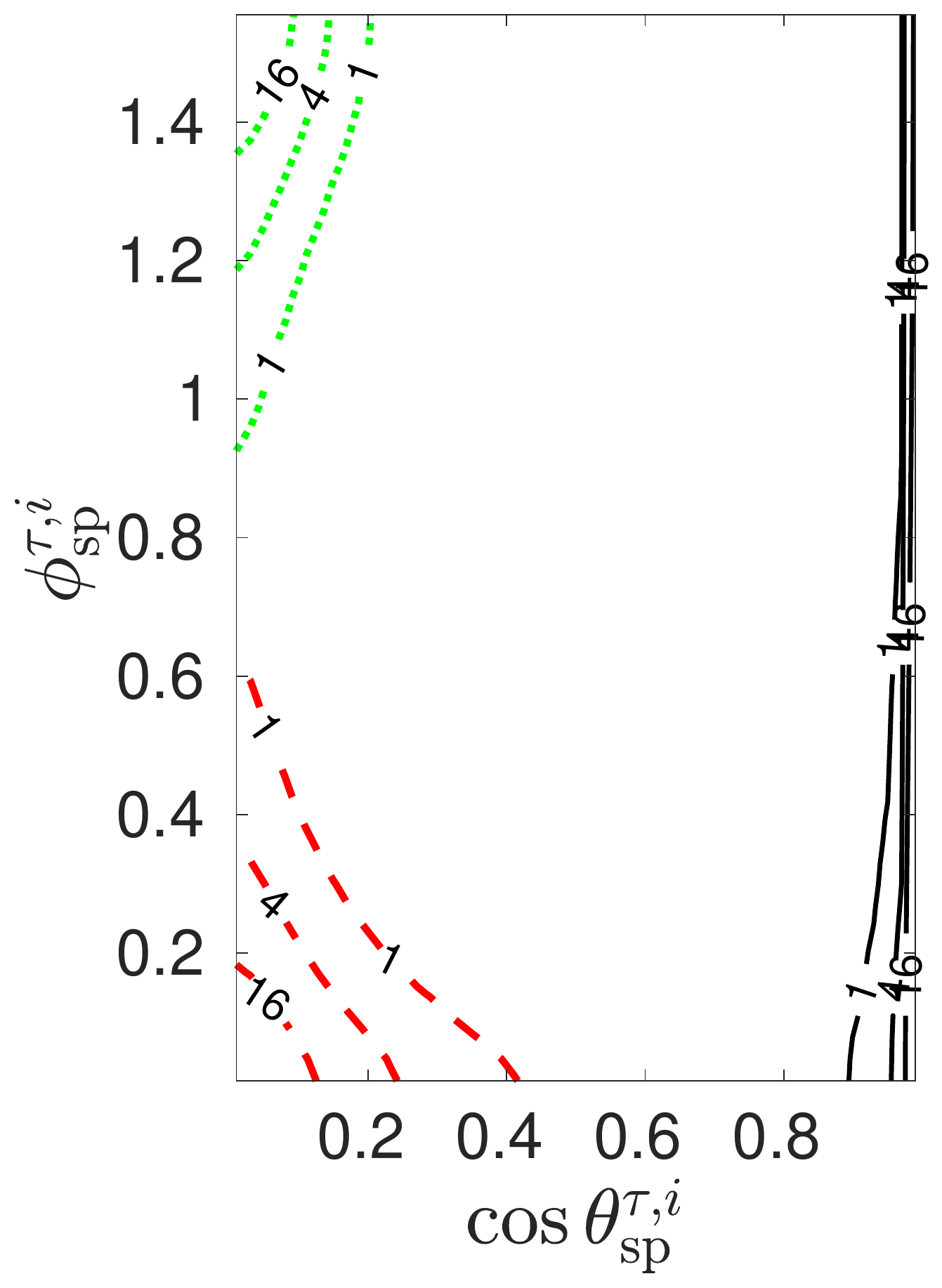}}
\emn %
\bmn{0.33\lnw}
  \centerline{\includegraphics[width=\lnw]{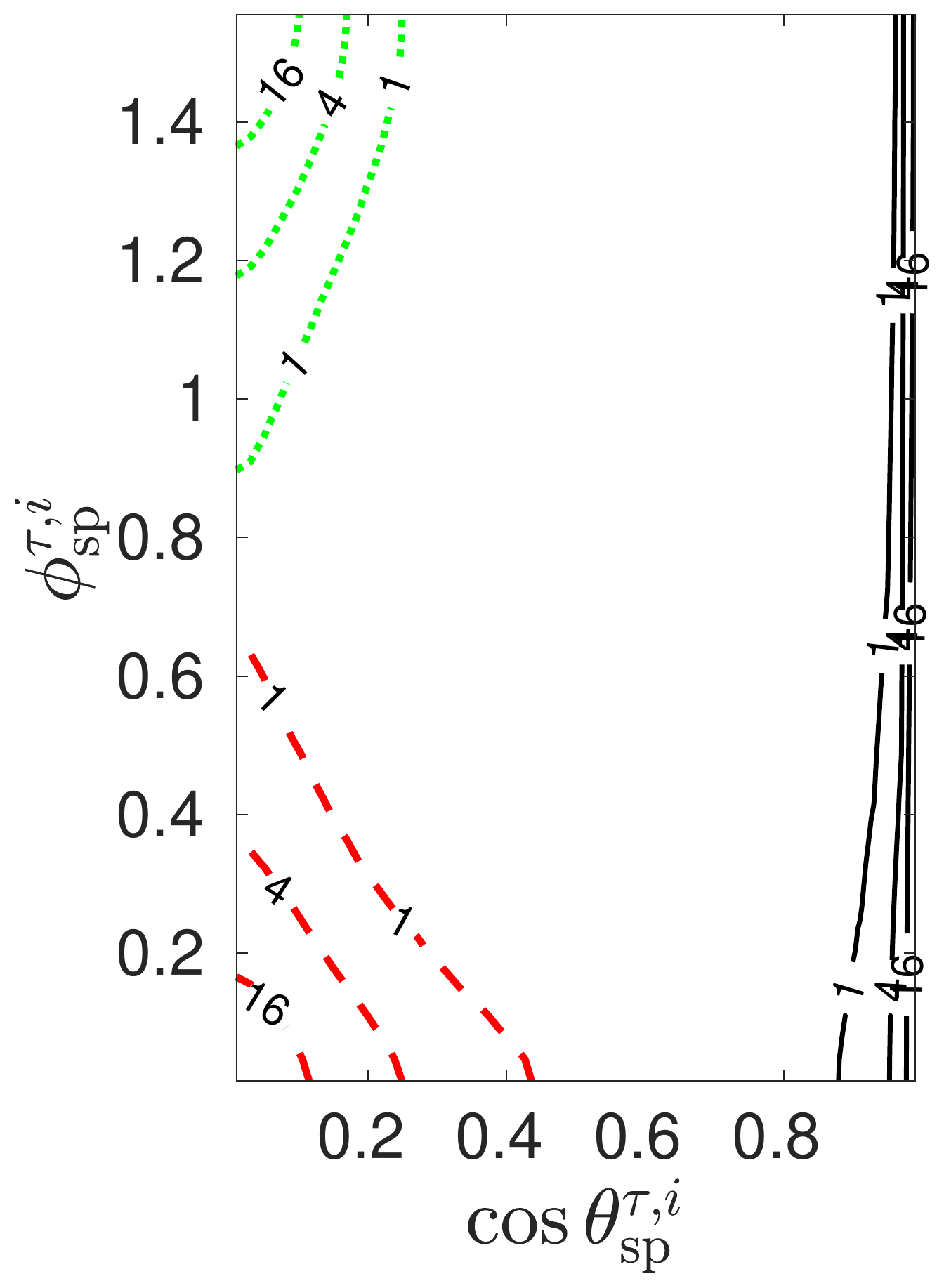}}
\emn %
\bmn{0.33\lnw}
  \centerline{\includegraphics[width=\lnw]{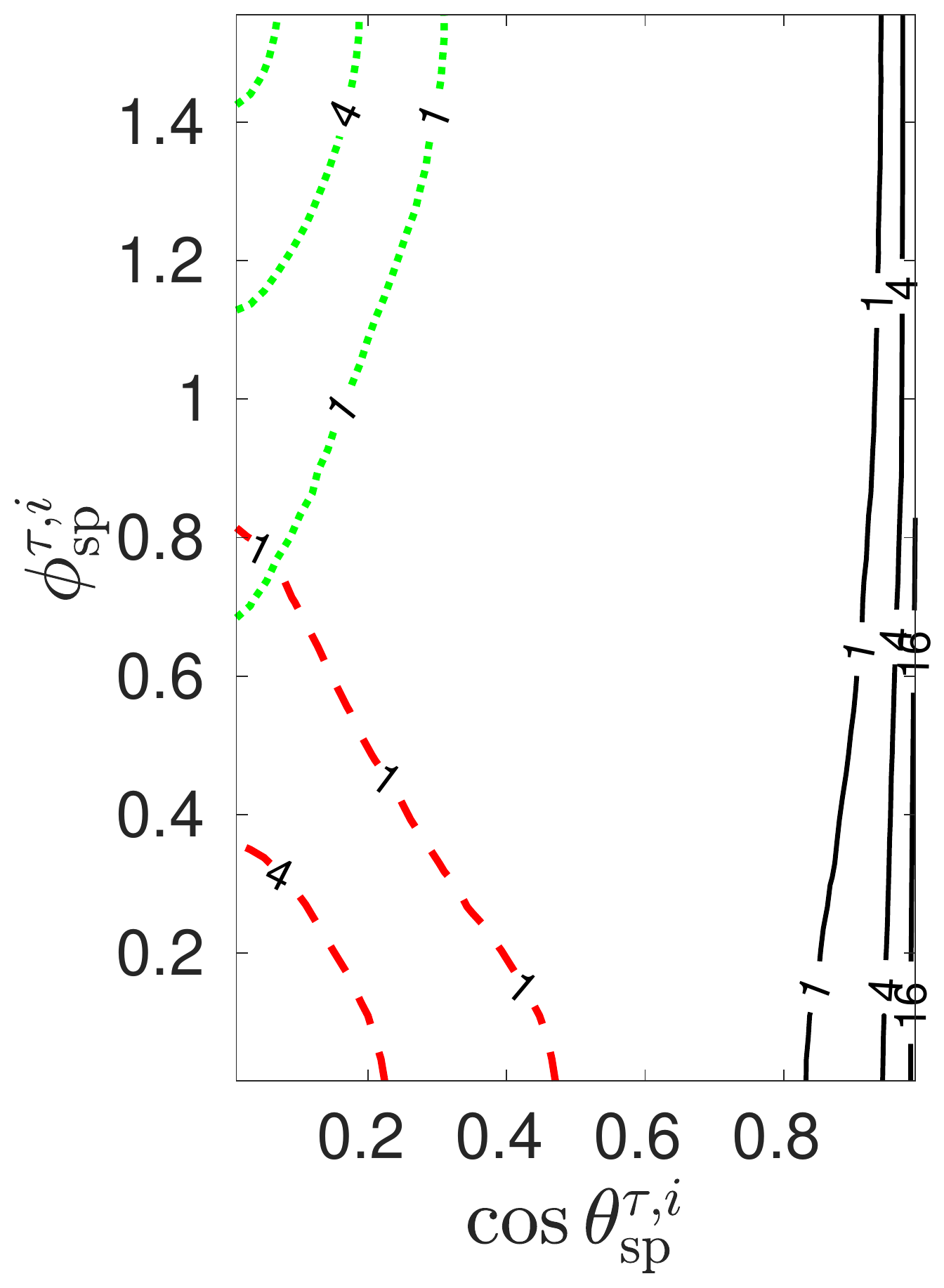}}
\emn
  \caption{The joint PDF of $(\cos\theta^{\tau,i}_{\rm sp},
  \phi^{\tau,i}_{\rm sp})$ for angles made by
  $\e^{\tau,\u}_i$ in the eigen-frame of $-\tau^{d,\v}_{ij}$. Left: $\Delta =
  \Delta_m/2$ and $t = T$ from case C1; middle: $\Delta = \Delta_m/4$ and
  $t = T$ from case C1; right: $\Delta = \Delta_m/2$ and $t = T$ from case C6. Solid lines: $i=\alpha$;
  dashed lines: $i=\beta$; dotted lines: $i=\gamma$. The contour
  levels are $1$, $4$, and $16$.}
\label{fig:aligntt}
\efig

The tensorial structures of $\ts^\u_{ij}$ and $\ts^\v_{ij}$
are compared using
the orientation of the eigenvectors of the former in the eigen-frame of the latter.
The alignment is given in the left panel of Fig. \ref{fig:alignss} for
$\Delta = \Delta_m/2$ and $T=\tau_L$,
in term of the joint PDFs of
$(\cos\theta^{s,i}_{\rm sp}, \phi^{s,i}_{\rm sp})$, where
$\theta^{s,i}_{\rm sp}$ and $\phi^{s,i}_{\rm sp}$ are the angles
made by $\e^{s,\u}_i$ in the eigen-frame of $\ts_{ij}^\v$ (c.f., Fig.
\ref{fig:eigenframe}).
According to the definitions of the angles, perfect alignment between
$\e^{s,\v}_\alpha$ and $\e^{s,\u}_\alpha$ corresponds to
$\cos\theta^{s,\alpha}_{\rm sp} = 1$ and an undetermined
$\phi^{s,\alpha}_{\rm sp}$. The solid lines in Fig.
\ref{fig:alignss} show a sharp ridge of peak values at
$\cos\theta^{s,i}_{\rm sp} = 1$, which indicates that $\e^{s,\v}_\alpha$
and $\e^{s,\u}_\alpha$ are strongly aligned.
Perfect
alignment between $\e^{s,\v}_\beta$ and $\e^{s,\u}_\beta$ corresponds to
a peak at $(0,0)$, whereas the perfect alignment between $\e^{s,\v}_\gamma$ and
$\e^{s,\u}_\gamma$ corresponds to a peak at $(0,\pi/2)$. The joint PDFs in these two cases
(shown with dashed and dotted lines, respectively) indeed display high
peak probabilities around these points. Therefore, these joint
PDFs demonstrate excellent agreement between the orientations of
$\ts^\u_{ij}$ and $\ts^\v_{ij}$.
The results at $\Delta =
\Delta_m/4$ given in the middle shows that strong alignment persists to smaller
filter scales. The joint PDFs obtained with $T=0.5\tau_L$, shown on the
right, have much weaker peaks, indicating a larger discrepancy in the
orientation of the two tensors.

Fig. \ref{fig:aligntt} plots the same statistics as those in Fig.
\ref{fig:alignss}, but for the SGS
stress tensors $-\tau^{d,\v}_{ij}$ and $-\tau^{d,\u}_{ij}$. The
observations
made from Fig. \ref{fig:alignss} are also observed in Fig. \ref{fig:aligntt},
except that the alignment shown in the latter is
in general somewhat weaker.

\subsubsection{Correlations}

We will look into the covariance and the correlation coefficients in
this subsection.
For two random
variables $X$ and $Y$, they are
denoted by $C(X,Y)$ and
$\rho(X,Y)$, respectively. By standard definitions,
\be
C(X,Y) \equiv  \lal (X - \lal X \ral)(Y- \lal Y \ral ) \ral,
\q \rho(X,Y) \equiv \frac{C(X,Y)}{\sigma_X \sigma_Y},
\ee
where $\sigma_X \equiv \sqrt{C(X,X)}$ is the RMS value of $X$ and $\sigma_Y$ is defined
similarly.
The physical quantities in $\u(\x,t)$ and $\v(\x,t)$ will be
denoted with superscripts $\u$ and $\v$, respectively.
The difference in $X$ is quantified by the RMS
value of $X^\u - X^\v$, $\sigma_{X^\u - X^\v}$, which will be normalized by the RMS
value of $X$ in the target fields $\sigma_{X^\v}$.

\bfig
\bmn{0.5\lnw}
  \centerline{\includegraphics[width=\lnw]{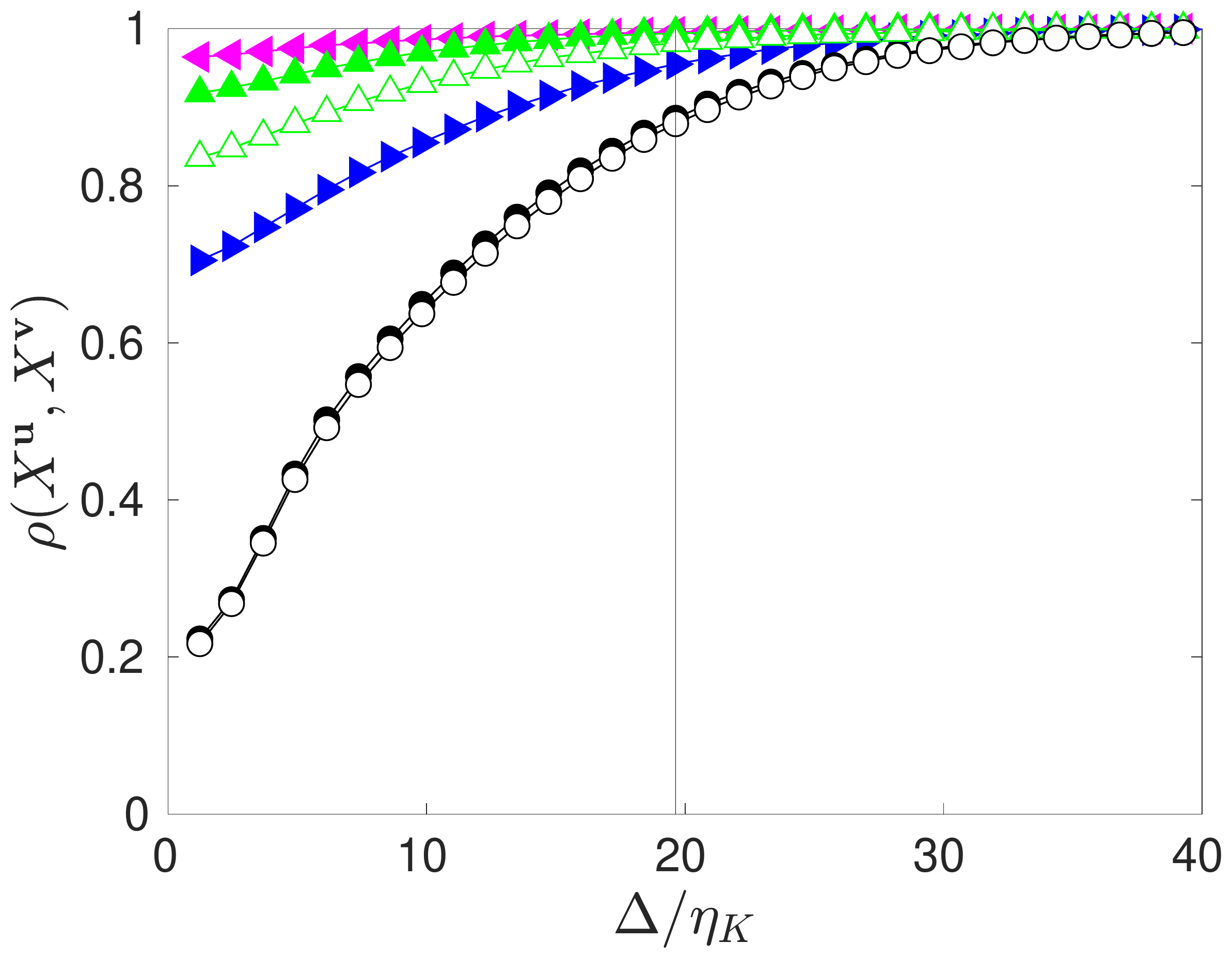}}
  \emn %
  \bmn{0.5\lnw}
  \centerline{\includegraphics[width=\lnw]{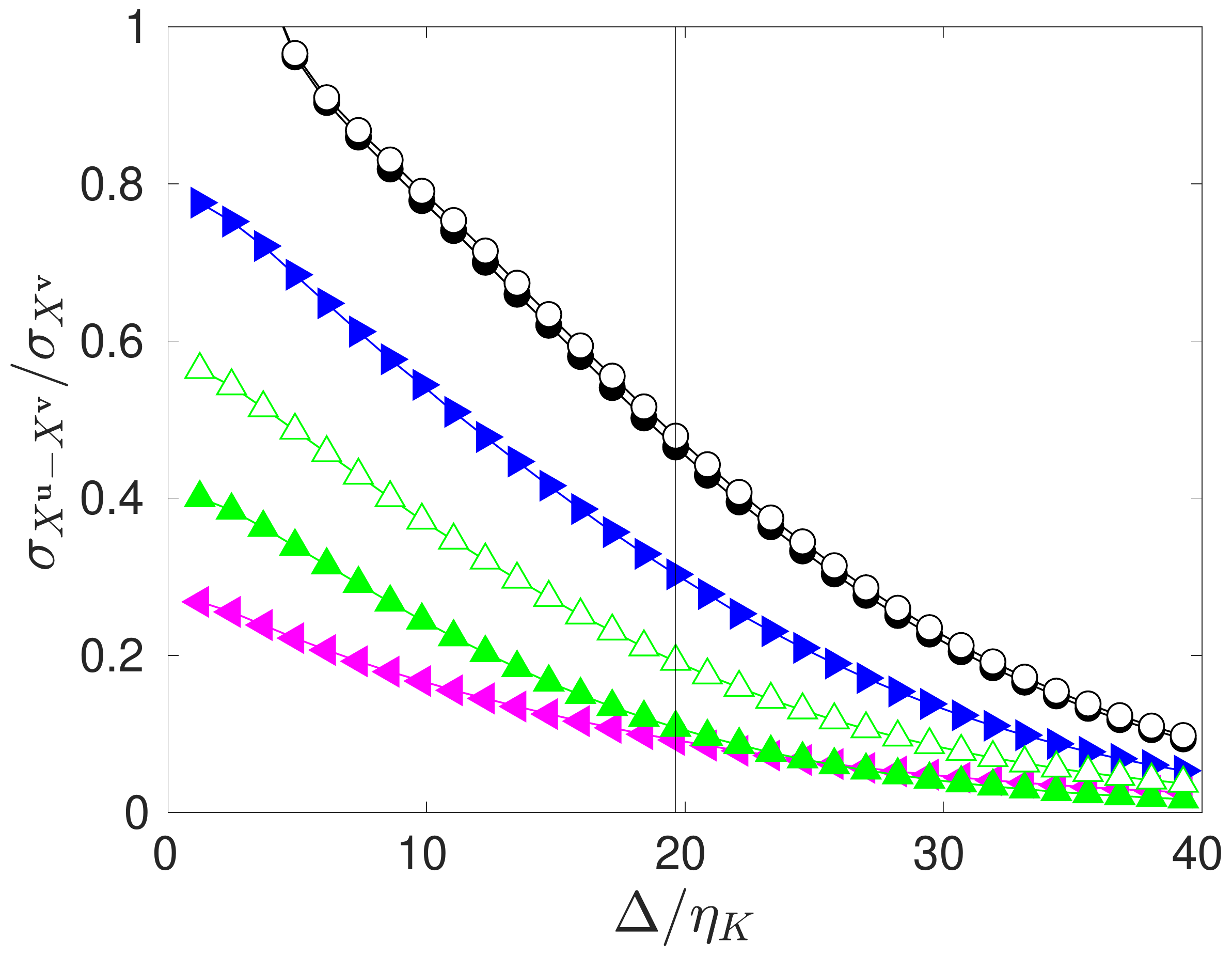}}
  \emn
  \caption{The correlation coefficient $\rho(X^\u, X^\v)$ (left)
and the normalized RMS value of $X^\u-X^\v$
(right) as a function of $\Delta/\eta_K$, where
  $X=\ts_{ij}\ts_{ij}$.
  Solid symbols: from case C1; open symbols: from case C6.
  Circles: $t=0$; upright triangles: $t=0.5\tau_L$; left-pointing
  triangles: $t=\tau_L$ (for case C1 only). Right-pointing solid triangles: from the DS method at
  $t=\tau_L$ for $k_m=4$. The vertical line
  marks the scale $\Delta = \Delta_m$.}
\label{fig:ss-covar}
\efig

The left panel in Fig. \ref{fig:ss-covar} shows
$\rho(X^\u, X^\v)$ where $X\equiv \ts_{ij}\ts_{ij}$.
The correlation deteriorates with decreasing filter
scales, but improves with $t/T$ for $t$ within the optimization
horizon. Excellent correlation across all scales is achieved at
$t/T=1$ for $T=\tau_L$.
The difference in
$X = \ts_{ij}\ts_{ij}$ is also examined in the right panel in terms of the RMS value of
$X^\u - X^\v$ normalized by the RMS value of $X^\v$.
For $t=T=\tau_L$,
the value can be reduced to about $27\%$ at the Kolmogorov scale where
$\Delta/\eta_K \approx 1$.
For results from the DS method (right-pointing solid triangles),
the correlation $\rho$ at the smallest $\Delta/\eta_K$ is about
$70\%$ and the RMS difference is about $80\%$, indicating significantly higher discrepancy.
This contrast is consistently observed in other statistics. Therefore, we will not present more
results from the DS method.

\bfig
\bmn{0.5\lnw}
  \centerline{\includegraphics[width=\lnw]{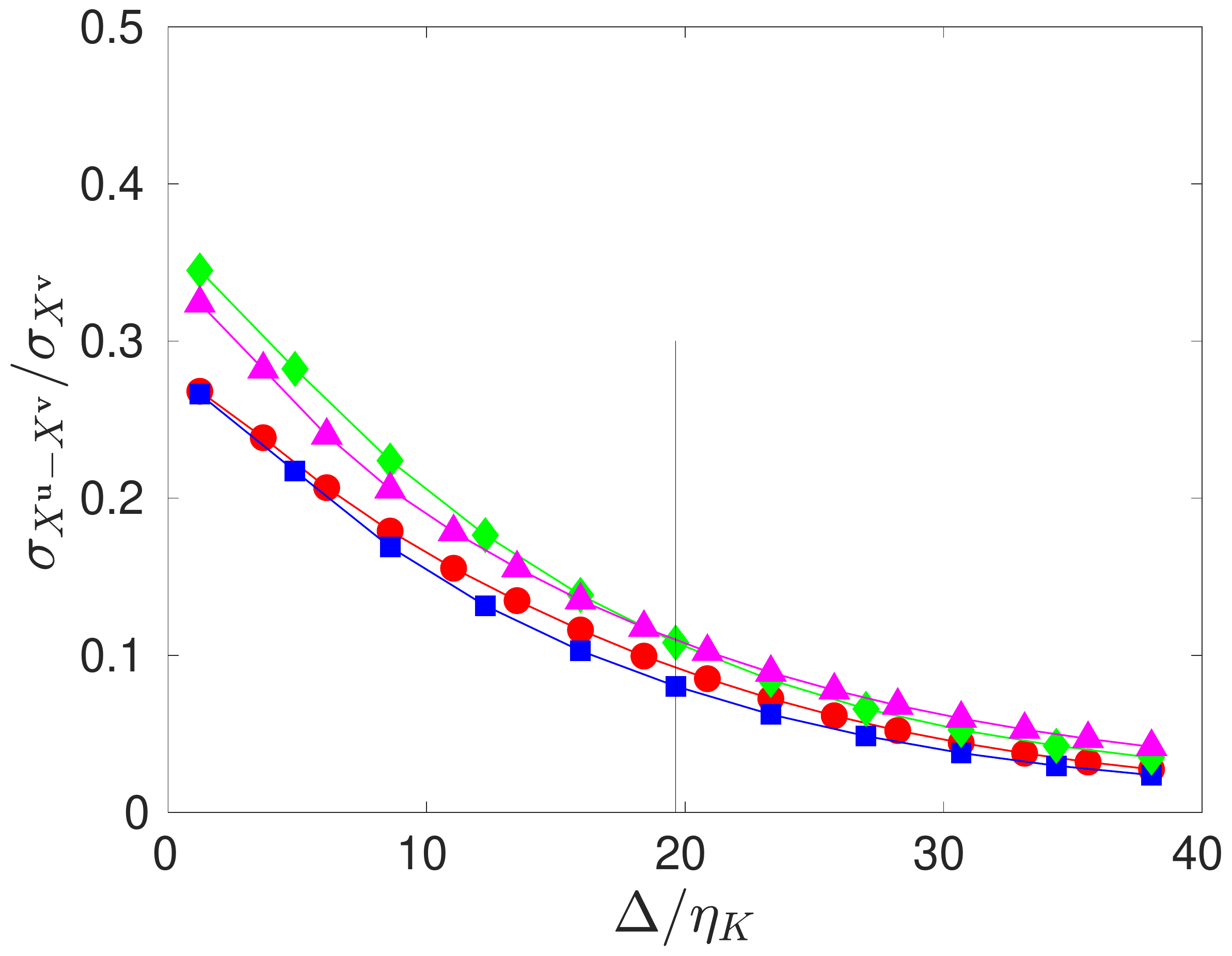}}
  \emn %
  \bmn{0.5\lnw}
  \centerline{\includegraphics[width=\lnw]{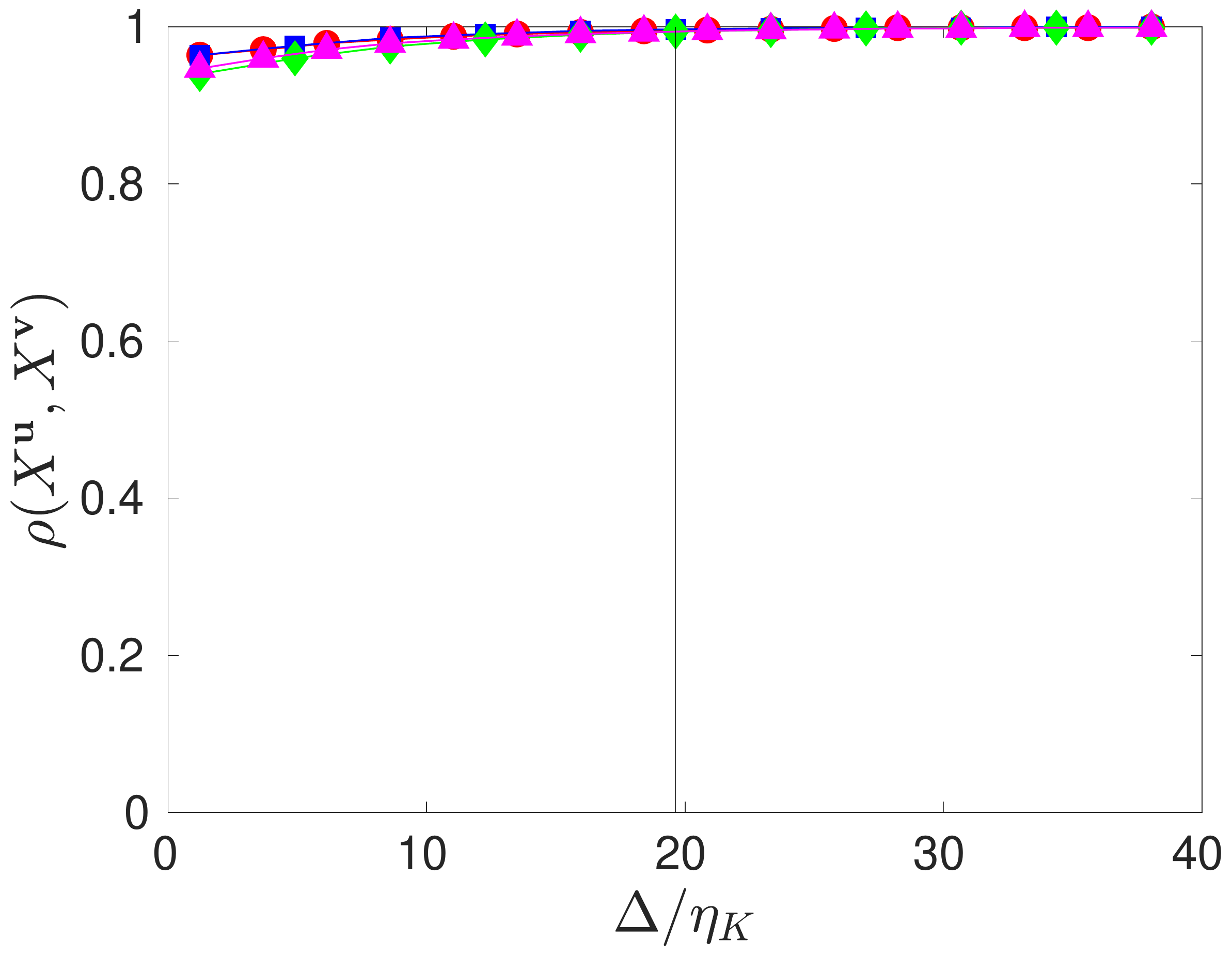}}
  \emn
  \caption{Left: The normalized
  RMS values of the differences $X^\u-X^\v$ from case C1, where $X\equiv \ts_{ij}\ts_{ij}$ (circles),
  $\tog_i\tog_i$ (squares), $P_\omega(\equiv \ts_{ij}\tog_i \tog_j)$ (diamonds), and
  $\Pi_\Delta$ (triangles), respectively. Right: the correlation
  coefficients $\rho(X^\u,X^\v)$ from case C1. The vertical line
  marks the scale $\Delta = \Delta_m$.}
\label{fig:m-diff-relvar}
\efig

$\sigma_{X^\u-X^\v}/\sigma_{X^\v}$ and $\rho(X^\u, X^\v)$ for
$X = \tog_i\tog_i$, $P_\omega$ and $\Pi_\Delta$ are plotted in
Fig. \ref{fig:m-diff-relvar}, where the results for $\ts_{ij}\ts_{ij}$
are also given as comparison. For all quantities, almost perfect
correlation is observed. The RMS difference for
$\tog_i \tog_i$
is similar to $\ts_{ij}\ts_{ij}$, both reaching about $27\%$ at the
Kolmogorov scale where $\Delta/\eta_K \approx 1$.
The discrepancy for $\Pi_\Delta$ and $P_\om $ is somewhat larger with the normalized RMS
difference found at about $35\%$ at the smallest filter scale.

%TODO: SGS dissipation
%Difference in SGS ED: we have correlations; conditional average; joint pdf of the ED; conditional average of
%increments; joint PDF of increment; averaged difference over scales; cross calculation.

\subsubsection{Joint PDFs and conditional statistics}
\bfig
  \centerline{\includegraphics[width=0.6\lnw]{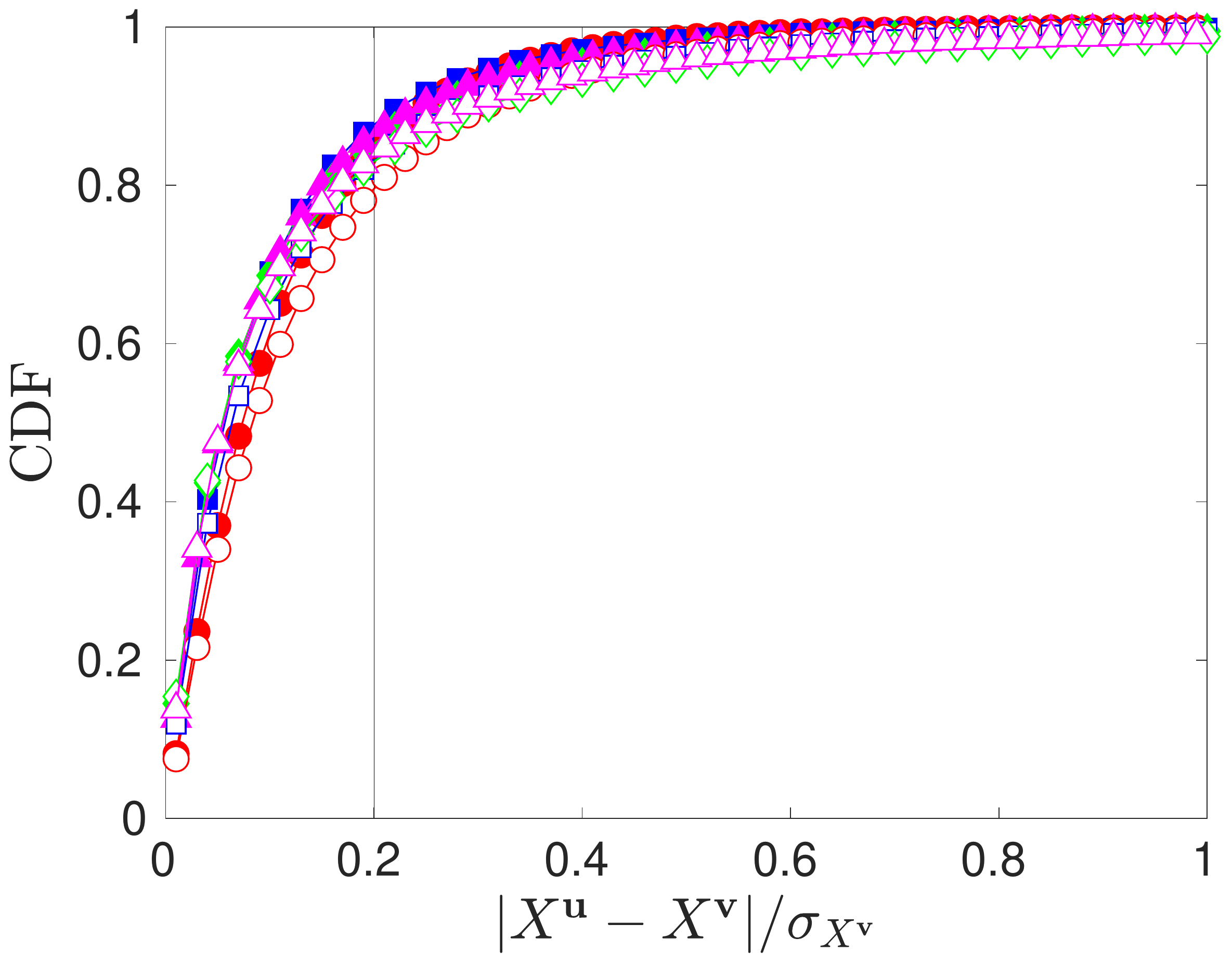}}
  \caption{The cumulative distribution functions for the normalized differences
  $|X^\u-X^\v|$ from case C1, where $X = \ts_{ij}\ts_{ij}$ (circles),
  $\tog_i\tog_i$ (squares), $P_\omega\equiv \ts_{ij}\tog_i\tog_j$ (diamonds), and
  $\Pi_\Delta$ (triangles). Filled symbols: $\Delta = \Delta_m/2$;
  empty symbols: $\Delta = \Delta_m/4$. The vertical line
  marks the $20\%$ threshold.}
\label{fig:cdf-absdiff-dx-cs400}
\efig

To examine the large fluctuations in the 
differences between the target and reconstructed parameters, we look into the
joint PDFs and conditional statistics in this subsection. 
Fig. \ref{fig:cdf-absdiff-dx-cs400} plots the cumulative distribution
functions for the normalized differences.
The results for both $\Delta = \Delta_m/2$ and $\Delta_m/4$ are given.
The vertical line indicates an arbitrarily chosen $20\%$
threshold.
The probabilities at $\Delta = \Delta_m/4$ are slightly lower than those at
$\Delta_m/2$, and those for $\ts_{ij}\ts_{ij}$ are slightly
lower than those for the other quantities.
The probability to observe a less than $20\%$ difference ranges
from $75\%$ to $90\%$ for different quantities.
%The results for $P_\omega$ and
%$\Pi_\Delta$ (shown in diamonds and triangles, respectively) essentially fall on top of each other.
%For both, there is a $65\%$ chance to find
%the normalized difference to be smaller than $20\%$, and the probability remains the same for the
%two filter scales.
%For $\ts_{ij} \ts_{ij}$ and $\tog_i\tog_i$, the probabilities to
%observe small fluctuations are slightly smaller, and they decrease
%with the filter scales.
%The normalized difference for $\ts_{ij}\ts_{ij}$ is less than $20\%$
%for about $55\%$ of the time for $\Delta = \Delta_m/2$. At $\Delta =
%\Delta_m/4$, the probability decreases to about $45\%$.

\begin{figure}
  \bmn{0.5\lnw}
  \centerline{\includegraphics[width=\lnw]{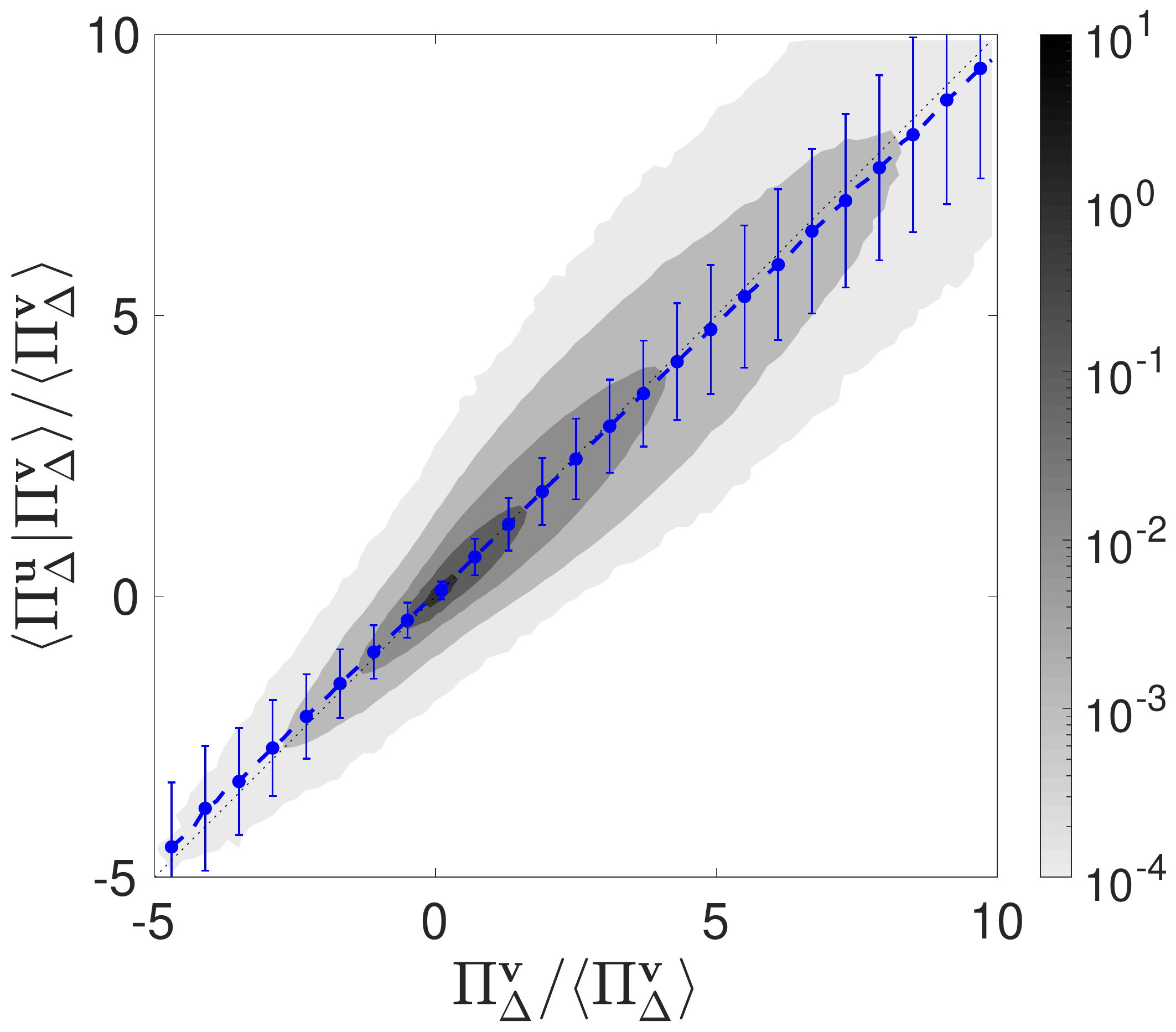}}
  \emn%
  \bmn{0.5\lnw}
  \centerline{\includegraphics[width=\lnw]{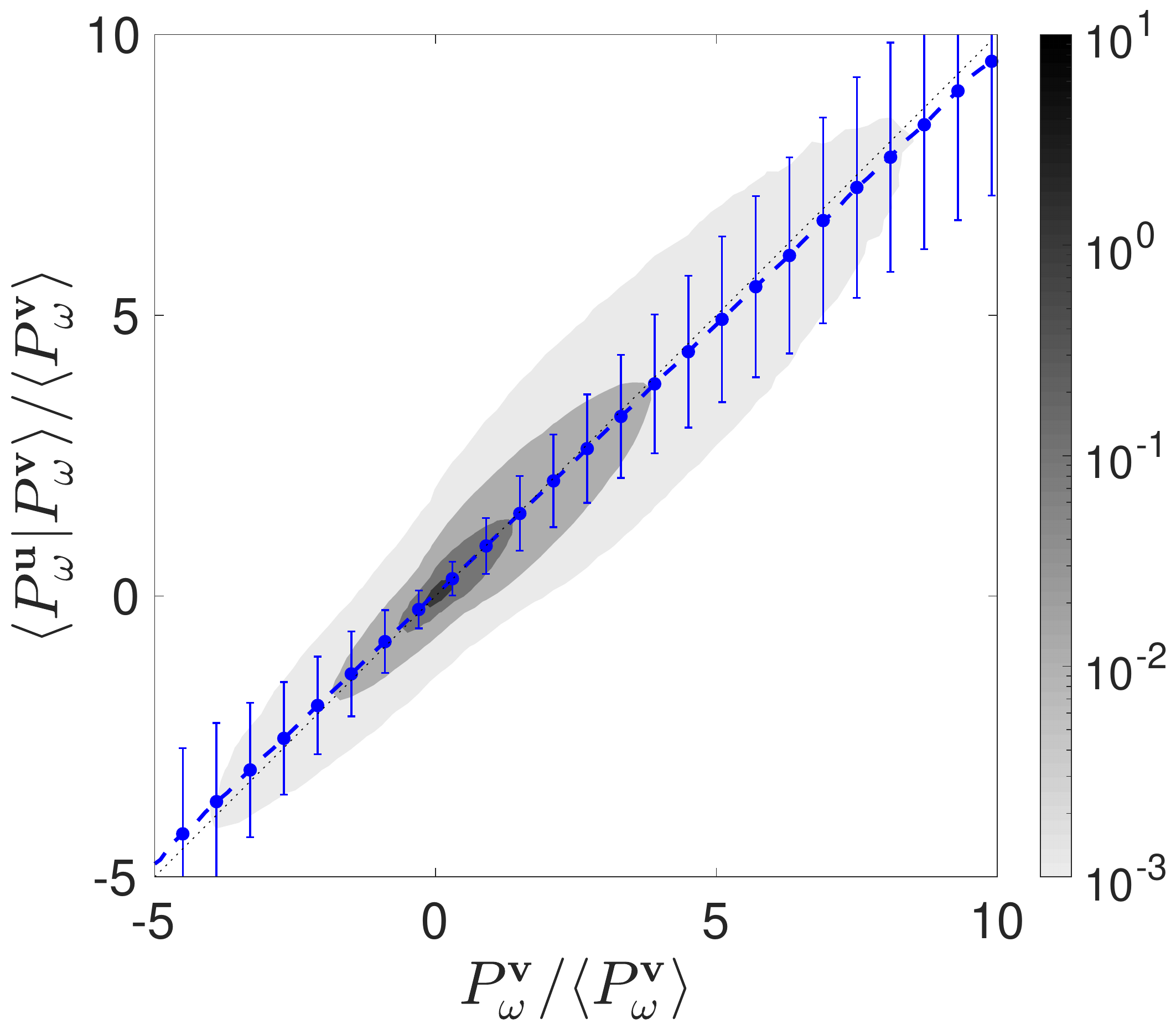}}
  \emn
  \caption{Left: the averaged $\Pi_\Delta$ in $\u(\x,t)$
  conditioned on $\Pi_\Delta$ in $\v(\x,t)$ (circles) and the joint PDF of the two (contours). 
  The error bars are the conditional RMS values of $\Pi_\Delta^\u /\lal
  \Pi^\v_\Delta\ral$. 
  Right: the same for $P_\omega$. For case C1 with 
  $\Delta = \Delta_m/4$ and $t=T$. }
\label{fig:jpdfsgsed-cnd-tgt-400cs400-8dx}
\end{figure}

\begin{figure}
  \bmn{0.5\lnw}
  \centerline{\includegraphics[width=\lnw]{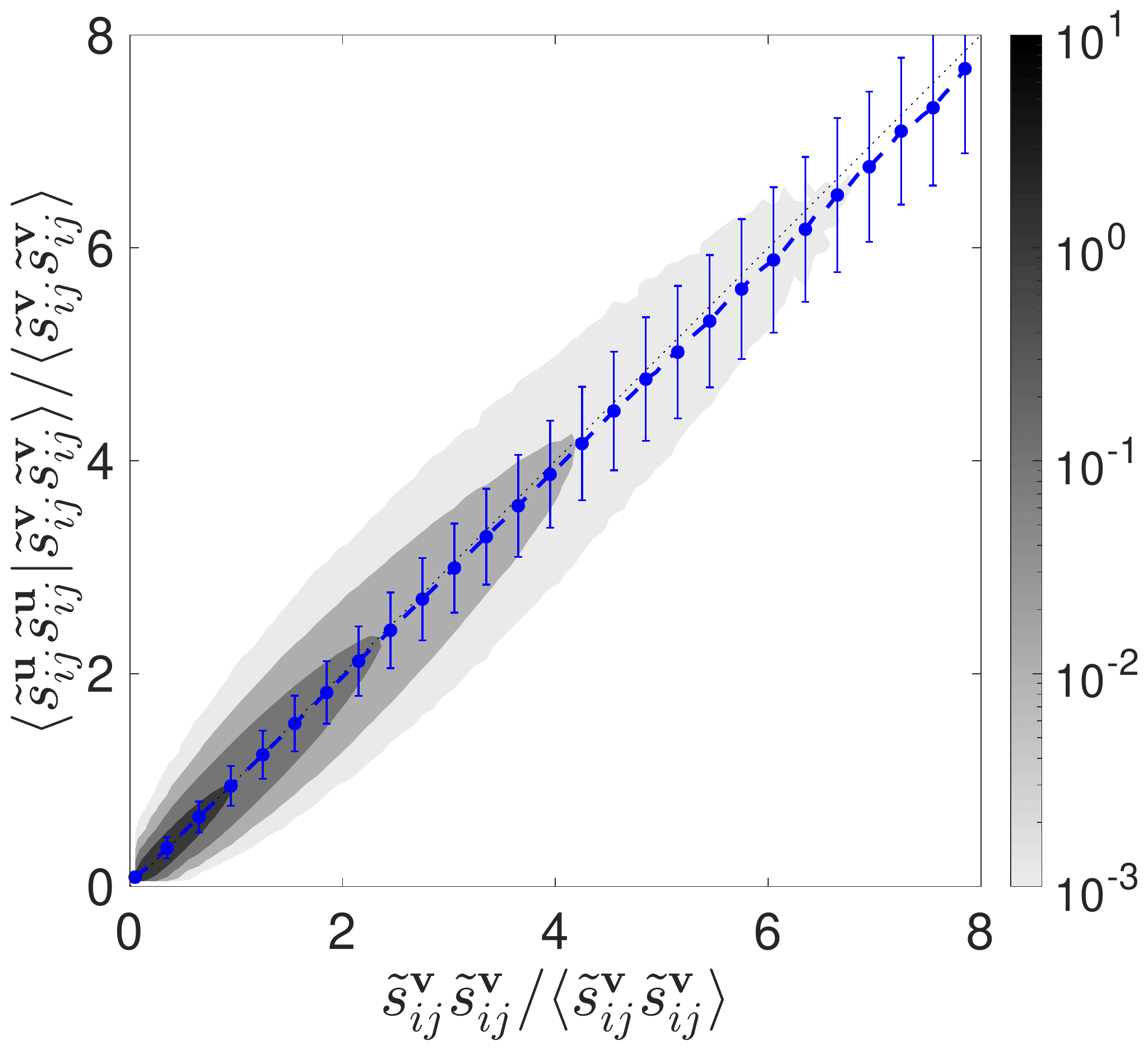}}
  \emn%
  \bmn{0.5\lnw}
  \centerline{\includegraphics[width=\lnw]{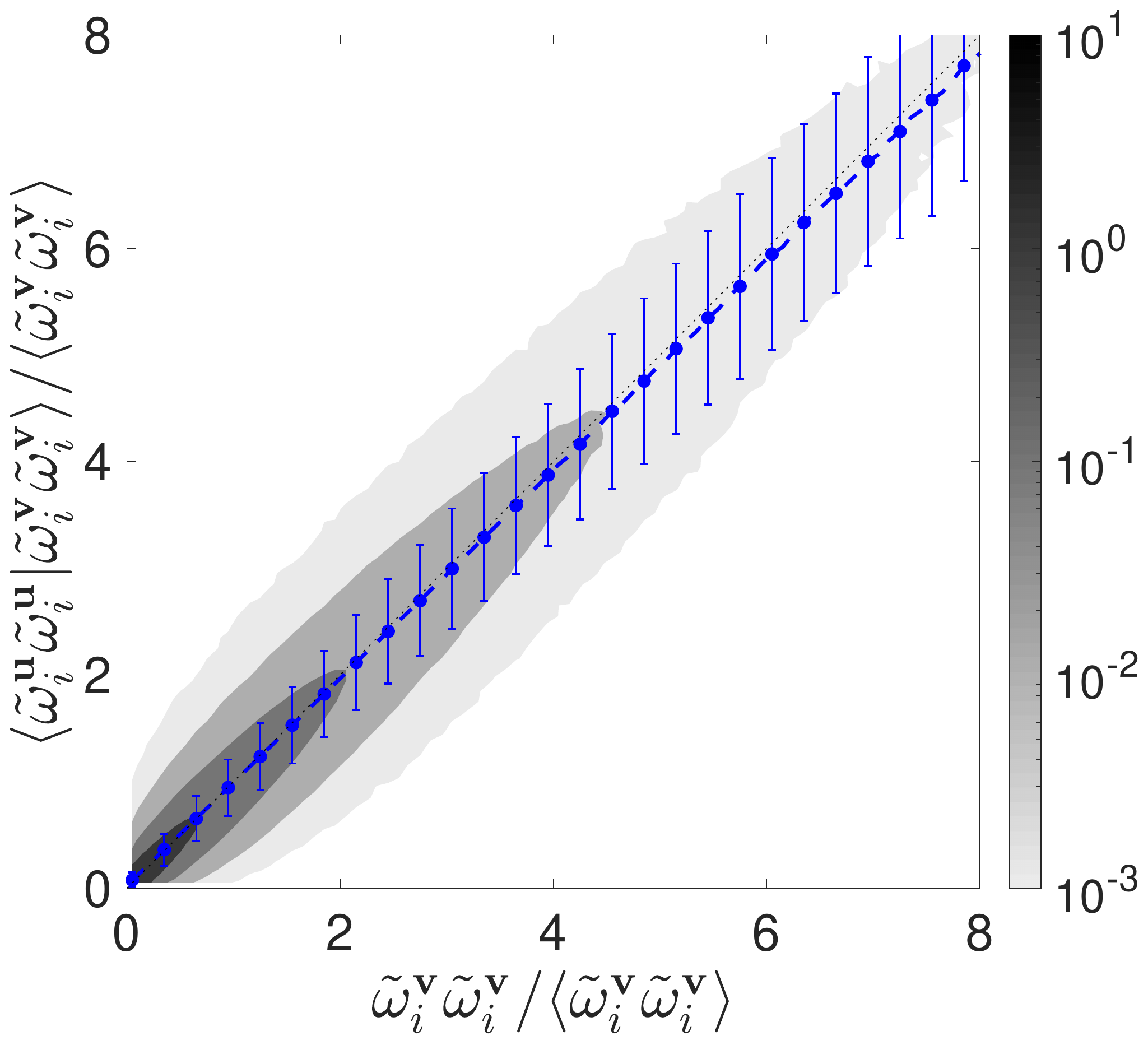}}
  \emn
  \caption{Same as Fig. \ref{fig:jpdfsgsed-cnd-tgt-400cs400-8dx} but for $\ts_{ij}\ts_{ij}$ (left)
  and $\tog_i \tog_i$ (right).
  }
\label{fig:mooss-cnd-tgt-400cs400-8dx}
\end{figure}

The conditional and joint statistics, e.g., the average of $X^\u$ conditioned on $X^\v$ and
the joint PDF of the two, are presented next.
The left panel of Fig. \ref{fig:jpdfsgsed-cnd-tgt-400cs400-8dx} is the results for $X \equiv
\Pi_\Delta$ with $\Delta = \Delta_m/4$. For a perfectly reconstructed field, the
conditional average will fall on the diagonal (dotted
line).
Therefore, the figure shows that $\Pi_\Delta^\u$ slightly underestimates
the magnitude of $\Pi^\v_\Delta$, and the underestimate increases slightly with the magnitude of
$\Pi_\Delta^\v$.
Nevertheless, the agreement is very good. The joint PDF does concentrate around the diagonal,
although there is significant scattering. The scattering is indicated by the error bars, which represent $\pm
\sigma_{\Pi_\Delta^\u|\Pi_\Delta^\v}/\lal
\Pi_\Delta^\v\ral$ with $\sigma_{\Pi_\Delta^\u | \Pi_\Delta^\v} \equiv [\lal
(\Pi_\Delta^\u)^2|\Pi_\Delta^\v\ral - \lal \Pi_\Delta^\u|\Pi_\Delta^\v\ral^2]^{1/2}$ 
being the conditional RMS value of $\Pi_\Delta^\u$. The error bar increases mildly with the magnitude of
$\Pi^\v_\Delta$.  
The right panel of Fig. \ref{fig:jpdfsgsed-cnd-tgt-400cs400-8dx} 
plots the same results for $P_\omega\equiv \ts_{ij}\tog_i\tog_j$. 
The results for enstrophy and strain rate are shown in Fig.
\ref{fig:mooss-cnd-tgt-400cs400-8dx}. Good agreement is also observed for these quantities. 
The error bars are slightly larger for $P^\u_\omega$ and $\tog^\u_i\tog^\u_i$, compared with the
other two quantities. 

\subsection{Reconstruction of non-local structures}

Fig. \ref{fig:oo-contour} has shown that
the shape,
orientation and locations
of the strong vortices in the reconstructed field can be remarkably similar to those in the target
field.
We quantify the similarity of the vortical and other structures using MVEEs and MVEE trees in
this subsection.

\subsubsection{Vortical structures}

The calculation of MVEEs have been explained in Section \ref{sect:cs_intro}.
Let the directions of the axes of an MVEE be denoted by $\e_i$ ($i = \al, \beta,
\gamma$), whereas the
length of axis $i$ be denoted by $\ell_i$. The indices $\alpha$, $\beta$
and $\gamma$ 
denote the major, intermediate and the minor axes, respectively, so that $\ell_\al \ge \ell_\beta \ge
\ell_\gamma$. As before, the parameters in the target fields are indicated by
superscript/subscript $\v$ whereas those in
the reconstructed fields by $\u$.

We now consider strong vortical structures where
$\tog \ge 3 \lal \tog_i\tog_i\ral^{1/2}$ with
filter scale $\Delta = \Delta_m/2$ and $\Delta_m/4$.
As it turns out, the number of isolated structures in a field may be too large
to process. If this happens, only the largest $N_s \equiv 10$ structures in the
target $\v$ and their matching structures in $\u$ are used.
Only the data in the five realizations of case C1 are
analyzed. There are thus in total
$N_e\equiv5N_s = 50$ pairs of matching MVEEs at each time step.
As the number of samples is modest,
medians instead of means are used to delineate the average behaviors,
because the former is less sensitive to extraneous fluctuations.
Additional details are given by percentiles and histograms.

\bfig
\bmn {0.5\lnw}
  \centerline{\includegraphics[width=\lnw]{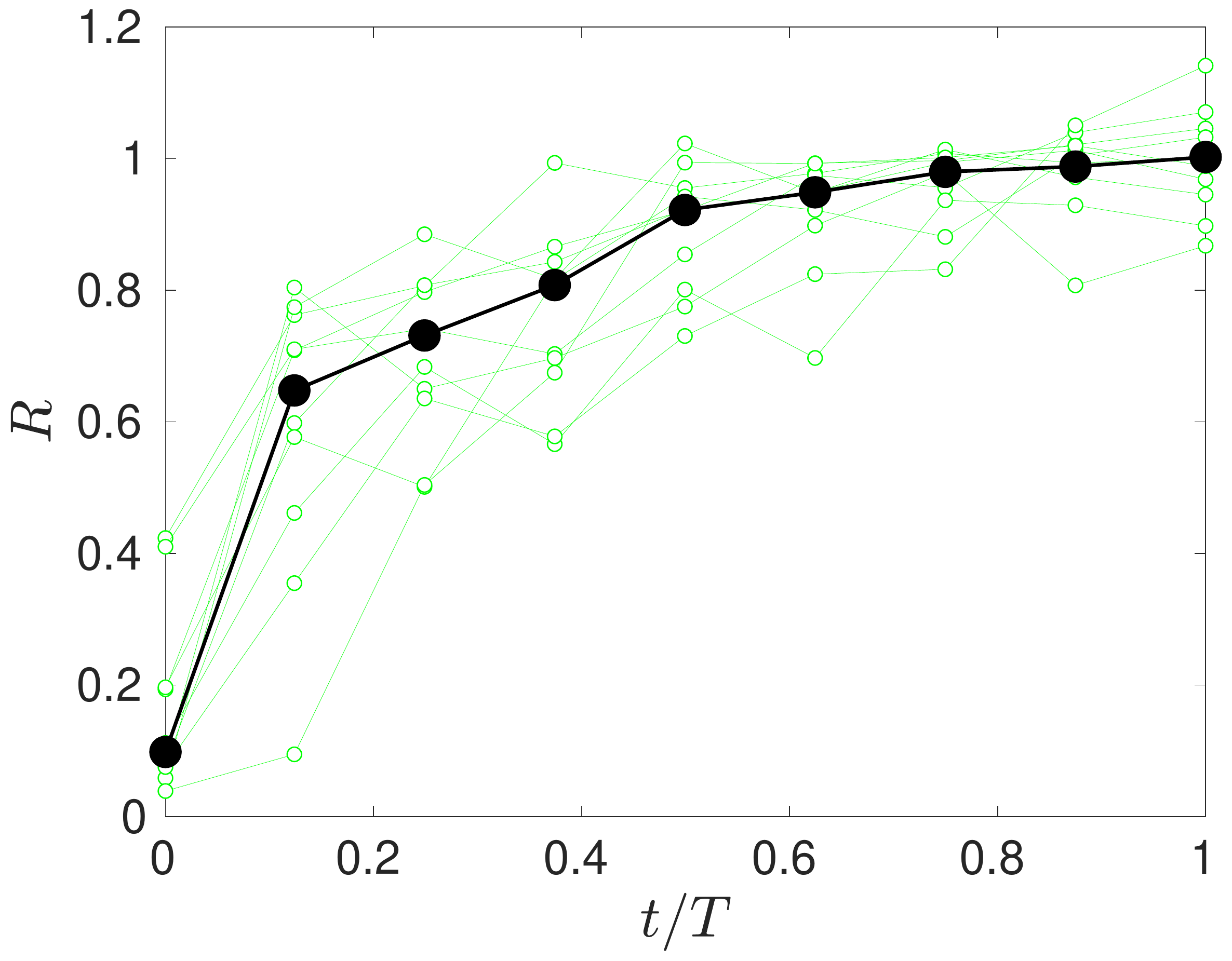}}
\emn %
\bmn {0.5\lnw}
  \centerline{\includegraphics[width=\lnw]{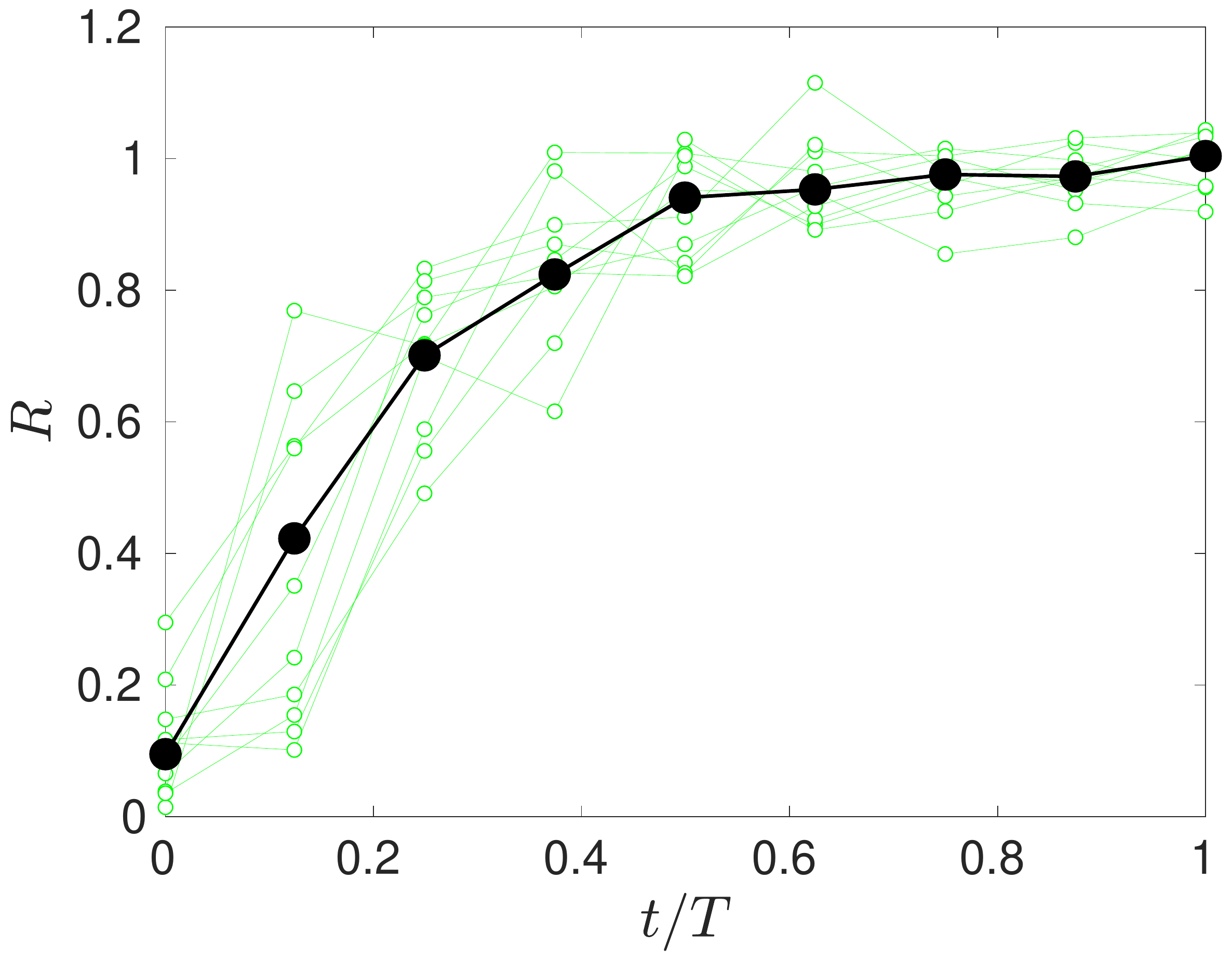}}
\emn
  \caption{The median $R$ of the ratio between the numbers of grid points in two matching
  structures (circles with thick line). The green thin lines show the medians in different size groups. Left: $\Delta =
  \Delta_m/2$; right: $\Delta = \Delta_m/4$. From case C1. }
\label{fig:sizes8dx}
\efig
\bfig
\centering
\bmn {0.33\lnw}
  \centerline{\includegraphics[width=\lnw]{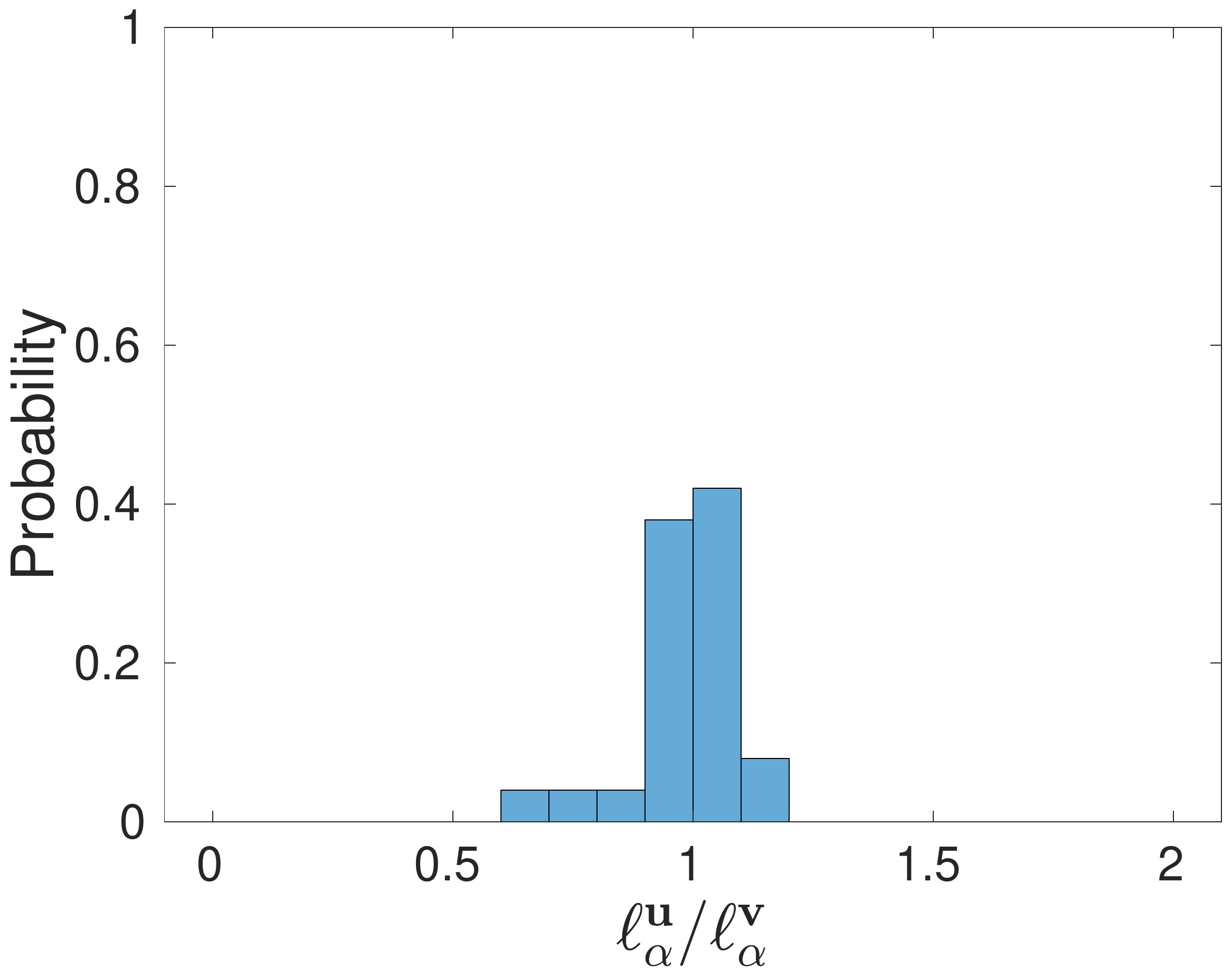}}
\emn %
\bmn {0.33\lnw}
  \centerline{\includegraphics[width=\lnw]{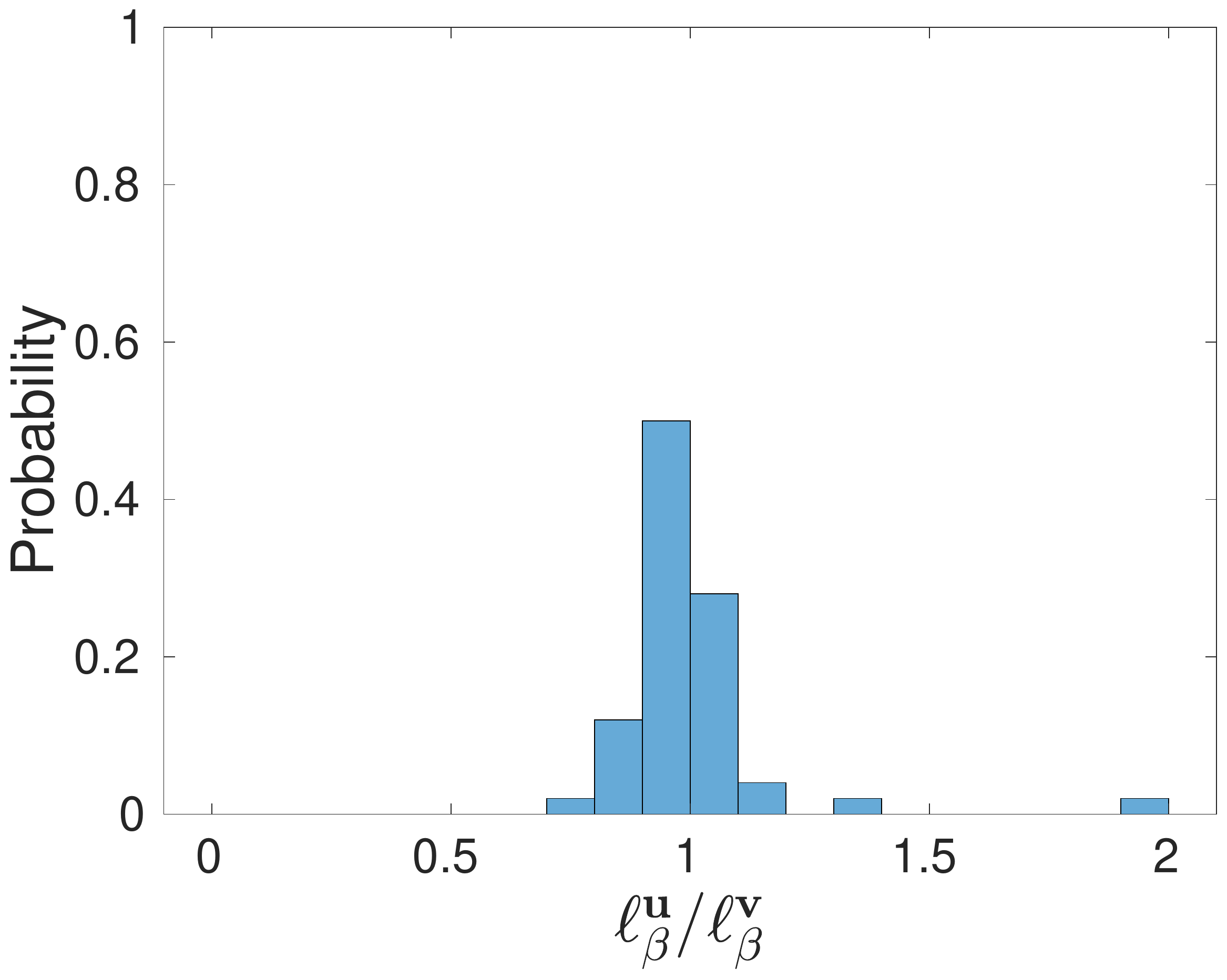}}
\emn %
\bmn {0.33\lnw}
  \centerline{\includegraphics[width=\lnw]{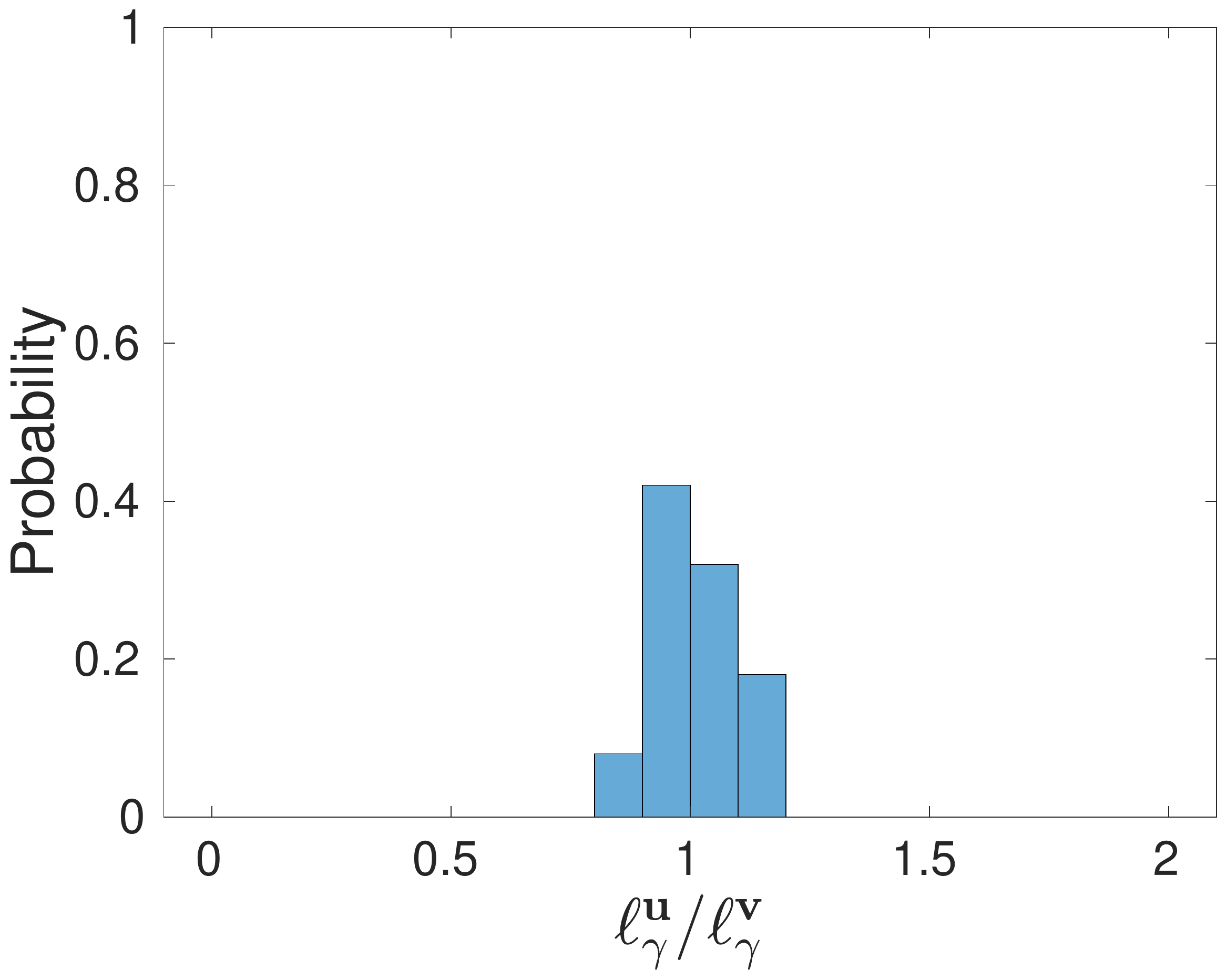}}
\emn

\bmn {0.33\lnw}
  \centerline{\includegraphics[width=\lnw]{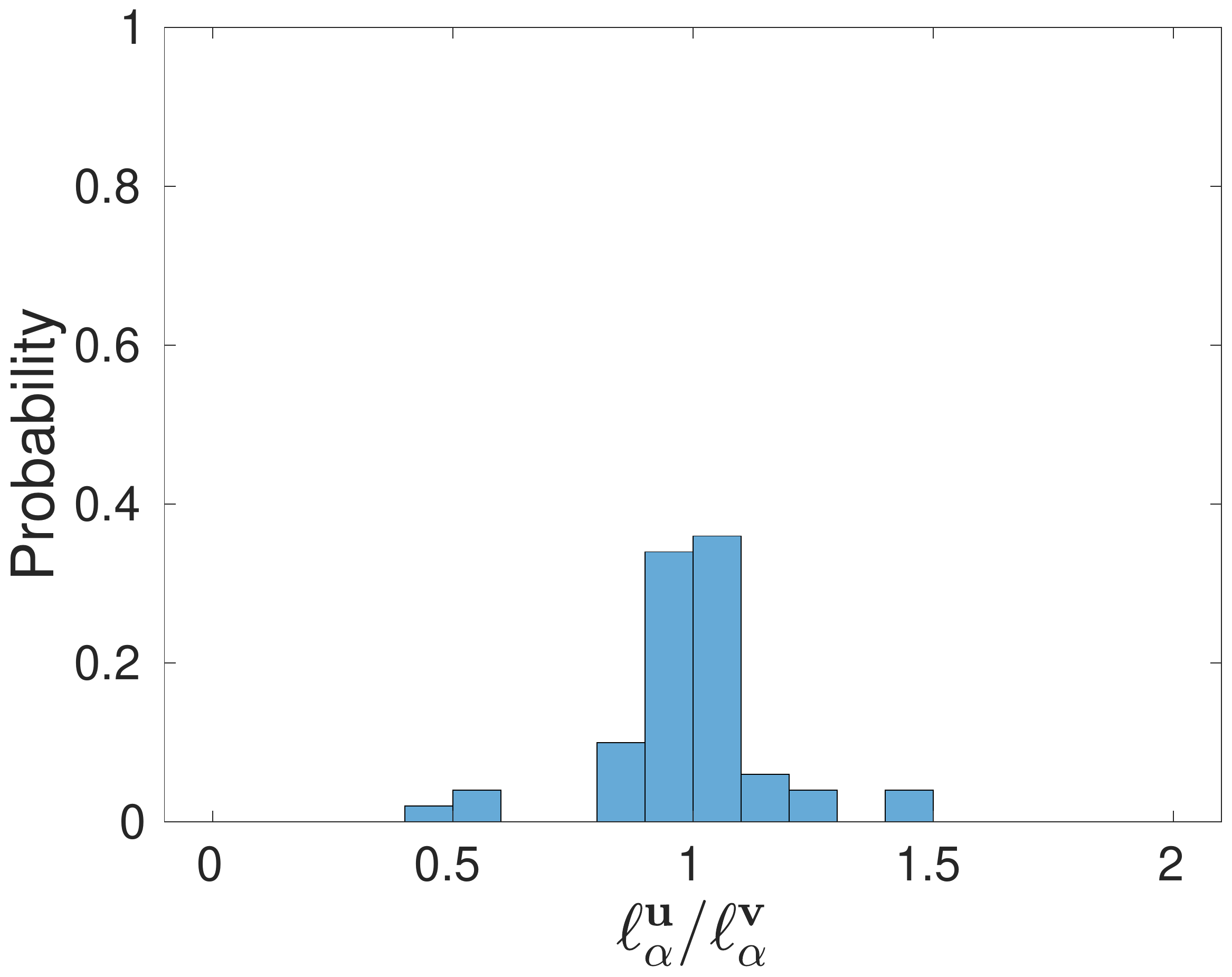}}
\emn %
\bmn {0.33\lnw}
  \centerline{\includegraphics[width=\lnw]{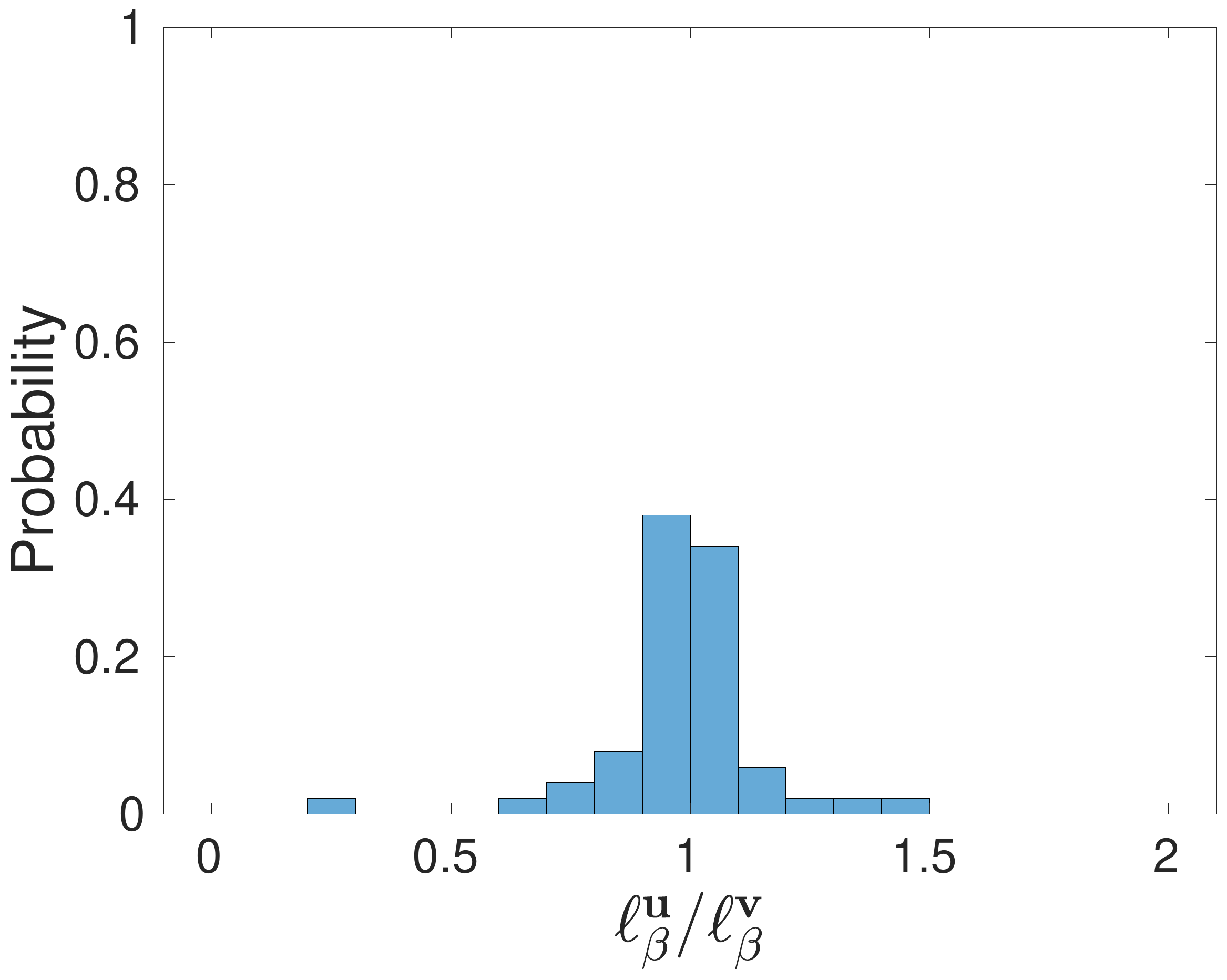}}
\emn %
\bmn {0.33\lnw}
  \centerline{\includegraphics[width=\lnw]{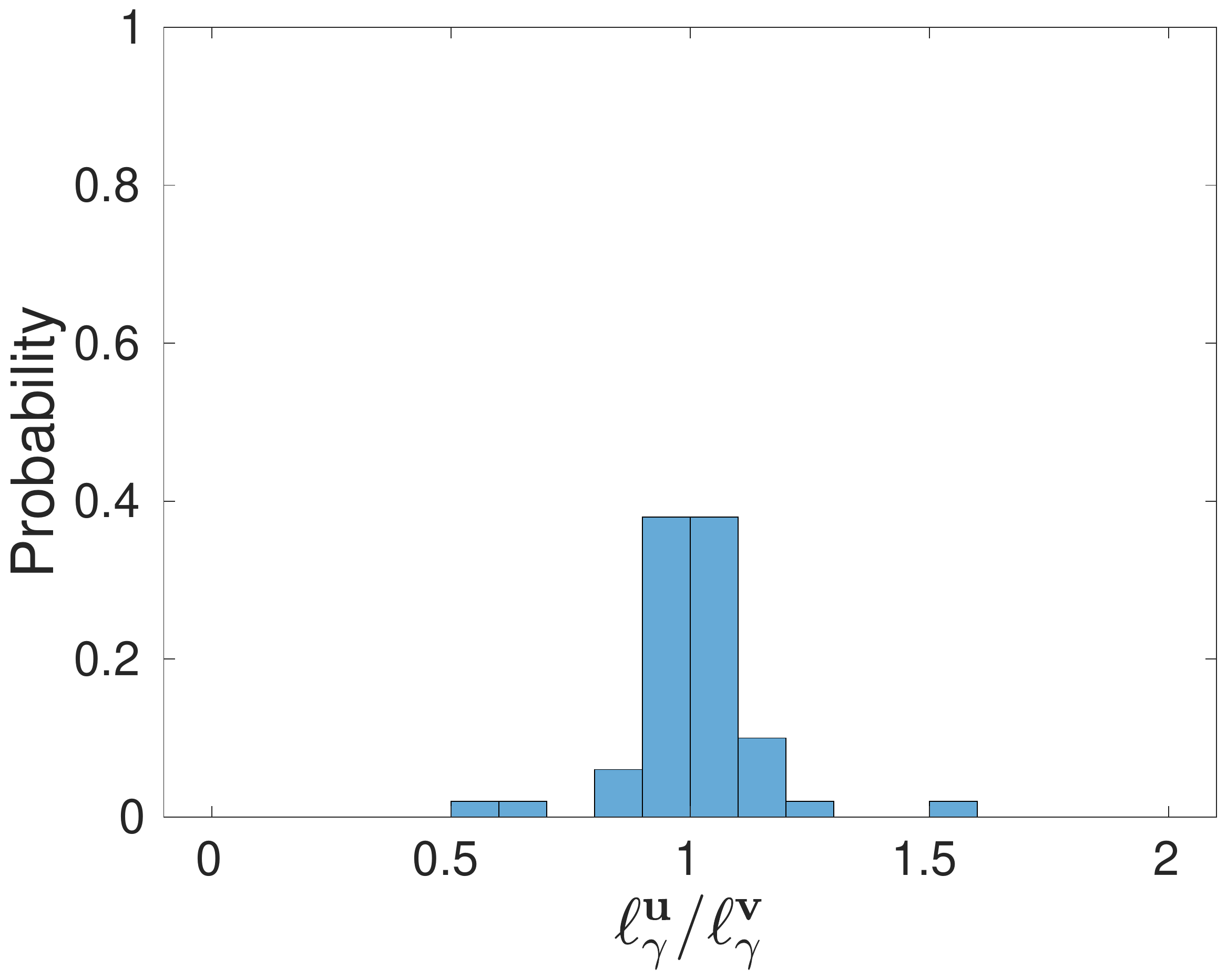}}
\emn
\caption{The normalized histograms of $\ell^{\u}_i/\ell^\v_i$ for
$i=\alpha$ (left column), $i=\beta$ (middle column),
and $i = \gamma$ (right column), with $\Delta = \Delta_m/2$ (top row)
and $\Delta = \Delta_m/4$ (bottom row). From case C1 with $t=T$.}
\label{fig:mvee-lenrat-hist}
\efig

\begin{figure}
\bmn{0.5\lnw}
  \centerline{\includegraphics[width=\lnw]{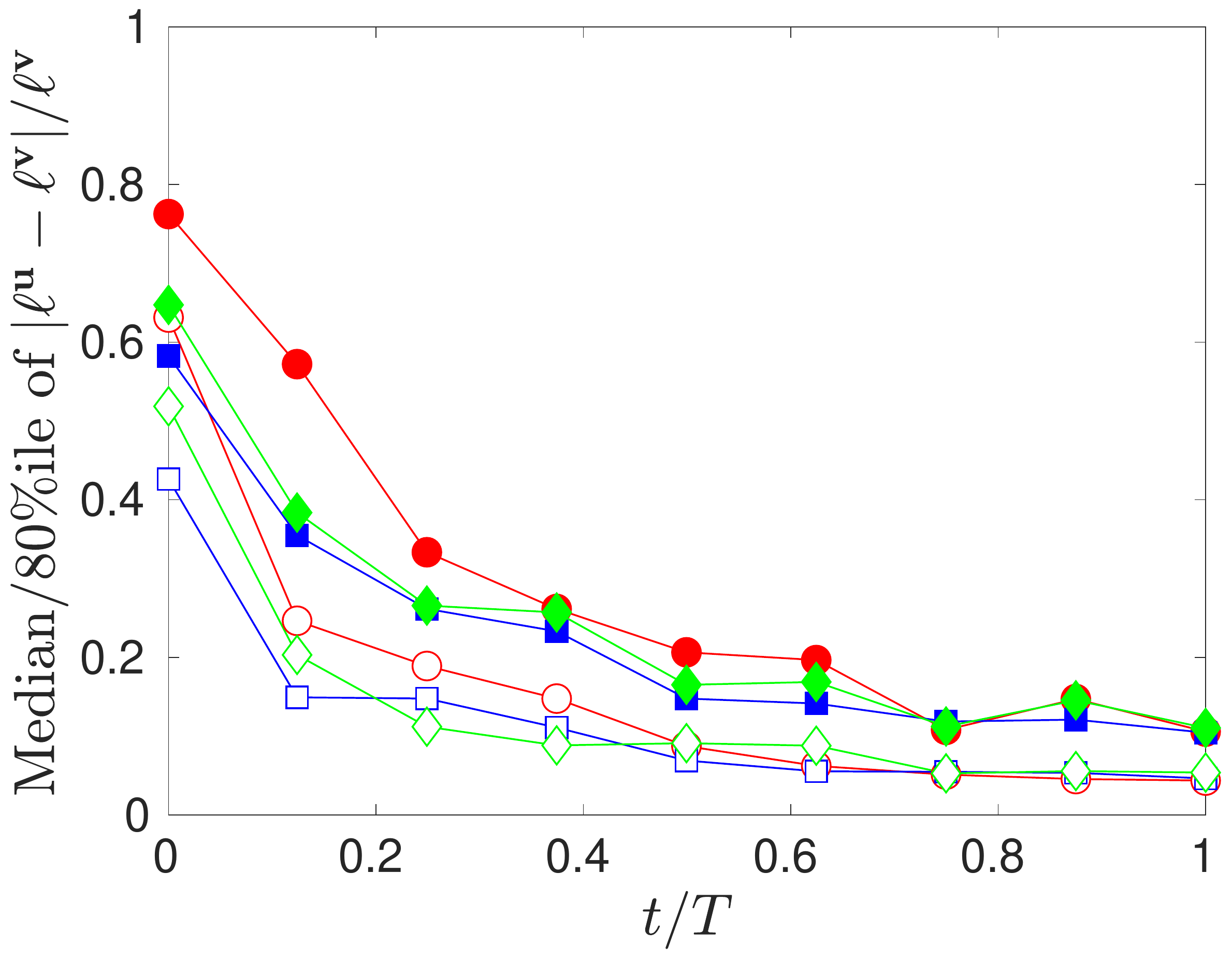}}
\emn %
\bmn{0.5\lnw}
  \centerline{\includegraphics[width=\lnw]{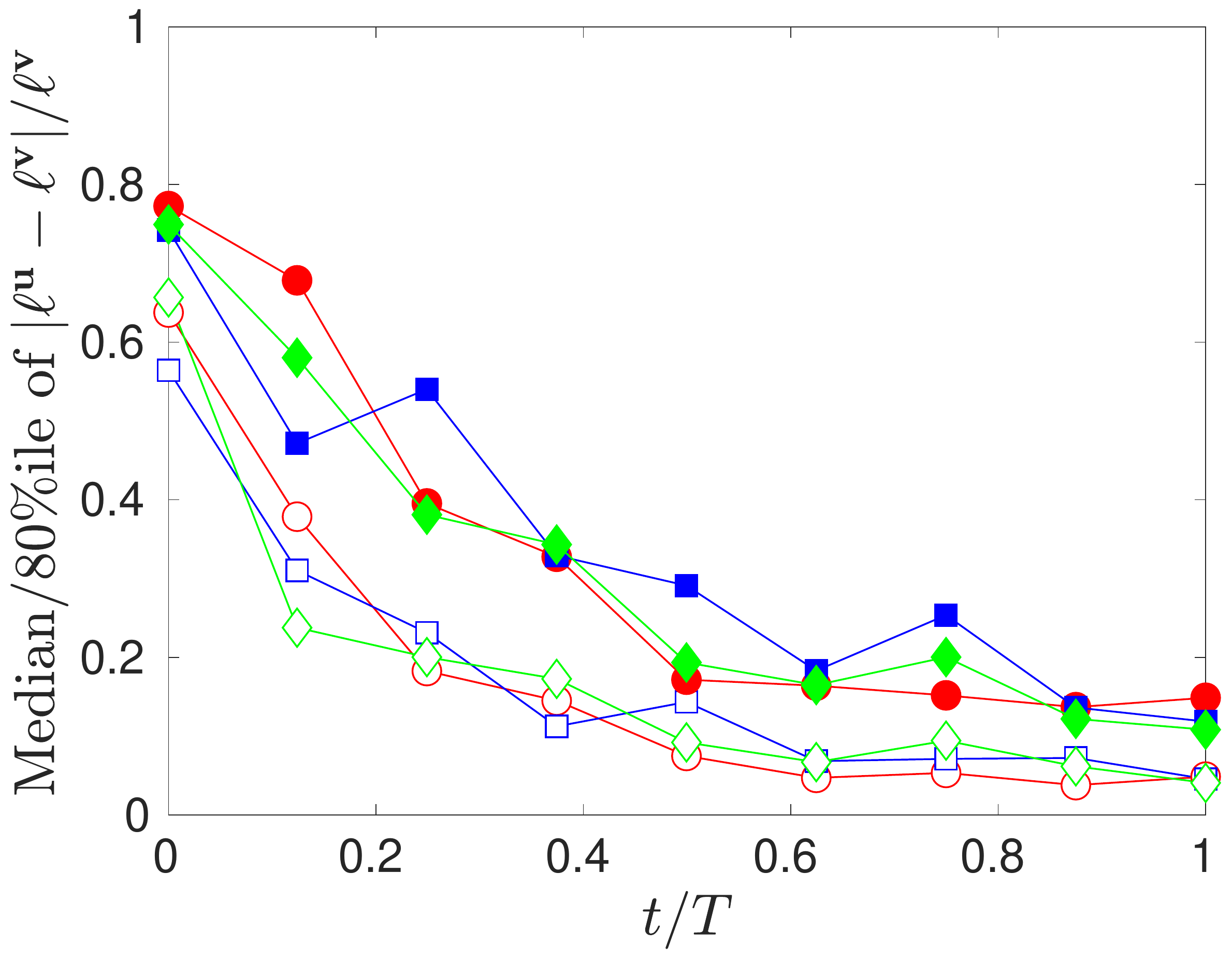}}
\emn
  \caption{The medians (empty symbols) and the $80$th percentiles (solid symbols)
  of $|\ell_i^\u-\ell^\v_i|/\ell^\v_i$ $(i=\alpha, \beta, \gamma)$ for the MVEEs of
  matching vortices. Circles: $i=\alpha$; squares: $i=\beta$;
  diamonds: $i=\gamma$. Left: $\Delta = \Delta_m/2$; right: $\Delta =
  \Delta_m/4$. From case C1.}
\label{fig:mvee-lenrat-median}
\end{figure}

\bfig
\centering
\bmn {0.33\lnw}
  \centerline{\includegraphics[width=\lnw]{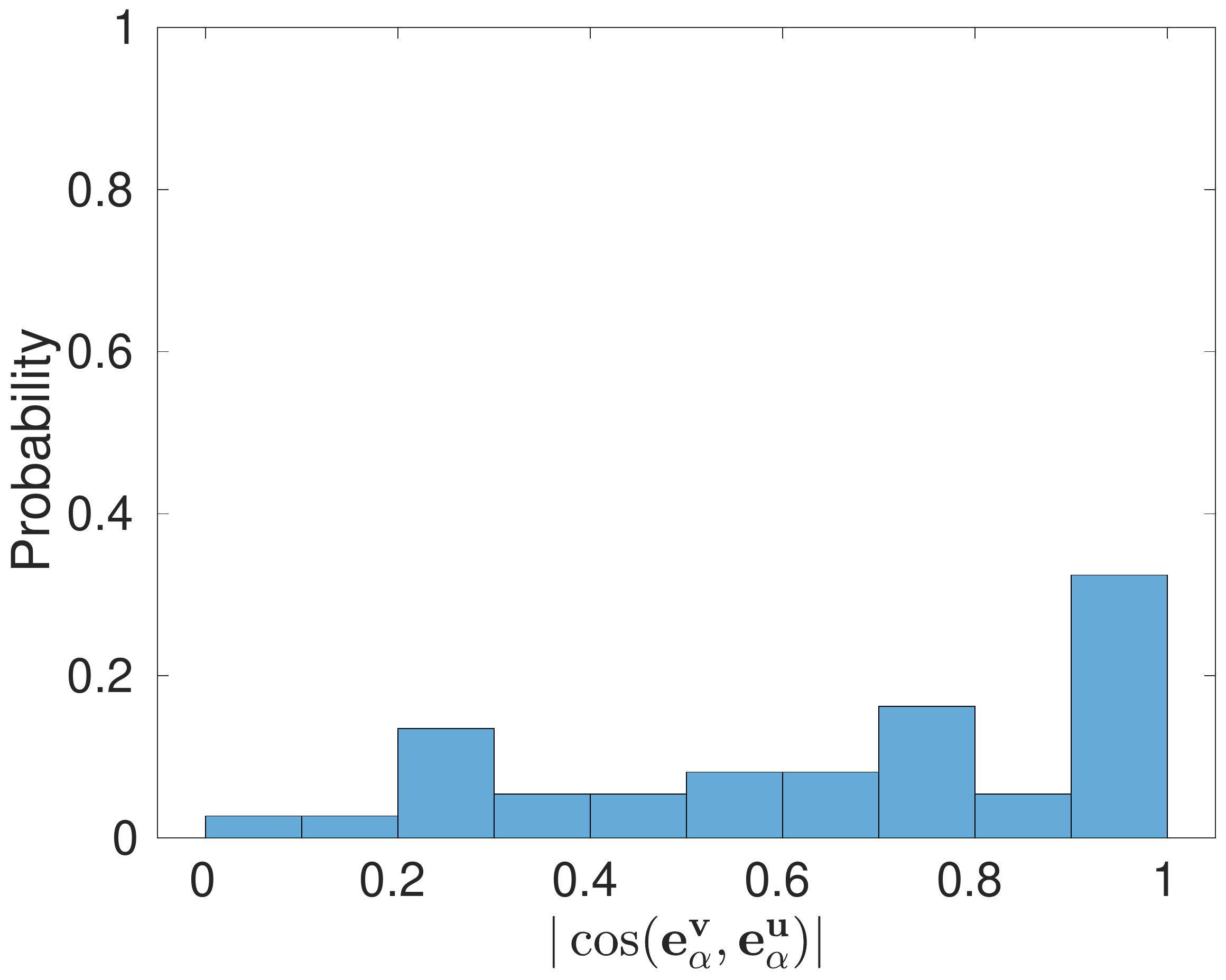}}
\emn %
\bmn {0.33\lnw}
  \centerline{\includegraphics[width=\lnw]{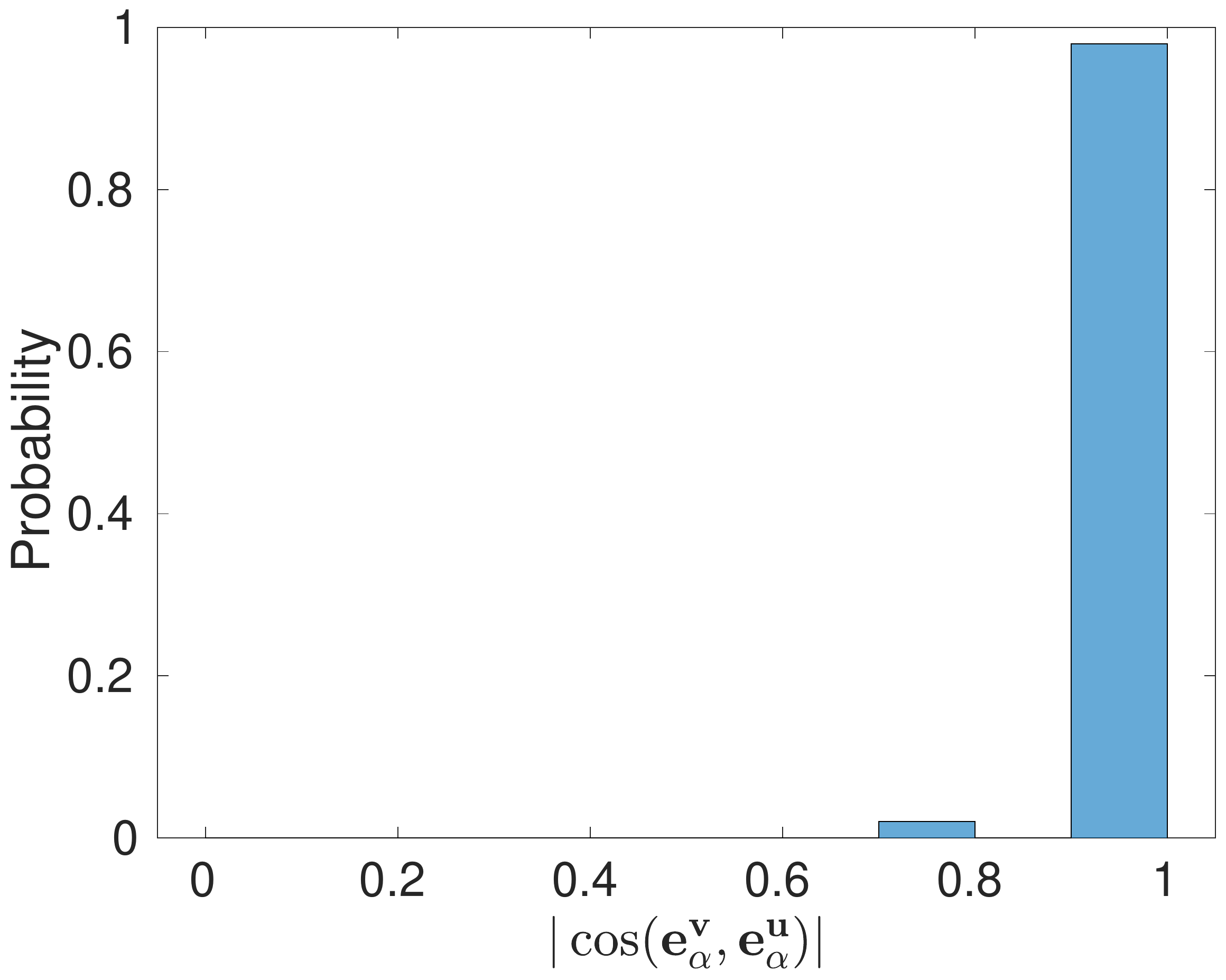}}
\emn %
\bmn {0.33\lnw}
  \centerline{\includegraphics[width=\lnw]{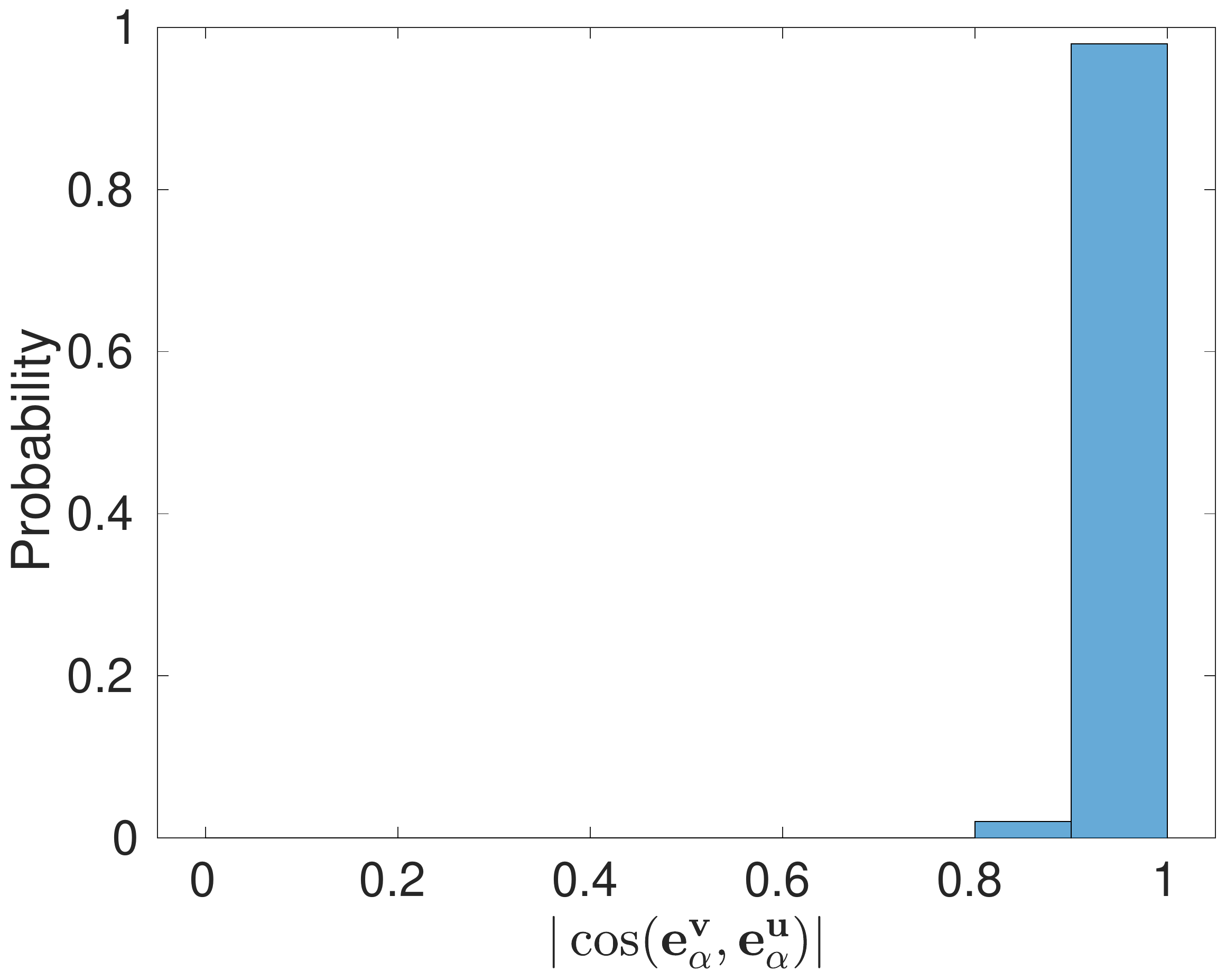}}
\emn \\
\bmn {0.33\lnw}
  \centerline{\includegraphics[width=\lnw]{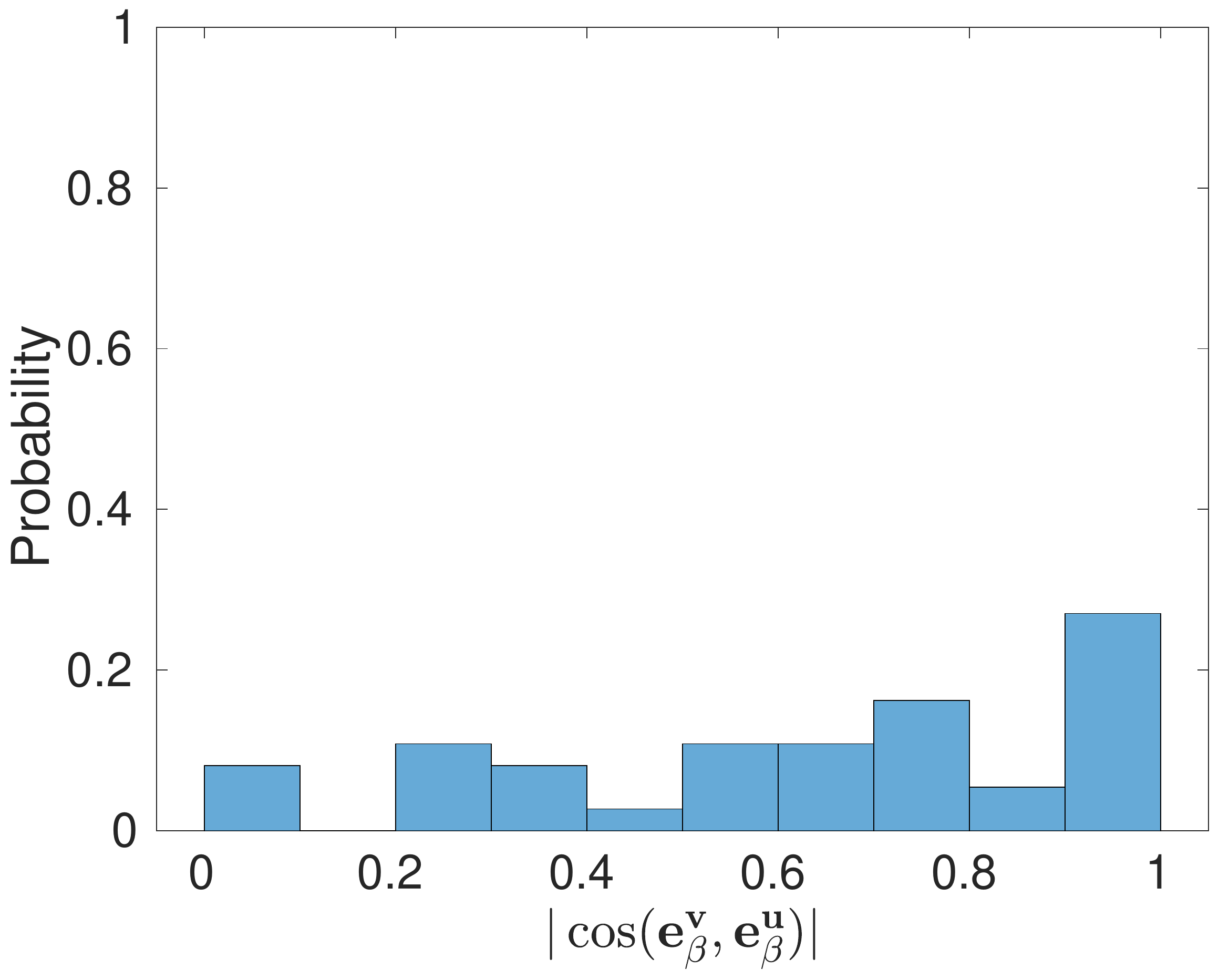}}
\emn %
\bmn {0.33\lnw}
  \centerline{\includegraphics[width=\lnw]{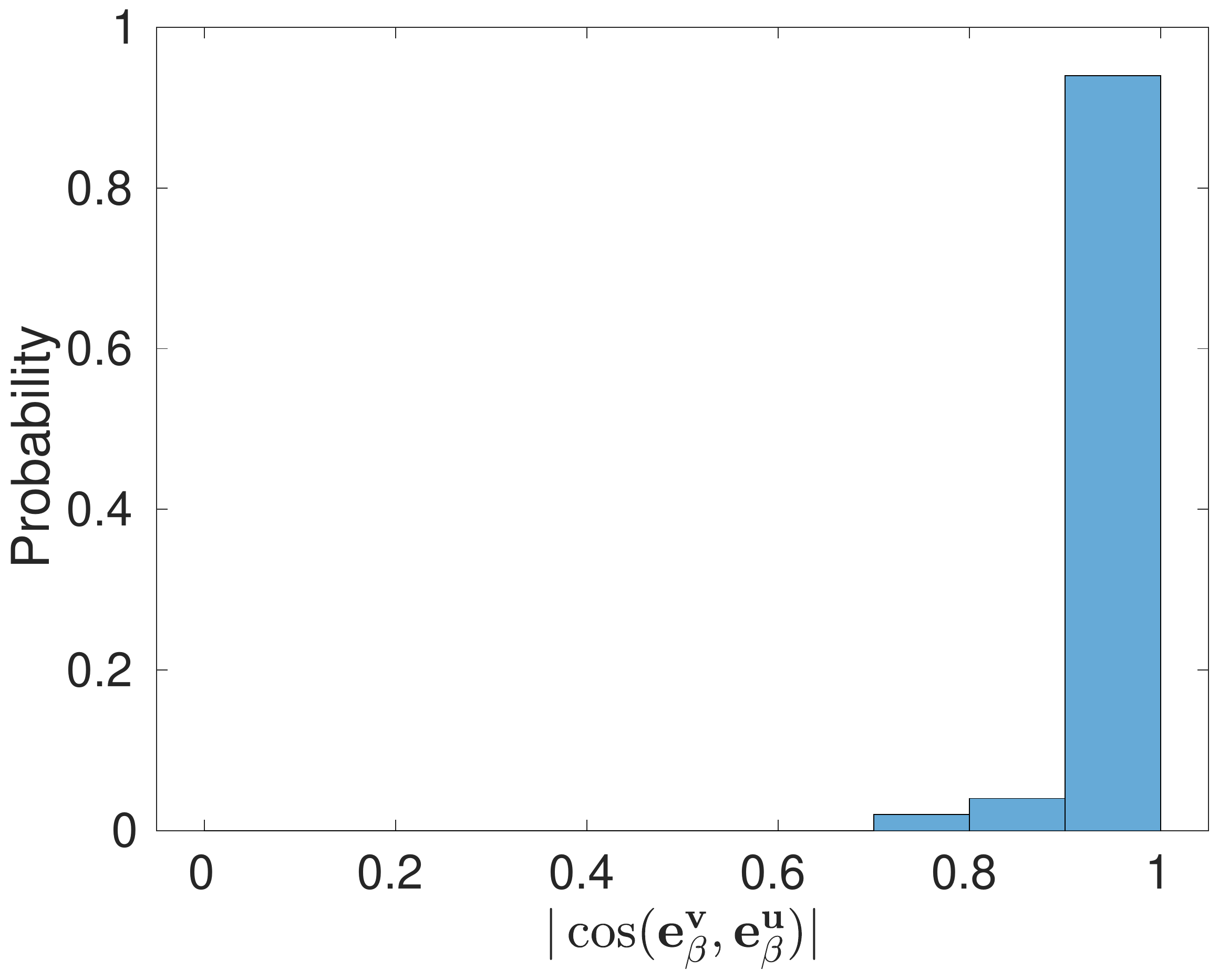}}
\emn %
\bmn {0.33\lnw}
  \centerline{\includegraphics[width=\lnw]{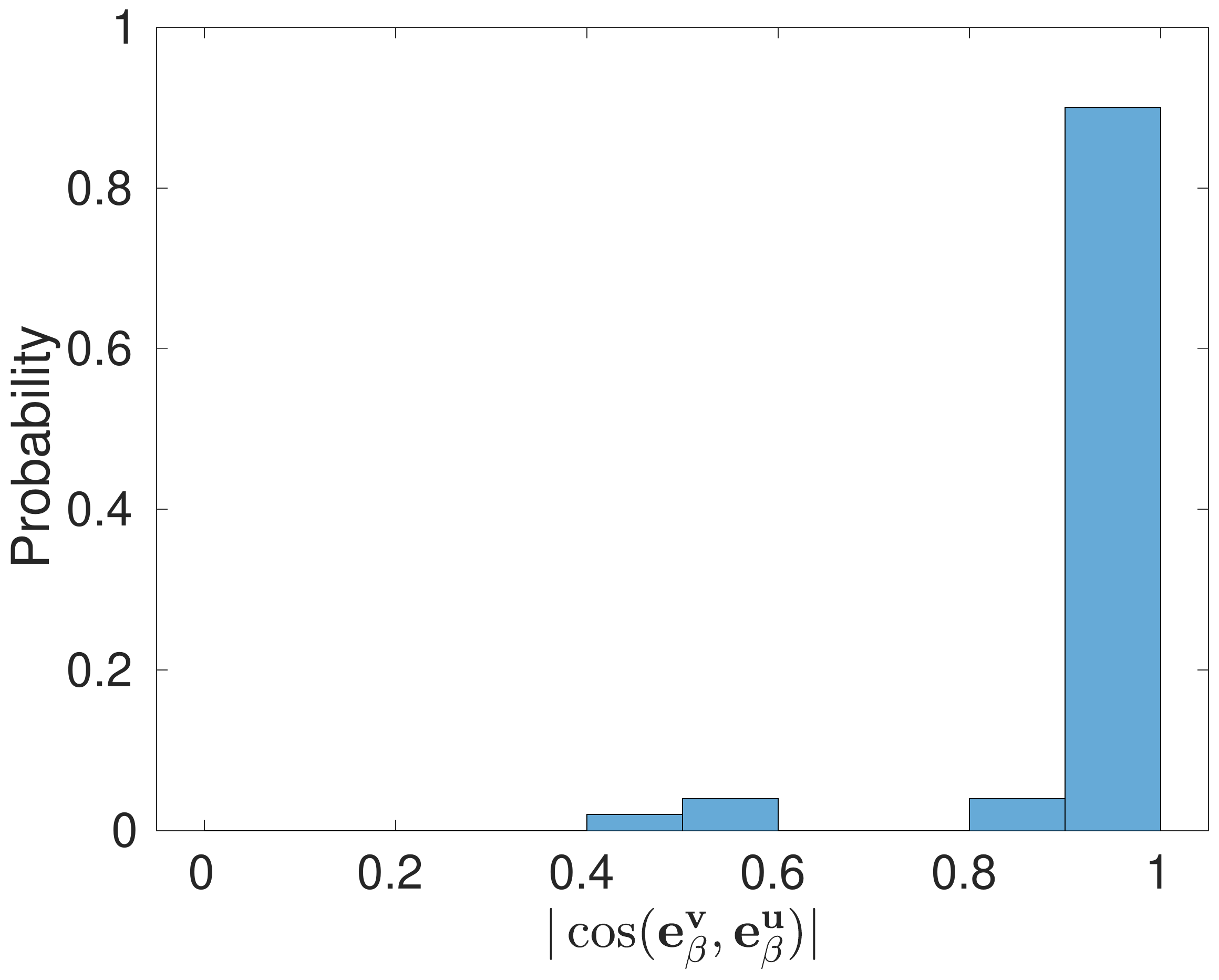}}
\emn \\
\bmn {0.33\lnw}
  \centerline{\includegraphics[width=\lnw]{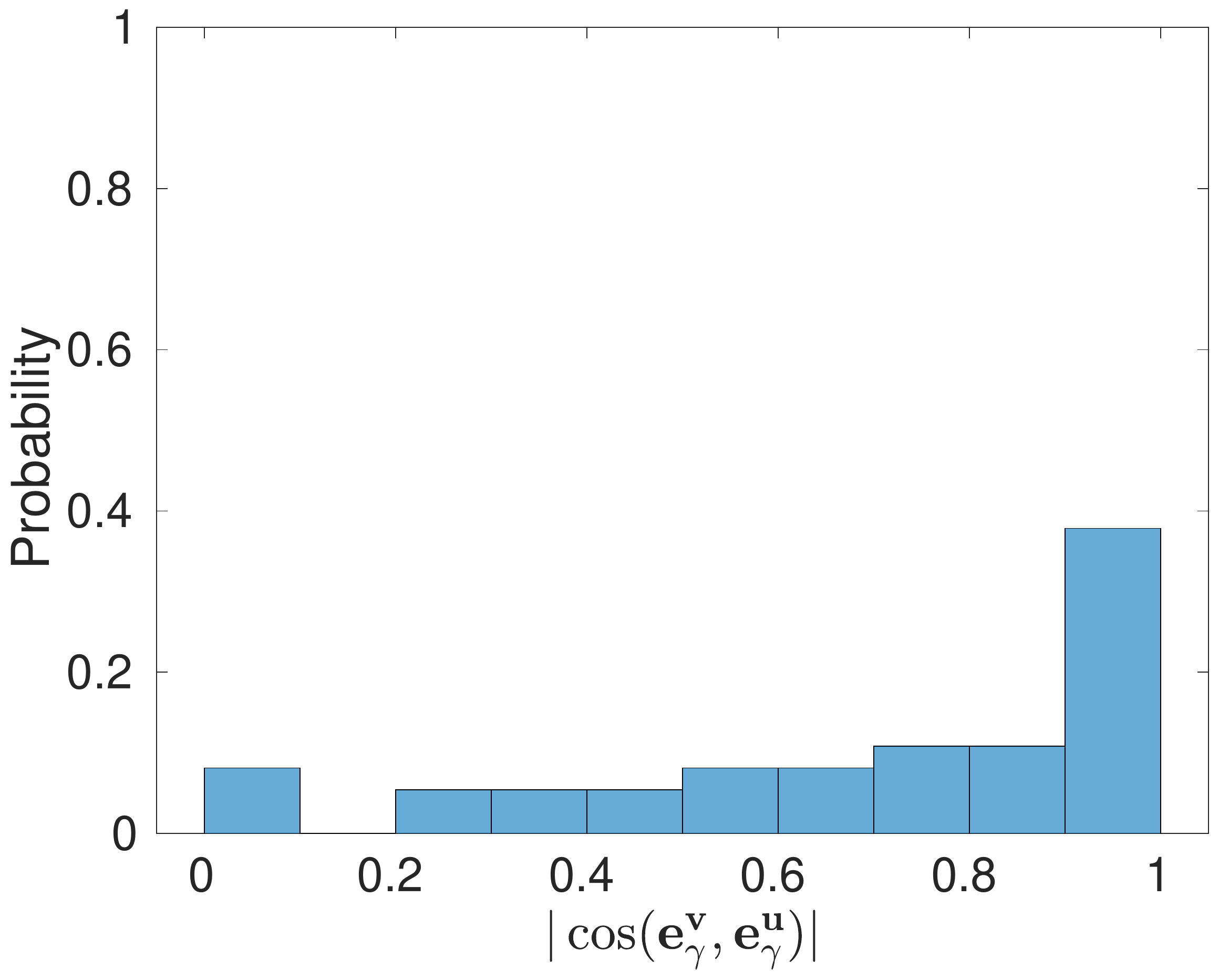}}
\emn %
\bmn {0.33\lnw}
  \centerline{\includegraphics[width=\lnw]{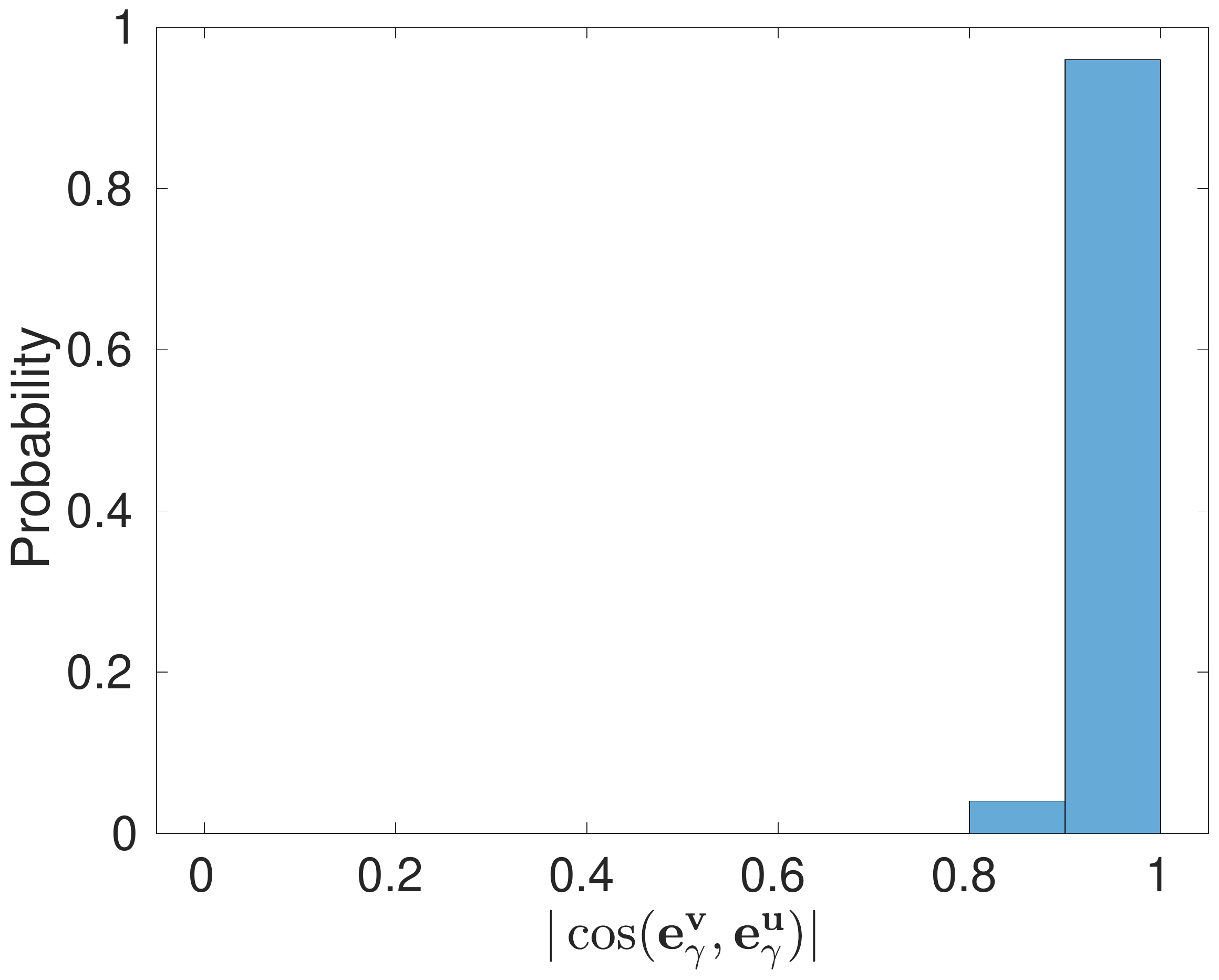}}
\emn %
\bmn {0.33\lnw}
  \centerline{\includegraphics[width=\lnw]{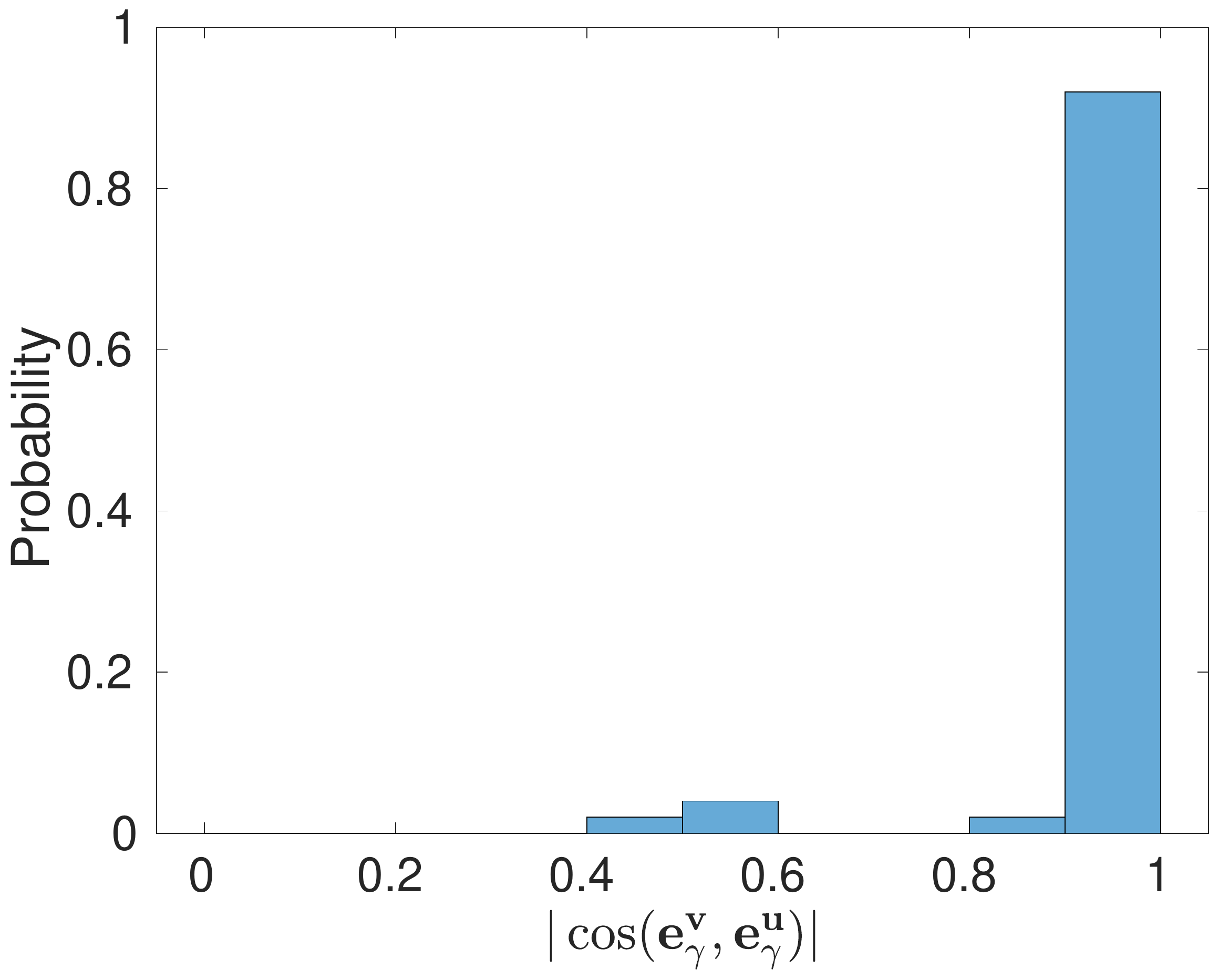}}
\emn
\caption{The histograms of $|\cos(\e_i^\v, \e_i^\u)|$ for $i=\alpha$ (top row), $i=\beta$ (middle
row), and $i = \gamma$ (bottom row), at $t/T=0$ (left column), $1/2$ (middle column), and $1$ (right
column). From case C1 with $\Delta = \Delta_m/2$.}
\label{fig:mvee-align-hist}
\efig

\begin{figure}
\bmn {0.5\lnw}
  \centerline{\includegraphics[width=\lnw]{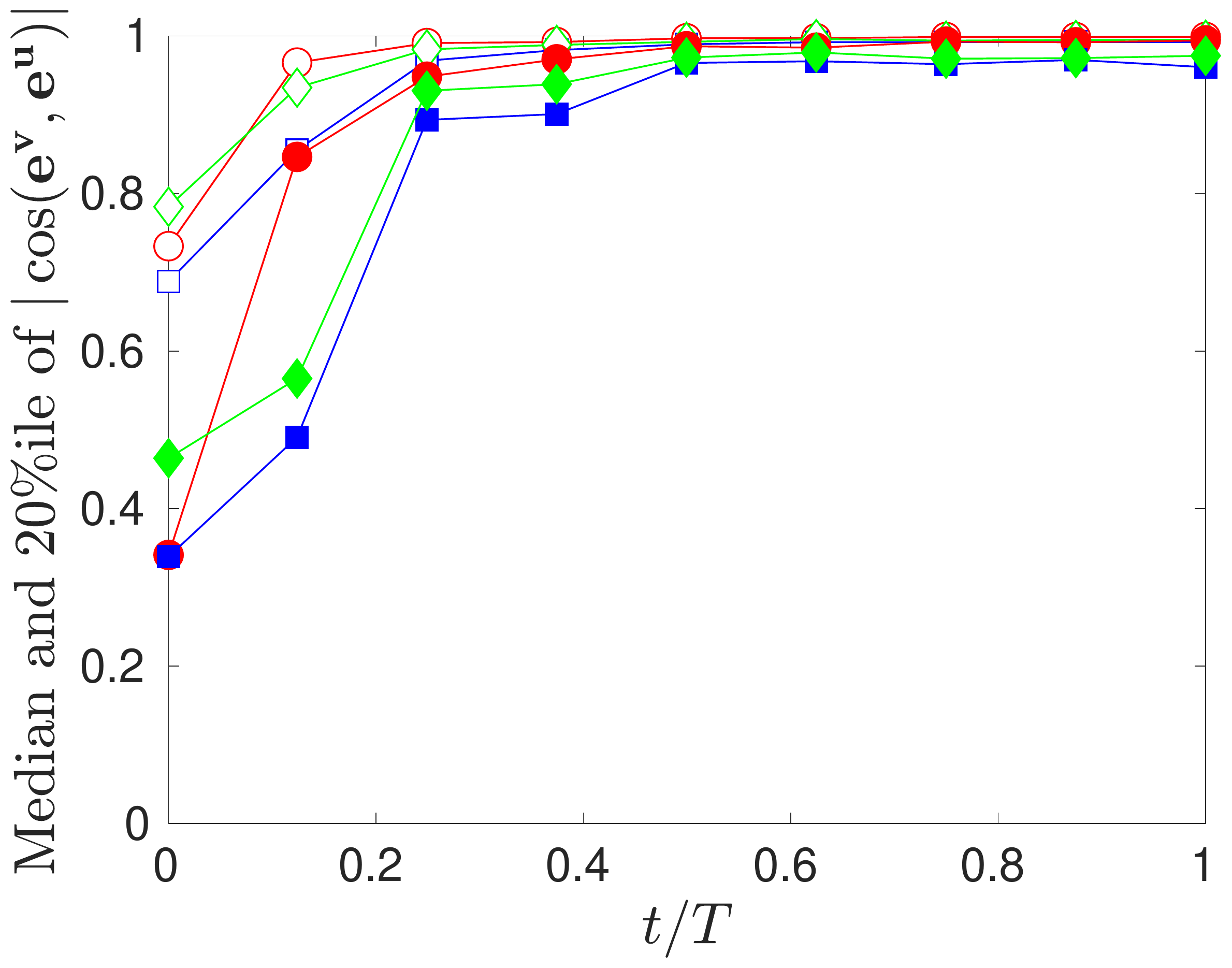}}
  \emn %
\bmn {0.5\lnw}
  \centerline{\includegraphics[width=\lnw]{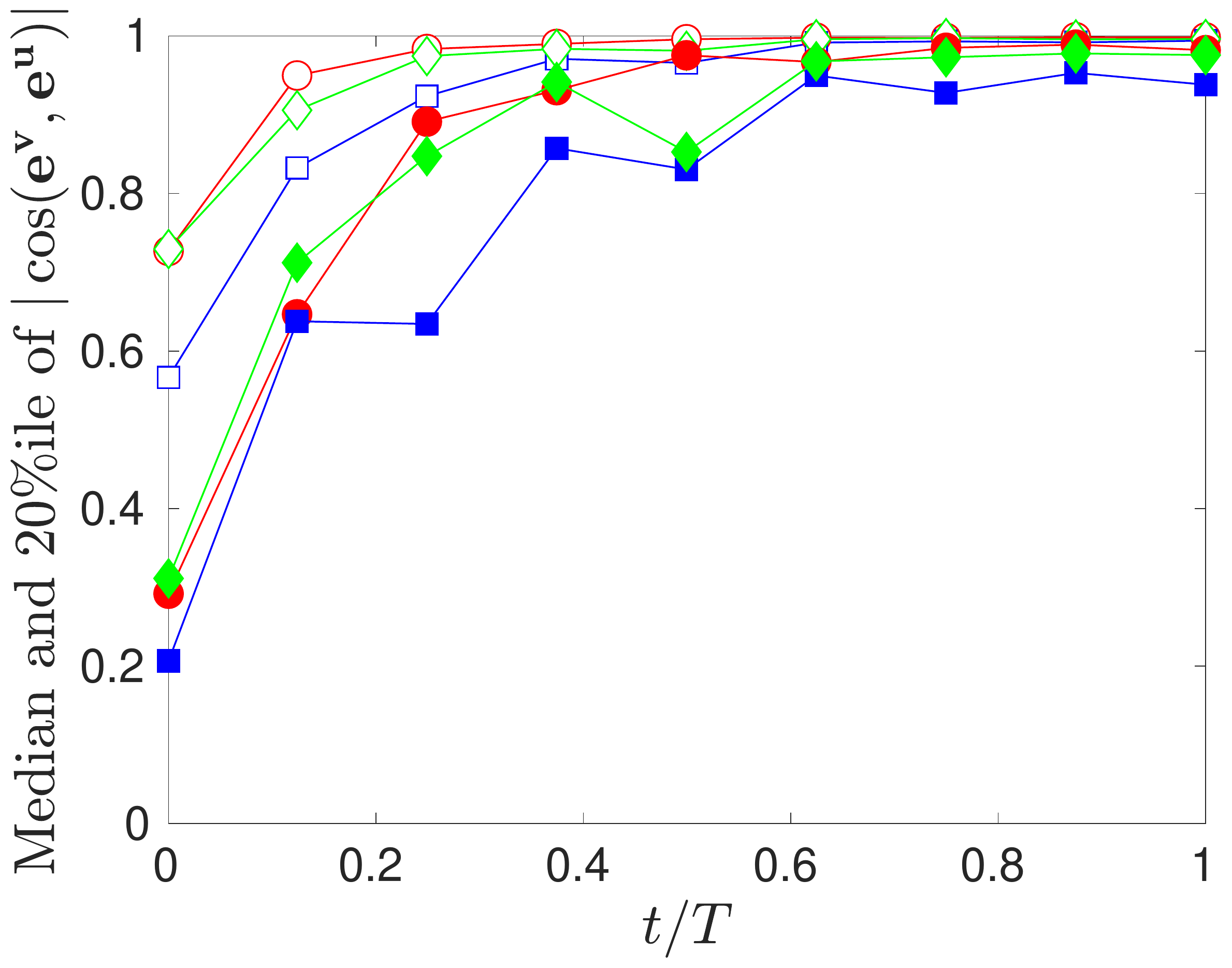}}
  \emn
  \caption{The medians (empty symbols) and the $20$th percentiles (solid symbols)
  of $|\cos(\e_i^\v, \e_i^\u)|$ $(i=\alpha, \beta, \gamma)$ for the MVEEs of
  matching vortices. Circles: $i=\alpha$; squares: $i=\beta$;
  diamonds: $i=\gamma$. Left: $\Delta = \Delta_m/2$; right: $\Delta =
  \Delta_m/4$. From case C1.}
\label{fig:median-align}
\end{figure}

\begin{figure}
\bmn {0.5\lnw}
  \centerline{\includegraphics[width=\lnw]{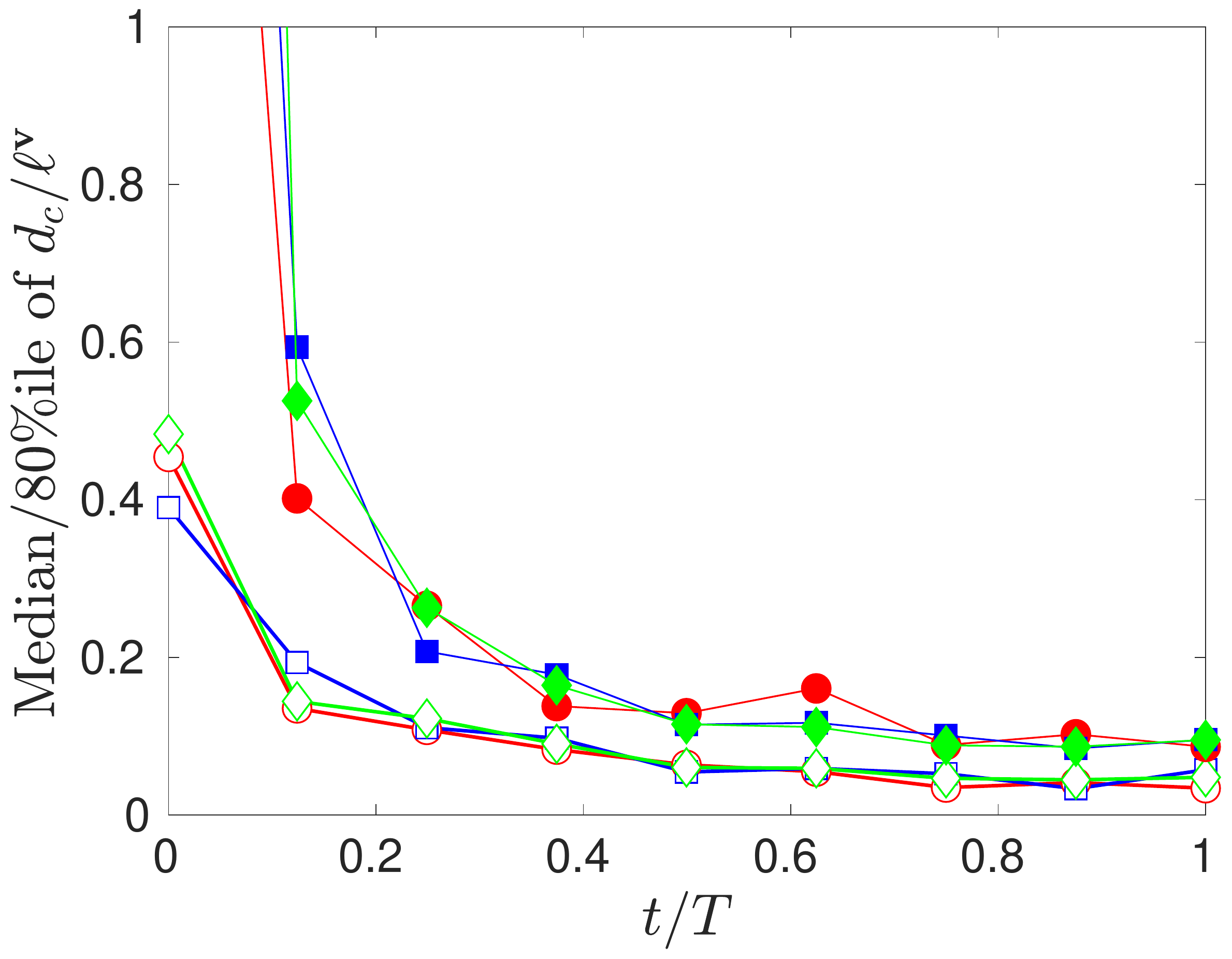}}
  \emn
\bmn {0.5\lnw}
  \centerline{\includegraphics[width=\lnw]{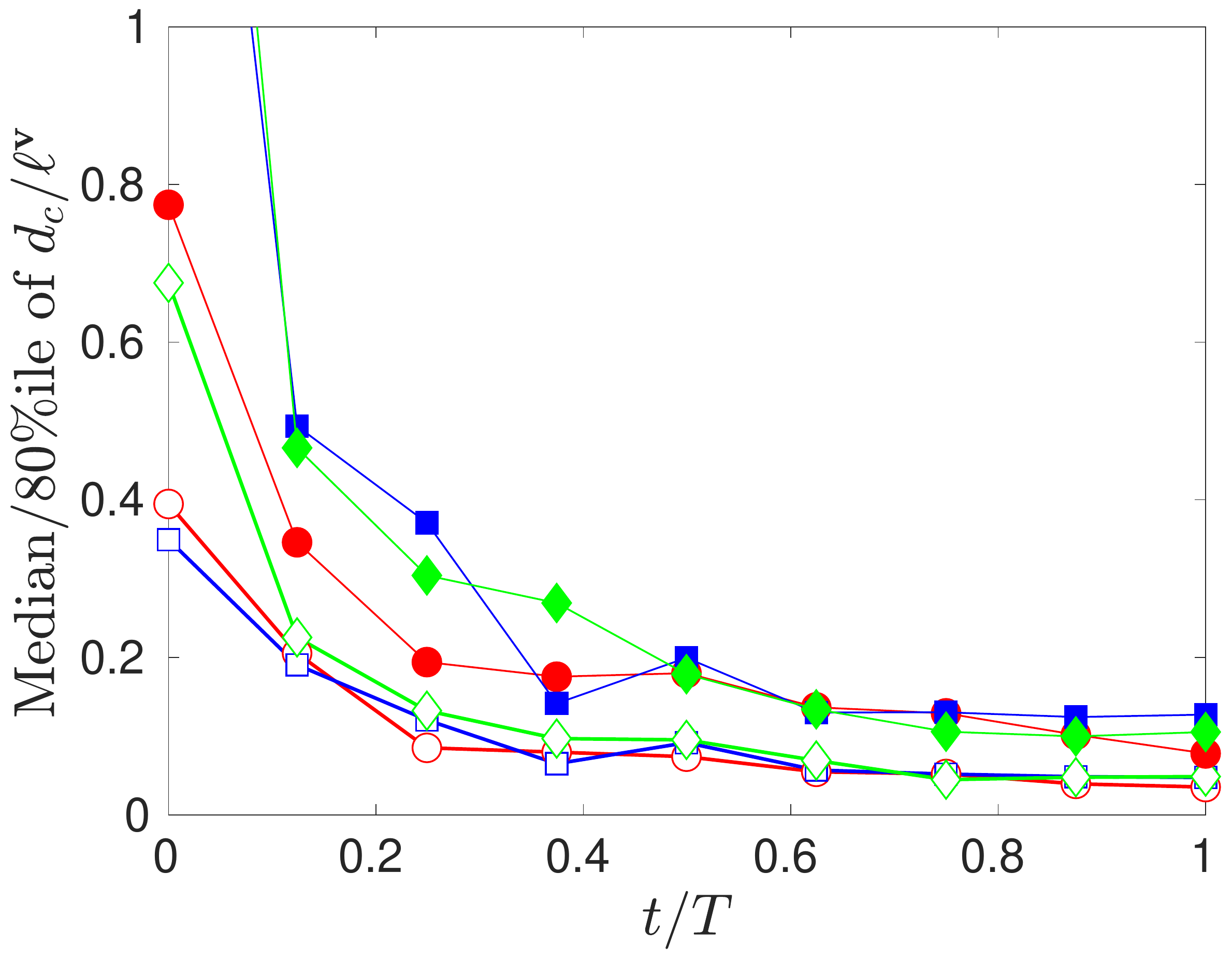}}
  \emn
  \caption{The medians (empty symbols) and the $80$th percentiles (solid symbols)
for $d_{c,i}/\ell^\v_i$ $(i=\alpha, \beta, \gamma)$ for the MVEEs of
  matching vortices. Circles: $i=\alpha$; squares: $i=\beta$;
  diamonds: $i=\gamma$. Left: $\Delta = \Delta_m/2$; right: $\Delta =
  \Delta_m/4$. From case C1.}
\label{fig:mvee-dist-median}
\end{figure}

\bfig
\centering
\bmn {0.33\lnw}
  \centerline{\includegraphics[width=\lnw]{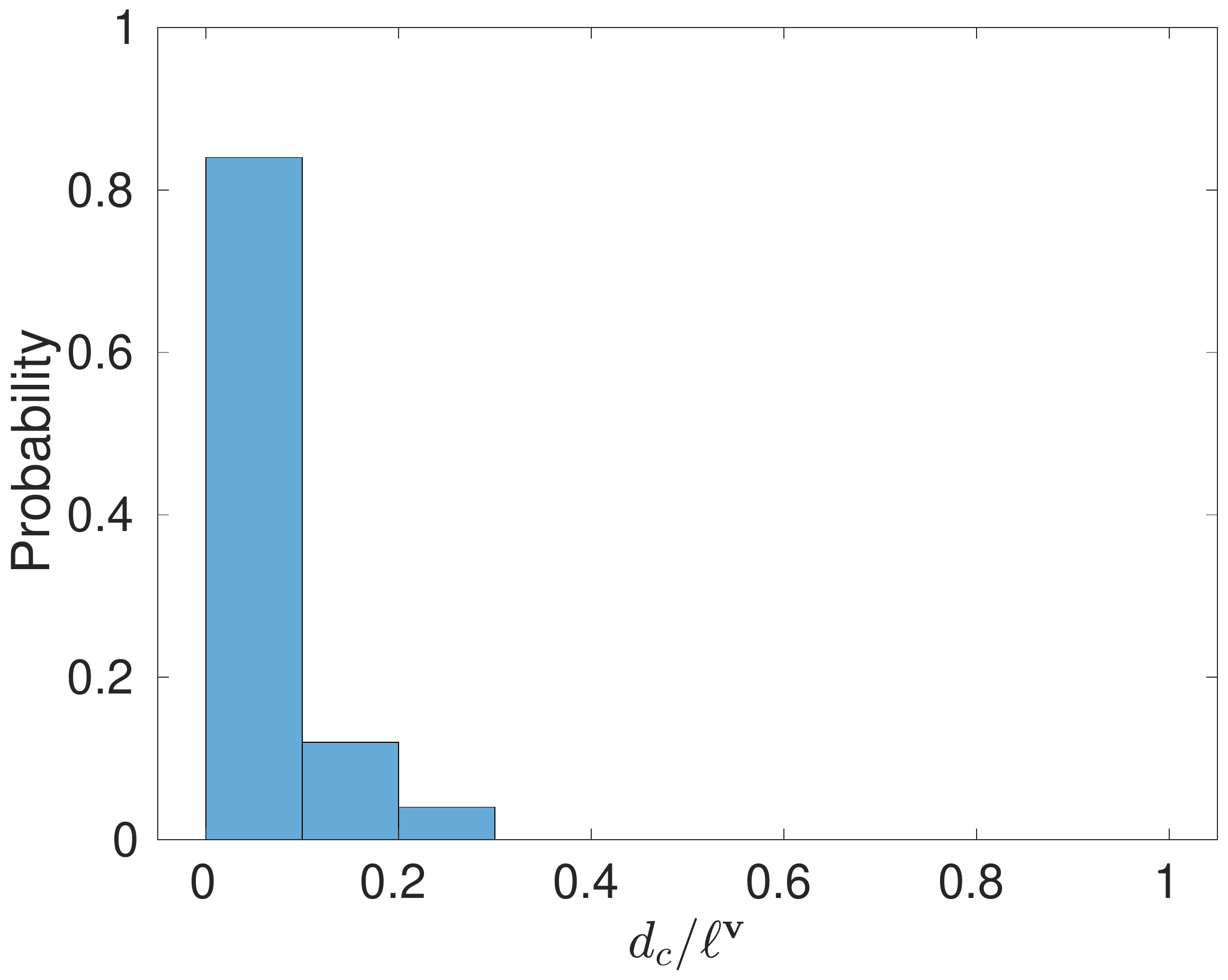}}
\emn %
\bmn {0.33\lnw}
  \centerline{\includegraphics[width=\lnw]{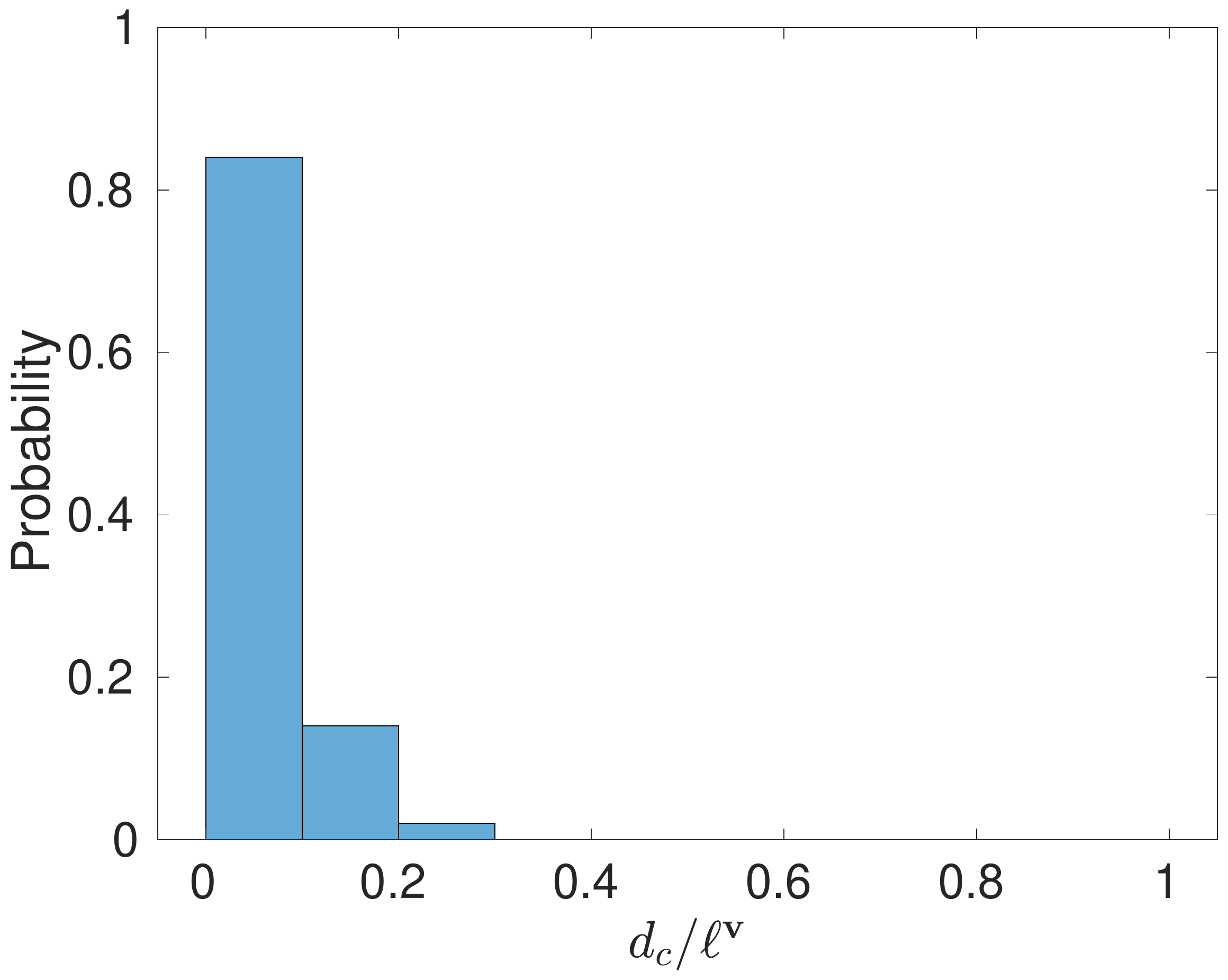}}
\emn %
\bmn {0.33\lnw}
  \centerline{\includegraphics[width=\lnw]{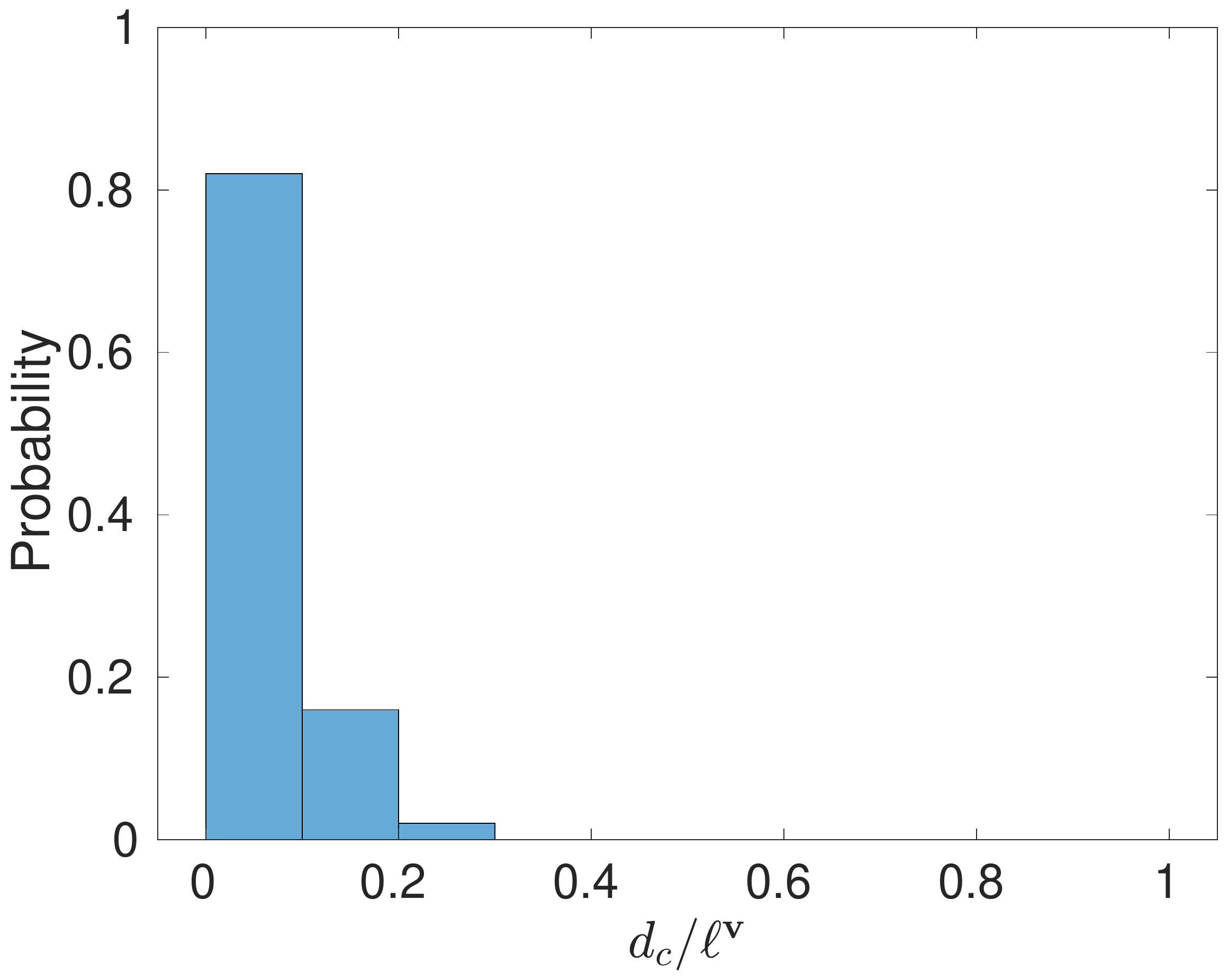}}
\emn
\caption{The normalized histograms of $d_{c,i}/\ell^\v_i$ for $i=\alpha$ (left), $i=\beta$ (middle),
and $i = \gamma$ (right), from case C1 with $t/T=1$ and $\Delta
= \Delta_m/2$.}
\label{fig:mvee-dist-hist}
\efig

Fig. \ref{fig:sizes8dx} compares the sizes of the vortical structures
in $\u$ and $\v$ in terms of
$R$, which is the median of the ratio between the numbers of grid
points in the two matching
vortices.
There are $N_s = 10$ groups of vortices of different sizes. The thick line with filled circles
shows $R$ calculated over all $N_s = 10$ groups, whereas the thin
lines with
empty circles show $R$ in each group separately.
The figure shows that the
vortices in $\u$ tend to be smaller initially, the median ratio
being only about $10\%$. However, it increases quickly, reaching $90\%$
at $t/T\approx 0.5$ and approximately $1$ at the end of the
optimization horizon. The same behavior is observed at two filter
scales. The thin lines show that there is significant
statistical variation between different groups.

The sizes can also be measured with the lengths of the axes of the
MVEEs. Fig.
\ref{fig:mvee-lenrat-hist} plots the histograms of the ratios of the axes in $\u$
to those in $\v$ at
$t/T=1$ for case C1. The
results are consistent with Fig. \ref{fig:sizes8dx}. The
peaks of the histograms show that
$80\%$ of the axes of the MVEEs in
$\u$ fall within $\pm 10\%$ of those in $\v$.

The median and $80$th percentile for
$|\ell_i^\u-\ell_i^\v|/\ell_i^\v$ (i.e., the relative error in the lengths of
the axes) are shown in Fig.
\ref{fig:mvee-lenrat-median} for two filter scales. Both values
decrease over time. At
$t/T=1$, the median is reduced down to somewhere between $5\%$ and
$10\%$, i.e., for about half of MVEEs, the relative error
in the length of the axes is
less than $5\sim 10\%$. The result for the $80$th percentile shows
that, for $80\%$ MVEEs the relative error is below $10\sim 15\%$. The
difference between different axes is very small.

The alignment between $\e^\v_i$ and $\e^\u_i$ is examined in Fig.
\ref{fig:mvee-align-hist}, where the
histograms of $|\cos(\e_i^\v, \e_i^\u)|$ are
plotted ($t/T=0,1/2,1$ from left to right, respectively).
The $i$th axes of the matching MVEEs would align perfectly if $|\cos(\e_i^\v,\e_i^\u)| = 1$.
Initially, the correct alignment between the axes is visible, but the peaks are not
strong. Perfect alignment develops quickly. The
histograms at $t/T=1/2$ and $1$ basically concentrate around $1$.
In Fig. \ref{fig:median-align}, the
medians and the $20$th percentiles of the cosines
are plotted for two filter scales.
In both cases,
$|\cos(\e_i^\v, \e_i^\u)|\ge 0.95$ is observed at $t/T=1$ for the $20$th percentiles for all axes, 
which corresponding to angles smaller than $18^\circ$.
Therefore, only $20\%$ samples have mis-alignment angle bigger than
$18^\circ$.

Fig. \ref{fig:mvee-dist-median} and \ref{fig:mvee-dist-hist} compare the locations of the MVEEs.
The quantity being examined is $d_{c,i}/\ell^\v_i$ ($i=\alpha, \beta, \gamma$),
where
$d_{c,i}\equiv |(\c^\v-\c^\u)\cdot \e^\v_i|$ is the displacement between the
centers of the MVEEs in
the direction of $\e^\v_i$. Thus, $d_{c,i}/\ell^\v_i$ is the
displacement relative to the lengths of the axes of the MVEE in $\v$.
The median and the $80$th percentile of $d_{c,i}/\ell^\v_i$ are given in Fig.
\ref{fig:mvee-dist-median}. The $80$th percentile is
quite large initially, but decreases quickly over time. For $\Delta =
\Delta_m/2$, the relative
displacement at $t/T=1$ is less than $10\%$ for the $80$th percentiles,
and less than $5\%$ for the medians. For $\Delta = \Delta_m/4$, the
displacement is slightly larger, but still is less than $15\%$ for the
$80$th percentile..

\subsubsection{Straining structures and SGS energy dissipation structures}

\bfig
  \bmn{0.33\lnw}
  \centerline{\includegraphics[width=\lnw]{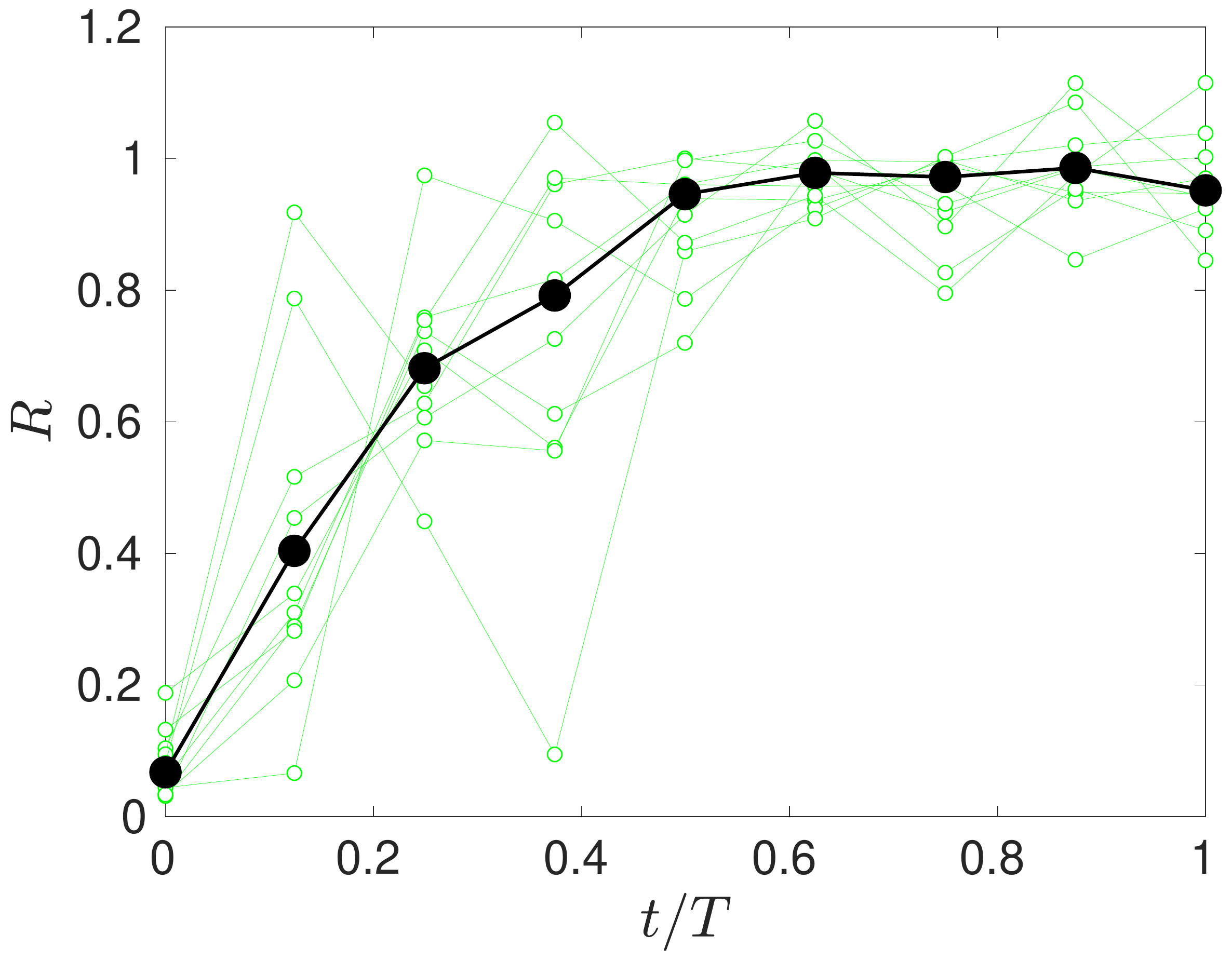}}
  \emn %
  \bmn{0.33\lnw}
  \centerline{\includegraphics[width=\lnw]{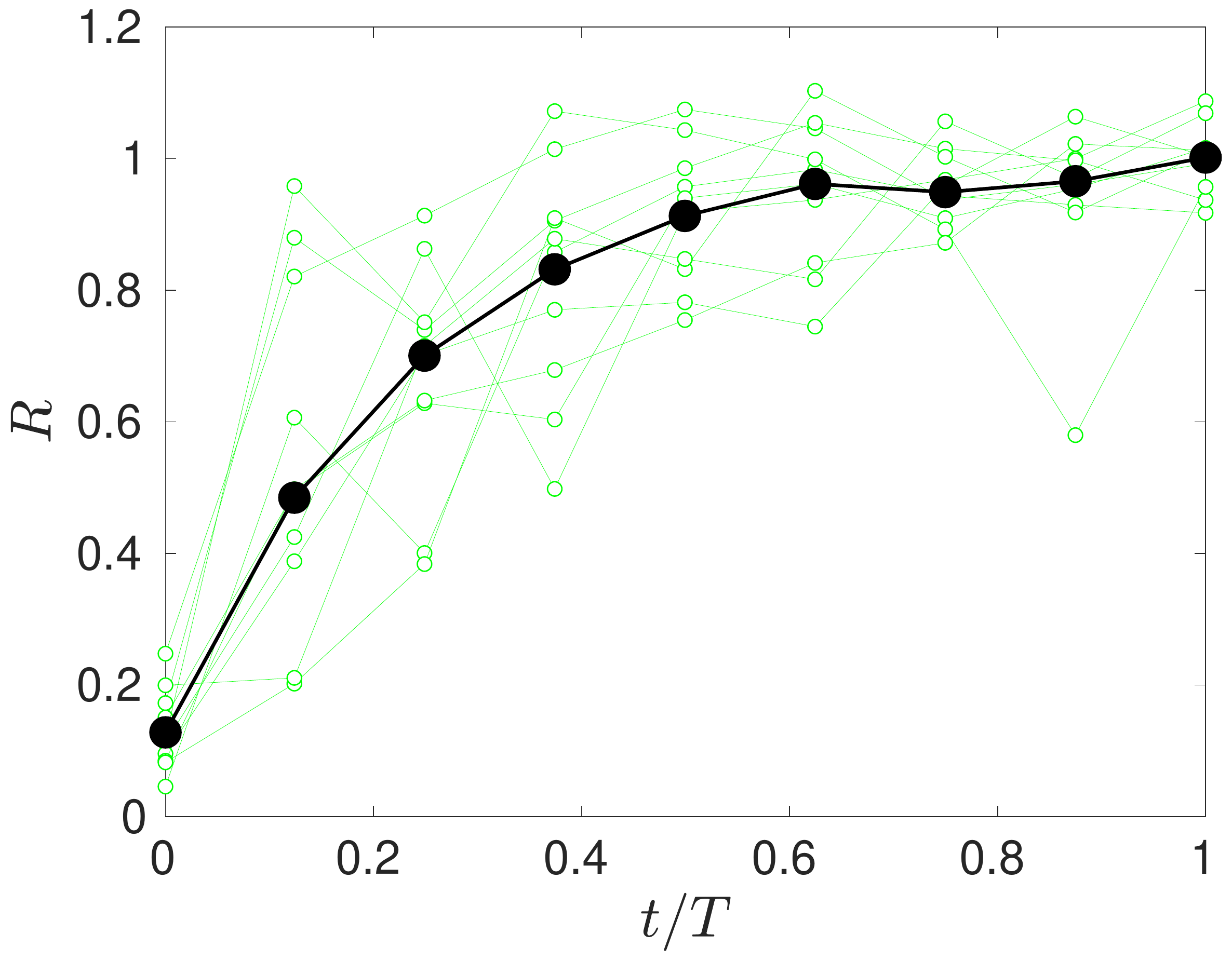}}
  \emn %
  \bmn{0.33\lnw}
  \centerline{\includegraphics[width=\lnw]{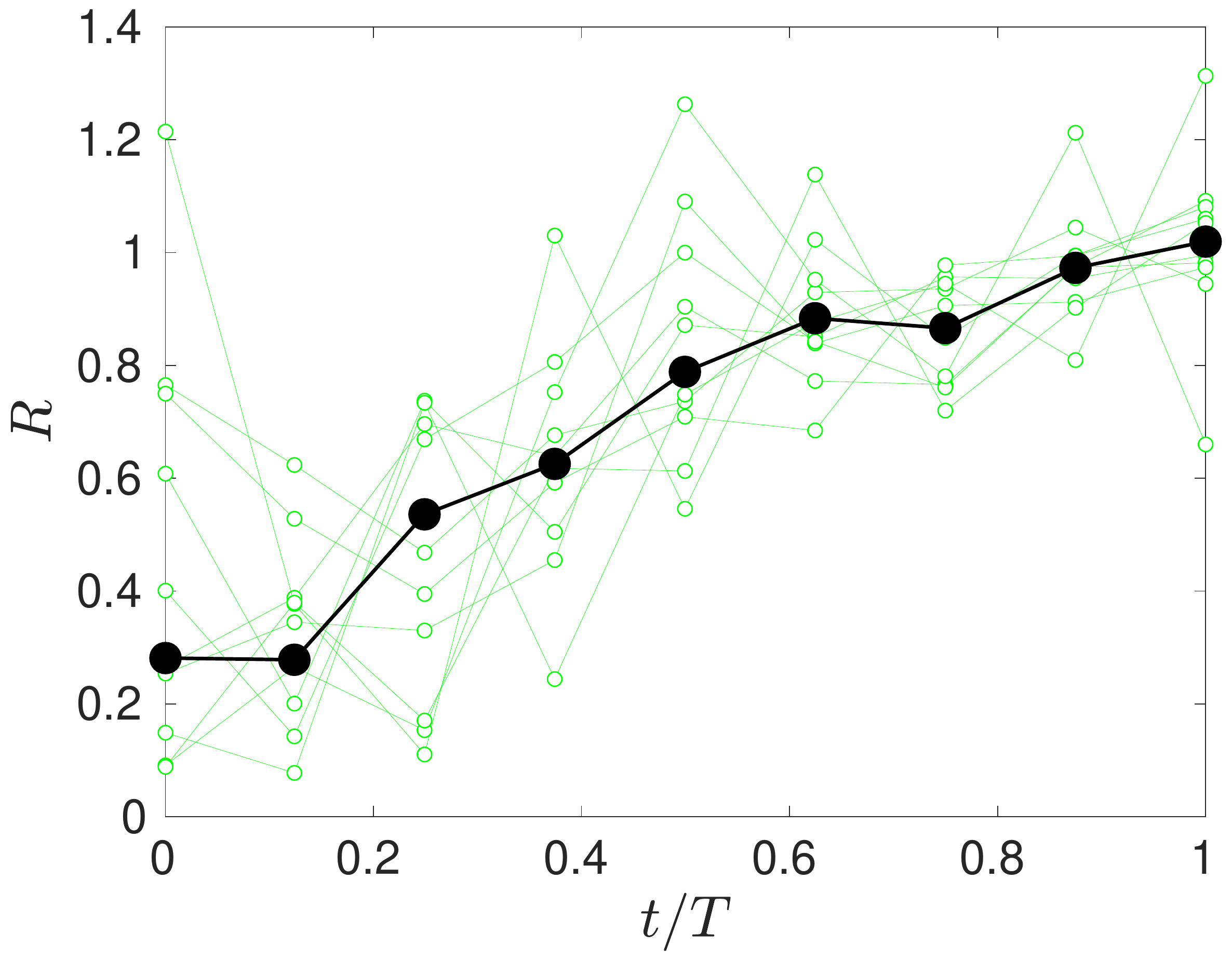}}
  \emn
  \caption{The median $R$ of the ratio between the numbers of grid points in two matching
  straining structures (circles with thick line). The green thin lines show the medians
  in different size groups. Left: for straining structures; middle: for
  structures with large positive
  $\Pi_\Delta$; right: for structures with large negative
  $\Pi_\Delta$. From case C1 with $\Delta = \Delta_m/4$. }
\label{fig:all-sizes8dx}
\efig

\begin{figure}
  \bmn {0.33\lnw}
  \centerline{\includegraphics[width=\lnw]{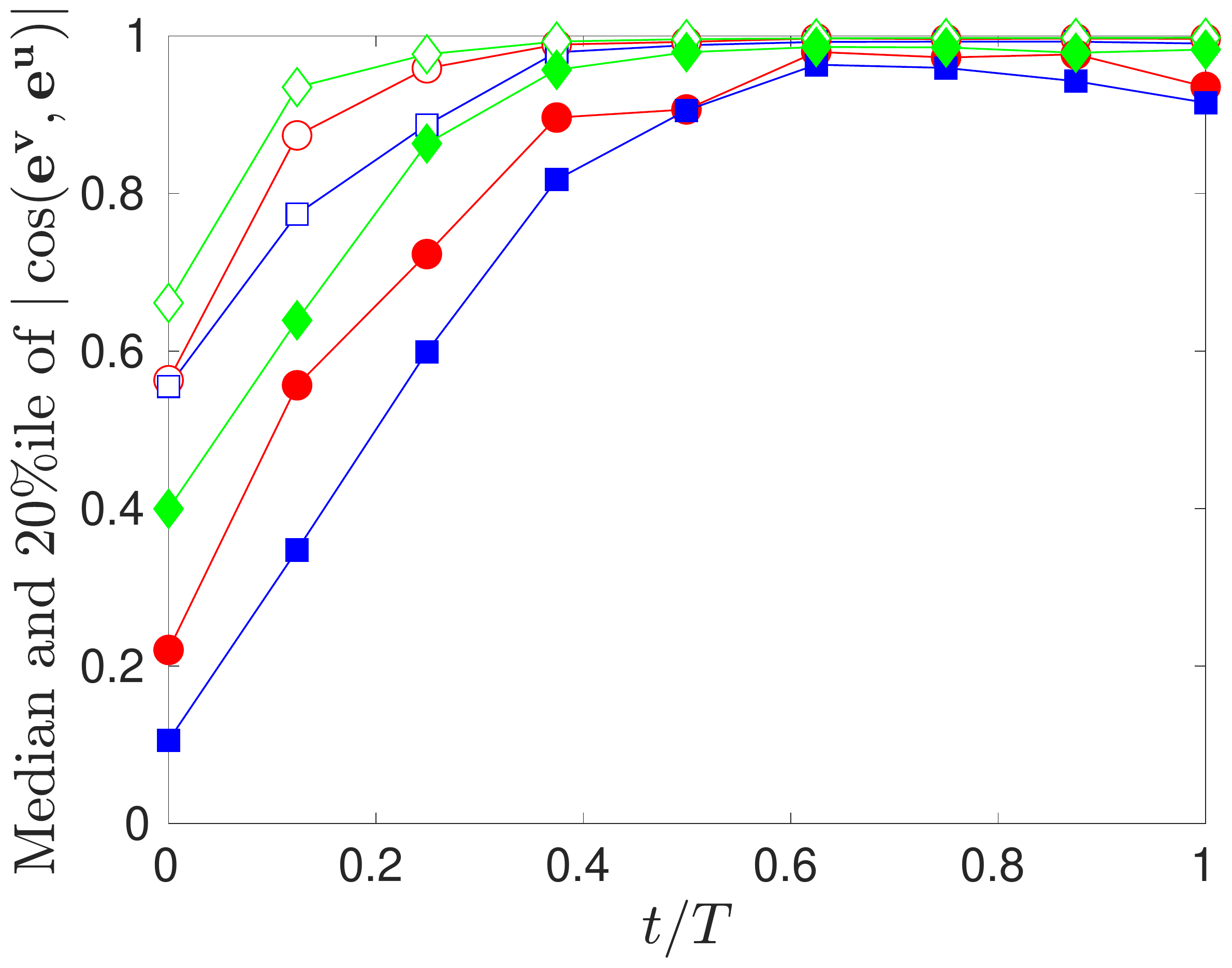}}
  \emn %
  \bmn {0.33\lnw}
  \centerline{\includegraphics[width=\lnw]{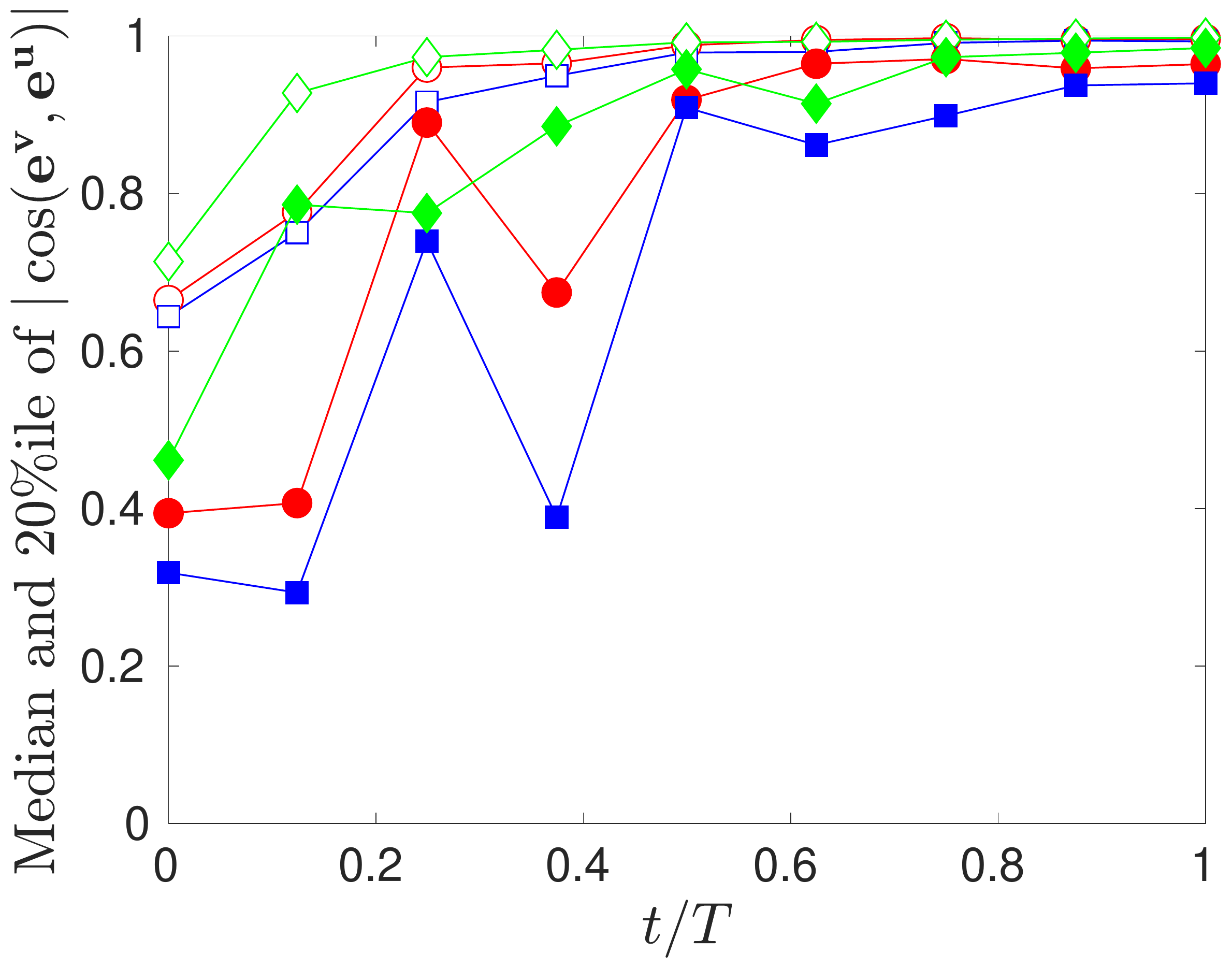}}
  \emn %
  \bmn {0.33\lnw}
  \centerline{\includegraphics[width=\lnw]{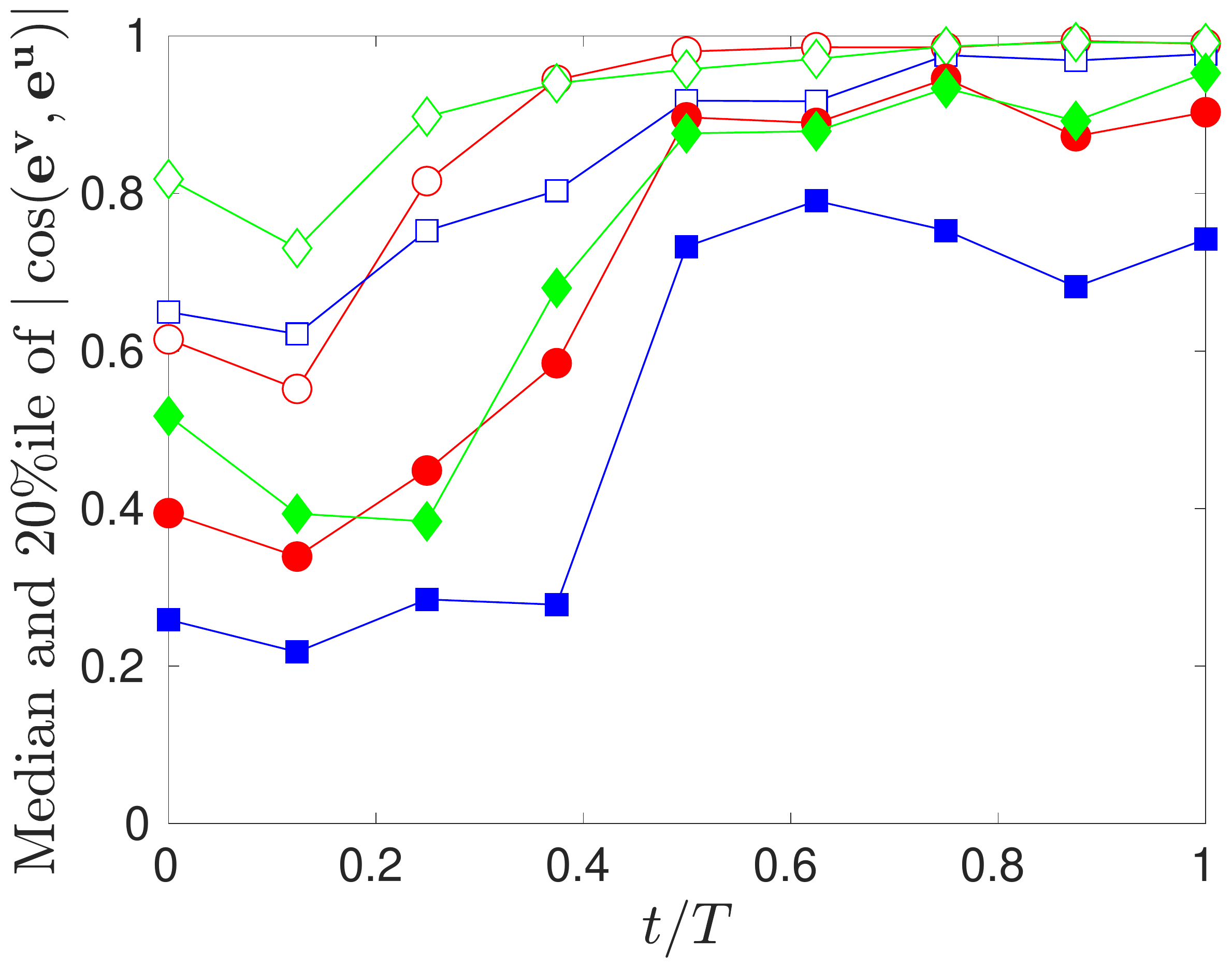}}
  \emn %
  \caption{The medians (empty symbols) and the $20$th percentiles (solid symbols)
  of $|\cos(\e_i^\v, \e_i^\u)|$ for
  straining structures (left), structures with large positive $\Pi_\Delta$ (middle), and structures
  with large negative $\Pi_\Delta$
  (right). Circles: $i=\alpha$; squares: $i=\beta$; diamonds:
  $i=\gamma$. From case C1 with $\Delta = \Delta_m/4$.}
\label{fig:median-align-all}
\end{figure}

\begin{figure}
  \bmn {0.33\lnw}
  \centerline{\includegraphics[width=\lnw]{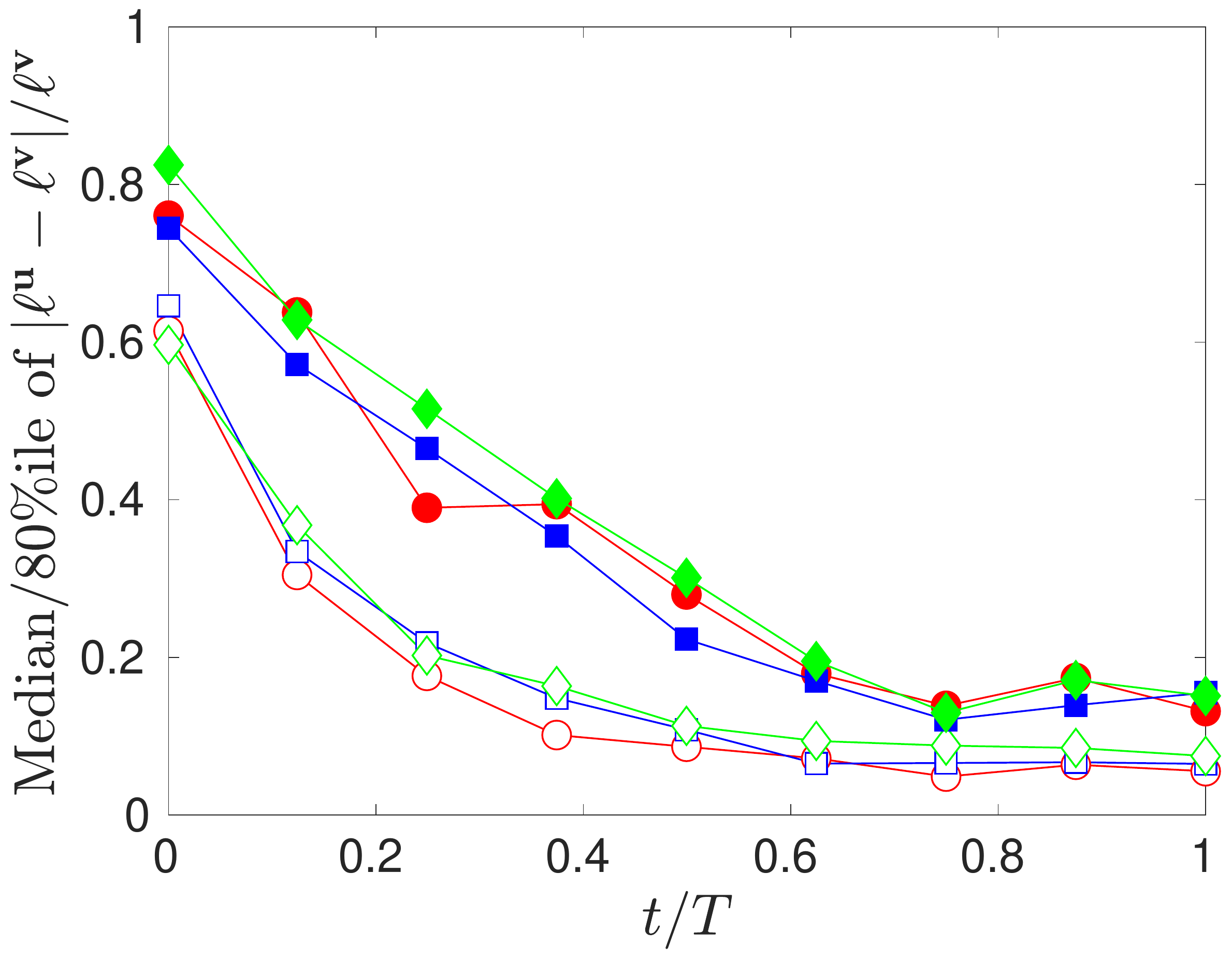}}
  \emn %
  \bmn {0.33\lnw}
  \centerline{\includegraphics[width=\lnw]{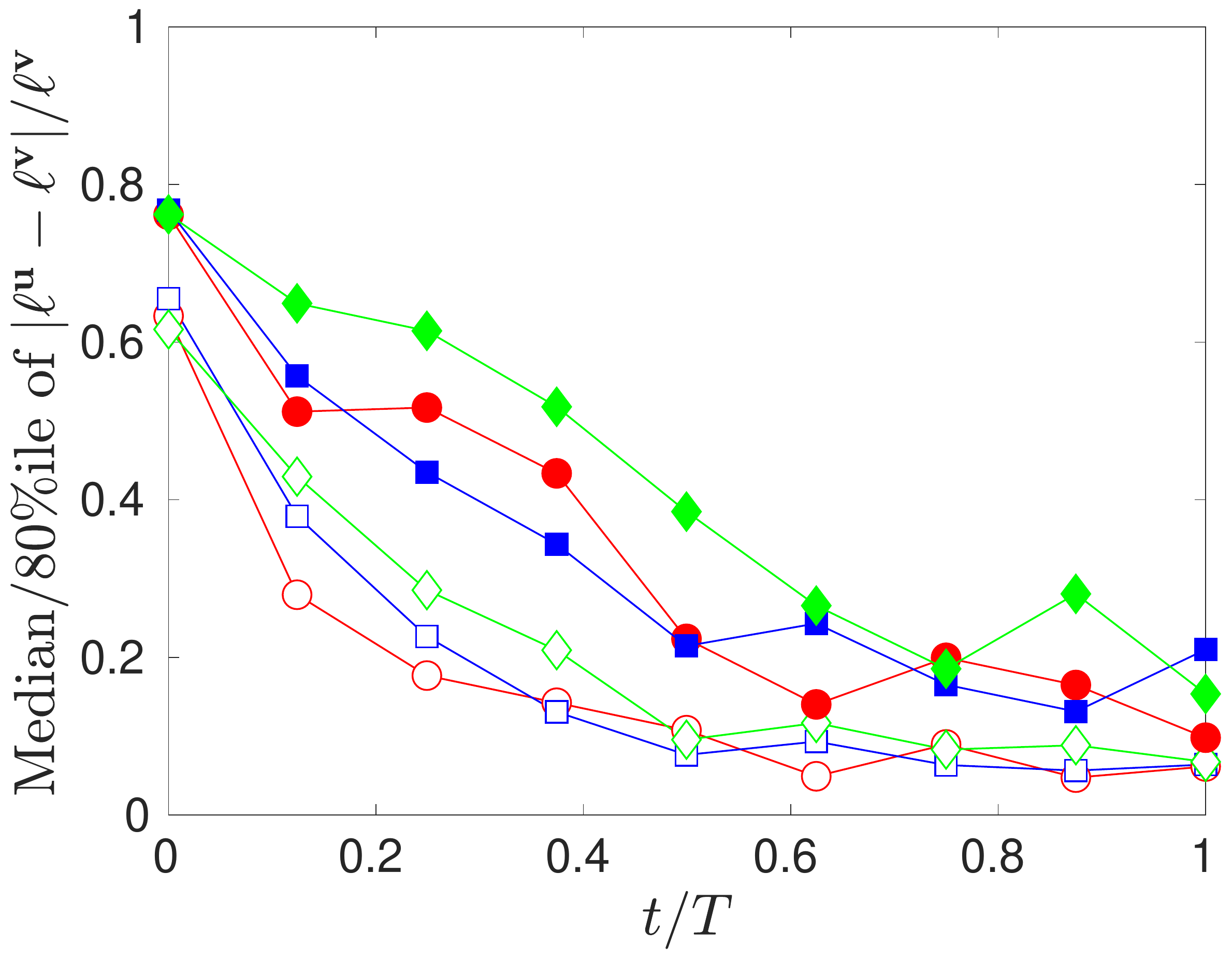}}
  \emn %
  \bmn {0.33\lnw}
  \centerline{\includegraphics[width=\lnw]{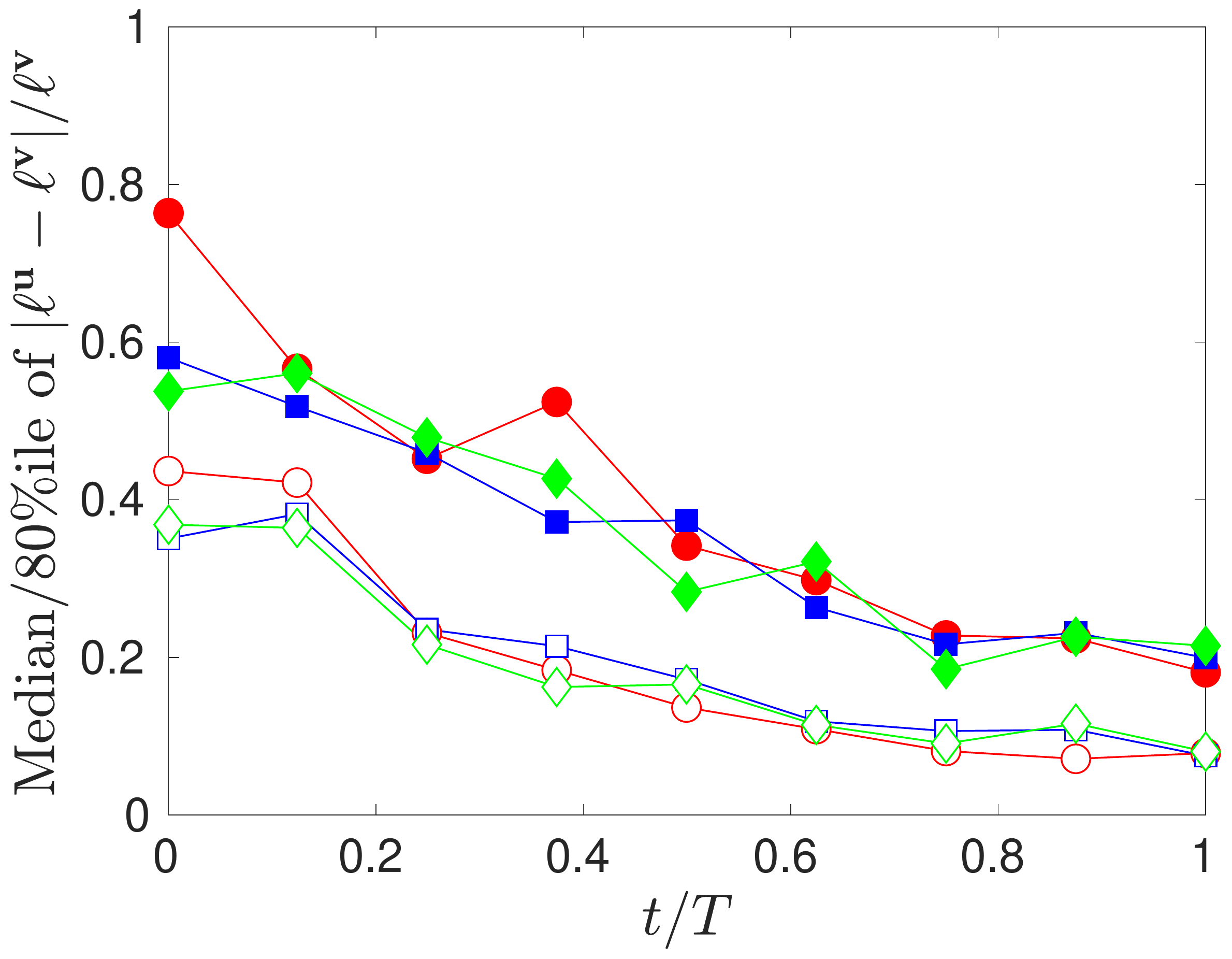}}
  \emn %
  \caption{The medians (empty symbols) and the $80$th percentiles (solid symbols)
  of $|\ell_i^\v-\ell_i^\u|/\ell^\v_i$ for
  straining structures (left), structures with large positive $\Pi_\Delta$ (middle), and structures
  with large negative $\Pi_\Delta$
  (right). Circles: $i=\alpha$; squares: $i=\beta$; diamonds:
  $i=\gamma$. From case C1 with $\Delta = \Delta_m/4$.}
\label{fig:median-axlenrat-all}
\end{figure}

\begin{figure}
  \bmn {0.33\lnw}
  \centerline{\includegraphics[width=\lnw]{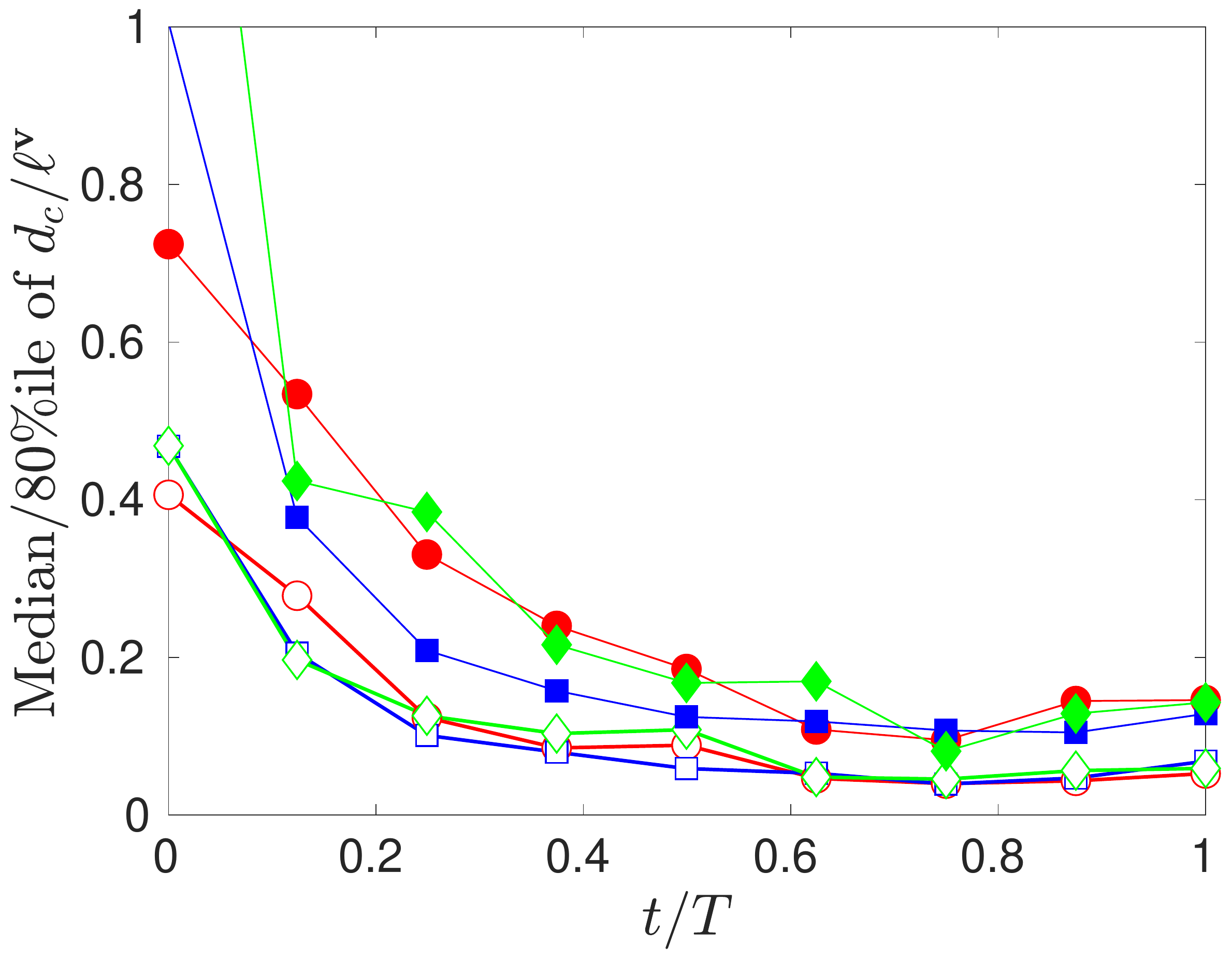}}
  \emn %
  \bmn {0.33\lnw}
  \centerline{\includegraphics[width=\lnw]{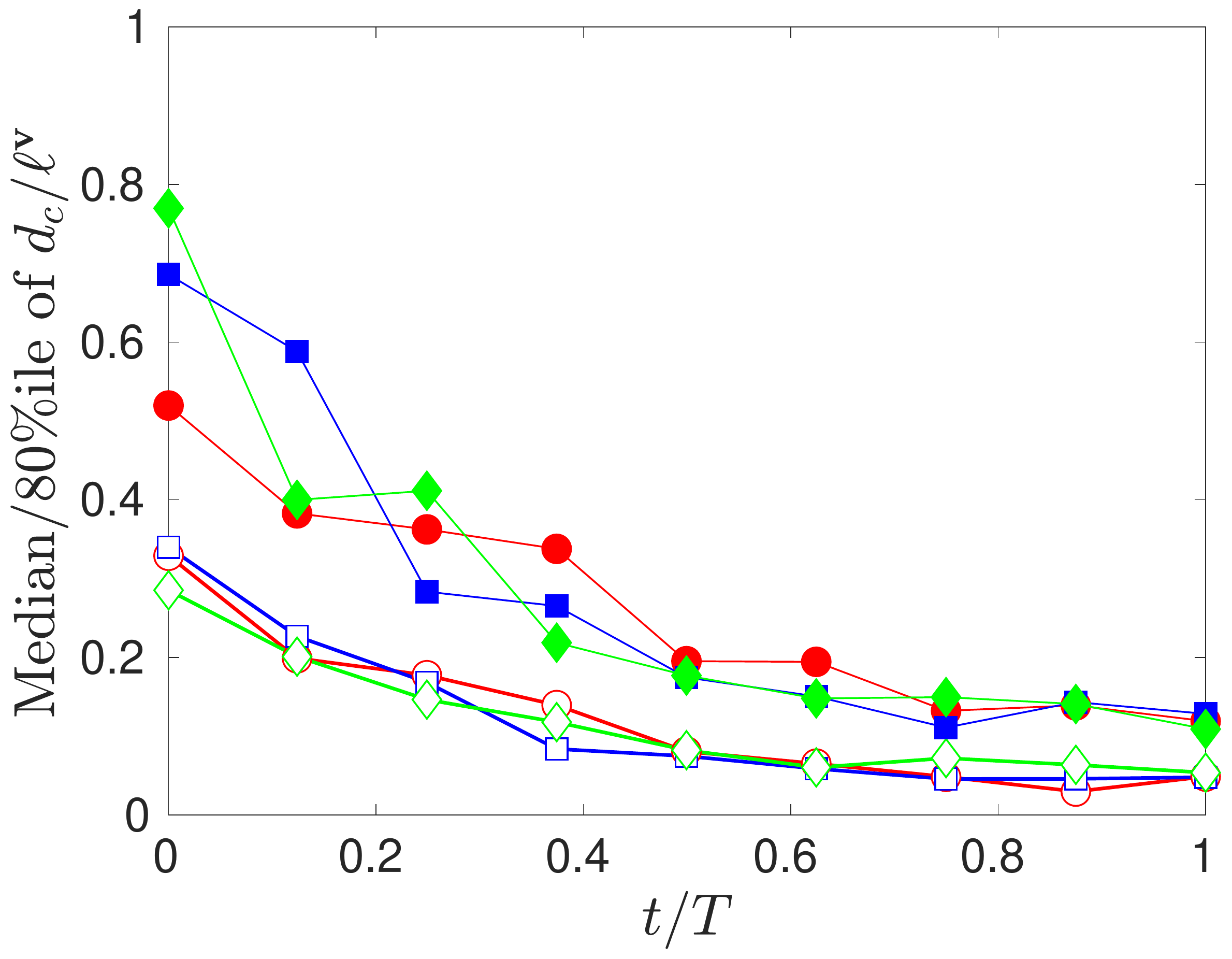}}
  \emn %
  \bmn {0.33\lnw}
  \centerline{\includegraphics[width=\lnw]{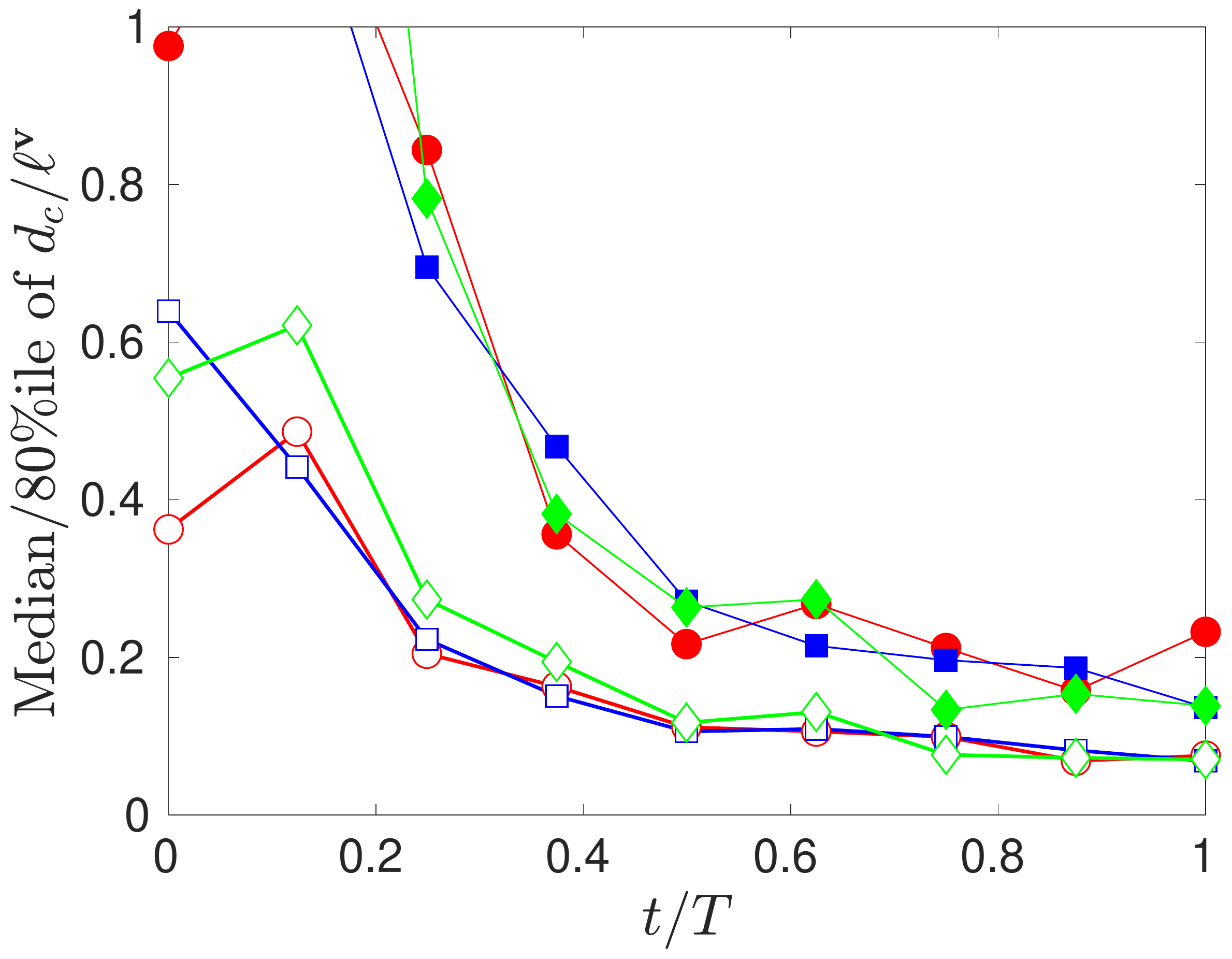}}
  \emn %
  \caption{
Same as Fig. \ref{fig:median-axlenrat-all}, but for the relative
  displacement $d_{c,i}/\ell^\v_i$.}
\label{fig:median-dist-all}
\end{figure}

The MVEEs are used in this subsection to investigate the strain
rate and the SGS energy dissipation rate. As the results show behaviors similar to those for the
vortical structures, only selected results are presented.   
For the strain rate, the structures
with $(\ts_{ij}\ts_{ij})^{1/2}\ge 2.5 \lal \ts_{ij}\ts_{ij}\ral^{1/2}$
are considered. For $\Pi_\Delta$, we examine both structures with large positive values
and those with large negative values.
The latter
represents the
regions where backscattering happens.
These structures are defined by $\Pi_\Delta \ge 10 \lal \Pi_\Delta
\ral$ and $\Pi_\Delta \le -2.5 \lal \Pi_\Delta \ral$,
and referred to as $P^+$ and $P^-$ structures, respectively.

Fig. \ref{fig:all-sizes8dx} shows the sizes of the reconstructed structures using the
median ratio $R$.
The reconstructed structures tend to be quite small initially, but the
discrepancy decreases over time quickly, similar to the results for
vortical structures in Fig. \ref{fig:sizes8dx}.
At $t/T=1$, all curves level off at about $1$.
The sizes for the $P^-$ structures need longer time to improve.

The medians and selected percentiles for alignment cosine
$|\cos(\e^\v, \e^\u)|$, relative
difference in axis length $|\ell^\u-\ell^\v|/\ell^\v$, and relative
displacement $d_c/\ell^\v$ are plotted in
Figs. \ref{fig:median-align-all} to \ref{fig:median-dist-all}. For both straining and $P^+$ structures,
excellent alignment is observed at $t/T=1$, with the medians and the
$20$th percentile being mostly above 0.9 (Fig.
\ref{fig:median-align-all}).
The medians and $80$th percentiles for $|\ell^\u-\ell^\v|/\ell^\v$ and $d_c/\ell^\v$ are
found to be around $5\%$ and $15\%$, respectively, at $t/T=1$, similar to those for the vortical
structures (Fig.
\ref{fig:median-axlenrat-all} and \ref{fig:median-dist-all}).
For $P^-$ structures, however, the results show
stronger discrepancies between $\v$ and $\u$.
For the alignment at $t/T=1$, the medians still reach
above $0.9$. However, the $20$th percentile result for the intermediate
axis is only about $0.7$ (c.f. Fig. \ref{fig:median-align-all}).
The medians and $80$th percentiles for $|\ell^\u-\ell^\v|/\ell^\v$ and $d_c/\ell^\v$ are found
at $10\%$ and $20\%$, respectively. These values are slightly higher than those for the other
two structures (see Fig. \ref{fig:median-axlenrat-all} and
\ref{fig:median-dist-all}).

\subsubsection{MVEE Trees for vortical structures}

\begin{figure}
\centering
\bmn {0.5\lnw}
  \centerline{\includegraphics[width=\lnw]{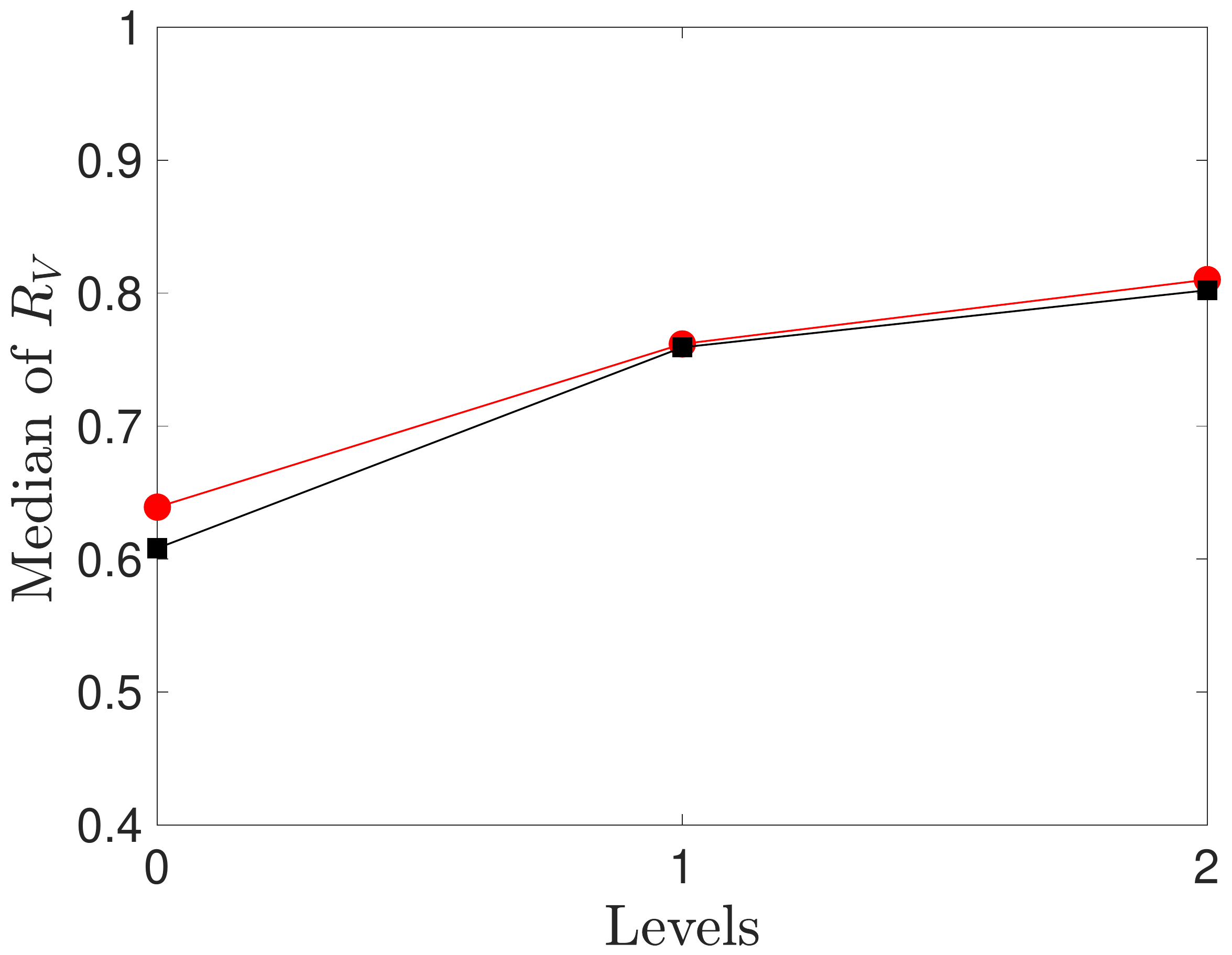}}
\emn %
\bmn {0.5\lnw}
  \centerline{\includegraphics[width=\lnw]{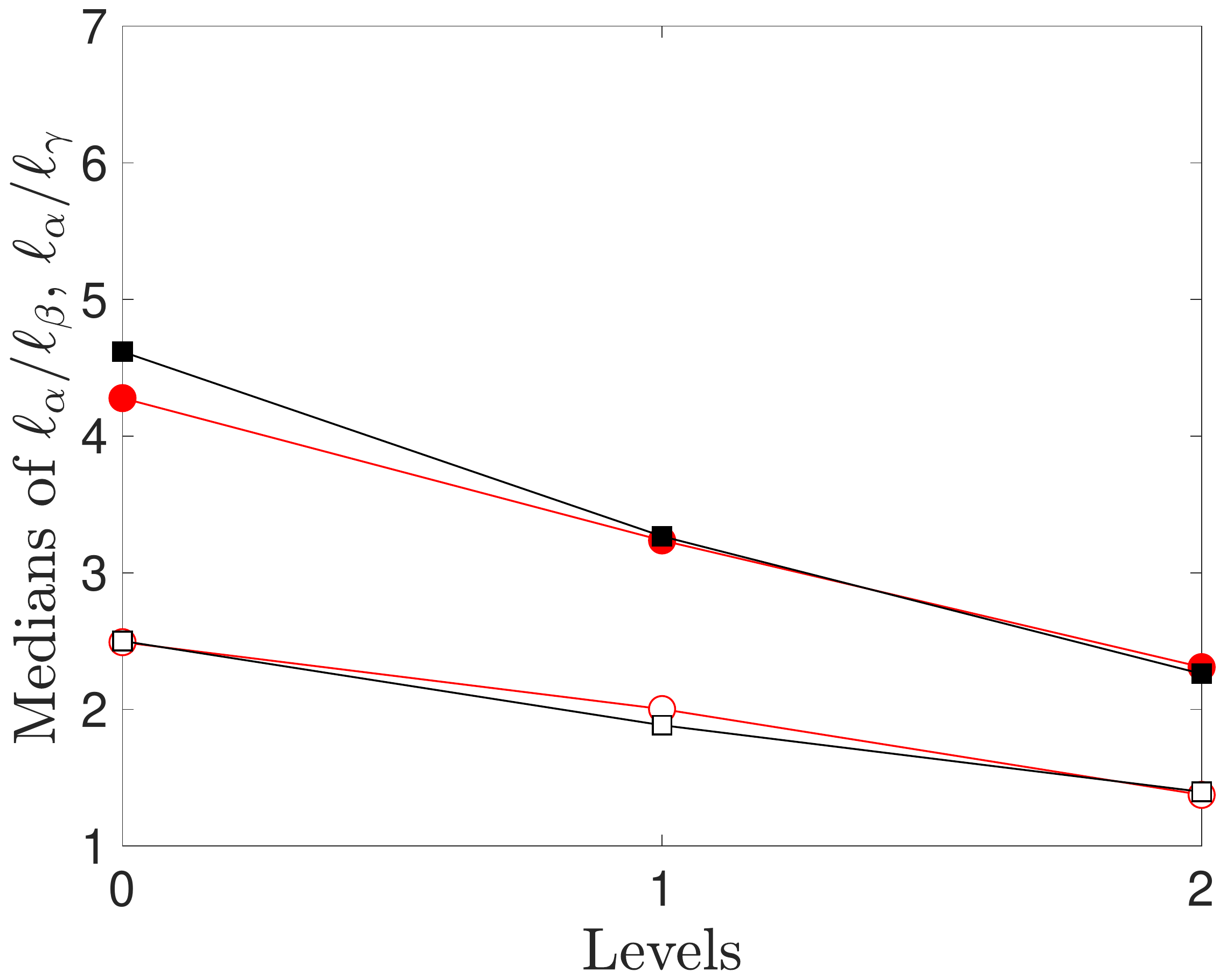}}
\emn
  \caption{Left: median of $R_V$, the ratio of the volume of a structure to the volume of its convex hull at different
  levels. Circle: for $\u$; squares: for $\v$. Right: the median of
  $\ell_\alpha/\ell_\gamma$ (solid symbols) and $\ell_\alpha/\ell_\beta$ (empty symbols) at different levels.
  Circle: for $\u$; squares: for $\v$.
  For strongly non-convex vortical structures only.
  For case C1 with $t/T=1$ and $\Delta = \Delta_m/2$. }
\label{fig:mveetree-shape}
\end{figure}

\begin{figure}
  \centerline{\includegraphics[width=\lnw]{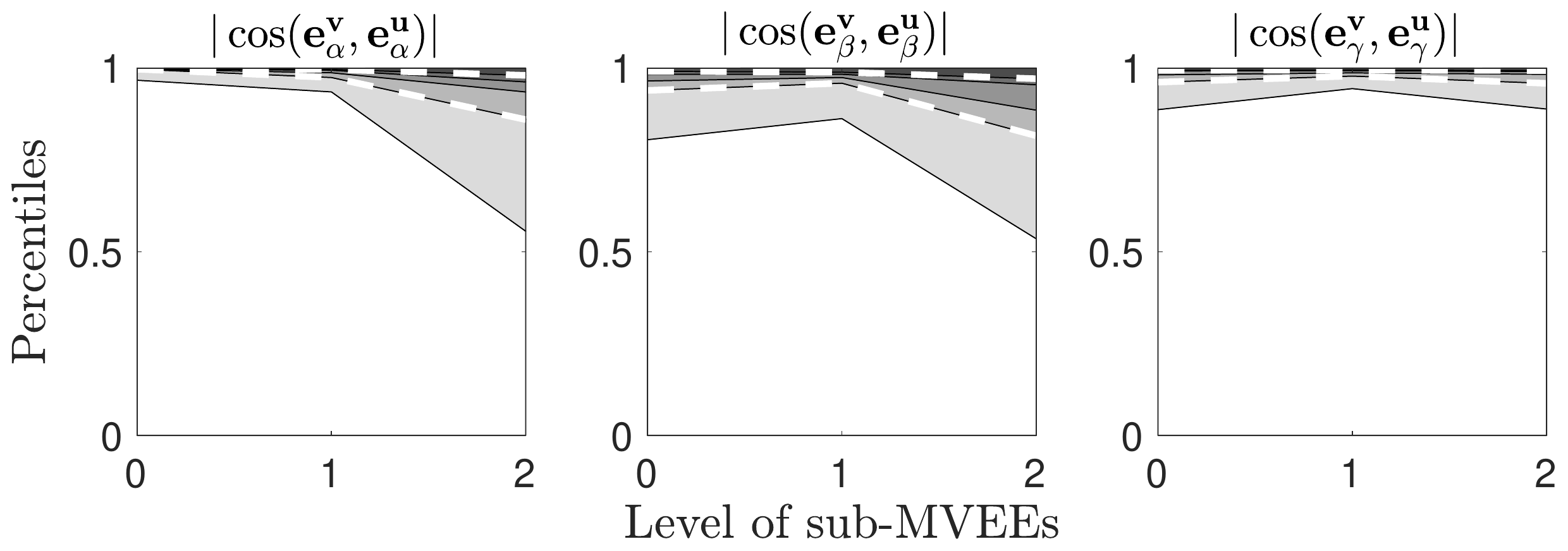}}
  \caption{The percentiles for $|\cos(\e_i^\v, \e_i^\u)|$ $(i=\alpha, \beta, \gamma)$ for MVEEs of
  matching vortices and their sub-MVEEs.
  From bottom: the $10$th, $20$th, $30$th, $40$th, and $50$th percentiles. The dashed lines 
  highlight the $20$th and $50$th percentiles. From case C1 with $t/T=1$ and $\Delta = \Delta_m/2$. }
\label{fig:mveetree-align-median}
\end{figure}

\begin{figure}
  \centerline{\includegraphics[width=\lnw]{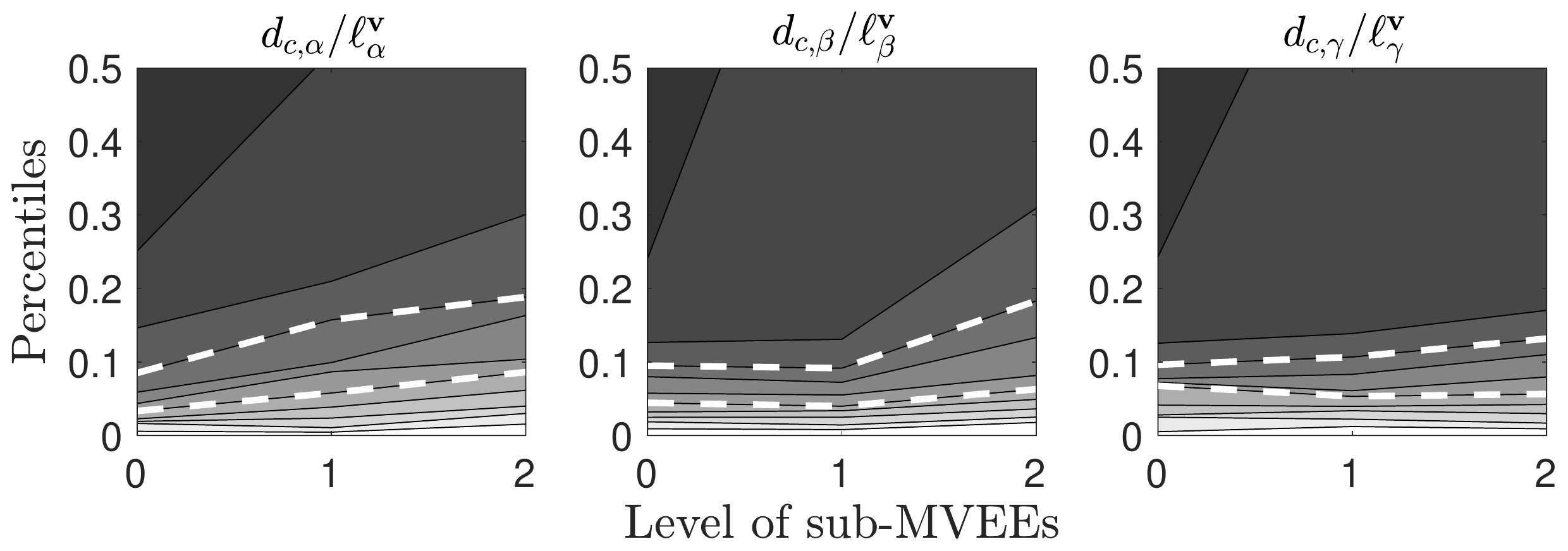}}
  \caption{The percentiles for $d_{c,i}/\ell^\v_i$ $(i=\alpha, \beta, \gamma)$ for MVEEs of
  matching vortices and their sub-MVEEs. From bottom: the $10$th to $100$th percentiles with
  increment of $10$ percent. 
  The dashed lines highlight the 50th and 80th percentiles.
  For case C1 with $t/T=1$ and $\Delta = \Delta_m/2$.
  }
\label{fig:mveetree-dist-median}
\end{figure}

\begin{figure}
  \centerline{\includegraphics[width=\lnw]{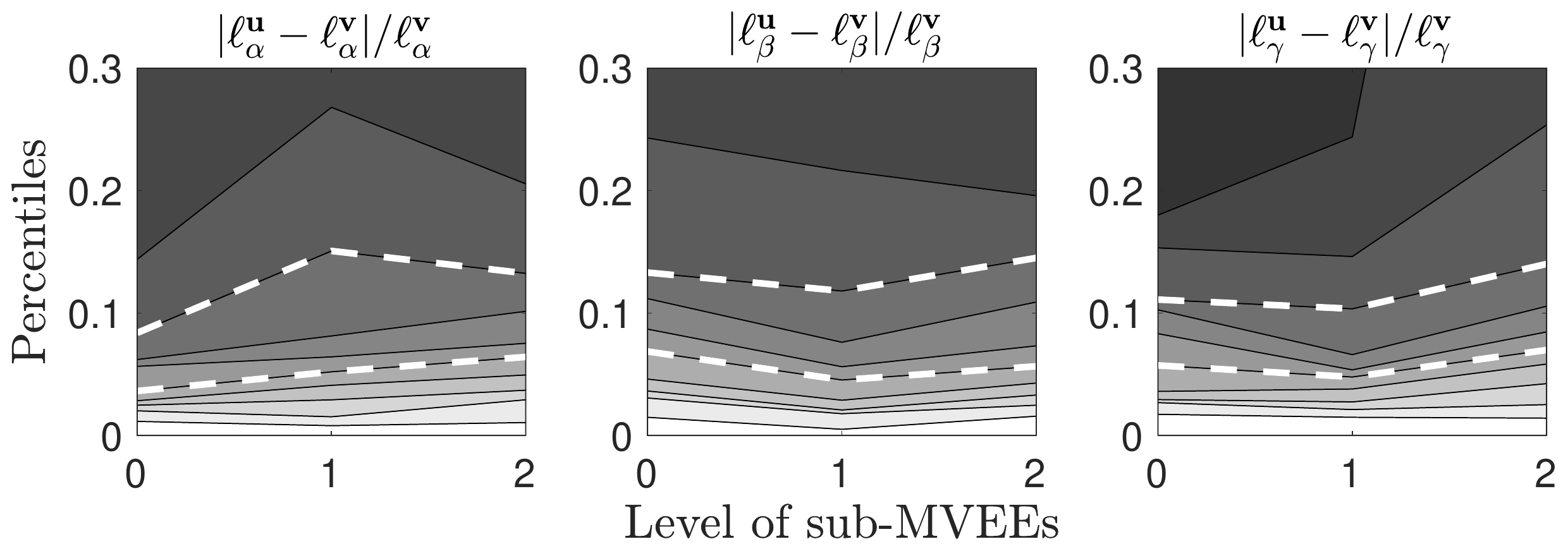}}
  \caption{Same as Fig. \ref{fig:mveetree-dist-median} but for the percentiles of
  $|\ell^\v_i - \ell^\u_i|/\ell^\v_i$ $(i=\alpha, \beta, \gamma)$. }
\label{fig:mveetree-ratLen-median}
\end{figure}

The analyses so far have examined the overall sizes, orientations and locations of the
non-local structures as defined by their MVEEs.
For strongly non-convex structures, the MVEEs do not provide full
descriptions. 
These structures are further examined with the MVEE trees in this sub-section (c.f. Section \ref{sect:cs_intro}).
Note that, if two MVEEs are significantly different
(e.g., seriously mis-aligned), the difference between their sub-MVEEs will only be bigger.
Therefore, this analysis is meaningful mainly for matching structures whose MVEEs are in good agreement; it
provides a more detailed comparison for these structures.
We will limit our discussion to vortical structures, as the results for other structures are
similar, and only the structures at $t=T$ for the base line case C1 are investigated. 
The analysis is applied to the non-convex vortical structures with
$R_V\le 0.8$ where $R_V$ is the ratio of the volume of a structure to the volume of its 
convex hull (c.f. Section \ref{sect:cs_intro}). 
Five pairs of matching vortical 
structures are found in each realization, and $25$ pairs in total from 5 realizations.
Two splitting are used to find the sub-MVEEs (c.f. Section \ref{sect:cs_intro}) so that
$3$ levels of sub-MVEEs are calculated: level 0, 1, and 2, which will be referred to as L0, L1, and
L2 sub-MVEEs. L0
sub-MVEEs are the same as the MVEEs already presented.

The geometry of the sub-structures at different levels is different.
Fig. \ref{fig:mveetree-shape}
displays the difference in two aspects. The median of $R_V$ (left) shows
that $R_V$ is larger at higher levels.
Therefore, the
sub-structures at higher levels are closer to their convex hulls. The right panel shows the median of
the ratio of the lengths of the
axes for the sub-MVEEs. The ratio is smaller for a 
higher level. The implication is that the sub-structures are rounder than their parent
structures. 

The alignment results for the sub-MVEEs are given in Fig.
\ref{fig:mveetree-align-median}. The five curves in the figure (the spaces between the curves
are filled with different shades of gray) are the $10$th, $20$th, $30$th, $40$th
and $50$th percentiles. 
The values for the L0 sub-MVEEs
are close to those in Fig. \ref{fig:median-align}, although they are not exactly the same
since here only strongly non-convex structures are considered.
For the major axes (shown in the left panel), 
the alignment becomes weaker at higher levels, since a given percentile decreases as the level number increases.
For
example, the 20th percentile is nearly 1 at L0 and approximately 0.85 at L2. It means that
80\% matching L0 sub-MVEEs align almost perfectly, while
we are only able to say the major axes of $80\%$
matching L2 sub-MVEEs are aligned to  
within $32^\circ$, corresponding to $|\cos(\e_\al^\u, \e_\al^\v)| \ge 0.85$. 
The $50$th percentile for L2 is also nearly 1, indicating a near perfect alignment is
still observed for more than a half matching sub-MVEEs. On the other hand, the $10$th percentile
drops quite significantly at L2. This shows that, although the MVEEs for these matching structures 
indicate very good agreement in their overall shapes, the compositional parts of these structures 
do differ quite substantially. This happens for about $10\%$ of the matching structures. 
For the intermediate (middle panel) and minor axes (right panel), similar behaviours can be
observed, although the variation with respect to the
levels is somewhat more complicated. However, even at the highest level (L2), the 20th 
percentile is still almost 1 for the minor axes,
and 0.8 for the intermediate axes. Thus, very good
alignment is observed for more than $80\%$ samples.

There are several competing
factors that affect the alignment of the sub-MVEEs. The higher level sub-structures are smaller and are closer to
being convex (c.f., the left panel of Fig. \ref{fig:mveetree-shape}). As such, the
orientations of the sub-MVEEs are
less sensitive to the details of the shapes, so
that it is easier to obtain better alignment.
On the other hand, the mis-alignment has a cumulative effect. That is,
the mis-alignment between the sub-MVEEs at a lower level naturally affects the alignment between
those at higher levels.
Furthermore, higher level structures tend to be rounder (c.f., the right panel
of Fig. \ref{fig:mveetree-shape}).
This too increases the chance of mis-alignment at higher levels,
because small perturbation to a sphere could lead to ellipsoids with very different axis
directions.
The competition of these mechanisms may lead to the observed non-monotonic behavior.
For the same reason,  
mis-alignment between the MVEEs does not necessarily imply large deviation in the shapes of the
structures. 

The relative displacement of the sub-MVEEs is presented in Fig. \ref{fig:mveetree-dist-median},
which shows higher percentiles increase quicker with the level
numbers (see, e.g., the 80th percentiles). This is likely due to the cumulative effect of
displacement. 
The curves for smaller percentiles (which correspond to cases with small relative
displacements) are rather flat.
Therefore, if the matching MVEEs locates closely to each other, so do their sub-MVEEs.
Quantitatively, the 50th
percentiles at L2 are approximately 0.08, 0.07 and 0.06 for the major, intermediate and minor axes,
respectively. Thus, in about 50\% cases, the relative displacement 
is less than 8\%, 7\%, and 6\% in the major, intermediate and minor axis directions, respectively.
Fig. \ref{fig:mveetree-ratLen-median} plots the relative difference in the lengths of the
axes. For this quantity, the relative differences at
L2 and L0 are not much different; the median value is found around 5\%.
It is probably not surprising since there is no cumulative
effect for the lengths of the axes.

To sum up, we observe that, if the
matching MVEEs already are close to each other, 
their sub-MVEEs also show good agreement in orientation, sizes and locations. 
Therefore, the description provided by
the MVEEs might already be enough in some applications.

\section{Discussion and conclusions \label{sect:conclusion}}

The four dimensional variational (4DVAR) method has been used to reconstruct the small scales of a
sequence of 3D Kolmogorov turbulent velocity fields with a moderate Reynolds number.
Velocity on a coarse grid from a time sequence
of DNS velocity fields is used as the
measurement data. We focus on the reconstruction
of instantaneous distributions of the small scales,
and present detailed assessment at scales one and two octaves smaller than
the measurement grid. In terms of Fourier modes, the number of modes contained down to
these scales are eight and sixty-four times of those in the measurement grid, respectively. 
The filtered vorticity $\tog_i$, filtered strain rate tensor
$\ts_{ij}$, the vortex stretching term $\ts_{ij} \tog_i \tog_j$, and the subgrid-scale energy
dissipation $P_\Delta = -\tau_{ij}\ts_{ij}$ are used to characterize the small scales. 
In order to quantitatively assess the reconstruction of non-local structures,
minimum volume enclosing ellipsoids (MVEE) and MVEE trees are introduced. A method to apply these
objects in the analysis is proposed and applied to the 4DVAR data.

Different parameters are tested, including the resolution of the measurement data $k_m$, the
tolerance for the convergence test $e_{\rm tot}$, and the optimization horizon 
$T$. $k_m$ has the most significant effects on the reconstruction. In
reference to the threshold value $k_c = 0.2\eta_K^{-1}$ found previously in the literature with a
direct substitution method, 
the reconstruction failed when $k_m$ is significantly smaller than $k_c$. Accurate reconstruction
is found for $k_m$ slightly smaller than $k_c$. Although more research is needed to prove that 
the threshold value is the same in 4DVAR, these results 
demonstrate that reconstruction from a finite time sequence of data using 4DVAR
can be accomplished for $k_m$ at or above the
threshold value $k_c$. 

The reconstruction 
at the initial time $t=0$ and at the end of the optimization horizon $t=T$ 
is examined separately for the successful cases. 
Larger $T$ and smaller $e_{\rm tot}$ improve the reconstruction at $t=T$. 
For $T$ equal the large eddy turnover time scale, accurate reconstruction at scales at least two
octaves smaller than the measurement grids
is obtained for a range of quantities.
The main findings are summarized as follows,
where approximate numbers are quoted
to give an estimate of the efficacy of the method:
\begin{enumerate}
\item
The reconstruction achieves higher than 95\% point-wise correlation 
    for the filtered enstrophy and the other parameters. 
    The spectral correlation for velocity is higher than 80\% across the
    spectrum. 
The normalized RMS point-wise difference and the normalized spectral difference 
    increase towards the small scales, and 
    reaches around 30\% at the Kolmogorov scale. 
%\item 
%Large fluctuations in the difference between the reconstruction and target are investigated using
%    joint PDFs and conditional statistics.  
%Geometrical structures, such as the orientation of the eigen-frames of $\ts_{ij}$ and $\tau_{ij}$,
%    can be reconstructed accurately. 
\item
  The non-local structures are reconstructed accurately. For most quantities,
    the misalignment is less than $15^\circ$ for the majority of the samples, the sizes differing within
    $\pm 10\%$, and the locations displaced by less than 15\% of the lengths of the axes. The structures
    with strong negative SGS energy dissipation have slightly larger errors. 
\item
A direct substitution scheme
    provides much poorer reconstruction,
    where the RMS point-wise difference is about three times as
    big for the filtered enstrophy, and is almost ten times as big in terms of the spectral 
    difference of the velocity fields. 
    Meanwhile, Lagrangian interpolation is simply infeasible.  
\end{enumerate}

The reconstruction at $t=0$ can capture the qualitative features of the small scales in turbulence, including 
qualitatively correct energy spectrum and mean SGS energy dissipation. 
However, significant discrepancy exists and the agreement is improved very little by tuning the optimization
horizon or the tolerance. 
A plausible explanation is that our optimal solutions have not truly converged. 
Due to the high computational costs, we have not been able to further improve the solution by 
using even smaller $e_{\rm tot}$. 
Given the chaotic nature of the flow, it is challenging
to obtain convergent results if $T$ is significantly larger or $e_{\rm tot}$ is much smaller.

This study confirms the efficacy
of the 4DVAR scheme, and demonstrates that MVEEs and MVEE trees can provide useful
information about non-local structures for CFD with data assimilation
when the prediction of instantaneous velocity field is one of the main interests.
As the application of data assimilation in CFD is still in an early stage, many questions remain open.
Most importantly, the Reynolds number for the flow considered here is relatively low. 
The natural next step is to investigate the quality of reconstruction when coarse-grained models
(such as large eddy simulations) are
used to model the data. This will allow us to investigate flows with higher Reynolds numbers.
%Preliminary results show that accurate reconstruction for a range of small scales are also possible. 
To obtain better converged results,
a better understanding of the behavior of the adjoint fields is useful.
These questions are the focus of our on-going research.

\section*{Acknowledgement}
Y. Li gratefully acknowledges Dr. Ashley Willis for his help with the computer code.
N. S. Abdullah wishes to acknowledge the financial support from the Iraqi government via the Higher
Committee for Education Development and the College of Education at the University of Mosul.

\bibliographystyle{jfm}
% Note the spaces between the initials
\bibliography{../../turbref}

\end{document}